\documentclass[revtex4]{emulateapj}
\usepackage{amsmath}
\usepackage{graphicx}
\usepackage{subfigure}

\shorttitle{Massive Star Cluster Formation and Destruction in LIRGs}
\shortauthors{Linden et al. 2017}

\begin{document}

\title{Massive Star Cluster Formation and Destruction in Luminous Infrared Galaxies in GOALS}

\author{ S. T. Linden\altaffilmark{1}, A. S. Evans\altaffilmark{1,2}, J. Rich\altaffilmark{3}, K. Larson\altaffilmark{3}, L. Armus\altaffilmark{3}, T. D\'iaz-Santos\altaffilmark{4}, G. C. Privon\altaffilmark{5}, J. Howell\altaffilmark{3}, H. Inami\altaffilmark{6}, D.-C. Kim\altaffilmark{2}, L.-H. Chien\altaffilmark{7}, T. Vavilkin\altaffilmark{8}, J. M. Mazzarella\altaffilmark{9}, J. A. Surace\altaffilmark{10}, S. Manning\altaffilmark{11}, A. Abdullah\altaffilmark{12}, A. Blake\altaffilmark{12}, A. Yarber\altaffilmark{12}, T. Lambert\altaffilmark{13}}
\altaffiltext{1}{Astronomy Department, University of Virginia, 530 McCormick Road, Charlottesville, VA 22904 USA: stl7ey@virginia.edu}
\altaffiltext{2}{National Radio Astronomy Observatory, 520 Edgemont Road, Charlottesville, VA 22903 USA}
\altaffiltext{3}{Infrared Processing and Analysis Center, California Institute of Technology, MS 100-22, Pasadena, CA 91125 USA}
\altaffiltext{4}{N\'ucleo de Astronom\'ia de la Facultad de Ingenier\'ia, Universidad Diego Portales, Av. Ej\'ercito Libertador 441, Santiago, Chile}
\altaffiltext{5}{Instituto de Astrof\'sica, Facultad de F\'isica, Pontificia Universidad Cat\'olica de Chile, Casilla 306, Santiago 22, Chile}
\altaffiltext{6}{Centre de Recherche Astrophysique de Lyon (CRAL), Observatoire de Lyon, 9 avenue Charles Andr\'{e}, 69230 Saint-Genis-Laval France}
\altaffiltext{7}{Space Telescope Science Institute, 3700 San Martin Drive, Baltimore, MD 21218, USA}
\altaffiltext{8}{Department of Physics \& Astronomy, Stony Brook University, Stony Brook, NY, 11794-3800 USA}
\altaffiltext{9}{Spitzer Science Center, Pasadena, CA 91125 USA}
\altaffiltext{10}{Eureka Scientific, Inc. 2452 Delmer Street Suite 100 Oakland, CA 94602-3017 USA}
\altaffiltext{11}{Department of Astronomy, The University of Texas at Austin, 2515 Speedway Boulevard Stop C1400, Austin, TX 78712, USA}
\altaffiltext{12}{Department of Physics and Astronomy, Howard university, 2355 6th St. NW, Washington, DC 20059}
\altaffiltext{13}{Department of Astronomy, University of Maryland, College Park, MD 20742}


\begin{abstract}

We present the results of a {\it Hubble Space Telescope} ACS/HRC FUV, ACS/WFC optical study into the cluster populations of a sample of 22 Luminous Infrared Galaxies in the Great Observatories All-Sky LIRG Survey. Through integrated broadband photometry we have derived ages and masses for a total of 484 star clusters contained within these systems. This allows us to examine the properties of star clusters found in the extreme environments of LIRGs relative to lower luminosity star-forming galaxies in the local Universe. We find that by adopting a Bruzual \& Charlot simple stellar population (SSP) model and Salpeter initial mass function, the age distribution of clusters declines as $dN/d\tau = \tau^{-0.9 +/- 0.3}$, consistent with the age distribution derived for the Antennae Galaxies, and interpreted as evidence for rapid cluster disruption occuring in the strong tidal fields of merging galaxies. The large number of $10^{6} M_{\odot}$ young clusters identified in the sample also suggests that LIRGs are capable of producing more high-mass clusters than what is observed to date in any lower luminosity star-forming galaxy in the local Universe. The observed cluster mass distribution of $dN/dM = M^{-1.95 +/- 0.11}$ is consistent with the canonical -2 power law used to describe the underlying initial cluster mass function (ICMF) for a wide range of galactic environments. We interpret this as evidence against mass-dependent cluster disruption, which would flatten the observed CMF relative to the underlying ICMF distribution.


\end{abstract}



\section{Introduction}

Galaxies with high infrared (IR) luminosities, e.g., luminous infrared galaxies (LIRGs: $L_{\rm IR} [8 - 1000 \mu{\rm m}] > 10^{11.0}$ L$_\odot$),  are rare in the local Universe, yet they are a cosmologically important class of objects because they dominate the infrared luminosity density at redshifts $z=1-2$ (Magnelli et al. 2013). Their high bolometric luminosities emanate from energetic star-formation (SF) regions, and sometimes active galactic nuclei (AGN), which are primarily triggered by interactions and mergers of gas-rich galaxies (e.g., Sanders \& Mirabel, 1996). Further, the complex structure of these dynamically evolving systems and the presence of both dust-obscured and un-obscured activity necessitates the need for high-resolution observations that sample as much of the electromagnetic spectrum as possible to best identify and reconstruct the distribution and luminosity of star-formation and AGN-related phenomena, and to probe the connection between merger stage and the observed activity. Understandably, the ultraviolet (UV) properties of these very IR-luminous galaxies have received far less scrutiny. However, the small fraction of the UV radiation 
from super star clusters, AGN, and diffuse stellar emission that escapes can nonetheless make LIRGs powerful sources of UV radiation (e.g., Evans et. al. 2008; Armus et al. 2009; Howell et al. 2010; Inami et al. 2010). 

Of interest for the present study of LIRGs are the luminous star clusters (SCs), which track basic information regarding the formation and fate of star formation in a variety of different environments.
The {\it Hubble Space Telescope} (HST) has been instrumental in the detection of numerous star clusters ($\gtrsim 1000$) in gas-rich mergers (e.g. NGC 3256: Zepf et al. 1999; NGC 4038/9: Whitmore \& Schweizer 1995, Whitmore et al. 1999) and recent merger remnants (e.g. NGC 3921: Schweizer et al. 1996; NGC 7252: Miller et al. 1997, Schweizer \& Seitzer 1998; NGC 3610: Whitmore et al. 1997). The presence of young ($\lesssim 10$ Myr) and intermediate age (100 - 500 Myr) star cluster populations in late stage mergers such as the Antennae galaxies (NGC 4038/4039; Whitmore et al. 1999), Arp 220 (Wilson et al. 2006), the Mice galaxies (NGC 4676 A/B; Chien et al. 2007) is consistent with the description of these galaxies as experiencing powerful starbursts triggered by the interaction and merger of pairs of gas-rich galaxies. However, optical studies of other late stage mergers such as NGC 6240 (Pasquali, de Grijs \& Gallagher 2003) and NGC 7673 (Homeier, Gallagher \& Pasquali 2002) reveal only young star clusters, indicating that older star clusters, which would have formed earlier on in the merger, are either undetected or rare. In contrast, the lack of young star clusters in the tidal tails of NGC 520 and NGC 2623 (Mulia et al. 2015) relative to what is observed for NGC 3256 (Trancho et al. 2007) suggests that the remaining reservoirs of predominately neutral hydrogen (HI) gas in the tails cannot always form new clusters.

Many studies have been devoted to understanding the long-term stability of the youngest clusters in mergers (e.g. Fall et al. 2009; Whitmore et al. 2007). It appears that only those which survive the disruption processes and are still dense and gravitationally bound are likely to become the globular clusters (GCs) we observe today (Zhang \& Fall 1999). The relative contributions from various cluster disruption mechanisms such as infant mortality (Fall et al. 2005; Chandar et al. 2010a), two-body relaxation (Fall et al. 2009), and tidal shocks (Gnedin \& Ostriker 1997) as a function of galactic environment continues to be the subject of much work. Infant mortality or rapid disruption, is caused by mass-loss during the early gas expulsion phase of cluster evolution and is expected to work on timescales of $\leq 10$ Myr. In contrast, disruption from large scale shocks is expected to be important over roughly $10^{8}$ yr timescales, and two body relaxation will cause disruption on even longer timescales (on the order of a Hubble time). Ultimately, the manner in which these young massive clusters (YMCs) evolve is crucial to connecting them to present day globular clusters. If YMCs are indeed local analogues to present-day GCs, then by understanding their formation and evolution, it is possible to gain insight into the formation of the earliest most massive clusters in the Universe (Kruijssen 2014).


In addition to understanding the fate of clusters, it is important to understand to what degree their environment affects where and how they form, as well as what their collective properties are -- e.g., the distribution of massive clusters (Initial Cluster Mass Function: ICMF) and the efficiency with which bound star clusters form (Larsen \& Richtler 2000; Bastian 2008). Although the low mass end of the ICMF appears to be universal (de Grijs et al. 2003; Fall \& Chandar 2012), the formation conditions of the highest-mass clusters are still subject to debate. 

One idea is that the formation mechanism of the most massive clusters is independent of environment (Whitmore et al. 2007, Chandar et al. 2015), and thus the total number and maximum cluster mass scale linearly with the star formation rate of the galaxy (Hunter et al. 2003, Whitmore et al. 2010, Vavilkin et al. 2016). Alternatively, the formation of the most massive clusters may require special physical conditions, such as high ambient pressure or enhanced gas densities. Kruijssen et al. (2012) predicts that the formation of bound stellar clusters takes place in the highest-density peaks of the ISM. Therefore, YMCs should form more efficiently at high gas pressures (and hence gas surface densities), because these conditions lead to higher density peaks. This leads to a non-linear scaling of the maximum cluster observed and the star formation rate surface density $(\Sigma_{SFR})$ of the galaxy.

To really quantify the role of galactic environment in shaping massive cluster formation and destruction, we need to study the properties of star clusters in a statistically larger sample of Luminous Infrared Galaxies which represent the most extreme star-forming systems observed in the local Universe. The Great Observatories All-Sky LIRG Survey (GOALS), is a multi-wavelength imaging and spectroscopic study of a complete flux density-limited ($S_{\rm 60\mu m} > 5.24$ Jy) sample of the 202 LIRGs in the {\it IRAS} Revised Bright Galaxy Sample (RBGS; Sanders et al. 2003, GOALS; Armus et al. 2009). The proximity, size, and completeness of the sample, combined with broad wavelength coverage, makes GOALS the definitive sample for studying star clusters in local, luminous star forming galaxies. The present study makes use of HST UV and optical images from GOALS to estimate the cluster age distribution,the cluster mass function, and the cluster formation efficiency in a sample of 22 LIRGs.

The paper is organized as follows: In \S 2, the sample selection is summarized. In \S 3, the observations and data reduction are described, as well as our method for identifying clusters. In \S 4, the manner in which the cluster ages are estimated is described. In \S 5, the age distribution, the mass function and the cluster efficiency are discussed within the context of lower luminosity star-forming galaxies. Section 6 is a summary of the results. 

Throughout this paper, we adopt a WMAP Cosmology of $H_0 = 70$ km s$^{-1}$ Mpc$^{-1}$, $\Omega _{\rm matter} = 0.28$, and $\Omega _{\Lambda} = 0.72$ (e.g., see Armus et al. 2009).

\section{Sample Selection}
Within GOALS, there are HST B- and I-band observations of all 88 LIRGs with $L_{\rm IR} \geq 10^{11.4}$ L$_\odot$. Of those, we select the 22 LIRGs observed to have greater than 100 B-band luminous clusters ($m_{\rm B} \sim 21-23$ mag) within the central $30\times30\arcsec$ of the galaxy (i.e, a limit imposed by our far-UV imaging field of view -- see below). In total we observed 9131 B-band luminous star clusters from galaxies in the sample.

\begin{deluxetable*}{lrrcccccll}
\tabletypesize{\footnotesize}
\tablewidth{0pt}
\tablecaption {Properties of the 27 GOALS Galaxies in the sample}
\tablehead{
\colhead{Name} & \colhead{RA} & \colhead{Dec} & \colhead{Log(LIR)} & \colhead{D(Mpc)} & \colhead{SFR\tablenotemark{a}} & \colhead{IR/UV\tablenotemark{a}} & \colhead{$f_\nu$(FUV)\tablenotemark{a}} & \colhead{MS\tablenotemark{b}} & \colhead{$A_v$\tablenotemark{c}}} \\
\startdata
NGC 0017 & 00:11:06.5000 & -12:06:26.00 & 11.49 & 83 & 55.25 & 31.2 & 1.53e-14 & 5 & 3.0 \\
Arp 256S & 00:18:50.9000 & -10:22:37.00 & 11.45 & 110 & 48.63 & 7.6 & 1.42e-14 & 3 & 1.7 \\
Arp 256N & 00:18:50.0430 & -10:21:43.62 & 10.36 & 110 & 3.95 & 7.6 & 1.24e-14 & 3 & 1.7 \\
NGC 0695 & 01:51:14.2000 & +22:34:57.00 & 11.68 & 130 & 84.64 & 37.1 & 2.50e-16 & 0 & 2.8 \\
UGC 02369 & 02:54:01.8000 & +14:58:25.00 & 11.60 & 132 & 50.11 & 39.81 & - & 2 & 2.3 \\
NGC 1614 & 04:33:59.8000 & -08:34:44.00 & 11.60 & 67 & 51.28 & 15.13 & - & 5 & 4.0 \\
2MASX J06094582-2140234 & 06:09:45.8000 & -21:40:24.00 & 11.60 & 165 & - & - & - & 3 & 1.0 \\
2MASX J08370182-4954302 & 08:37:01.8000 & -49:54:30.00 & 11.60 & 115 & - & - & - & 3 & 3.7 \\
NGC 2623 & 08:38:24.1000 & +25:45:17.00 & 11.60  & 84 & 69.19 & 95.6 & 5.44e-15 & 5 & 1.5 \\
UGC 04881 & 09:15:55.1000 & +44:19:55.00 & 11.74 & 178 & 97.13 & 52.3 & 2.52e-15 & 2 & 1.9 \\
IC 2545 & 10:06:04.5810 & -33:53:05.55 & 11.70 & 150 & - & - & - & 4 & 4.0 \\
NGC 3256 & 10:27:51.3000 & -43:54:14.00 & 11.64 & 38 & 76.46 & 71.6 & 3.10e-14 & 5 & 3.7 \\
Arp 148 & 11:03:53.2000 & +40:50:57.00 & 11.60 & 160 & - & - & - & 2 & 2.1 \\
NGC 3690E & 11:28:33.4470 & +58:33:46.08 & 11.41 & 45.2 & 45.19 & 29.4 & 3.59e-14 & 3 & 3.4 \\
NGC 3690W & 11:28:30.3390 & +58:33:39.48 & 11.77 & 45.2 & 101.44 & 29.4 & 8.32e-14 & 3 & 3.9 \\
NGC 5257E & 13:39:57.6830 & +00:49:49.80 & 11.32 & 99 & 36.06 & 9.1 & 2.84e-14 & 2 & 2.6 \\
NGC 5257W & 13:39:52.9530 & +00:50:23.10 & 11.31 & 99 & 35.66 & 9.1 & 1.10e-14 & 2 & 1.8 \\
NGC 5331S & 13:52:16.2140 & +02:06:03.28 & 11.54 & 139 & 60.78 & 32.7 & 1.32e-15 & 3 & 3.6 \\
NGC 5331N & 13:52:16.3810 & +02:06:29.88 & 11.02 & 139 & 18.10 & 32.7 & 3.89e-15 & 3 & 1.8 \\
UGC 09618NED02 & 14:57:00.8000 & +24:37:04.00 & 11.70 & 150 & 65.56 & 10.47 & - & 1 & 2.4 \\
IC 4687N & 18:13:39.7490 & -57:43:29.20 & 11.32 & 77 & 38.51 & 35.3 & 4.11e-15 & 2 & 2.8 \\
IC 4687S & 18:13:40.4750 & -57:44:53.95 & 11.02 & 77 & 15.49 & 35.3 & 1.92e-15 & 2 & 3.7 \\
NGC 6786 & 19:10:53.9000 & +73:24:37.00 & 11.40 & 101 & - & - & - & 2 & 1.0 \\
IRAS 20351+2521 & 20:37:17.8000 & +25:31:38.00 & 11.50 & 15 & - & - & - & 1 & 9.4 \\
II ZW 096 & 20:57:23.3000 & +17:07:34.00 & 11.94 & 150 & 156.77 & 23.9 & 1.64e-14 & 2 & 3.0 \\
ESO 148-IG002 & 23:15:46.8000 & -59:03:16.00 & 12.06 & 190 & 204.60 & 48.8 & 5.88e-15 & 4 & 2.5 \\
NGC 7674 & 23:27:56.7000 & +08:46:45.00 & 11.56 & 120 & 61.26 & 16.4 & 1.42e-14 & 2 & 2.0
\enddata
\tablenotetext{a}{SFRs calculated using IR+UV data taken from Howell et al. 2010 and U et al. (2012)}
\tablenotetext{b}{Merger Stages taken from Haan et al. (2013) and Stierwalt et al. (2013)} 
\tablenotetext{c}{The maximum $A_v$ adopted for each galaxy taken from the literature. See Appendix for more details.} 
\end{deluxetable*}

\section{Observations, Data Reduction, and Cluster Selection}

The HST B- (F435W) and I- (F814W) band images were obtained with the Wide Field Camera (WFC) on the Advanced Camera for Surveys (ACS) during the period 2005 August to 2007 January (PI: A. Evans; PID 10592). In all but a few cases, the wide field-of-view of the WFC ($202\arcsec \times  202\arcsec$) enabled the full extent of each LIRG to be observed. Each galaxy was observed in both filters per orbit, with two and three dithered exposures in ACCUM mode in the F814W filter and F435W filters, respectively. The approximate integration times for each filter were 21 minutes in F435W and 12 minutes in F814W. The ACS data were reduced with the Multidrizzle software included in $IRAF/STSDAS$ provided by STScI, to identify and reject cosmic rays and bad pixels, remove geometric distortion, and to combine the images into mosaics. Because of the limited number of dithers, additional cosmic rays rejection routines were run on each image prior to drizzling (see Kim et al. 2013 for a detailed description).

The HST far-UV (F140LP) and optical images in the sample were obtained with the Solar Blind Channel (SBC) on the Advanced Camera for Surveys (ACS) during the period 2008 April -- 2009 August (PID 11196; PI: A. Evans). The field of view of the SBC is $\sim 30\arcsec\times30\arcsec$ -- this placed a limit on the area within each LIRG over which the clusters could be analyzed. The data were taken in the ACCUM mode using the PARALLELOGRAM four-position dither pattern for a total integration time per galaxy of 40--45 minutes. We further reduced the SBC data with the Multidrizzle software included in $IRAF/STSDAS$ provided by STScI, to identify and reject cosmic rays and bad pixels, remove geometric distortion, and to combine the images into mosaics.

Before an automated routine for cluster identification could be applied to the images, contamination from foreground stars and distant background galaxies outside of the area of each image subtended by the LIRG (i.e., the ``sky'' area) had to be minimized. Masks of each image were made by first creating a median-smoothed version of the F435W and F814W images. The effect of this filtering is to minimize structures in the sky region with spatial extents significantly smaller than the filter size (i.e., faint stars and distant background galaxies). The backgrounds, containing low pixel values, were then set to zero, while the high pixels corresponding to the LIRG were set to one. Finally, pixels associated with any bright stars in the image were set to zero. The original reduced image  was then multiplied by the final mask of the galaxy to set the regions outside of the galaxy equal to zero.

Star clusters in all three bands were selected using the program SExtractor (Bertin \& Arnouts 1996). The identification of clusters and the extraction of photometry is complicated by the non-uniform surface brightness of the underlying galaxy. To estimate and subtract the underlying galaxy, Source Extractor iteratively computes the median and standard deviation of the pixels within a mesh of $n \times n$ pixels. During each iteration, outlier pixels are discarded until all of the pixels within each mesh are within $3\sigma$ of the median value. Several mesh sizes were tested, and for each mesh the photometry of several of the clusters was separately computed via the IPAC image display and analysis program Skyview and compared to values estimated from the original image (Skyview allows users to manually size and place apertures on clusters, and it allows for local background around the aperture to be subtracted). The mesh sizes varied between 9 and 14 pixels, and overall did an efficient job of removing the underlying galaxy and minimizing the creation of negative value holes surrounding clusters created through over-subtraction of the local background. Cluster photometry across all background-subtracted images was then calculated using the IDL package APER (originally modified from DAOPHOT). We used an aperture of radius 6.0 pixels for the HRC images and 3.0 pixels for the WFC images (= 0.15$\arcsec$ in both cases). An annulus with radius 4 pixels and a thickness of 5 pixels was used to measure the local background in the WFC images; the radii and thickness of the annulus was adjusted accordingly for the SBC images. Aperture corrections were calculated based on the flux calibrations of unresolved sources by Sirianni et al. (2005). We corrected the photometry for foreground Galactic extinction, using the Schlafly \& Finkbeiner (2011) dust model combined with the empirical reddening law of Fitzpatrick et al. (1999) available through the NASA Extragalactic Database (NED).

In the process of doing the photometry, we filtered out all sources which had a signal-to-noise ratio, $S/N < 5$ and which were not visible in all three filters. This left us with a total of 1186 cluster candidates identified in the sample. We then used ISHAPE (Larsen 1999) to measure the FWHM values for all remaining sources in all three wavelengths; this was done in order to separate stars and background galaxies from clusters. ISHAPE measures FWHMs by de-convolving the HST instrumental point spread function with a King profile, then performing a $\chi^{2}$ calculation to test the goodness of fit to each individual cluster (King 1966). ISHAPE iterates through different values for the effective radius until a minimum $\chi^{2}$ is found. Similiar to the approach in Mulia et al. (2015), we find that a conservative cut of 2 pixels FWHM effectively removes extended sources in both the nearest and furthest galaxies in the sample. Additionally, we made a cut of $M_{B} \leq -9.5$ mag, corresponding to the Humphreys \& Davidson (1979) limit, where we might expect contamination of the cluster sample from single bright yellow supergiants in the Milky Way. This was shown in Whitmore et al. 2010 to be an effecive way to remove foreground stars by their luminosity alone. A total of 665 clusters across all 22 LIRGs (27 nuclei) meet the above criteria.

One remaining concern with this approach was that at the average distance of the galaxies in our sample (115 Mpc), our size estimates would not correspond to physically relevant values for individual clusters. Indeed a 2 pixel FWHM at the resolution of WFC gives an average cluster size of $R_{eff} \sim 24$~pc. For the most nearby galaxies in the sample, we derive consistent results with the established cluster size in the Antennae of $ R_{eff} \sim 5-10$ pc (Anders et al. 2007). However, for the most distant galaxies in the sample our size estimates are nearly three times larger ($\sim 37$ pc), which is an effect we must take into consideration when interpreting our results (see \S 5). Importantly, the measured cluster sizes are all still well below the average size of an entire cluster complex or OB association ($R_{eff} \sim 100-200$ pc:  Bastian et al. 2006), where the application of simple stellar population (SSP) models would be questionable.



\section{Age-Dating Clusters}

\subsection{Model Fitting}

For each cluster in each galaxy, the measured colors were compared with the evolutionary tracks from GALAXEV (version 2003), a library of evolutionary stellar population synthesis models that were computed using the isochrone synthesis code of Bruzual \& Charlot (2003), hereafter referred to as BC03. This code computes the spectral evolution of a stellar population based on a stellar evolution prescription (Padova 1994) and a library of observed stellar spectra. The output of the model SED was multiplied by the ACS F435W, F814W, and SBC 140LP filter response functions in order to obtain magnitudes and colors in these filters. We first estimate the age and the extinction $A_{V}$ by performing a $\chi^{2}$ fit assuming an instantaneous burst simple stellar population (SSP), a Salpeter IMF (Salpeter 1955), and both solar and sub-solar metallicities as suggested for LIRGs by Kewley et al. (2010). We also apply a Calzetti extinction law of the form $k(\lambda) = A(\lambda)/E(B-V)_{*} = a + b/\lambda + c/\lambda^{2} + d/\lambda^{3}$, where a, b, c, and d are constants in a given wavelength range, and $A(\lambda)$ is the attenuation in magnitudes. The total attenuation of the stellar continuum, $R_{V} = A(V)/E(B-V)_{*} = 4.05 \pm 0.8$, is calibrated specifically for starburst galaxies and differs from the typical Milky Way value of $R_{V} \sim 3.1$ (Calzetti et al. 2000). It has been shown empirically that clusters and HII regions are more heavily attenuated than the underlying stellar continuum, due to the fact that these objects are often found near dusty regions of ongoing star formation (Calzetti 1994). From galaxy to galaxy, there can be considerable variations in the detailed dust distributions, but Calzetti et al. (2000) points out that in all the cases they studied, the empirical law recovers the total dust optical depth of UV-bright starburst galaxies within a factor of two.


It is worth noting here that a major concern in estimating cluster ages is the effect of stochasticity which affects clusters with low masses. Such clusters have too low a mass to adequately produce a sufficient of number of stars in all mass ranges, and thus any age-dating prescription making use of a standard IMF fails to predict the correct cluster age. Given the distance of the LIRGs in the sample and thus the brightness of clusters detected by our HST observations, the detected clusters are unlikely to have low masses. Indeed, stochastic fluctuations are relatively minor for clusters with masses greater than $10^{4} M_{\odot}$ (Fouesneau et al. 2012), which in our case is the lower limit of clusters we can observe.

Another factor affecting the age estimates is the metallicity. LIRGs are known to have gas-phase metallicities within 0.2 dex of solar in $12+\log{\rm [O/H]}$ (Relano et al. 2007; Rupke et al. 2012). Thus we consider both a solar ($z=0.02$) and sub-solar ($z=0.008$) BC03 model for each galaxy. Rich et al. 2012 also finds that the metallicity gradients in LIRGs are flattened by the merging process, allowing us to parameterize the metallicity of clusters with a single value for each galaxy. 

The mass of each cluster is estimated from the observed B-band extinction-corrected luminosity and the mass-to-light ratios ($M (M_{\odot}) = L_{B}\times(M/L)$) predicted by the un-extincted model at the fitted age. The models assume that the stellar IMF for each cluster is fully sampled. The largest contribution to the uncertainty in the mass estimates are the uncertainties in the estimated ages, which are typically on the order of 0.3 dex in $\log(\tau)$. These translate to similar uncertainties of 0.3 and 2 in $\log (M)$ and $M$, respectively. The derived masses of the clusters depend on the IMF assumed in the stellar population models. For example, if a Chabrier IMF is adopted, the estimated mass of each cluster would decrease by a near constant 40 percent (although the shape of the mass function would not change). The average fractional uncertainty in the distances of each galaxy taken from NED are $\sim 7\%$. This would introduce uncertainties in the cluster mass estimates of roughly $13\%$, which is less than the error contribution from our cluster age fitting procedure.

The age and mass estimations using color-color diagrams together with evolutionary tracks suffer from age--reddening degeneracy. As pointed out in Maoz et. al. (2001), the use of a UV filter when examining the colors of star clusters does help to avoid the issue of ``backtracking,'' whereby the reddening shifts the models in a direction nearly parallel to the aging direction. However, a cluster that appears red in the FUV-B, B-I color space can still be either very old, or young and heavily obscured by dust. In particular, young star clusters are assumed to be embedded in dust that is present in the star-forming region. Despite the fact that a fraction of the dust can be cleared away from young star-forming regions in as little as a few Myr (Larsen 2010), $0.5 \leq A_{V} \leq 2.5$ mag extinction has been reported for 4 Myr old clusters in nearby, lower luminosity galaxies (Whitmore \& Zhang, 2002; Reines et al. 2008). Since our analysis involves the use of three filters, we cannot break this degeneracy with our photometry alone. Thus, the ages of clusters in our LIRG sample are solved for by creating a suite of SSPs within the FUV-B, B-I color space, incrementing by 0.1 in $A_{V}$ as input to the extinction law, then solving for the age--reddening of each cluster based on the best $\chi ^2$ fit to an individual model within the suite. Further, it is important to note that because FUV light can accurately trace the ages of star clusters over two orders of magnitude (Meurer et al. 1995) our analysis of cluster ages is not biased by the requirement to detect a cluster in the F140LP SBC filter.

In order to better refine the age--reddening estimates for each cluster, two additional constraints were applied: First, we required that the extinction of any given cluster could not exceed estimates for the $A_V$ of its host galaxy taken from the literature. Considering the fact that our F140LP cluster detections often span the entire SBC field-of-view, the average galaxy $A_V$ is a good proxy for the amount of reddening one would expect each cluster could have before we are unable to detect it. It is important to note that only $5\%$ of clusters in the final sample have extinctions which are equal to the maximum allowed for their host galaxy based on our fits, meaning that our choice of $A_{V}$ is not systematically biasing our final derived values. This constraint additionally prevents our model from obtaining cluster properties with arbitrarily high extinctions and therefore cluster masses, which exceed what is possible for bound stellar clusters so far observed in extragalactic systems (Maraston et al. 2004). Second, we constructed B-I color images in order get a visual clue of where the projected dust lanes are in each galaxy. The reasoning is that by making a manual assessment of each image we can distinguish globular clusters, which have much redder colors and are often found in uncrowded regions away from sites of recent star formation (e.g., see Whitmore et al. 2014). One complicating factor is that a YMC that forms behind a projected dust lane can appear to have a color similar to these old GCs. By overlaying the cluster centroids, we identified which clusters had no obvious dust lanes in a surrounding annulus of $4-9$ pixels. These clusters are therefore more likely to be young and extincted as opposed to relatively old and dust-free clusters. The results can be seen in the false-color images shown in the appendix. In total, only $10\%$ of the clusters modeled had ages which differ by 0.6 dex (roughly twice the expected uncertainty) when including or excluding the additional dust-lane constraints. Whitmore et al. (2014) used this additional constraint when looking at the cluster populations of 20 star-forming galaxies in the Local Universe, and found it to be effective regardless of the detailed galaxy morphologies seen in the color images.


We consider here how these constraints can be understood based on the $F435W-F814W$ value of each cluster: Clusters designated with $(F435W-F814W) < 0.51$ mag can be reliably age-dated as being younger than 7 Myr, because the old-age track of the model never reach that part of the parameter space. Clusters with $(F435W-F814W) = 0.51 - 1.0$ mag have a wide range of possible ages ($7 - 500$ Myr), but if the cluster resides in a dustier region of the galaxy, then it is either an unreddened to moderately reddened old cluster or a young, heavily reddened cluster. This color bin covers the widest range of cluster ages and therefore contains the largest number of SCs. Finally, any SCs with $(F435W-F814W) = 1.0 - 1.5$ mag that do not reside in a more heavily extincted region of the galaxy are old, with ages between 500 Myr and 1 Gyr. The ages of clusters in these last two regions that lie in and around dust lanes are the ones most affected by our above criteria for solving the age--reddening degeneracy. Clusters with $(F435W-F814W) > 1.5$ have ages older than 1 Gyr assuming reasonable values for the internal extinction within the galaxy. 

By examining the distributions of internal visual extinction and age for each cluster derived from the model, we see that nearly $1/3$ of all young clusters in the sample have a relatively small dust correction ($A_{V} \leq 1$), and nearly 80\% of all young clusters have an $A_{V} \leq 2$ correction. Thus the majority of all clusters in the sample need only a relatively modest dust correction, compared to a galaxies global average, to properly derive young ages.
 

\subsection{Consistency Checks}

\subsubsection{Comparison with Direct SED-Fitting}
In order to account for the effect our chosen filter set has on the derived cluster properties as described above, we compare the results of anchoring each color to the F435W measurement, with the results from fitting the three broadband photometric measurements (F140LP, F435W, and F814W) simultaneously, as was similarly done in Maoz et al. (2001), and shown to be an effective way to further improve our ability to separate the effects of age and extinction. To perform this full ``SED-based'' fitting we use the same galaxy evolution code, extinction model (minus the additional dust-lane constraints in both cases), IMF, and metallicity. From our sample of 665 clusters, we further remove from the final analysis any clusters for which the method described in Section 4.1 and this SED fitting method do not produce ages which agree within 0.6 dex of each other. These clusters are almost always ones for which their is nearly equal probability of the cluster being young and highly-extincted or old and less heavily extincted. These highly degenerate cases are therefore removed due to their uncertain contribution to the overall shape of the age and mass distributions to be derived. This leaves us with a final sample of 484 ($\sim 83\%$ of verified clusters) clusters that have age and mass estimates independent of the fitting method chosen for deriving cluster properties. We also note that of the original 67 clusters which provide inconsistent age results in our own dust-lane vs. no-dust-lane analysis, 48 ($\sim 83\%$) are kept when comparing to the results of the full SED-fit. This again shows that our additional dust-lane constraints did not systematically bias the estimation of cluster ages.

\subsubsection{Comparisons with Spectroscopic-Derived Ages}

Chien (2010) measure Balmer line-derived cluster ages for a sample of GOALS LIRGs. Three of the systems in their sample overlap with our present study (NGC 2623, Arp 256, and Arp 299). Figure 1 is a comparison of our photometrically derived ages and the Balmer line-derived ages. Approximately 77\% (17 of 22) of the clusters have ages that agree to within $\pm0.3$ dex, and 91\% (20 of 22) have ages that agree to within $\pm0.6$ dex. This means that the majority of our 3-band cluster ages agree with the spectroscopic ages within the uncertainty of the BC03 models. Further, it is important to note that we derive young ages for all seven of the star clusters in our sample with identified Wolf-Rayet spectral features from Chien (2010). Wolf-Rayet features are very sensitive probes of young cluster ages since they only exist for clusters with ages of $3-7$ Myr (Leitherer et al. 1999; Chien 2010).

It is potentially not surprising that the older clusters in the sample have more uncertain spectroscopic age measurements. In particular, as a cluster ages, the strength of the Balmer lines is significantly decreased (Gonzalez Delgado et al. 2005). Finally, the most discrepant age estimates come from NGC 2623. This could be due to the fact that the galaxy has a complicated morphology (Evans et al. 2008). All the young clusters identified come from a single ``pie-wedge'' structure to the right of the nucleus (see Appendix), while all the older clusters come from the nuclear regions. This makes using a simple prescription for an $A_{V}$ correction over the entire FOV more uncertain.

\begin{figure}
\centering
\includegraphics[scale=0.4]{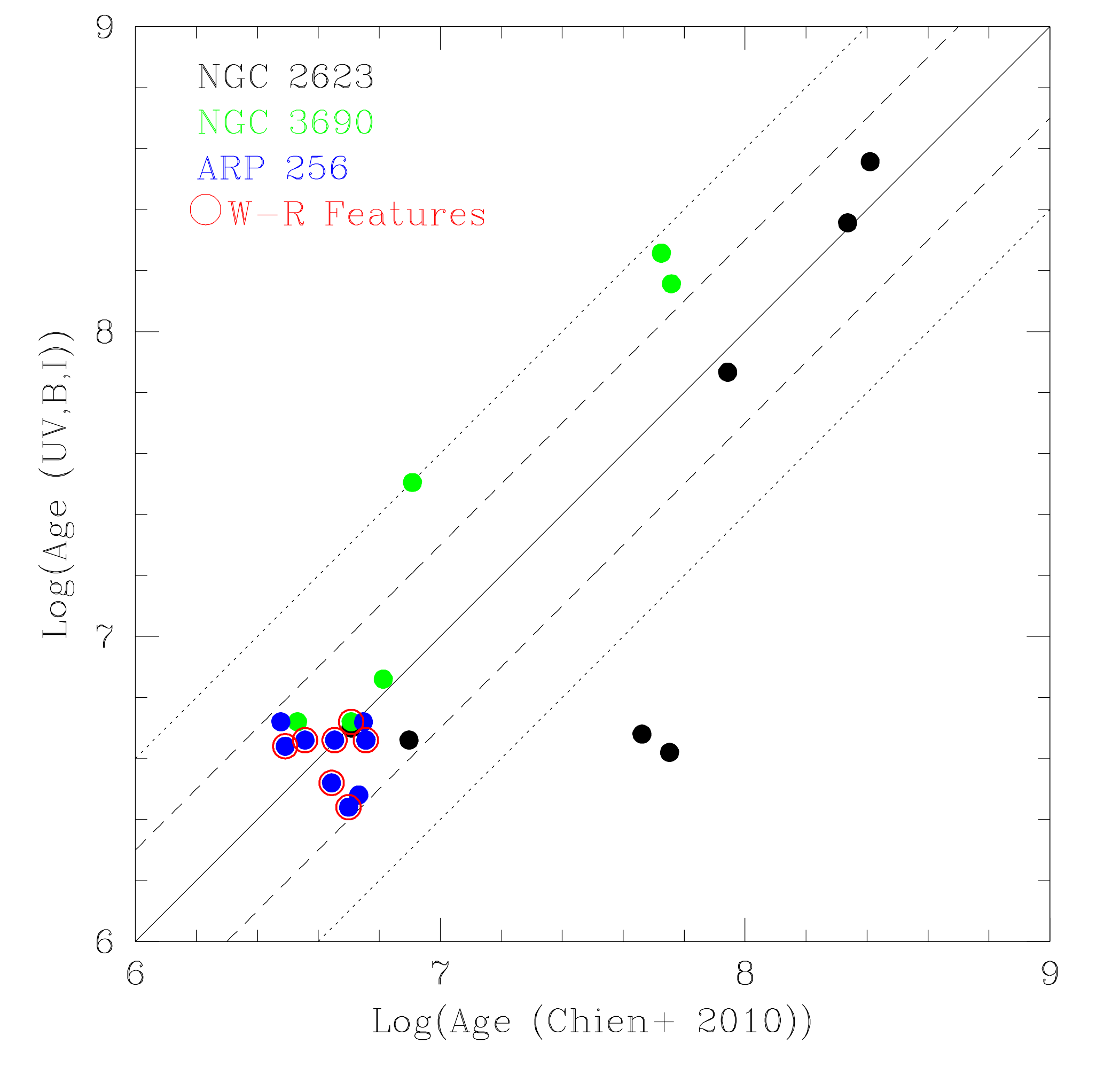}
\caption{A comparison between the spectroscopically derived ages from Chien 2010 to to our UV, B, I broadband age estimtates for NGC 2623, NGC 3690E/W, and ARP 256N/S. The red circles denote star clusters which have Wolf-Rayet spectral features as identified in Chien (2010). The solid line represents the $1:1$ correlation, whereas the dashed and dotted lines are within 0.3 and 0.6 dex of the $1:1$ correlation.}
\end{figure}

\subsubsection{Comparisons with Paschen-$\beta$ Equivalent Widths Derived from WFC3 Imaging}

Larson et al. (in prep) obtained Paschen-$\alpha$ and Paschen-$\beta$ imaging for a subset of the GOALS sample, with 6 LIRGs (9 galaxies) overlapping our present HST sample. For any B-band cluster centroid that is spatially coincident with a high density clump in the Pa$\beta$ images we can directly compare our cluster ages to ages derived via the equivalent width (in Angstroms) of the Pa$\beta$ emission line. For an instantaneous burst SSP and a Salpeter IMF, the presence of Pa$\beta$ emission constrains the burst age to less than 20 Myr because stars with masses greater than 10 M$_\odot$ are required for significant production of ionizing photons. We utilize Starburst99 models of Pa$\beta$ equivalent width as a function of clump age to independently derive ages for 27 clusters in the sample (Leitherer et al. 1999).

\begin{figure}
\centering
\includegraphics[scale=0.4]{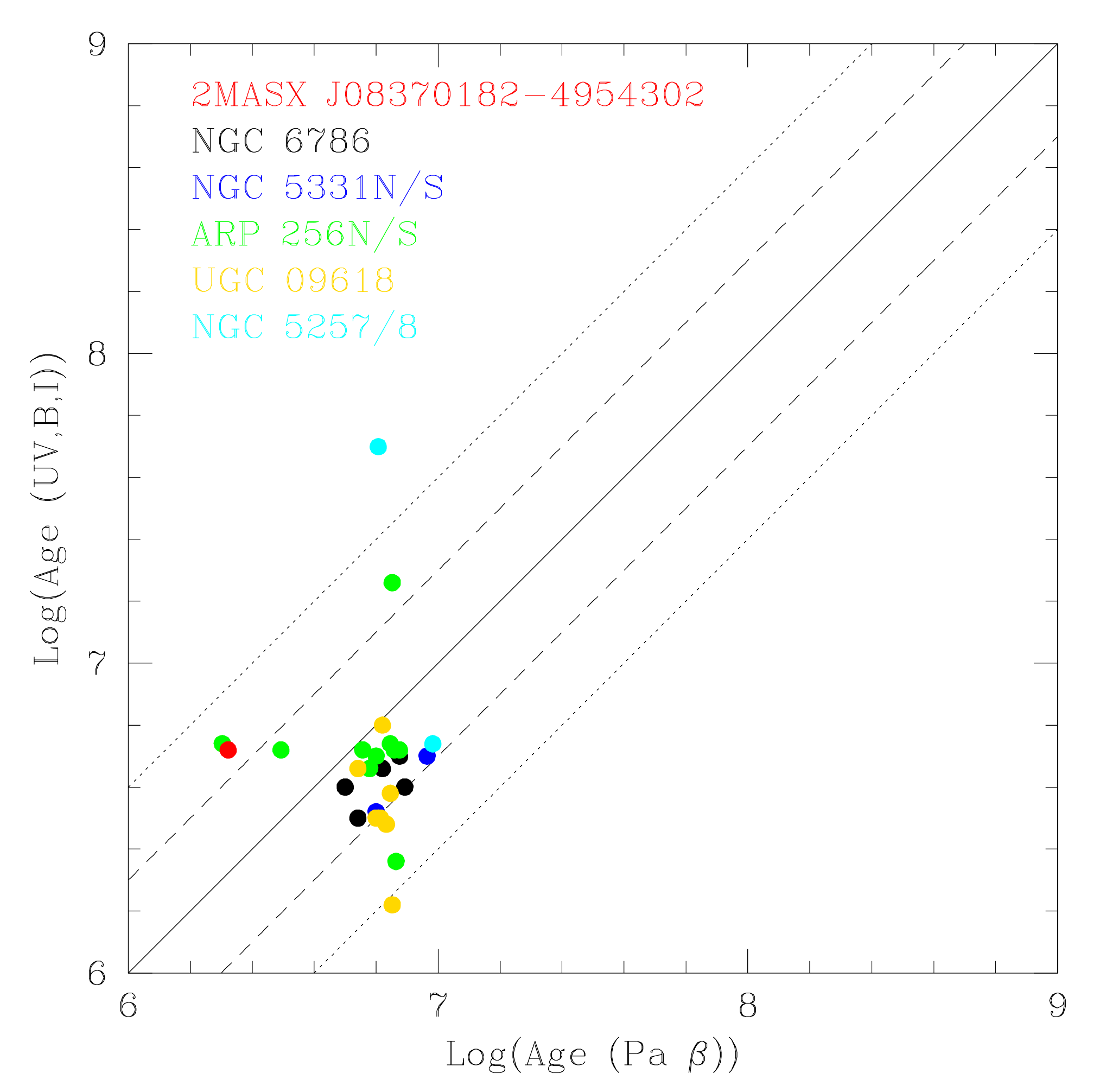}
\caption{The $1:1$ comparison of cluster ages derived using our UV, B, and I photometry and the equivalent width of the Pa$\beta$ emission line associated with the cluster centroid from Larson et al. (in prep). The solid line represents the $1:1$ correlation, whereas the dashed and dotted lines are within 0.3 and 0.6 dex of the $1:1$ correlation.}
\end{figure}
 
From Figure 2 we find that approximately 78\% (21 of 27) of the clusters have ages that agree to within $\pm0.3$ dex, and 96\% (26 of 27) have ages that agree to within $\pm0.6$ dex. This shows us that the majority of all clusters we identify as having bright Pa$\beta$ counterparts are indeed young. Additionally, 89\% of the clusters (= 24 out of 27) which are photometrically identified as having ages less than 20 Myrs have a mean Pa$\beta$ equivalent width of log($W(Pa\beta) [\AA]$) $\sim 1.7$ or log($Age_{SB99} (yr)$) $\sim 6.8$. It is important to note that of the 142 young ($t \leq 10^{7}$ yr) star clusters photometrically identified in these 6 LIRGs, we only associated a strong Pa$\beta$ clump in the continuum-subtracted image with 19\% (27 of 142) of them. This fraction is likely low for two reasons:
\noindent
(1) Our clusters are located primarily in the central regions of the galaxies, where the continuum subtraction is much more uncertain due to the larger contribution of diffuse large-scale NIR emission. As a result, the minimum equivalent width of a marginal 3$\sigma$ Pa$\beta$ detection can vary by a factor of a few within a galaxy and by almost an order of magnitude on a galaxy-by-galaxy basis. This variation corresponds to $\sim 0.3$ dex change in the maximum derivable age using the SB99 model, which if we assume a $1:1$ correlation, changes the age of the oldest cluster for which we would expect a counterpart in FUV emission by the same amount.
\noindent
(2) The resolution of the NIR Pa$\beta$ images is $0.12''$/pixel, which is a factor of two lower than what we achieve in the FUV and optical imaging. This makes detecting bright compact sources of Pa$\beta$ line emission embedded in a larger diffuse GMC cloud difficult at the distance of the galaxies in our sample. 

Ultimately, both the local background subtraction and resolution contribute to the lack of overlap we observe in the Pa$\beta$ and FUV emission. Regardless, this is an independent verification of our ability to derive accurate young ages for clusters in the sample, and shows us that our $A_{V}$ corrections can do a reasonable job at photometrically separating young and old clusters.


\subsection{Mass-Age Diagram and Completeness}

Figure 3 shows the derived age and corresponding mass of each cluster identified in the sample. An immediate observation one can make is the lack of low-mass, old clusters. This is due to the fact that clusters dim as they age and eventually become fainter than our UV detection limits. We also note the large number of clusters seen with ages below 10 Myr over the full range of masses. 

Although the cluster fitting method can create some observed structure in the mass-age diagram, it is unlikely to do so over all masses at young ages. In particular the lack of clusters with ages of $\sim 10^{7}$ Myr is a common feature of model-derived mass-age diagrams of star clusters in galaxies (Gieles et al. 2005, Goddard et al. 2010). This is due to the limited age resolution and overall degeneracy of the UV-B, B-I color track at these ages (See color-color diagrams in the Appendix). From the histograms in Figure 3, we conclude that there is a genuine over-density of clusters with ages below 10 Myr compared to above 10 Myr. 

In order to determine the completeness limit of the cluster sample, we used a similiar prescription to Whitmore et al. (1999), and set the limit for each galaxy as the magnitude at which $50\%$ of the clusters are detected at B and I, but are missed at FUV. The magnitude distributions for each band are corrected for foreground galactic extinction, and spatially matched to the FOV of the SBC. Of the 22 LIRGs in the sample, 19 have magnitude distributions which span the full range of observed cluster values ($M_{B} = -10 \sim -15$ mag), and have a mean completeness of $M_{B} \sim - 11.2$ mag. The three remaining sources have completeness limits which are shifted to higher magnitudes $M_{B} \sim -13$ mag, likely due to the fact that they are all further away than the mean distance of the galaxies in the sample ($115$ Mpc). It is important to note however, that there are several other galaxies for which a larger distance did not result in a shifted magnitude distribution, meaning that the actual $50\%$ limit for the sample is not a strong function of the mean distance to any galaxy. Additionally, these outliers represent only $7\%$ of the total cluster population. Therefore, to minimize their contributions to the final adopted limit for the entire sample, we calculated a cluster-weighted mean completeness limit, and found that the mean shifted only slightly to $M_{B} = -11.26$.

By applying this completeness limit to the BC03 model, we can define regions of this parameter space (both as a function of cluster ages over a mass range and masses over an age range) where we are observationally complete and thus working with a mass-limited sample of clusters. Mass-limited cluster samples have the advantage over luminosity-limited samples because they recover the underlying shape of the age distribution, and are thus not affected by the distance to each galaxy. However, the total number of clusters can be highly uncertain simply because the lower mass clusters are not included. We will discuss the implications for this fact in \S 5.

The four cuts were selected to sample distinct regions of the mass and age distribution for which we could maintain completeness. We define Region 1 to be:

\begin{figure*}
\centering
\includegraphics[scale=0.75]{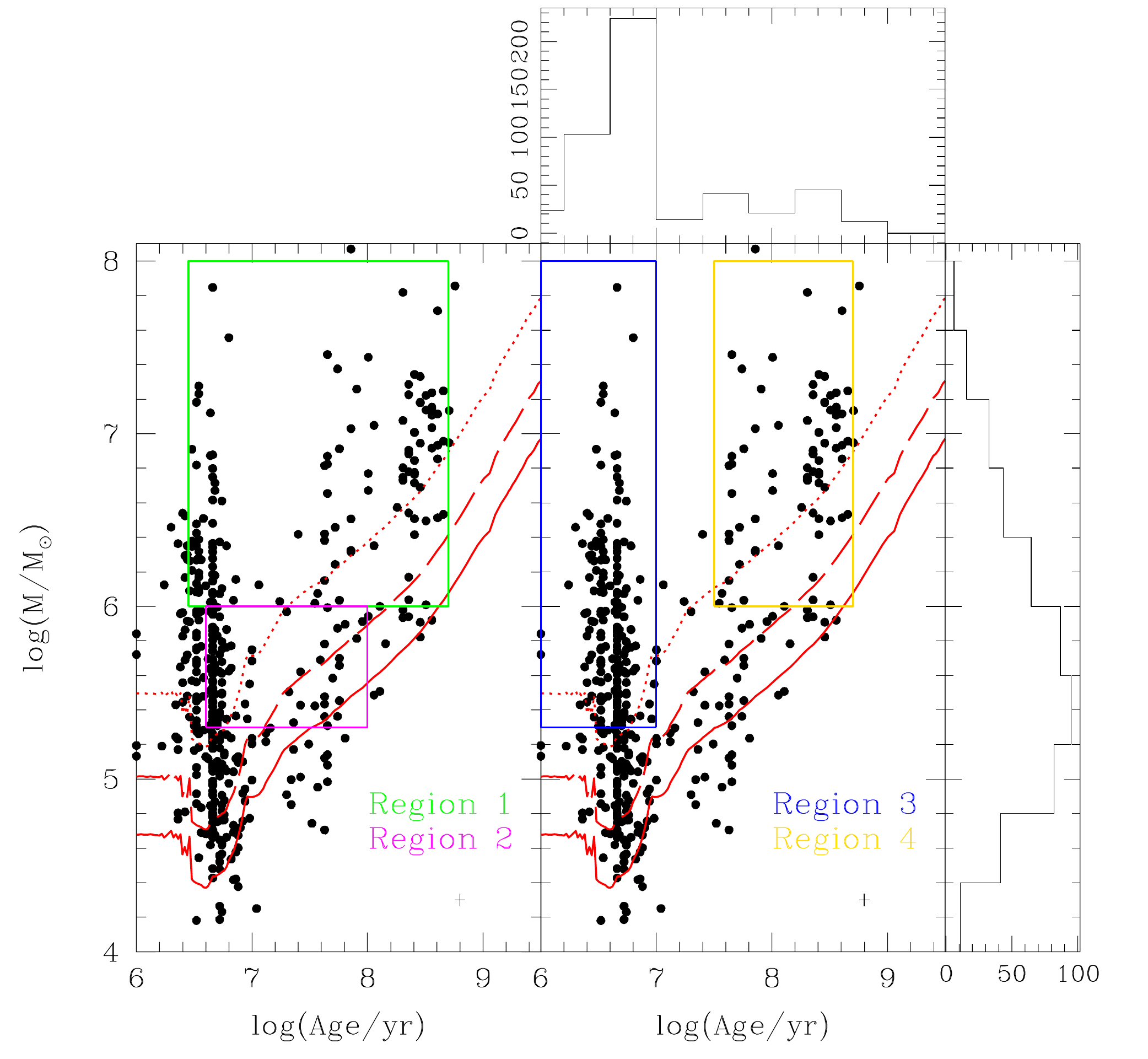}
\caption{The mass, age distribution of all 484 clusters found in the 27 galaxies. The solid, dashed, and dotted red curves represent mass-age tracks produced from the BC03 model with an input of $M_{B} = -11.26$, $M_{B} = -12.07$, $M_{B} = -13.31$ for the $50\%$, $75\%$, and $100\%$ completeness limits respectively. The green and purple boxes in the left panel represent Regions 1 and 2 respectively, and are used for the two mass-age cuts applied when analyzing the cluster age distribution. The blue and gold boxes in the middle panel represent Regions 3 and 4 respectively, and are used for the two mass-age cuts applied when analyzing the cluster mass distribution. The histograms show the distribution of cluster ages and masses for the full sample. The cross on the bottom right of each panel represents the median errors in cluster age and mass bootstrapped from our model.}
\end{figure*}

\begin{equation}
  6 < log(M/M_{\odot}) < 8
\end{equation}
\begin{equation}
  6.5 < log(\tau) < 8.7
\end{equation}

\noindent
Region 2 to be:

\begin{equation}
  5.3 < log(M/M_{\odot}) < 6
\end{equation}
\begin{equation}
  6.6 < log(\tau) < 8
\end{equation}

\noindent
Region 3 to be:

\begin{equation}
  log(\tau) < 7
\end{equation}
\begin{equation}
  5.3 < log(M/M_{\odot}) < 8
\end{equation}

\noindent
and Region 4 to be:

\begin{equation}
  7.5 < log(\tau) < 8.7
\end{equation}
\begin{equation}
  6 < log(M/M_{\odot}) < 8
\end{equation}

\noindent
The two mass cuts are marked as Regions 1 and 2 in the left panel of Figure 3. Since older clusters are intrinsically fainter, a higher mass limit will result in a cluster population that is mass-limited to a wider range of ages. 
Note that the chosen mass-regimes do not contain the youngest least massive clusters that are only observed in a subset of our galaxies, and thus would bias any estimate for the global mass and age distributions of all the galaxies combined. Region 2 is chosen to match the age and mass limits from Fall et al. (2005), allowing us to make accurate comparisons to the cluster population of the most well-studied nearby major merger, the Antennae Galaxy. Regions 3 and 4 are chosen to sample the young ($\leq 10$ Myr) and old ($\tau \geq 10^{7.5}$) clusters respectively within the completeness limit. When analyzing Regions 1, 3, and 4 we will exclude the largest mass bin of $log(M/M_{\odot}) = 8.0$. These very high masses are most likely the result of either an imperfect extinction correction or multiple star clusters in close proximity appearing as a single star cluster at the resolution of these images, resulting in a large derived total mass (See \S 5.2). While clusters of these masses have rarely been observed in abundance, we note that Bastian et al. (2013c) studied several young star clusters in NGC 7252 with masses greater than $10^{7} M_{\odot}$, including one cluster with a total mass of $\sim 10^{8} M_{\odot}$.


\section{Discussion}

After determining ages, masses, and extinctions for the entire cluster sample we directly compare these distributions with those of nearby normal and interacting galaxies. We focus on the interpretation of the derived cluster age distribution and mass function, and briefly discuss the implications for cluster formation efficiency. Ultimately, we discuss to what degree the differences observed in our cluster population can be attributed to the extreme star-forming environment unique to LIRGs in the local Universe. Individual cluster age and mass functions for the most 'cluster-rich' (i.e. greater than 25 detected clusters) galaxies are computed in Table 2.

\subsection{Age Distribution}

\begin{figure*}
\centering
\includegraphics[scale=0.7]{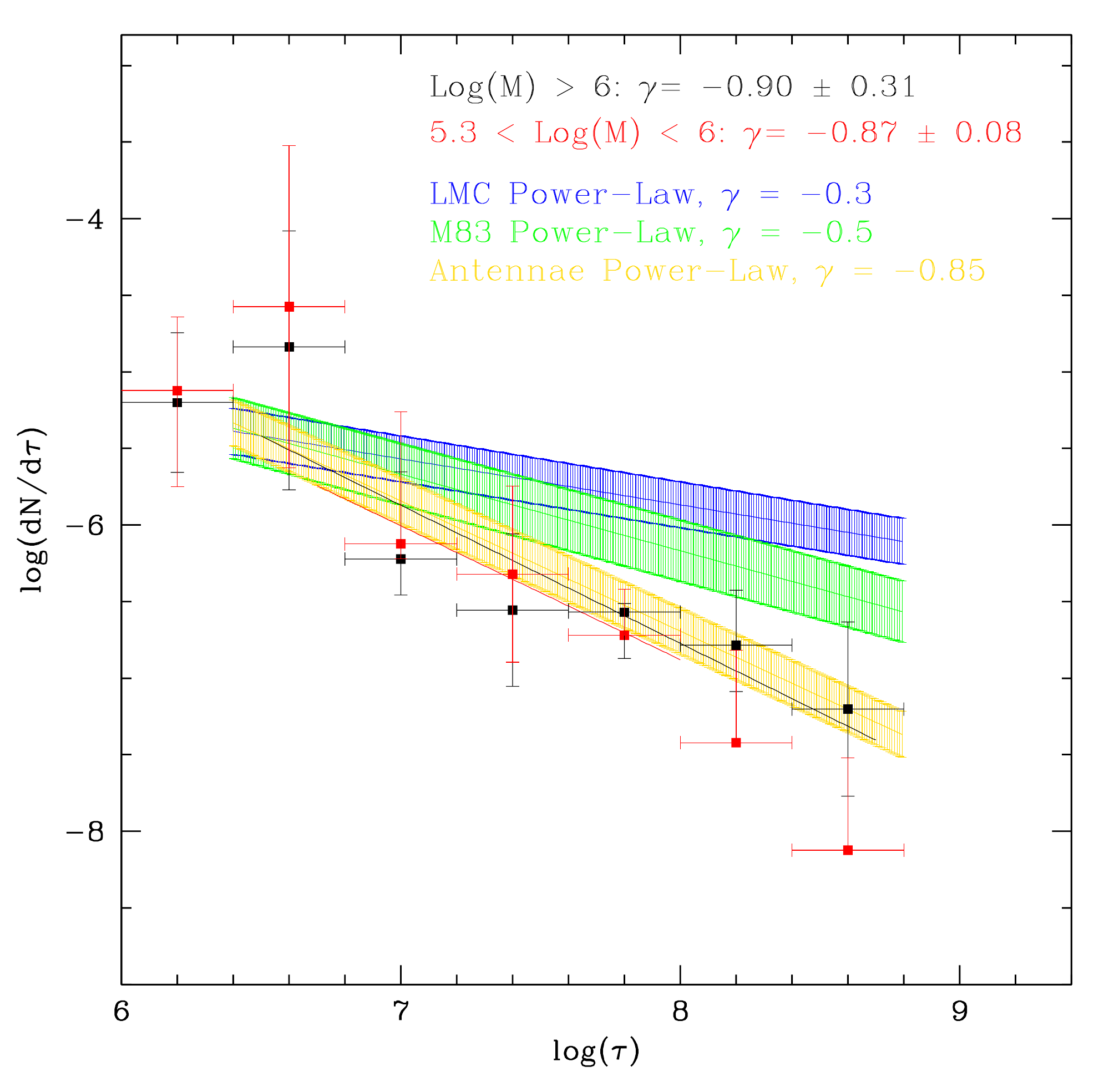}
\caption{The stacked age distribution functions for all 27 galaxies. We have broken our age distribution up into the two age-mass ranges described in Equations 1 and 2, and shown as Regions 1 and 2 in the left panel of Figure 3. The red and black lines represent weighted linear least squares fits to the data. The blue, green, and yellow age functions of the LMC, M83, and the Antennae respectively, are taken from Adamo \& Bastian 2016, and are normalized to the total number of clusters in our sample to best compare the slope for each galaxy.}
\end{figure*}

We consider the age distribution of clusters in our complete LIRG sample over the two mass ranges (i.e., Regions 1 and 2) described in \S 4.3. Specifically, we are interested in measuring the power law index $\gamma$, where $dN/d\tau = \tau^{\gamma}$. Figure 4 is a plot of the logarithm of the number of clusters per time interval, $\log (dN / d\tau)$, versus the logarithm of the cluster age, $\log (\tau)$. The plotted data are binned by 0.4 in $\log (\tau)$ so as to fully encapsulate the model errors of 0.3 in $\log (\tau)$ discussed in \S 4.1. We see that that a large fraction ($\sim 30\%$) of the clusters have ages less than 7.5 Myr. For the youngest most massive clusters in the sample (contained in Region 1), a weighted linear least-squares fit to the cluster age distribution gives a power law index of $\gamma = -0.9 \pm 0.3$, consistent with the derived power law index for the Antennae Galaxies within $1\sigma$ (Fall et al. 2005; Figure 2).

The distribution of the lower mass clusters (Region 2) can be fit with a power law index of $\gamma = -0.87 \pm 0.1$, also consistent with the derived power law index for the Antennae Galaxies within $1\sigma$. The change in $\gamma$ ($\sim 0.04$) for the solar and sub-solar models was less than the uncertainty in the fit to the data in Figure 4. The similarity in the slope of the power-law index between the two mass-cuts is also further confirmation that we are working in a mass-limited regime, where the slope of the age distribution does not get systematically flatter with increasing cluster mass or distance to the host galaxy (Bastian 2016).


Also plotted in Figure 4 are the age distributions for M83 and the LMC, normalized to the fitted-number of clusters in the youngest age bin. As can be seen, $\gamma$ for the LIRG sample is steeper than what is measured for these lower mass, normal star-forming systems. In addition, Adamo \& Bastian (2016) provide a Table summary of $\gamma$ for several local galaxies; in all cases, $\gamma$ is flatter than -1. 

There are two possible interpretations of this plot:

\noindent
(1) If a continuous (or near continuous) cluster formation rate is assumed during the merging process for each LIRG, then the index of $\gamma = -1$ is an indication that 90\% of the clusters formed are disappearing every age dex. In the case of the Antennae Galaxies, Fall et al. (2005) concluded that the majority of the clusters are rapidly disrupted within the merger via 'infant mortality'. This scenario not only seems to fit into the nature of the violent environments of galaxy mergers, but may also explain the negative value of $\gamma$ (albeit, not as negative as measured for mergers) observed in lower-mass, less star-forming, quiescent nearby spirals. 

We note that when discussing 'infant mortality', it is important to mention that the rapid decrease in the number of clusters as a function of age could be due to the inclusion of young, low-density, unbound OB associations in cluster catalogues (e.g., Bastian et al.2012; Silva-Villa et al. 2014). When these associations are removed, the age distributions for local star-forming galaxies appear to flatten. Kruijssen et al. (2015) point out that these effects can be minimized by selecting slightly older clusters ($10-50$~Myr), so that associations will have already been dispersed into the field. If this were a dominant effect in our sample, we would expect the age distribution of Region 2 to be much flatter and inconsistent with the Antennae value. Further, while we cannot verify the amount of contamination from OB associations for the youngest clusters ($t < 10$~Myr) in our sample, the high mass cut-off for Region 1, ensures that the this effect is minimized.

\noindent
(2) The star formation rate has increased such that the bulk of the star formation, and cluster formation, has happened fairly recently as a result of the interaction of the two galaxies. This seems unlikely due to the fact that many of the galaxies within the sample have been interacting for a few hundred million years, whereas the median age of clusters for the whole sample is only $\sim 10^{7}$ years. Hopkins et al. (2013) finds that when simulations use realistic prescriptions for galaxy feedback, the star formation in a galaxy merger can in fact be time-variable and drops between each passage. Therefore, the average SF enhancement is only ever a factor of a few during the course of a merger, which is not enough to explain a 90\% decrease in the number of clusters at each age dex (Karl et al. 2011). We could assume that all of the galaxies across the various merger stages are being viewed at these bursty peaks in the star formation rate, but we also consider this an unlikely scenario.

Under this framework, we would also be forced to accept that the star formation rates in nearby normal galaxies (which have negative $\gamma$ values - though, note the above discussion of possible OB association contamination) are also increasing. In well-studied star-forming galaxies like the Milky Way and the Magellanic Clouds, the SFR is observed to have been nearly constant over the last Gyr, which argues strongly for the fact that the decline in $dN/d\tau$ is primarily a consequence of disruption in the MW and Magellanic Clouds (Harris \& Zaritsky 2009; Chandar et al. 2010a).

Given the above, the most plausible explanation is that clusters are being rapidly destroyed in luminous galaxy mergers at a rate that exceeds the cluster destruction process occurring in nearby normal galaxies.

\subsection{Mass Function}

The cluster mass function (CMF) has the form $dN/dM \sim M^{\beta}$. For star clusters in our sample, this was derived by stacking the mass distributions of each galaxy, keeping the binning constant (0.4 in $\log(M)$), and then performing a cluster-weighted linear least-square fit as a function of derived mass. For clusters with ages $t \leq 10^{7}$ years and $t \sim 10^8$ years, we derive a mass function with a $\beta = -1.95 \pm 0.11 $ and $-1.67 \pm 0.33$, respectively (see Figure 5). By comparison, $\beta$ is commonly measured to be $-2$ for the majority of lower luminosity star-forming galaxies, as well as the Antennae Galaxies (Larsen 2010). The change in $\beta$ for the solar and sub-solar models was less than the uncertainty in the fit (i.e., $< 0.1$) to the data in Figure 5. 

An alternative approach to modeling the ICMF is with a two component Schechter function of the form $dN/dM = (M/M_{c})^{\alpha} e^{(M/M_{c})}$. For reference, the $M_{c}$, or characteristic mass, measured for the Milky Way is $\sim 10^{5} M_{\odot}$ (Bastian 2008). If we assumed that a star formation rate of $\sim 100 M_{\odot}$/yr went into forming only clusters, the number clusters with $M \geq 10^{7} M_{\odot}$ would still be negligible for $M_{c} = 10^{5} M_{\odot}$; even if these high SFRs could be sustained for $\sim 100$ Myrs. Thus, the mere presence of $10^{7} M_{\odot}$ clusters in our sample indicates that the cluster formation environment in more extreme systems is different than that observed in lower-luminosity spiral galaxies.

Larsen (2010) shows that a Schechter function with a canonical -2 power-law slope and $M_{c} = 10^{6.3} M_{\odot}$ can reproduce the observed distribution in the Antennae galaxies equally well. In Figure 5 it is clear that we cannot simply adopt these parameters to fit our observations. Instead, we require both a slightly shallower power-law slope and a slightly larger cut-off mass due primarily to the fact that we are observing clusters with masses greater than $10^{6.5} M_{\odot}$, which simply are not observed in the Antennae. It is important to note that our data (Region $3+4$) is consistent to within $1\sigma$ of a -2 power law in $dN/dM$ over the same mass range as the Antennae, but can also be fit at the high-mass end using a modified Schechter function with a cut-off mass of $10^{7} M_{\odot}$. This is clearly larger than what has been recently observed in M31, where the observed cut-off mass for the cluster sample is $M_{c} \sim 8 \times 10^{3} M_{\odot}$ (Johnson et al. 2017). Interestingly, in that work, the authors define a relationship for the expected $M_{c}$ as a function of $\Sigma_{SFR}$ as: log~$M_{c} = (1.07 \pm 0.10) \times log \Sigma_{SFR} + (6.82 \pm 0.20)$. If a typical value of $\Sigma_{SFR} $ for LIRGs in the GOALS sample is used (U et al. 2012), we expect an $M_{c} \sim 10^{7} M_{\odot}$, which is consistent with our derived fit, and indicates that high-mass clusters can indeed form more efficiently in higher star-forming environments.

When interpreting these results, it is important to consider several possible factors which could affect our derived mass functions:

\noindent
(1) If lower-mass star clusters are preferentially disrupted, the mass distribution of the surviving star clusters in a merger remnant will be shallower than what is observed in a quiescent spiral galaxy (Kruijssen et al. 2012; Li et al. 2016). We might also expect this to correspond to a steeper age distribution for the lower-mass cluster sample (Figure 4; Region 2), but given that our 'low-mass' clusters are still rather massive, the lack of a clear difference in $dN/d\tau$ is not surprising. Therefore, the cluster disruption in these galaxies appears to be mostly mass-independent (i.e., we find that $\gamma \sim -1$ over the mass range of $M_{\odot} = 10^{5} - 10^{6}$), a finding that Whitmore et al (2010) confirmed for the Antennae over the same range of cluster masses (Figure 5; yellow track). 

When we increase the lower limit cluster mass for Region 1 to $10^{6.5} M_{\odot}$ in Figure 4, we observe a disruption rate of $dN/d\tau \sim \tau^{-0.75 \pm 0.4}$. This leads us to conclude that cluster disruption in LIRGs appears largely consistent with what is seen in the Antennae up to $10^{6.5} M_{\odot}$. We note that the uncertainty on the measured slope is much larger than for Region 2, so in principle, gamma could be shallower than the Antennae Galaxies in this mass regime. However, if this were a strong effect in our data we would expect our observed CMF in Region 3 to be shallower than the -2 power-law used to represent the underlying ICMF.

\noindent
(2) The choice of bin size for our data could systematically flatten the measured $\beta$ (Maiz Apellaniz \& Ubeda 2005). We use bin sizes in mass and age of 0.4 dex in $Log(M)$ and $Log(t)$, chosen to fully encapsulate the typical uncertainty associated with our age and subsequent mass estimations. To test the effect this choice has on the measured slope, we explored two other bin sizes, 0.2 and 0.6 dex respectively. We found that the slopes derived for $dN/d\tau$ and $dN/dM$ change on average by $0.1-0.2$ dex. As this is comparable with the $1\sigma$ uncertainties on each slope measurement, we conclude our choice of bin size is not significantly affecting our determination of the shape of the cluster mass distribution.

\noindent
(3) At the resolution of our observations, multiple lower mass clusters may appear as one, massive cluster, and thus systematically flatten the CMF. To test this possibility, we ran Source Extractor on B-band and I-band WFC images of NGC4038/9 from the Hubble Legacy Archive (HLA) to identify star clusters. The distance used for NGC4038/9 is $\sim 24$Mpc, but the median distance of our sample is four times farther away. Since the pixel scale of the Drizzlepac output images is the same, we simply smoothed the HLA images with a boxcar function of 4 pixels. Source Extractor was then run on this smoothed image with the Source Extractor results from the original, pre-smoothed images as a reference. For this step, Source Extractor only outputs sources which are both identified in the smooth image and also match a source in the original list (within a search radius of 4 pixels, i.e., the same size as the smoothing). The ratio: $(N_{orig} - N_{smoothed})/N_{orig}$ should give a upper limit for the fraction of dual sources identified as 1 in the smoothed image. 

For the B-band and I-band image comparisons, this ratio is $0.3$ and $0.26$, respectively. Thus, roughly $30\%$ of `blended clusters' identified in our LIRGs with $D \geq 100$Mpc would actually be identified as a complex of single clusters at the resolution of the Antennae. By redistributing to the lower mass end this percentage of clusters, with masses greater than $5\times10^{6} M_{\odot}$, we observe a steepening of the mass function $\sim 0.1$ dex. Despite this fact, it is clear that the existence of young high-mass ($\geq 10^{7} M_{\odot}$) clusters in our sample cannot be solely attributed to a resolution limit. Finally, it is worth noting that cluster blending can effect the estimated cluster ages. The effect most likely pushes clusters toward the median cluster age, and thus if deblending randomly populates the young and old cluster parts of the age distribution, there will not be a dramatic effect on $\gamma$. 



\begin{figure*}
\centering
\includegraphics[scale=0.7]{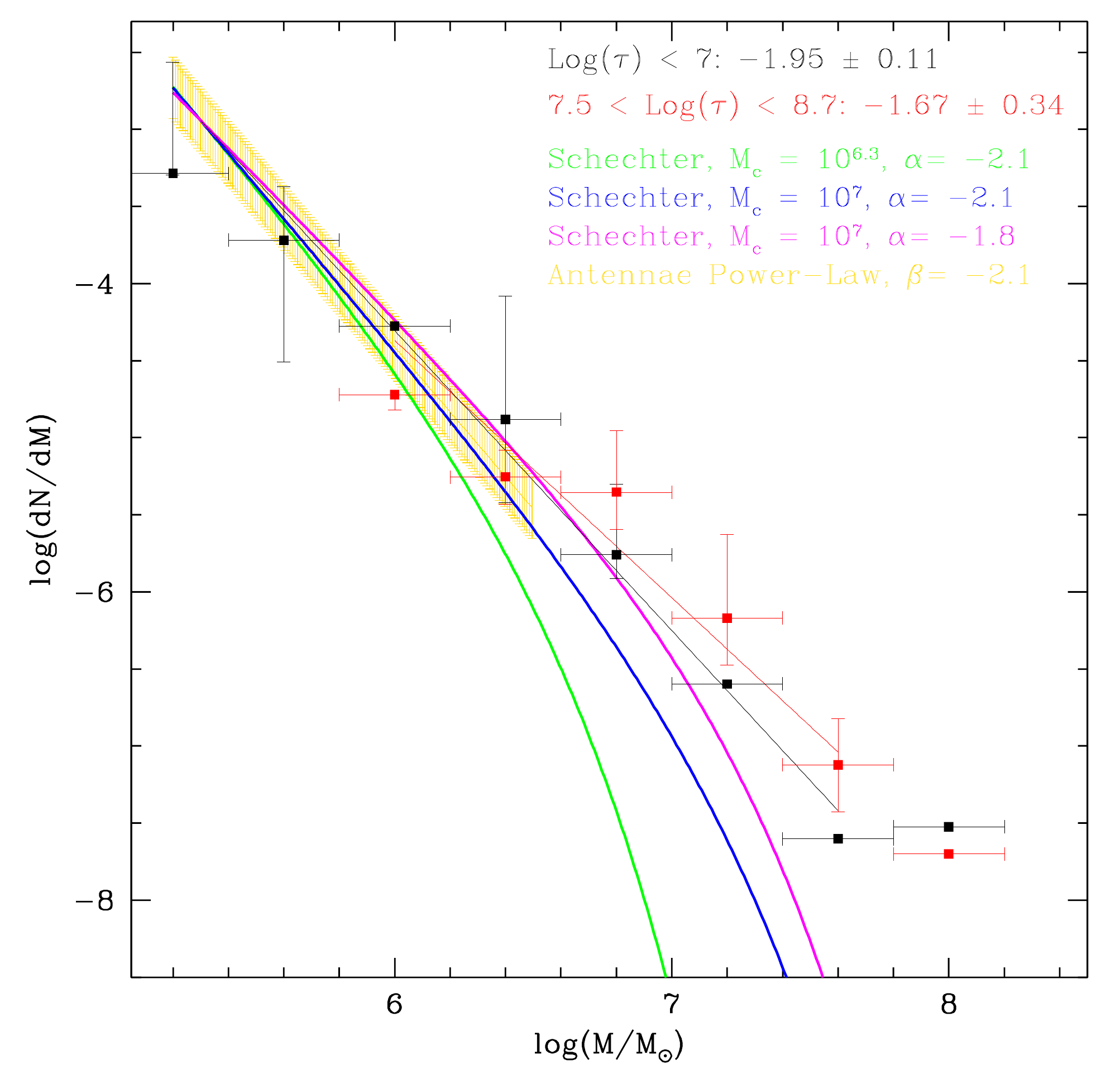}
\caption{The stacked mass distribution functions for all 27 galaxies. We have separated our mass distribution up into two mass-age ranges described in Equations 3 and 4, and shown as Regions 3 and 4 in the right panel of Figure 3. These cuts allow us to test the effects of our completeness limits and the mass dependence of cluster disruption in the sample. The red and black lines represent weighted linear least squares fits to the data. The yellow error bars represent the mass function of the Antennae taken Whitmore et al. (2010), and is normalized to the total number of clusters in our sample. The green, magenta, and blue lines represent three different analytic Schechter function fits to the empirical distribution.}
\end{figure*}


Given the above, it appears that the differences in the slope observed in the LIRG sample relative to the Antennae mass function is not caused by mass-dependent cluster disruption from $10^{5} - 10^{6.5} M_{\odot}$. When we consider the effect of a resolution limit on the high-mass end of the distribution, we can reconcile the small descrepancies in the observed slopes. Therefore, cluster formation in these galaxies can be explained with a universal -2 power law fit to the mass distribution up to at least $10^{6.5} M_{\odot}$. However, we emphasize that the prevalence of the most-massive clusters observed in the sample is compelling evidence that these clusters exist more predominately in the more extreme star-forming environments of LIRGs.

This idea is further supported by the fact that a Schechter function, with an $M_{c} \sim 10^{7} M_{\odot}$, can also fit our data over the full range of observed cluster masses relative to a simple power-law formalism. This implies that GMCs in LIRGs can have higher ISM pressures and densities than what has been seen in other galactic environments. Recently, Maji et al. (2016) used hydrodynamic simulations of two equal-mass MW-like merging galaxies to show that such ISM conditions are actually capable of producing clusters in the range of $10^{5.5-7.5} M_{\odot}$ (Figure 4), consistent with the mass-scales we observe in our LIRG sample.

\begin{deluxetable*}{lcccccccc}
\tabletypesize{\footnotesize}
\tablewidth{0pt}
\tablecaption {Derived Age and Mass Function Slopes}
\tablehead{
\colhead{Name} & \colhead{$\gamma_{0.02}$} & \colhead{$\sigma_{\gamma}$} & \colhead{$\gamma_{0.008}$} & \colhead{$\sigma_{\gamma}$} & \colhead{$\beta_{0.02}$} & \colhead{$\sigma_{\beta}$} & \colhead{$\beta_{0.008}$} & \colhead{$\sigma_{\beta}$}} \\
\startdata  
NGC 1614 & -0.96 & 0.18 & -1.16 & 0.17 & -1.35 & 0.23 & -1.60 & 0.10 \\
NGC 7674  & -1.67 & 0.46 & -0.78 & 0.28 & -1.15 & 0.12 & -1.32 & 0.29  \\
NGC 3690E & -0.62 & 0.54 & -1.01 & 0.44 & -1.44 & 0.14 & -1.31 & 0.23 \\
NGC 3690W & -1.26 & 0.12 & -1.24 & 0.14 & -1.92 & 0.24 & -1.45 & 0.26 \\
Arp 148 & -0.87 & 0.38 & -1.38 & 0.69 & -1.44 & 0.17 & -1.8 & 0.18 \\
IRAS 20351+2521 & -1.19 & 0.11 & -1.27 & 0.10 & -1.60 & 0.52 & -1.12 & 0.25 \\
NGC 6786 & -1.29 & 0.18 & -1.17 & 0.26 & -1.40 & 0.12 & -1.58 & 0.21 \\
UGC 09618NED02 & -1.18 & 0.23 & -1.42 & 0.12 & -2.13 & 0.47 & -1.52 & 0.31
\enddata
\end{deluxetable*}





\subsection{Merger Stage Dependence}

Since our LIRG sample spans the full range of merger stages, we can test if our explanation of cluster formation and destruction depends on the dynamical state of the galaxy. Haan et al (2013), Kim et al. (2013), and Stierwalt et al. (2013) have classified the merger stage of each U/LIRG in the GOALS sample based on their morphological appearance at multiple wavelengths. These merger classification schemes run from pre-first passage to single coalesced nuclei. We separated the sample into early (classes 0-2), middle (classes 3-4), and late-stage (classes 5-6) mergers. In order to quantify any differences in each age distribution we ran a KS-test comparing the normalized distributions of the early, middle, and late-stage mergers to the total sample. We find that within our sub-sample of GOALS LIRGs these individual merger stage distributions are drawn from the same parent distribution of ages with a 92\% probability or higher.

For galaxies classified across all merger stage bins we find that the most massive clusters in the sample (Region 1) are always consistent with a -1 power law in $dN/d\tau$, which is further justification for combining the cluster populations for each galaxy into a single sample, and indicates that disruption does not vary much, within uncertainty, throughout the merger. It also provides credence to the idea that the SFR of a merging galaxy is bursty, which given the large size of our age bins, is an effect on the age distribution we can safely ignore. This allows us to characterize each galaxy as having an elevated but roughly continuous SFR.

\begin{figure}
\centering
  \subfigure{\includegraphics[scale=0.4]{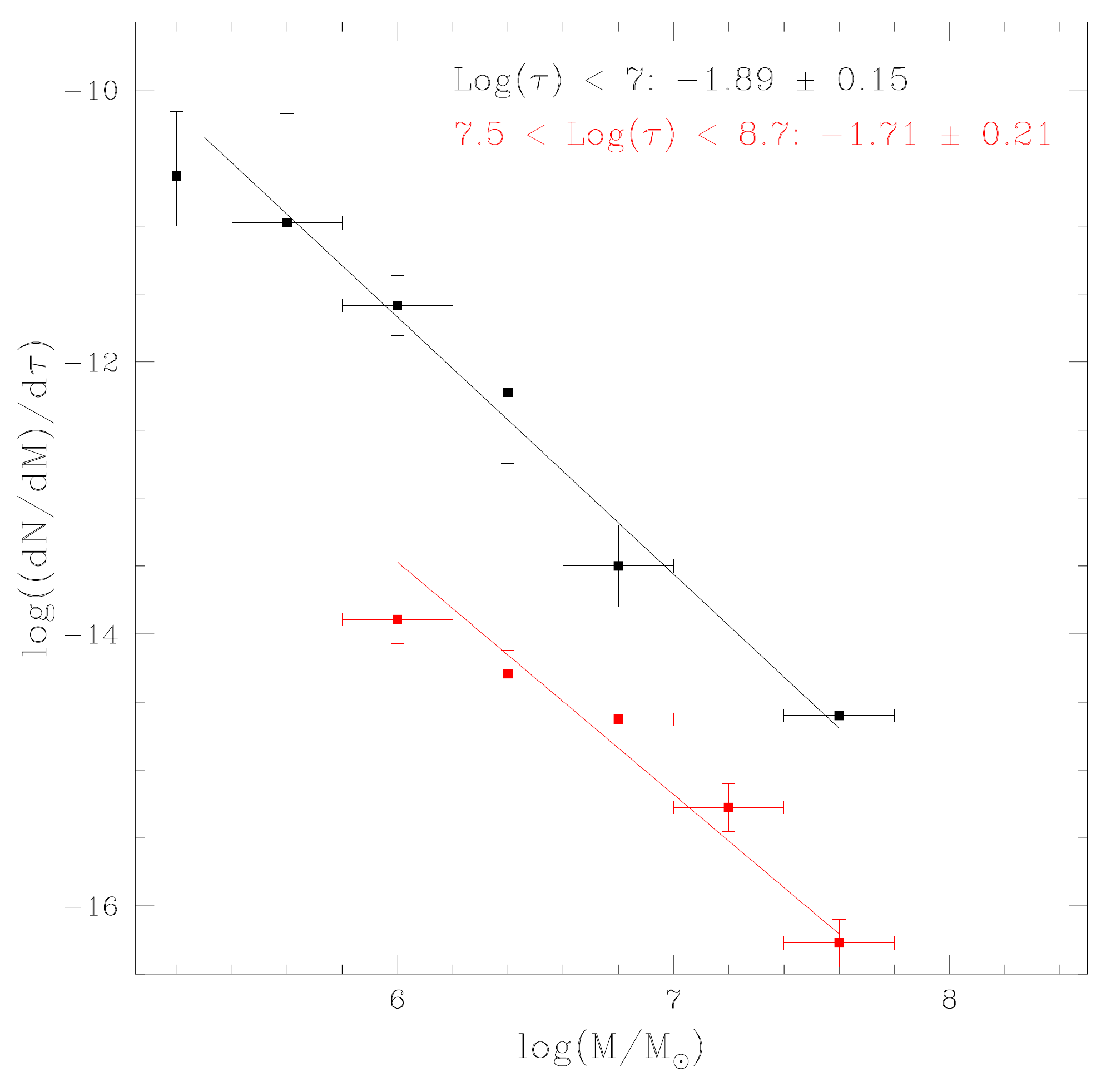}}
  \subfigure{\includegraphics[scale=0.4]{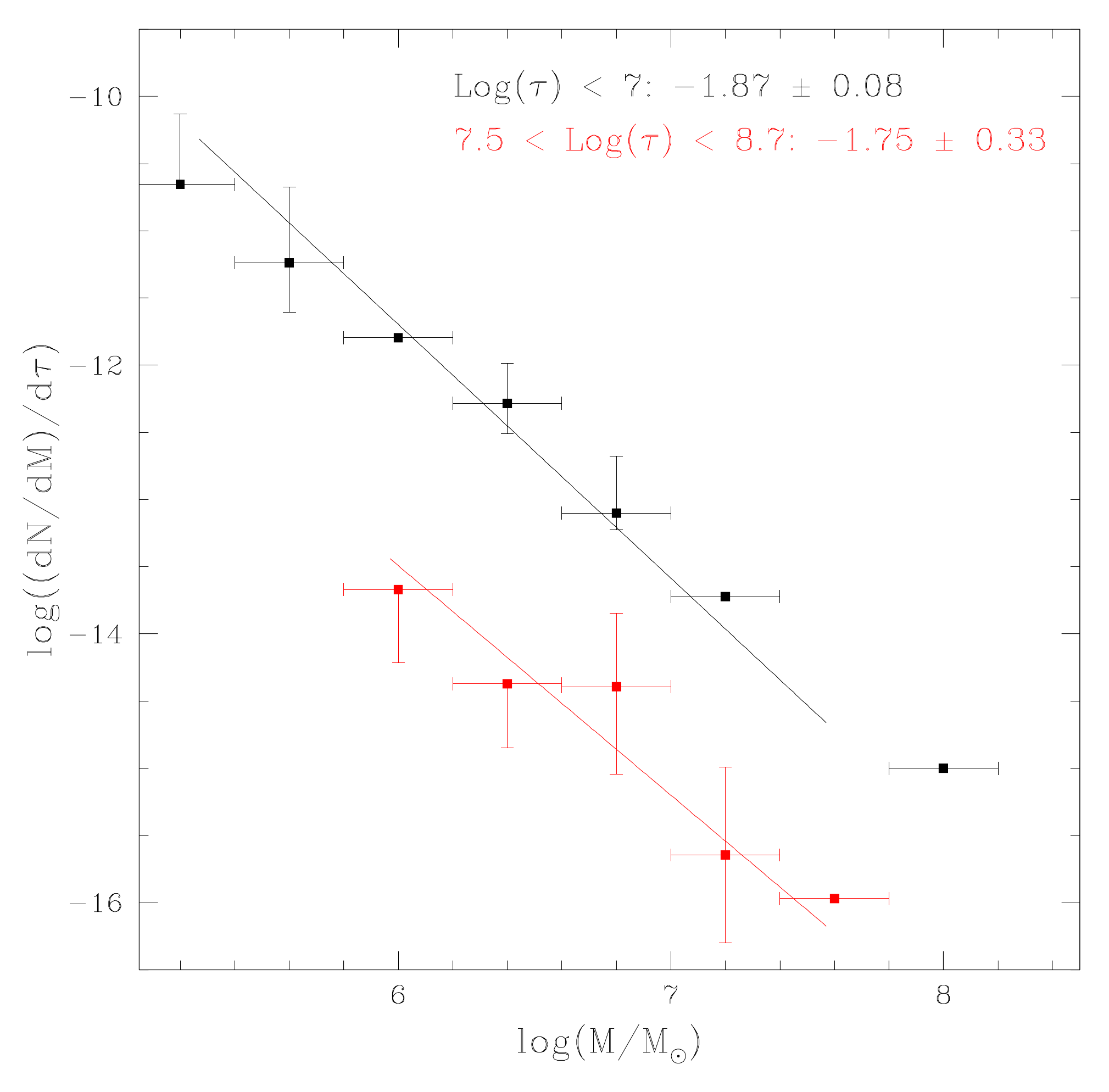}}
\caption{The stacked age-normalized mass distribution functions for the 11 galaxies with a merger class of 0-2 (Top Plot) and the 8 galaxies with a merger class 3-4 (Bottom Plot) identifying them as early-stage and mid-stage mergers respectively. The total number of clusters in each class is 260 and 154 for pre- and mid-stage mergers respectively. We have broken our mass distribution up into the same age ranges as in Figure 5}
\end{figure}

When breaking the sample down to early and mid-stage mergers in Figure 6 we find that star clusters in both early and mid-stage mergers show a power-law distribution of $dN/dM \sim M^{-1.8}$ across both age-regimes. Additionally, each mass function is normalized by the total duration within their respective age bin in order to remove any artifacts of the bins having different time ranges. This helps to emphasize that the number of clusters that survive decreases in absolute number and independent of mass from the pre- to ongoing-merger systems.

Under the assumption of a constant SFR, the youngest clusters in each galaxy merger class will show the same slope in $dN/dM$. Our results are consistent with the idea that the star formation history is not changing significantly between merger stages, and thus cannot be a dominant effect in driving the observed age distributions we see for our sample, when combing all galaxies together.

Additionally, when analyzing the cluster mass distribution we assumed that the formation conditions (i.e characteristic mass $M_{c}$ and slope $\alpha$) do not change significantly throughout the merging process. The similarity of the slopes between each merger class is consistent with simulations of merging disk galaxies, which find that the characteristic mass $M_{c}$ evolves at a rate of only $\sim 0.3-0.4$ dex/Gyr (Kruijssen et al. 2012).


\subsection{Cluster Formation Efficiency}

Finally, we consider the efficiency of cluster formation (CFE) within the high star formation rate environments of LIRGs. CFE, or $\Gamma$, is defined the ratio of the rate of stellar mass formation in bound clusters, $\dot{M}_{\rm SC}$, to the global star formation rate, $\dot{M}_{\rm SF}$, over the same time interval, i.e.,  

\begin{equation} 
\Gamma = {\dot{M}_{\rm SC} \over \dot{M}_{\rm SF}} \times 100\%.
\end{equation}

For our sample, the fact that we do not detect clusters well below $10^{5} M_{\odot}$, and that we have significant cluster disruption over all masses, makes the estimation of the cluster formation efficiency (CFE) highly uncertain.

This is compounded by the fact that our UV-Bright cluster population is not sampling the full SFR as traced by the total UV+IR based SFR measurements from Howell et al. (2010). Additionally, we cannot match our UV-based CFR to the total GALEX UV SFR estimation because the field of view of the SBC is $\sim 1/140$ that of GALEX, and thus a correction for the clusters we miss is uncertain. The large amount of obscuring dust also makes a completeness correction to derive a total mass and CFR based on our mass distributions difficult for our LIRG sample. Johnson et al. (2016) notes that CFE calculations are best done in dust-free environments that show little sign of significicant cluster disruption, a scenario we are simply not presented with in our sample. Therefore we leave a discussion about CFE in LIRGs to future studies involving deep IR-based observations that have both a larger FOV, and the ability to detect more dust-enshouded low-mass clusters.

\section{Summary}

{\it Hubble Space Telescope} ACS/HRC FUV (F140LP) and ACS/WFC optical (F435W and F814W) observations of a sample of 22 star cluster-rich LIRGs in the GOALS sample were obtained. 
These observations have been utilized to derive the ages and masses of the star clusters contained within these systems in order to examine the 
cluster properties in extreme starburst environments relative to those in nearby, lower luminosity star-forming galaxies. The following conclusions are reached:

\noindent
(1) We have detected 665 clusters within the inner $30\arcsec \times30\arcsec$ of these 22 LIRGs (27 nuclei). These clusters have $S/N \geq 5$ in all three filters and de-convolved FWHMs as measured by ISHAPE of $\leq 2$ pixels.

\noindent
(2) Cluster ages have been derived by assuming an instantaneous SSP, Salpeter IMF, and either a solar or sub-solar metallicity. By requiring the derived cluster ages to be consistent when using both a color-color and SED-based fitting technique, we obtain a final sample of 484 clusters whose properties are reliably constrained within the 1$\sigma$ uncertainties of the SSP models. The derived cluster ages imply a disruption rate of $dN/d\tau = \tau^{-0.9 +/- 0.3}$ for cluster masses $\geq 10^{6} M_{\odot}$, and $dN/d\tau = \tau^{-0.87 +/- 0.08}$ for cluster masses $10^{5.3} < M < 10^{6} M_{\odot}$. This is consistent with what is seen in the Antennae,  and indicates the general influence mergers have on the creation and destruction of star clusters. The measured $\gamma$ is steeper than that measured for lower mass, less star-forming systems in the local Universe, implying that the merging process produces a fundamentally different cluster disruption law.

\noindent
(3) We have identified a large number of $M \geq 10^{6} M_{\odot}$ clusters in the sample, which indicates that the more extreme star-forming environments of LIRGs are capable of producing more high-mass clusters than what is observed in galaxies like the Milky Way or even the Antennae (Larsen 2009; Bastian et al. 2012; Whitmore et al. 2010). The derived cluster masses also imply a CMF for the sample of $dN/dM = M^{-1.95 +/- 0.11}$, which is consistent with a -2 power law in $dN/dM$. Together with the fact that we do not see a significant change in the age distribution slope as a function of mass, we interpret our mass function slope as evidence against mass-dependent cluster disruption at $M \geq 10^{5.3} M_{\odot}$ which would flatten the observed CMF relative to a canonical -2 power law in this regime.

\acknowledgements
The authors thank B. Whitmore, R. Chandar, A. Mulia, and G. Soutchkova for useful discussions and assistance. The authors also thank the referee for detailed comments and suggestions which have improved the manuscript.
S.T.L. was supported by the NASA VSGC Graduate Fellowship. A.S.E., and D.C.K. were supported by NSF grant AST 1109475 and by NASA through grants HST-GO10592.01-A, HST-GO11196.01-A, and HST-GO13364 from the Space Telescope Science Institute, which is operated by the Association of Universities for Research in Astronomy, Inc., under NASA contract NAS5- 26555. G.C.P. was supported by a FONDECYT Postdoctoral Fellowship (No.\ 3150361). A.S.E. was also supported by the Taiwan, R.O.C. Ministry of Science and Technology grant MoST 102-2119-M-001-MY3. T.D-S. acknowledges support from ALMA-CONICYT project 31130005 and FONDECYT regular project 1151239.

Portions of this work were performed at the Aspen Center for Physics, which is supported by National Science Foundation grant PHY-1066293. This work was partially supported by a grant from the Simons Foundation. Finally, This research has made use of the NASA/IPAC Extragalactic Database (NED) which is operated by the Jet Propulsion Laboratory, California Institute of Technology, under contract with the National Aeronautics and Space Administration. 


\appendix
\section{Galaxy Descriptions}

In the following sections we give a brief description of the basic morphology and star cluster spatial distributions within each galaxy, as well as the adopted values for the maximum amount of visual extinction we use in our model. See Evans et al. 2017 for a detailed description of all 88 LIRGs in the GOALS sample that have been observed with HST.


\subsection{NGC 0017}

NGC 0017 is a late stage merger that contains a single resolved nucleus surrounded by dust lanes associated with spiral arms in the inner few kpc. Several bright star clusters are visible within this nuclear spiral region. The maximum $A_{V}$ adopted for this galaxy is 3.0 mags of visual extinction. (Dametto et al. 2014). 

\begin{figure*}[h]
\centering
\includegraphics[scale=0.25]{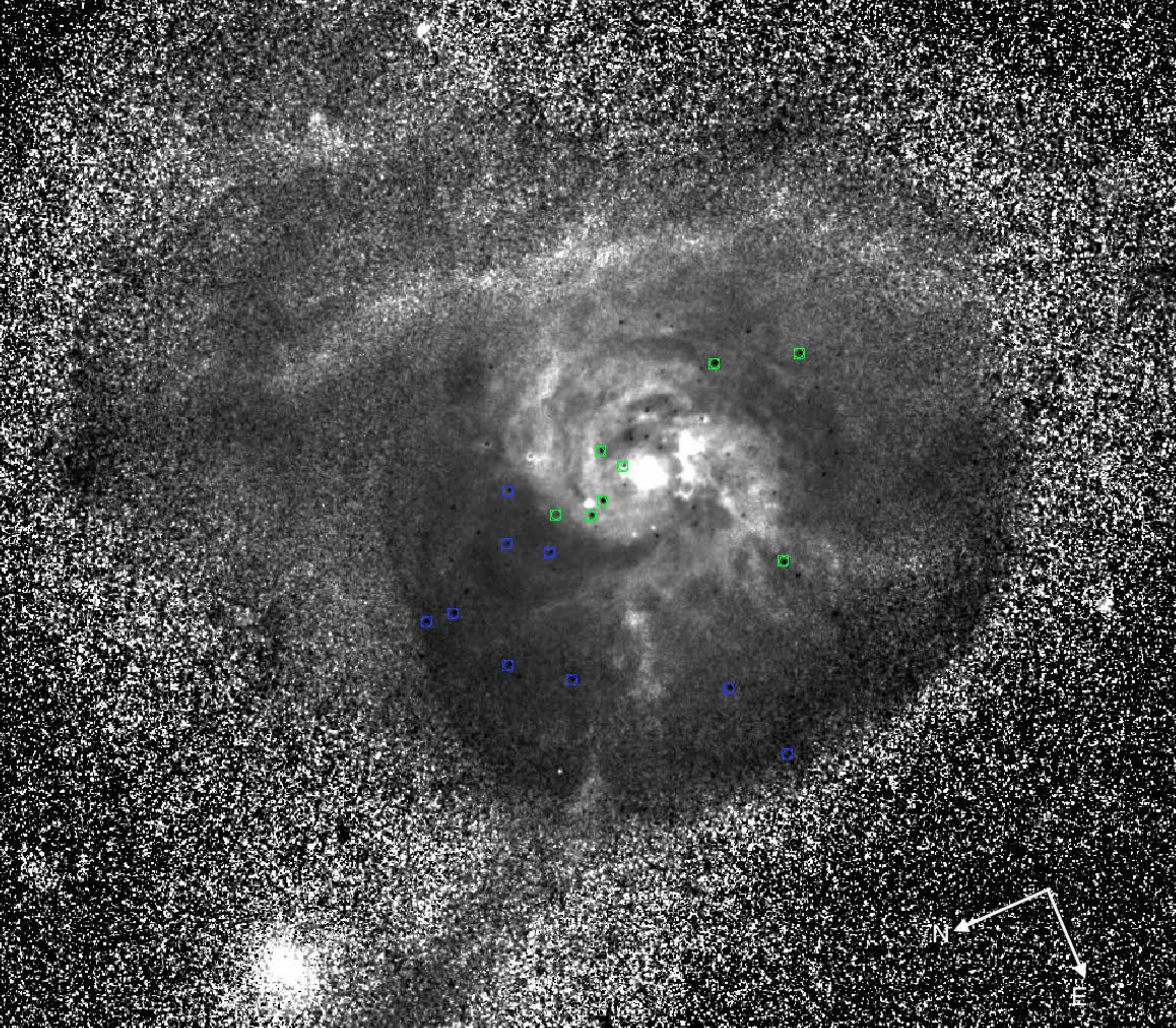}
\caption{Inverted black and white B-I image of NGC 0017 taken with HST ACS/WFC F814W and F435W. The bright emission corresponds to redder (i.e. dustier) regions of the galaxy. The blue centroids correspond to clusters found in relatively ``dust-free'' regions of these galaxies, whereas the green centroids correspond to clusters found in relatively dustier regions of the galaxy.}
\end{figure*}

\begin{figure*}
\centering
\includegraphics[scale=0.55]{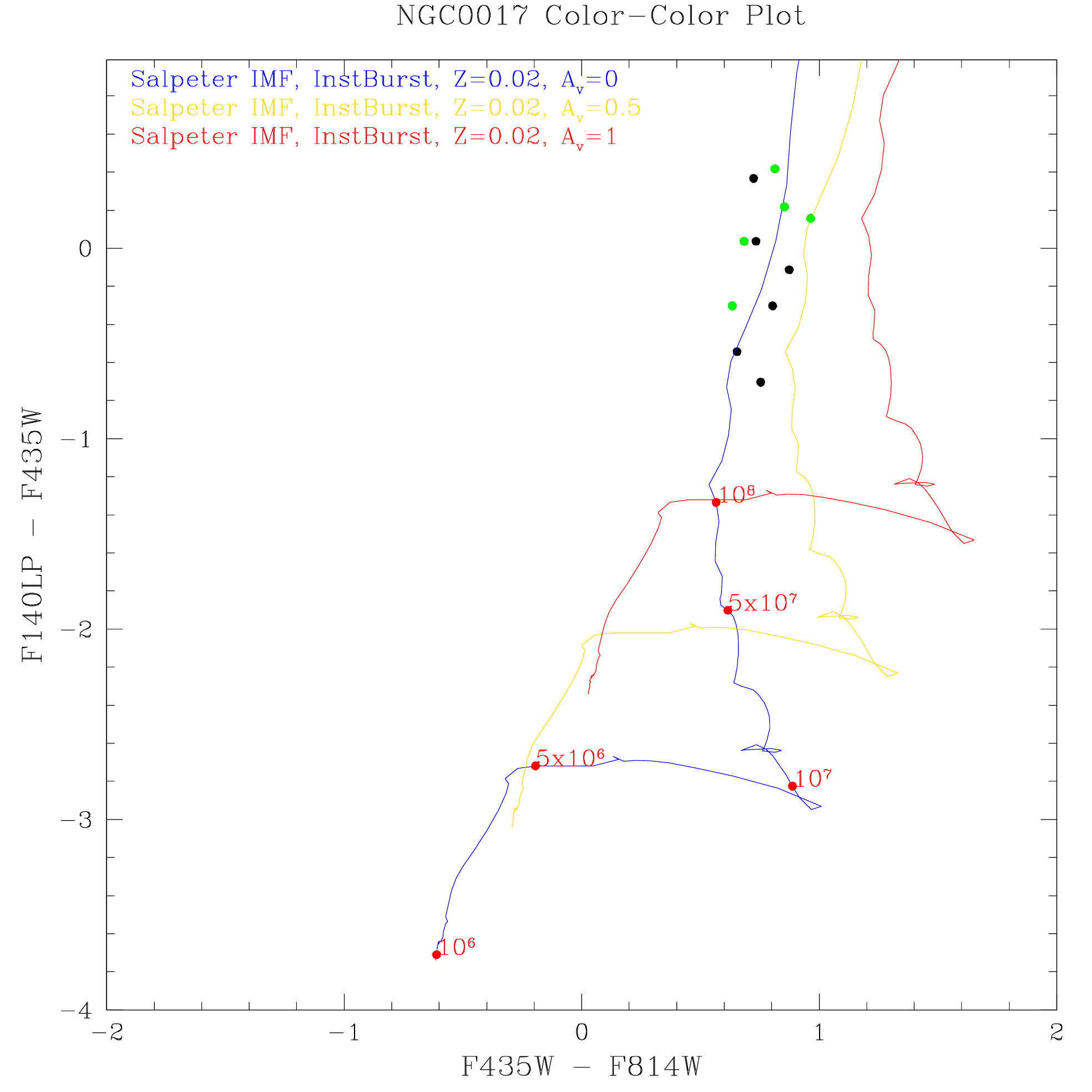}
\caption{Color-Color plot of all star clusters identified in NGC 0017 in F814W, F435W, and F140LP plotted against SSP models with various amount of visual extinction. The green points correspond to the clusters found in dustier regions of the galaxy in Figure 7}
\end{figure*}

\begin{deluxetable*}{ccccccccc}
\tabletypesize{\footnotesize}
\tablecolumns{10}
\tablewidth{0pt}
\tablecaption {Observed Properties of Star Clusters in NGC 0017}
\tablehead{
\colhead{ID} & \colhead{RA} & \colhead{Dec} & \colhead{$M_{B}$} & \colhead{$\sigma_{B}$} & \colhead{$M_{I}$} & \colhead{$\sigma_{I}$} & \colhead{$M_{FUV}$} & \colhead{$\sigma_{FUV}$}} \\
\startdata
1 & 2.776819641 & -12.10616498 & -13.81 & 0.01 & -14.87 & 0.01 & -12.67 & 0.03 \\
2 & 2.777320196 & -12.10713204 & -13.89 & 0.01 & -14.53 & 0.01 & -14.20 & 0.01 \\
3 & 2.777273773 & -12.10658526 & -11.05 & 0.06 & -11.91 & 0.05 & -10.84 & 0.18 \\
4 & 2.777432692 & -12.10695518 & -13.56 & 0.01 & -14.43 & 0.01 & -12.32 & 0.05 \\
5 & 2.777660645 & -12.10636335 & -12.56 & 0.01 & -13.44 & 0.01 & -12.68 & 0.03 \\
6 & 2.778745035 & -12.10880075 & -14.29 & 0.01 & -15.27 & 0.01 & -12.96 & 0.02 \\
7 & 2.777920896 & -12.10509066 & -12.03 & 0.01 & -12.76 & 0.01 & -11.67 & 0.08 \\
8 & 2.778717571 & -12.10546794 & -10.72 & 0.02 & -11.38 & 0.02 & -11.27 & 0.12 \\
9 & 2.779140091 & -12.10608216 & -15.05 & 0.01 & -15.86 & 0.01 & -15.36 & 0.01 \\
10 & 2.779887259 & -12.10772035 & -11.65 & 0.01 & -12.39 & 0.01 & -11.62 & 0.16 \\
11 & 2.780845501 & -12.10807908 & -10.35 & 0.03 & -11.11 & 0.03 & -11.06 & 0.02 \\
12 & 2.777023706 & -12.10749699 & -13.10 & 0.04 & -14.80 & 0.16 & -11.10 & 0.14 \\
13 & 2.776766427 & -12.10731248 & -13.23 & 0.02 & -14.20 & 0.02 & -13.08 & 0.02 \\
14 & 2.776532232 & -12.10983287 & -11.54 & 0.01 & -12.36 & 0.01 & -11.13 & 0.08 \\
15 & 2.776287946 & -12.10888848 & -13.35 & 0.01 & -14.04 & 0.01 & -13.32 & 0.02
\enddata
\end{deluxetable*}

\begin{deluxetable*}{ccccccc}
\tabletypesize{\footnotesize}
\tablewidth{0pt}
\tablecaption {Derived Properties of Star Clusters in NGC 0017}
\tablehead{
\colhead{ID} & \colhead{Log(Age)} & \colhead{$\sigma_{Age}$} & \colhead{Log($M/M_{\odot}$)} & \colhead{$\sigma_{M}$} & \colhead{$A_{V}$} & \colhead{$\sigma_{A_{V}}$}} \\
\startdata  
1 & 8.61 & 0.02 & 7.34 & 0.16 & 0.20 & 0.05 \\
2 & 8.36 & 0.02 & 7.11 & 0.17 & 0.10 & 0.06 \\
3 & 8.46 & 0.58 & 6.04 & 0.61 & 0.10 & 0.63 \\
4 & 8.66 & 0.02 & 7.18 & 0.16 & 0.01 & 0.04 \\
5 & 6.66 & 0.81 & 6.18 & 0.70 & 1.90 & 0.77 \\
6 & 8.66 & 0.69 & 7.47 & 0.16 & 0.01 & 5.27 \\
7 & 6.66 & 0.03 & 6.08 & 0.16 & 2.10 & 0.04 \\
8 & 6.66 & 0.79 & 5.29 & 0.69 & 1.60 & 0.74 \\
9 & 6.64 & 0.10 & 7.10 & 0.16 & 1.80 & 5.62 \\
10 & 6.66 & 2.37 & 5.87 & 0.16 & 2.00 & 4.73 \\
11 & 6.34 & 0.86 & 5.65 & 0.67 & 2.10 & 0.71 \\
12 & 8.46 & 0.18 & 7.55 & 0.52 & 1.40 & 0.51 \\
13 & 8.36 & 0.29 & 7.00 & 0.37 & 0.40 & 0.33 \\
14 & 8.51 & 1.93 & 6.23 & 0.16 & 0.01 & 0.03 \\
15 & 8.46 & 5.32 & 6.91 & 0.16 & 0.01 & 0.78
\enddata
\end{deluxetable*}

\subsection{Arp 256S}

Arp 256 is a mid-stage merger containing a southern (MCG-02-01-051) and northern (MCG-02-01-052) galaxy. Arp 256S has an elongated $\sim 1''$ (400 pc) nucleus, and the north and southwest tails contain the majority of the star clusters in the galaxy. The maximum $A_{V}$ adopted for this galaxy is 1.7 mags of visual extinction (Smith et al. 2014). 

\begin{figure*}
\centering
\includegraphics[scale=0.25]{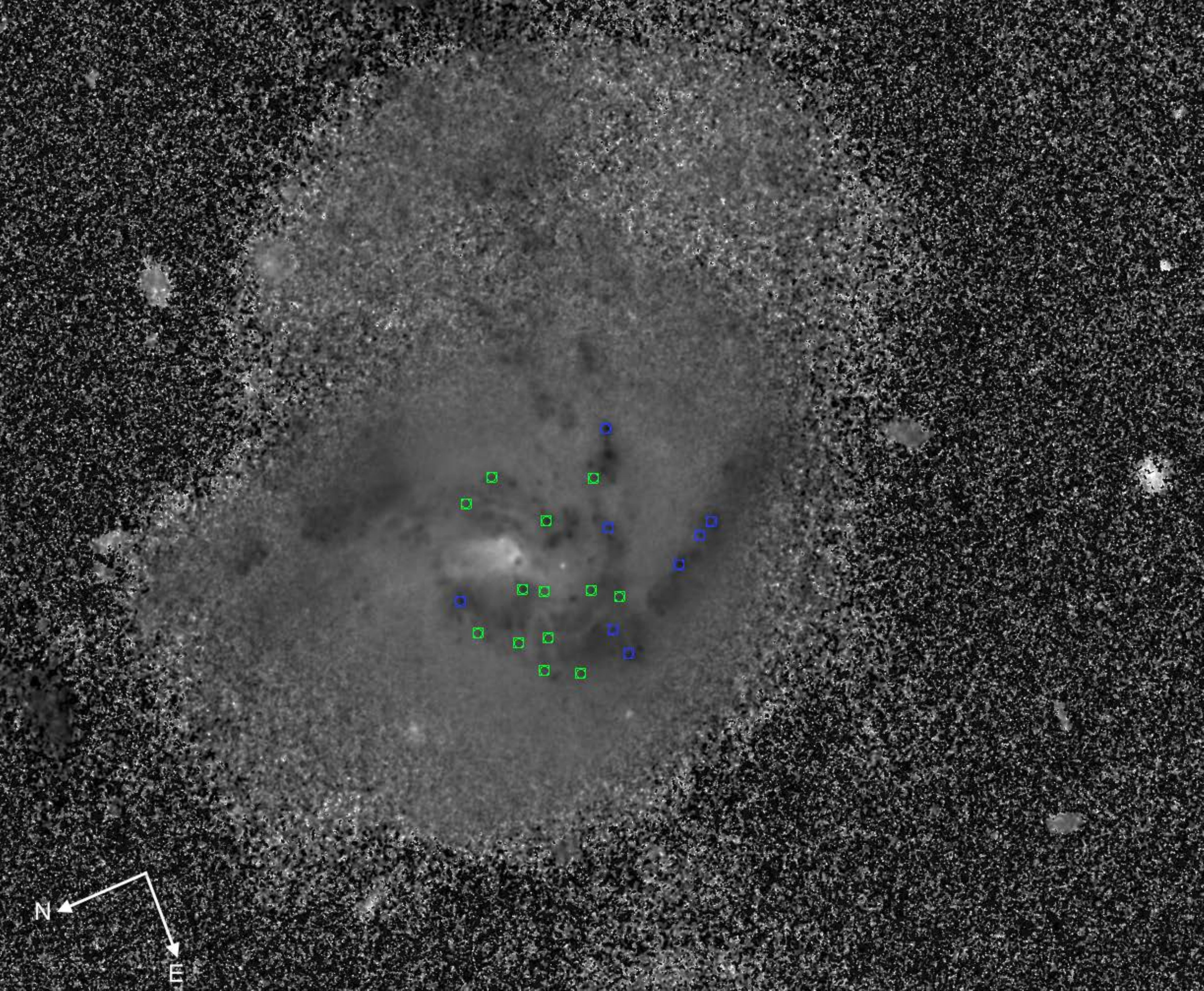}
\caption{Inverted black and white B-I image of Arp 256S taken with HST ACS/WFC F814W and F435W. The bright emission corresponds to redder (i.e. dustier) regions of the galaxy. The blue centroids correspond to clusters found in relatively ``dust-free'' regions of these galaxies, whereas the green centroids correspond to clusters found in relatively dustier regions of the galaxy.}
\end{figure*}

\begin{figure*}
\centering
\includegraphics[scale=0.55]{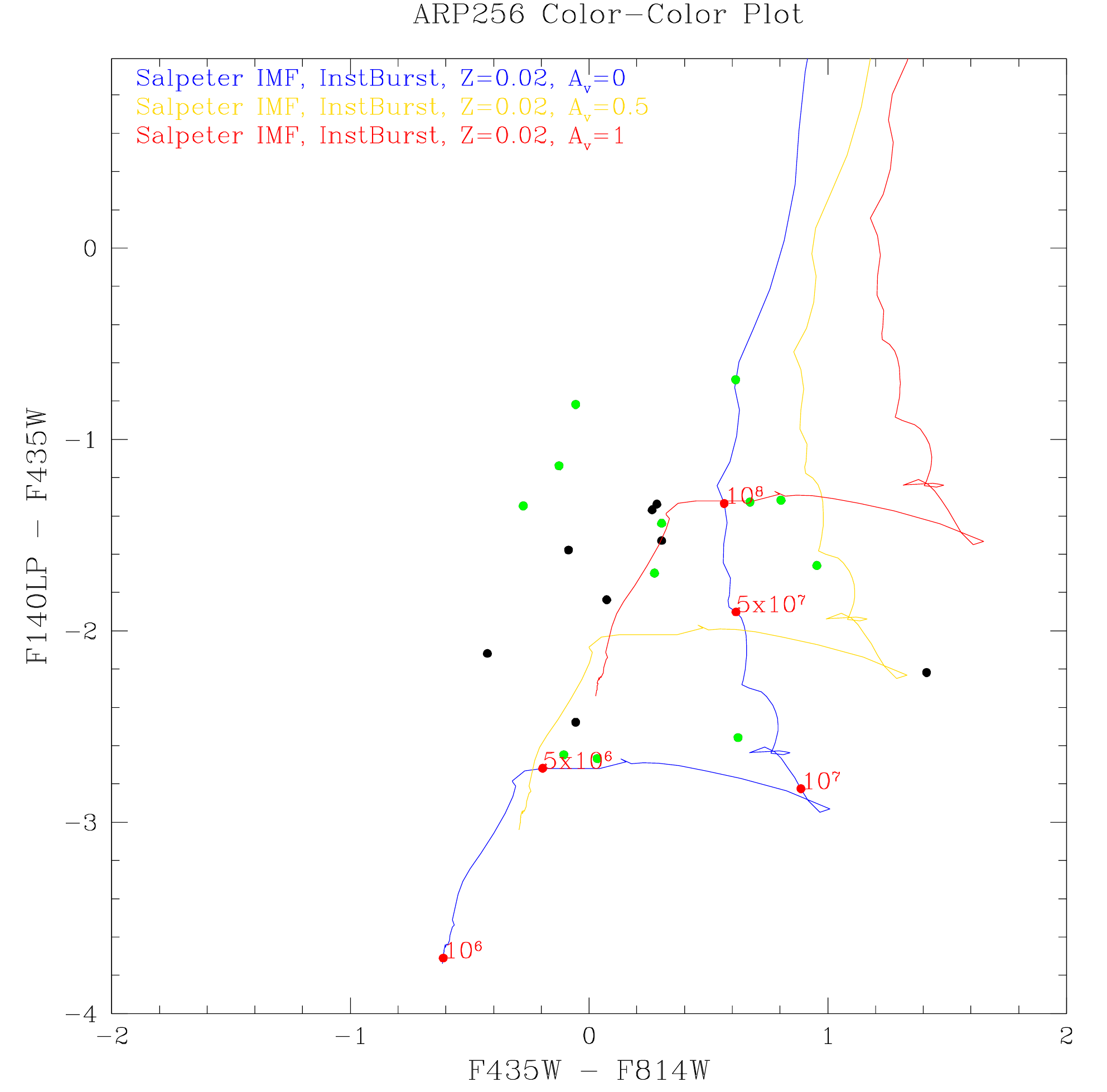}
\caption{Color-Color plot of all star clusters identified in Arp 256S in F814W, F435W, and F140LP plotted against SSP models with various amount of visual extinction. The green points correspond to the clusters found in dustier regions of the galaxy in Figure 9}
\end{figure*}

\begin{deluxetable*}{ccccccccc}
\tabletypesize{\footnotesize}
\tablewidth{0pt}
\tablecaption {Observed Properties of Star Clusters in Arp 256S}
\tablehead{
\colhead{ID} & \colhead{RA} & \colhead{Dec} & \colhead{$M_{B}$} & \colhead{$\sigma_{B}$} & \colhead{$M_{I}$} & \colhead{$\sigma_{I}$} & \colhead{$M_{FUV}$} & \colhead{$\sigma_{FUV}$}} \\
\startdata
1 & 4.713679548 & -10.37685826 & -10.41 & 0.05 & -11.03 & 0.07 & -12.97 & 0.05 \\
2 & 4.710809481 & -10.37888736 & -12.54 & 0.01 & -12.61 & 0.02 & -14.38 & 0.01 \\
3 & 4.710599656 & -10.377146 & -10.45 & 0.13 & -10.34 & 0.22 & -13.10 & 0.04 \\
4 & 4.711341331 & -10.37839176 & -11.99 & 0.02 & -12.79 & 0.02 & -13.31 & 0.03 \\
5 & 4.71074686 & -10.37664793 & -12.51 & 0.03 & -13.46 & 0.04 & -14.17 & 0.03 \\
6 & 4.711535325 & -10.37750998 & -13.53 & 0.02 & -13.80 & 0.02 & -15.23 & 0.02 \\
7 & 4.712724227 & -10.37952183 & -10.43 & 0.15 & -11.84 & 0.08 & -12.65 & 0.13 \\
8 & 4.712065091 & -10.37821862 & -12.93 & 0.03 & -13.21 & 0.04 & -14.27 & 0.03 \\
9 & 4.712819126 & -10.37928414 & -11.79 & 0.05 & -12.05 & 0.09 & -13.16 & 0.08 \\
10 & 4.713033719 & -10.37882884 & -12.69 & 0.02 & -12.63 & 0.05 & -15.17 & 0.02 \\
11 & 4.712218761 & -10.37674074 & -14.31 & 0.02 & -14.61 & 0.02 & -15.75 & 0.01 \\
12 & 4.712398738 & -10.3769871 & -12.19 & 0.03 & -12.13 & 0.21 & -13.01 & 0.05 \\
13 & 4.712719547 & -10.37756498 & -11.80 & 0.06 & -11.67 & 0.12 & -12.94 & 0.10 \\
14 & 4.71300344 & -10.37787256 & -12.42 & 0.03 & -13.03 & 0.03 & -13.11 & 0.04 \\
15 & 4.71191981 & -10.37588942 & -12.99 & 0.03 & -12.56 & 0.06 & -15.11 & 0.01 \\
16 & 4.713366435 & -10.37755202 & -13.26 & 0.05 & -13.56 & 0.07 & -14.79 & 0.03 \\
17 & 4.713006219 & -10.37670878 & -11.65 & 0.10 & -11.37 & 0.15 & -13.00 & 0.05 \\
18 & 4.712859587 & -10.37631389 & -11.93 & 0.03 & -12.60 & 0.05 & -13.26 & 0.04 \\
19 & 4.713776161 & -10.37757731 & -12.27 & 0.04 & -12.18 & 0.04 & -13.85 & 0.06 \\
20 & 4.713381893 & -10.37643331 & -10.68 & 0.07 & -10.71 & 0.14 & -13.35 & 0.03
\enddata
\end{deluxetable*}

\begin{deluxetable*}{ccccccc}
\tabletypesize{\footnotesize}
\tablewidth{0pt}
\tablecaption {Derived Properties of Star Clusters in Arp 256S}
\tablehead{
\colhead{ID} & \colhead{Log(Age)} & \colhead{$\sigma_{Age}$} & \colhead{Log($M/M_{\odot}$)} & \colhead{$\sigma_{M}$} & \colhead{$A_{V}$} & \colhead{$\sigma_{A_{V}}$}} \\
\startdata  
1 & 6.88 & 0.17 & 4.60 & 0.20 & 0.10 & 0.07 \\
2 & 6.64 & 0.02 & 5.51 & 0.18 & 0.70 & 0.05 \\
3 & 5.10 & 0.43 & 5.10 & 0.33 & 0.80 & 0.23 \\
4 & 7.81 & 0.42 & 6.12 & 0.34 & 0.30 & 0.28 \\
5 & 7.54 & 0.28 & 6.24 & 0.26 & 0.40 & 0.18 \\
6 & 6.46 & 0.27 & 6.49 & 0.33 & 1.30 & 0.27 \\
7 & 5.10 & 0.09 & 5.57 & 0.29 & 1.70 & 0.16 \\
8 & 6.66 & 0.04 & 5.85 & 0.20 & 1.00 & 0.09 \\
9 & 6.66 & 0.41 & 5.40 & 0.29 & 1.00 & 0.22 \\
10 & 6.40 & 0.64 & 5.78 & 0.35 & 0.80 & 0.30 \\
11 & 6.60 & 0.06 & 6.34 & 0.23 & 1.00 & 0.15 \\
12 & 6.66 & 0.25 & 5.66 & 0.28 & 1.20 & 0.21 \\
13 & 6.66 & 0.72 & 5.40 & 0.21 & 1.00 & 0.10 \\
14 & 6.66 & 0.81 & 5.92 & 0.68 & 1.50 & 0.73 \\
15 & 6.66 & 0.76 & 5.56 & 0.18 & 0.40 & 0.04 \\
16 & 6.56 & 0.17 & 5.99 & 0.35 & 1.10 & 0.30 \\
17 & 6.66 & 0.76 & 5.29 & 0.24 & 0.90 & 0.11 \\
18 & 6.74 & 0.53 & 5.48 & 0.43 & 1.00 & 0.39 \\
19 & 6.66 & 0.27 & 5.48 & 0.19 & 0.80 & 0.06 \\
20 & 6.74 & 0.38 & 4.45 & 0.31 & 0.01 & 0.24
\enddata
\end{deluxetable*}

\subsection{Arp 256N}

Arp 256N has a central, point-like nucleus. The majority of the star clusters are seen along the tidal tails in this galaxy. The maximum $A_{V}$ adopted for this galaxy is 1.7 mags of visual extinction (Smith et a. 2014). 

\begin{figure*}
\centering
\includegraphics[scale=0.25]{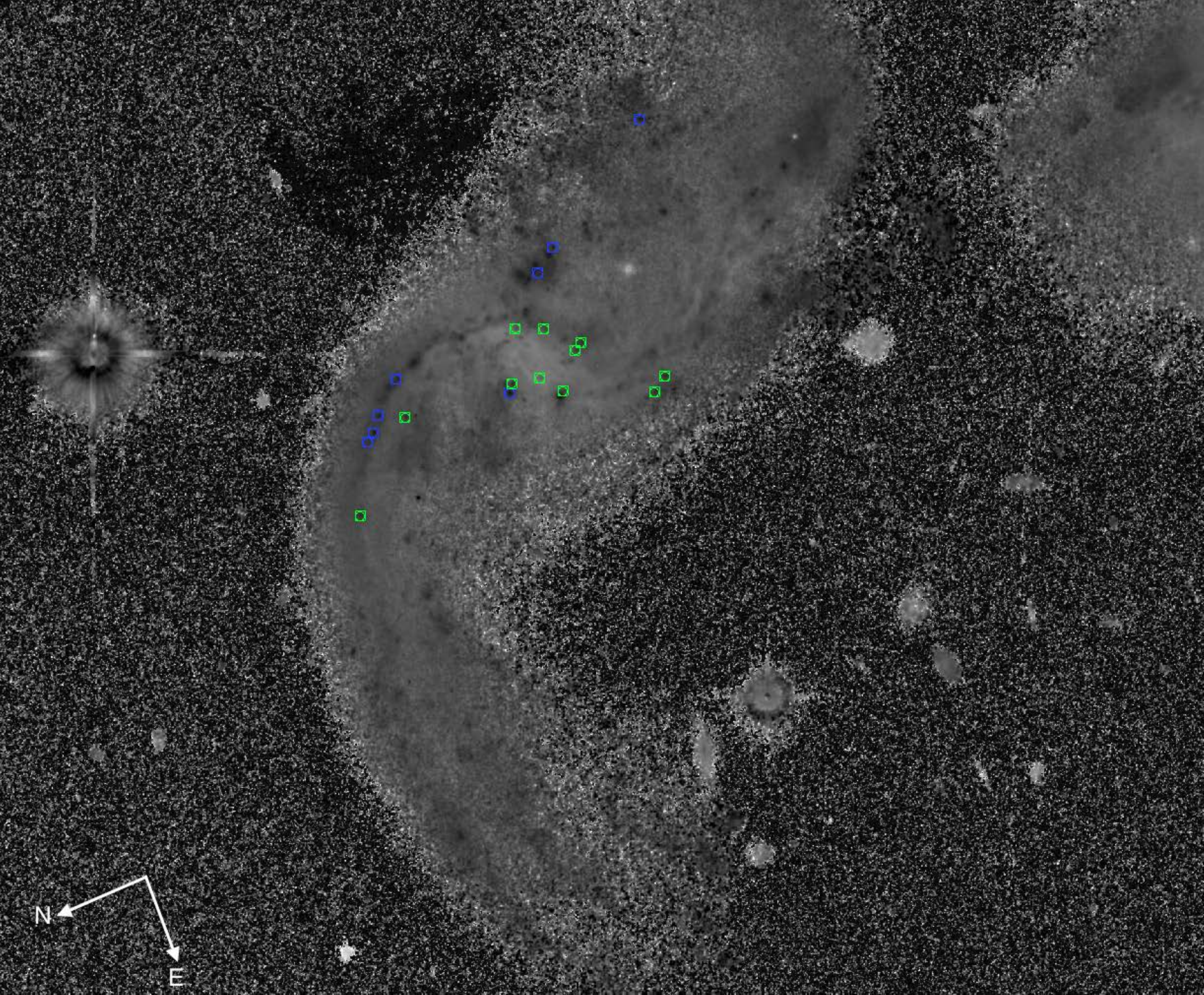}
\caption{Inverted black and white B-I image of Arp 256N taken with HST ACS/WFC F814W and F435W. The bright emission corresponds to redder (i.e. dustier) regions of the galaxy. The blue centroids correspond to clusters found in relatively ``dust-free'' regions of these galaxies, whereas the green centroids correspond to clusters found in relatively dustier regions of the galaxy.}
\end{figure*}

\begin{figure*}
\centering
\includegraphics[scale=0.55]{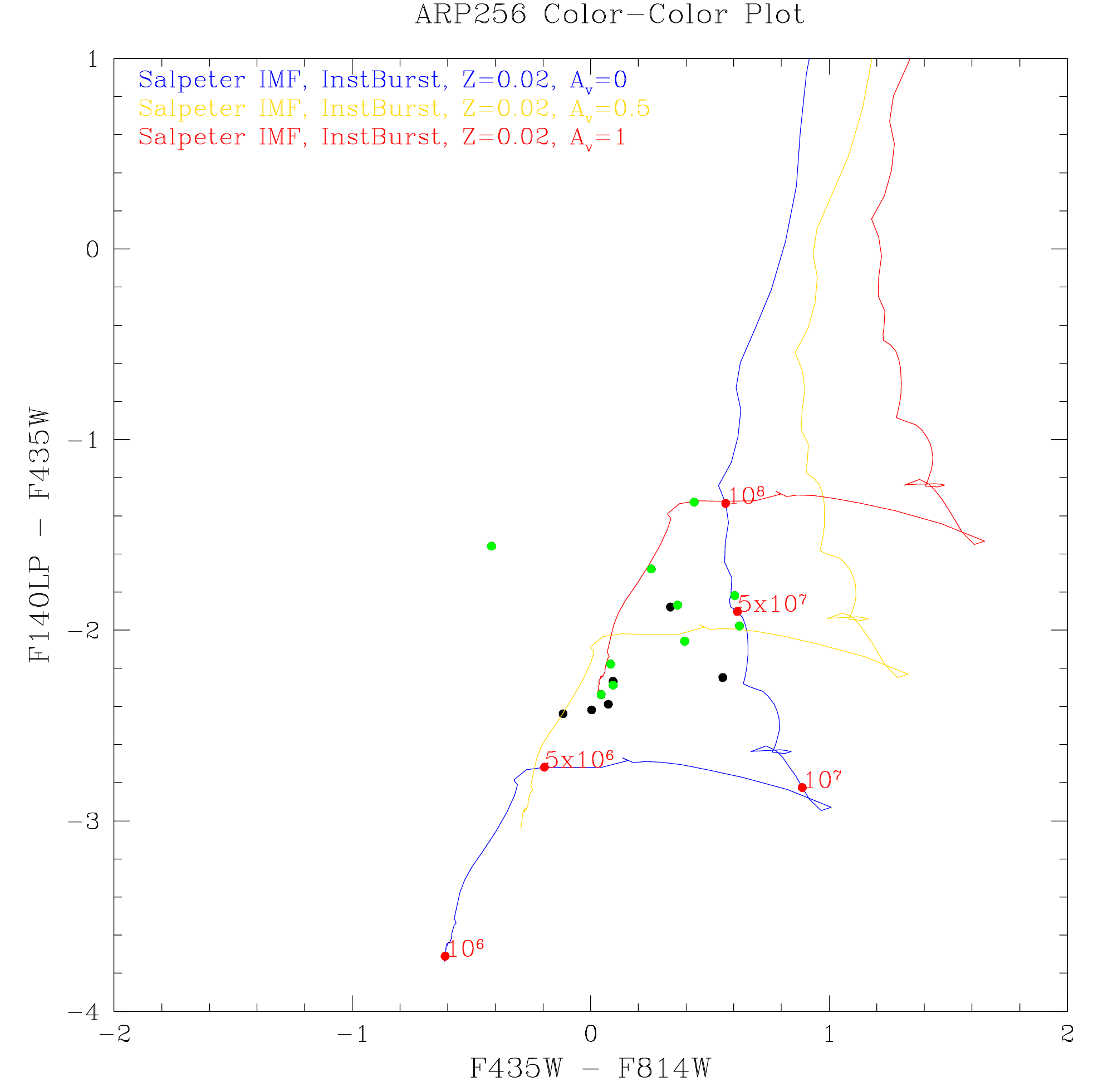}
\caption{Color-Color plot of all star clusters identified in Arp 256N in F814W, F435W, and F140LP plotted against SSP models with various amount of visual extinction. The green points correspond to the clusters found in dustier regions of the galaxy in Figure 11}
\end{figure*}

\begin{deluxetable*}{ccccccccc}
\tabletypesize{\footnotesize}
\tablewidth{0pt}
\tablecaption {Observed Properties of Star Clusters in Arp 256N}
\tablehead{
\colhead{ID} & \colhead{RA} & \colhead{Dec} & \colhead{$M_{B}$} & \colhead{$\sigma_{B}$} & \colhead{$M_{I}$} & \colhead{$\sigma_{I}$} & \colhead{$M_{FUV}$} & \colhead{$\sigma_{FUV}$}} \\
\startdata
1 & 4.705283874 & -10.36593186 & -10.52 & 0.03 & -10.40 & 0.07 & -12.96 & 0.03 \\
2 & 4.706681456 & -10.36309635 & -11.05 & 0.02 & -11.05 & 0.08 & -13.47 & 0.03 \\
3 & 4.706997794 & -10.36259164 & -12.33 & 0.02 & -12.88 & 0.02 & -14.58 & 0.04 \\
4 & 4.708060796 & -10.36212902 & -11.79 & 0.03 & -12.41 & 0.04 & -13.77 & 0.02 \\
5 & 4.707757416 & -10.3616235 & -11.16 & 0.06 & -10.74 & 0.22 & -12.72 & 0.06 \\
6 & 4.708684462 & -10.36263369 & -12.69 & 0.01 & -13.05 & 0.02 & -14.56 & 0.01 \\
7 & 4.708763384 & -10.36246337 & -11.99 & 0.02 & -12.38 & 0.03 & -14.05 & 0.03 \\
8 & 4.710151015 & -10.36378048 & -11.10 & 0.03 & -11.14 & 0.06 & -13.44 & 0.03 \\
9 & 4.708704027 & -10.36100592 & -12.08 & 0.05 & -12.17 & 0.03 & -14.37 & 0.04 \\
10 & 4.709365918 & -10.361838 & -12.00 & 0.04 & -12.08 & 0.13 & -14.18 & 0.03 \\
11 & 4.708851609 & -10.3608711 & -13.78 & 0.01 & -13.85 & 0.02 & -16.17 & 0.01 \\
12 & 4.710324609 & -10.36344198 & -11.80 & 0.02 & -12.23 & 0.02 & -13.13 & 0.06 \\
13 & 4.707903866 & -10.35832277 & -13.74 & 0.01 & -14.07 & 0.02 & -15.62 & 0.01 \\
14 & 4.708214025 & -10.3587839 & -11.98 & 0.02 & -12.23 & 0.04 & -13.66 & 0.02 \\
15 & 4.708155895 & -10.358066 & -13.40 & 0.01 & -13.49 & 0.02 & -15.67 & 0.01 \\
16 & 4.709517938 & -10.35699562 & -11.91 & 0.02 & -12.51 & 0.03 & -13.73 & 0.02
\enddata
\end{deluxetable*}

\begin{deluxetable*}{ccccccc}
\tabletypesize{\footnotesize}
\tablewidth{0pt}
\tablecaption {Derived Properties of Star Clusters in Arp 256N}
\tablehead{
\colhead{ID} & \colhead{Log(Age)} & \colhead{$\sigma_{Age}$} & \colhead{Log($M/M_{\odot}$)} & \colhead{$\sigma_{M}$} & \colhead{$A_{V}$} & \colhead{$\sigma_{A_{V}}$}} \\
\startdata  
1 & 6.54 & 0.08 & 4.77 & 0.22 & 0.70 & 0.13 \\
2 & 6.66 & 0.19 & 5.00 & 0.30 & 0.60 & 0.23 \\
3 & 7.42 & 0.21 & 5.84 & 0.21 & 0.10 & 0.11 \\
4 & 7.32 & 0.25 & 5.73 & 0.25 & 0.40 & 0.17 \\
5 & 6.52 & 0.01 & 5.29 & 0.21 & 1.20 & 0.10 \\
6 & 7.76 & 0.53 & 5.68 & 0.47 & 0.01 & 0.44 \\
7 & 7.63 & 0.45 & 5.34 & 0.36 & 0.01 & 0.31 \\
8 & 6.64 & 0.11 & 5.07 & 0.26 & 0.70 & 0.19 \\
9 & 6.66 & 0.27 & 5.47 & 0.29 & 0.70 & 0.22 \\
10 & 6.44 & 0.16 & 5.70 & 0.28 & 1.10 & 0.21 \\
11 & 6.68 & 0.26 & 6.09 & 0.24 & 0.60 & 0.15 \\
12 & 6.44 & 0.41 & 5.94 & 0.45 & 1.70 & 0.42 \\
13 & 6.72 & 0.49 & 6.16 & 0.42 & 0.80 & 0.39 \\
14 & 6.48 & 0.07 & 5.79 & 0.22 & 1.40 & 0.12 \\
15 & 5.70 & 0.44 & 6.52 & 0.29 & 1.10 & 0.23 \\
16 & 6.76 & 0.33 & 5.36 & 0.30 & 0.70 & 0.24
\enddata
\end{deluxetable*}

\subsection{NGC 0695}

NGC 0695 is a face-on spiral galaxy with a companion at a projected nuclear separation of $\sim 26''$ (16 kpc) to the northwest. There are multiple spiral arms on the northwestern half of the galaxy, and star clusters are distributed throughout disk. The maximum $A_{V}$ adopted for this galaxy is 2.8 mags of visual extinction (Kennicutt et al. 2009). 

\begin{figure*}
\centering
\includegraphics[scale=0.25]{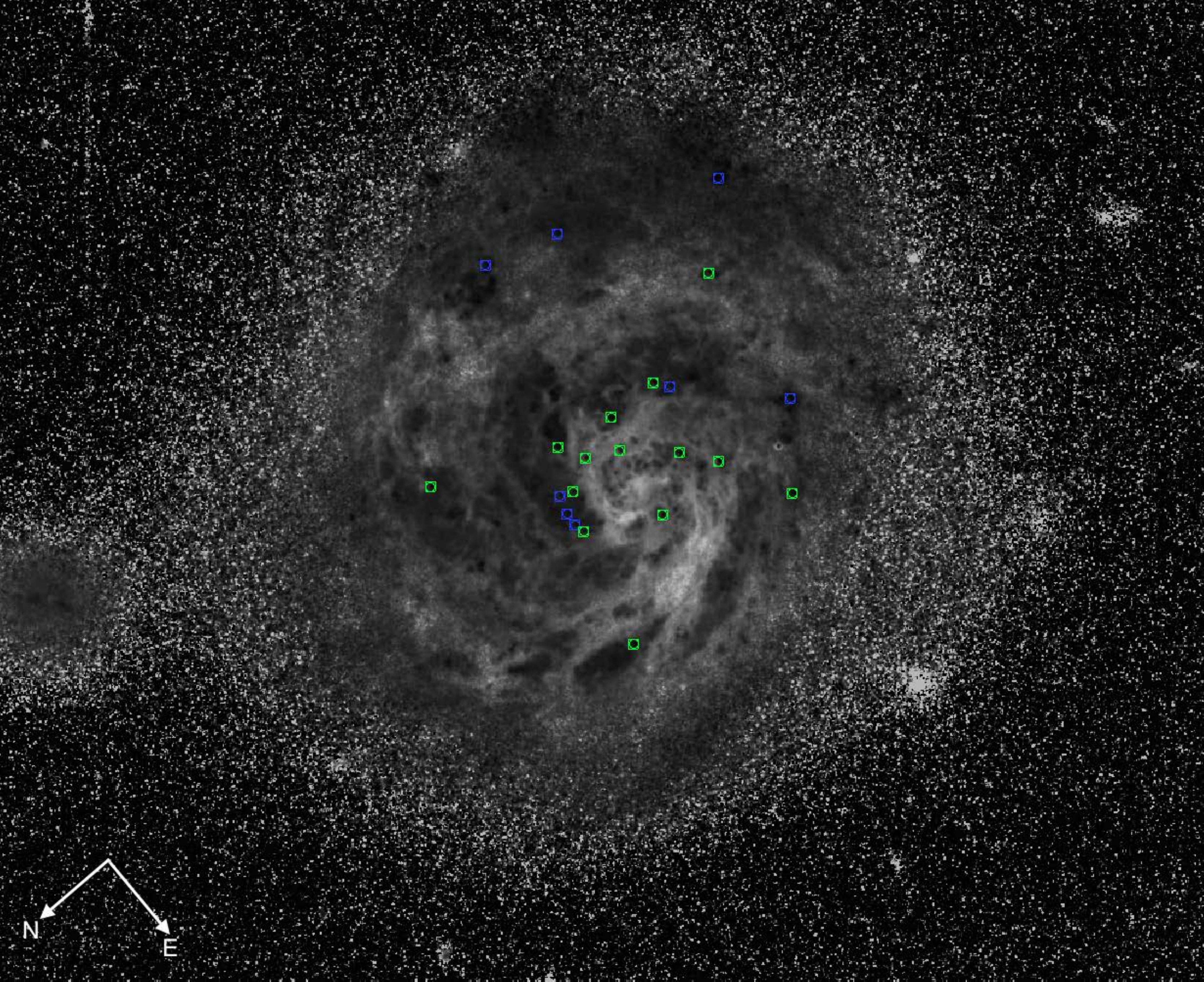}
\caption{Inverted black and white B-I image of NGC 0695 taken with HST ACS/WFC F814W and F435W. The bright emission corresponds to redder (i.e. dustier) regions of the galaxy. The blue centroids correspond to clusters found in relatively ``dust-free'' regions of these galaxies, whereas the green centroids correspond to clusters found in relatively dustier regions of the galaxy.}
\end{figure*}

\begin{figure*}
\centering
\includegraphics[scale=0.55]{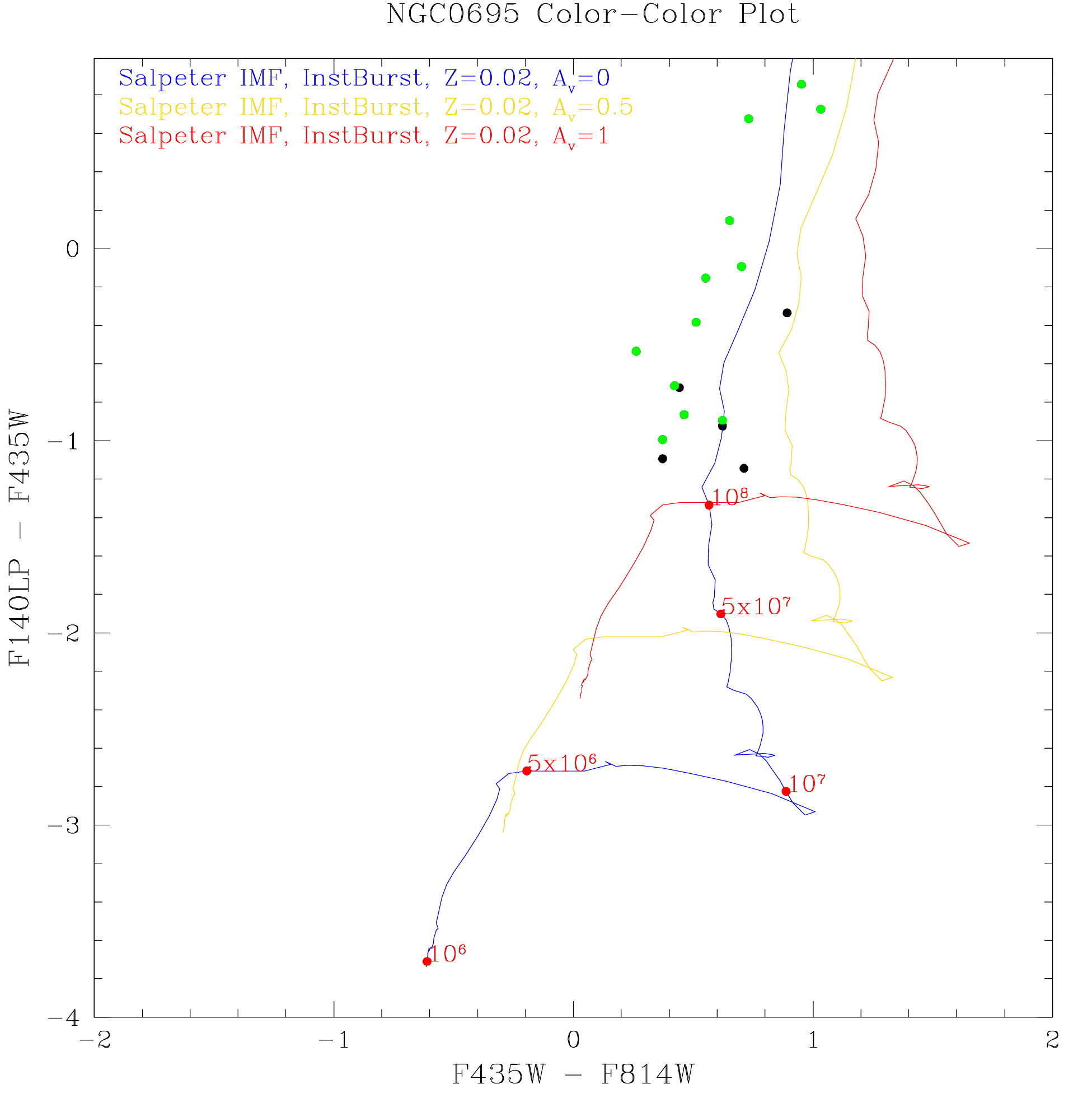}
\caption{Color-Color plot of all star clusters identified in NGC 0695 in F814W, F435W, and F140LP plotted against SSP models with various amount of visual extinction. The green points correspond to the clusters found in dustier regions of the galaxy in Figure 13}
\end{figure*}

\begin{deluxetable*}{ccccccccc}
\tabletypesize{\footnotesize}
\tablewidth{0pt}
\tablecaption {Observed Properties of Star Clusters in NGC 0695}
\tablehead{
\colhead{ID} & \colhead{RA} & \colhead{Dec} & \colhead{$M_{B}$} & \colhead{$\sigma_{B}$} & \colhead{$M_{I}$} & \colhead{$\sigma_{I}$} & \colhead{$M_{FUV}$} & \colhead{$\sigma_{FUV}$}} \\
\startdata
1 & 27.80620232 & 22.58288641 & -11.35 & 0.06 & -12.24 & 0.06 & -11.68 & 0.08 \\
2 & 27.8065178 & 22.57996324 & -11.21 & 0.02 & -11.65 & 0.03 & -11.93 & 0.06 \\
3 & 27.80625132 & 22.58196385 & -10.37 & 0.04 & -10.99 & 0.04 & -11.29 & 0.11 \\
4 & 27.80849949 & 22.58172092 & -10.96 & 0.10 & -11.58 & 0.10 & -11.85 & 0.06 \\
5 & 27.80866289 & 22.58234757 & -13.19 & 0.02 & -13.45 & 0.04 & -13.72 & 0.02 \\
6 & 27.808703 & 22.58307422 & -12.36 & 0.05 & -12.82 & 0.06 & -13.22 & 0.02 \\
7 & 27.80909168 & 22.58243561 & -12.16 & 0.02 & -12.89 & 0.06 & -11.48 & 0.10 \\
8 & 27.80945496 & 22.58181348 & -12.71 & 0.03 & -13.26 & 0.05 & -12.86 & 0.03 \\
9 & 27.80898694 & 22.58283593 & -12.19 & 0.03 & -13.14 & 0.04 & -11.33 & 0.03 \\
10 & 27.8097772 & 22.58144334 & -13.05 & 0.01 & -13.70 & 0.01 & -12.90 & 0.02 \\
11 & 27.80843983 & 22.58463146 & -10.70 & 0.07 & -11.21 & 0.08 & -11.08 & 0.14 \\
12 & 27.80929961 & 22.58314413 & -12.17 & 0.03 & -12.87 & 0.06 & -12.26 & 0.04 \\
13 & 27.8105601 & 22.5808227 & -10.89 & 0.02 & -11.31 & 0.03 & -11.60 & 0.05 \\
14 & 27.80927543 & 22.58330795 & -10.97 & 0.06 & -11.34 & 0.09 & -12.06 & 0.05 \\
15 & 27.80952221 & 22.58331847 & -11.14 & 0.05 & -11.85 & 0.13 & -12.28 & 0.04 \\
16 & 27.81139357 & 22.58329741 & -11.56 & 0.02 & -11.93 & 0.04 & -12.55 & 0.03 \\
17 & 27.80755568 & 22.58055984 & -12.43 & 0.02 & -13.46 & 0.04 & -11.70 & 0.07
\enddata
\end{deluxetable*}

\begin{deluxetable*}{ccccccc}
\tabletypesize{\footnotesize}
\tablewidth{0pt}
\tablecaption {Derived Properties of Star Clusters in NGC 0695}
\tablehead{
\colhead{ID} & \colhead{Log(Age)} & \colhead{$\sigma_{Age}$} & \colhead{Log($M/M_{\odot}$)} & \colhead{$\sigma_{M}$} & \colhead{$A_{V}$} & \colhead{$\sigma_{A_{V}}$}} \\
\startdata  
1 & 6.70 & 0.92 & 6.22 & 0.76 & 1.70 & 0.84 \\
2 & 6.66 & 0.82 & 5.99 & 0.66 & 1.40 & 0.70 \\
3 & 6.40 & 0.74 & 6.05 & 0.71 & 1.90 & 0.76 \\
4 & 6.70 & 0.86 & 5.85 & 0.63 & 1.30 & 0.65 \\
5 & 6.66 & 0.40 & 6.84 & 0.17 & 1.50 & 0.42 \\
6 & 6.68 & 0.24 & 6.40 & 0.31 & 1.30 & 0.25 \\
7 & 8.56 & 0.86 & 7.14 & 0.17 & 0.01 & 0.02 \\
8 & 8.41 & 0.01 & 7.23 & 0.18 & 0.01 & 0.03 \\
9 & 8.56 & 0.03 & 7.26 & 0.21 & 0.20 & 0.10 \\
10 & 8.46 & 0.48 & 7.40 & 0.17 & 0.01 & 0.02 \\
11 & 8.36 & 0.52 & 6.39 & 0.50 & 0.01 & 0.48 \\
12 & 8.41 & 0.04 & 7.07 & 0.20 & 0.10 & 0.10 \\
13 & 6.66 & 0.76 & 5.86 & 0.64 & 1.40 & 0.67 \\
14 & 6.66 & 0.40 & 5.79 & 0.28 & 1.20 & 0.21 \\
15 & 6.74 & 0.90 & 5.83 & 0.59 & 1.10 & 0.61 \\
16 & 6.66 & 0.49 & 6.03 & 0.18 & 1.20 & 0.05 \\
17 & 8.51 & 0.04 & 7.36 & 0.22 & 0.30 & 0.13
\enddata
\end{deluxetable*}

\subsection{UGC 02369}

UGC 02369 is a mid-stage merger consisting of a southern face on galaxy (MCG +02-08-029) and an inclined northern galaxy (MCG +02-08-030). The nuclei of the two galaxies are separated by $\sim 21''$ (13 kpc). A spiral arm containing multiple star clusters extends from the nucleus of the southern galaxy towards the northern galaxy. The maximum $A_{V}$ adopted for this galaxy is 2.3 mags of visual extinction (van Driel et al. 2001). 

\begin{figure*}
\centering
\includegraphics[scale=0.25]{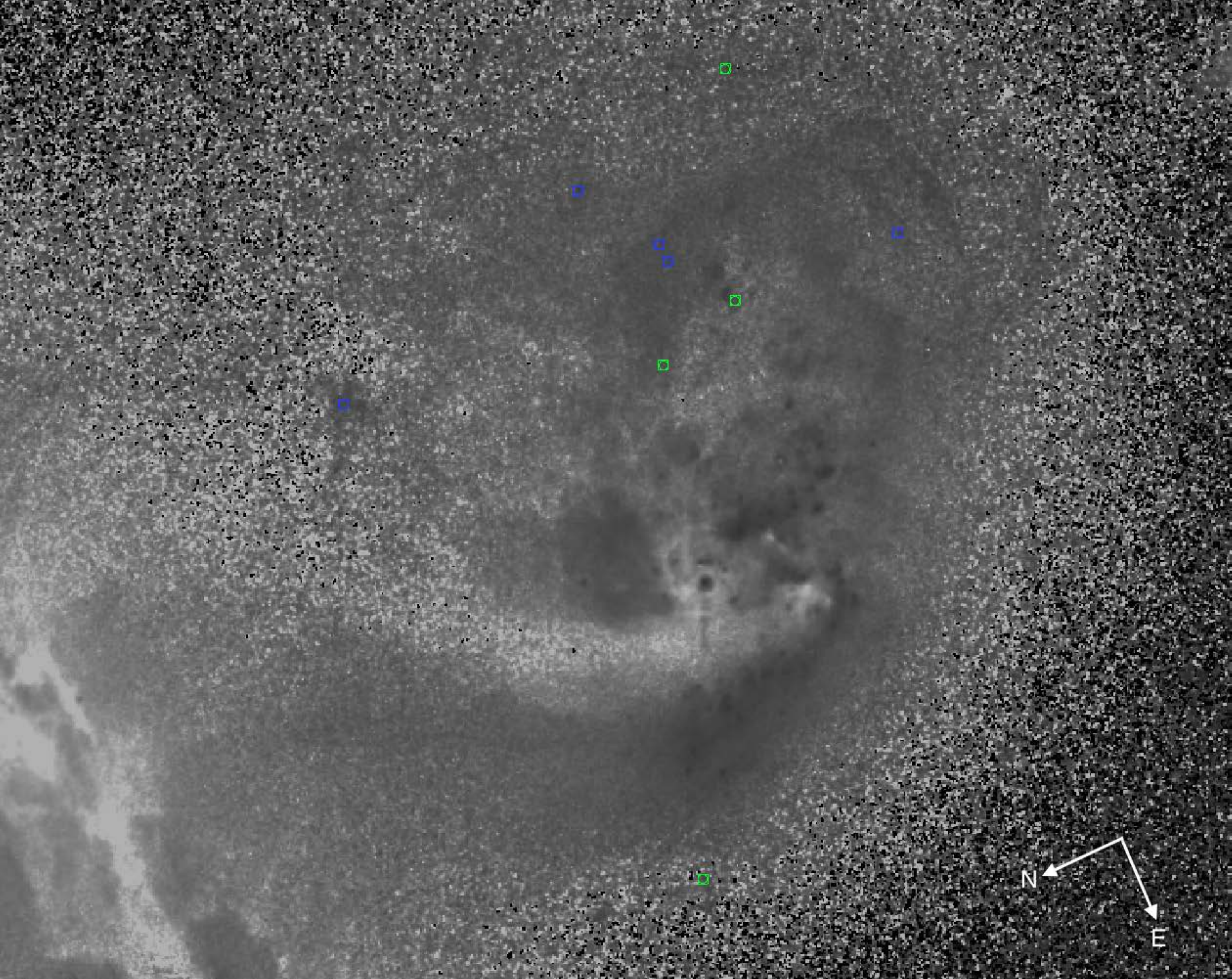}
\caption{Inverted black and white B-I image of UGC 02369 taken with HST ACS/WFC F814W and F435W. The bright emission corresponds to redder (i.e. dustier) regions of the galaxy. The blue centroids correspond to clusters found in relatively ``dust-free'' regions of these galaxies, whereas the green centroids correspond to clusters found in relatively dustier regions of the galaxy.}
\end{figure*}

\begin{figure*}
\centering
\includegraphics[scale=0.55]{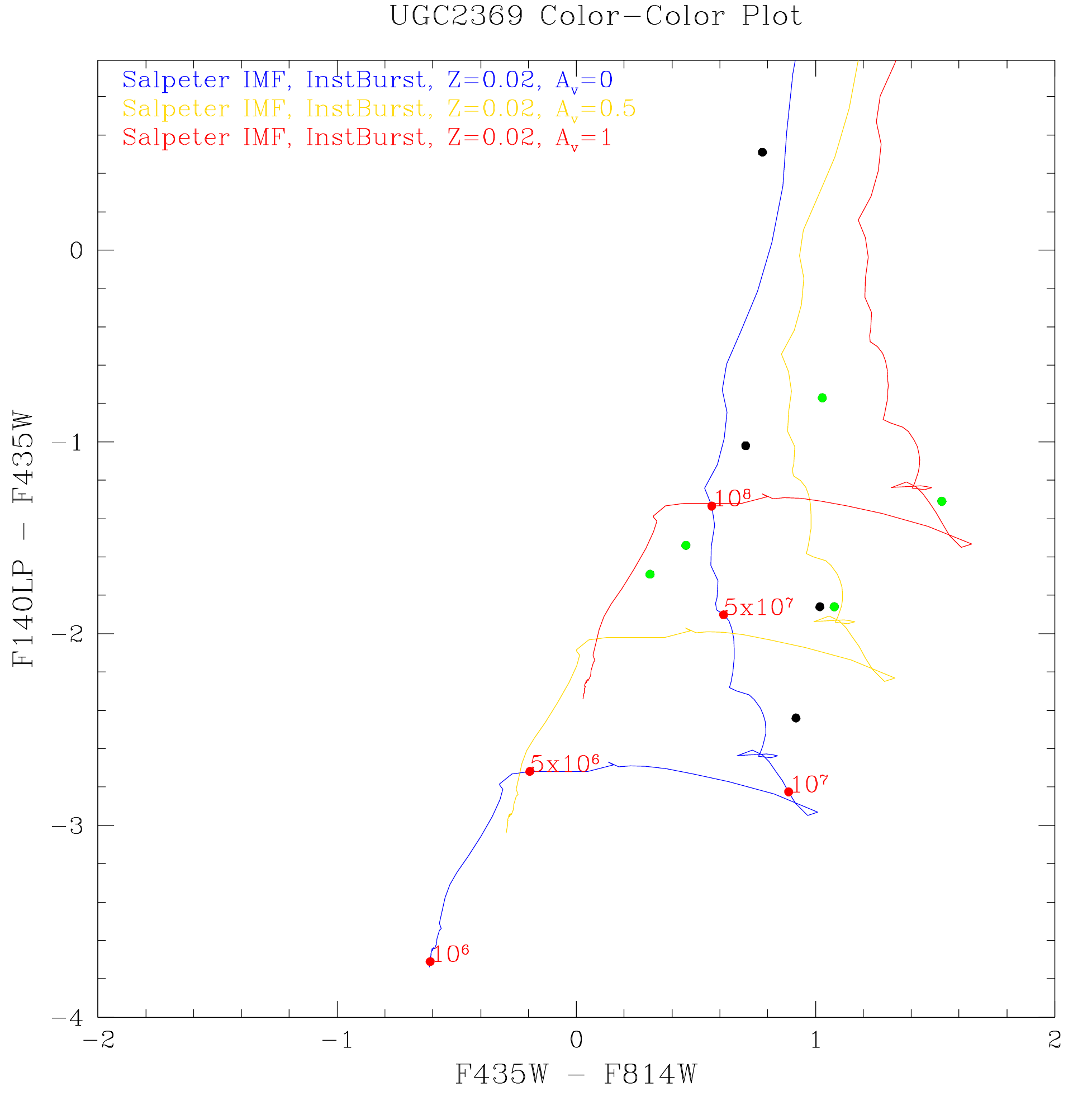}
\caption{Color-Color plot of all star clusters identified in UGC 02369 in F814W, F435W, and F140LP plotted against SSP models with various amount of visual extinction. The green points correspond to the clusters found in dustier regions of the galaxy in Figure 15}
\end{figure*}

\begin{deluxetable*}{ccccccccc}
\tabletypesize{\footnotesize}
\tablewidth{0pt}
\tablecaption {Observed Properties of Star Clusters in UGC 02369}
\tablehead{
\colhead{ID} & \colhead{RA} & \colhead{Dec} & \colhead{$M_{B}$} & \colhead{$\sigma_{B}$} & \colhead{$M_{I}$} & \colhead{$\sigma_{I}$} & \colhead{$M_{FUV}$} & \colhead{$\sigma_{FUV}$}} \\
\startdata
1 & 43.50963156 & 14.97660266 & -10.85 & 0.06 & -11.31 & 0.13 & -12.39 & 0.02 \\
2 & 43.50342276 & 14.96966218 & -9.69 & 0.10 & -10.01 & 0.15 & -11.38 & 0.05 \\
3 & 43.5048427 & 14.97058419 & -9.36 & 0.12 & -10.28 & 0.14 & -11.80 & 0.04 \\
4 & 43.50519206 & 14.96870793 & -9.54 & 0.10 & -10.56 & 0.09 & -11.40 & 0.05 \\
5 & 43.50404048 & 14.97111855 & -12.20 & 0.01 & -12.98 & 0.01 & -11.69 & 0.04 \\
6 & 43.50533035 & 14.97014987 & -9.73 & 0.17 & -11.26 & 0.10 & -11.04 & 0.05 \\
7 & 43.50518643 & 14.9734774 & -9.84 & 0.10 & -10.55 & 0.13 & -10.86 & 0.11 \\
8 & 43.50567241 & 14.97086958 & -9.81 & 0.11 & -10.84 & 0.10 & -10.58 & 0.11 \\
9 & 43.50995014 & 14.97180788 & -10.08 & 0.07 & -11.16 & 0.06 & -11.94 & 0.03
\enddata
\end{deluxetable*}

\begin{deluxetable*}{ccccccc}
\tabletypesize{\footnotesize}
\tablewidth{0pt}
\tablecaption {Derived Properties of Star Clusters in UGC 02369}
\tablehead{
\colhead{ID} & \colhead{Log(Age)} & \colhead{$\sigma_{Age}$} & \colhead{Log($M/M_{\odot}$)} & \colhead{$\sigma_{M}$} & \colhead{$A_{V}$} & \colhead{$\sigma_{A_{V}}$}} \\
\startdata  
1 & 5.10 & 0.77 & 5.68 & 0.52 & 1.60 & 0.50 \\
2 & 6.36 & 0.58 & 4.99 & 0.45 & 1.40 & 0.40 \\
3 & 7.04 & 0.20 & 4.47 & 0.26 & 0.20 & 0.15 \\
4 & 7.63 & 0.26 & 4.93 & 0.26 & 0.10 & 0.17 \\
5 & 6.66 & 0.03 & 6.20 & 0.17 & 2.20 & 0.05 \\
6 & 6.92 & 0.21 & 4.98 & 0.31 & 1.10 & 0.19 \\
7 & 6.72 & 0.74 & 4.74 & 0.59 & 1.20 & 0.59 \\
8 & 7.81 & 0.55 & 5.46 & 0.51 & 0.70 & 0.49 \\
9 & 7.34 & 0.18 & 5.22 & 0.23 & 0.50 & 0.12
\enddata
\end{deluxetable*}

\subsection{NGC 1614}

NGC 1614 is a late-stage merger with two resolved components in the nucleus separated by $\sim 0.8''$ (300 pc). Beyond the nucleus are two well defined spiral arms, with a significant number of bright clusters scattered throughout this region. The maximum $A_{V}$ adopted for this galaxy is 4.0 mags of visual extinction (Alonso-Herrero et al. 2001). 

\begin{figure*}
\centering
\includegraphics[scale=0.25]{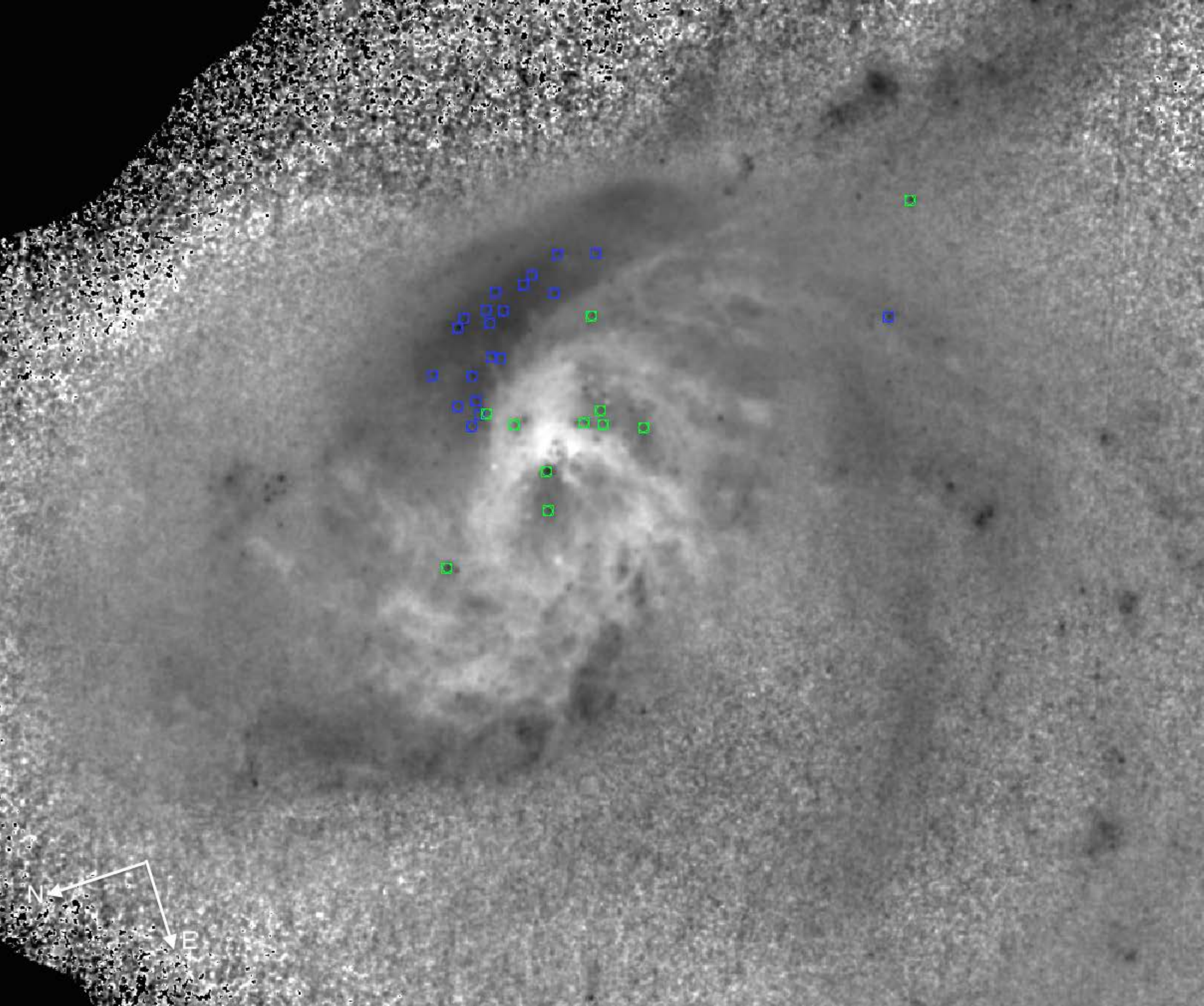}
\caption{Inverted black and white B-I image of NGC 1614 taken with HST ACS/WFC F814W and F435W. The bright emission corresponds to redder (i.e. dustier) regions of the galaxy. The blue centroids correspond to clusters found in relatively ``dust-free'' regions of these galaxies, whereas the green centroids correspond to clusters found in relatively dustier regions of the galaxy.}
\end{figure*}

\begin{figure*}
\centering
\includegraphics[scale=0.55]{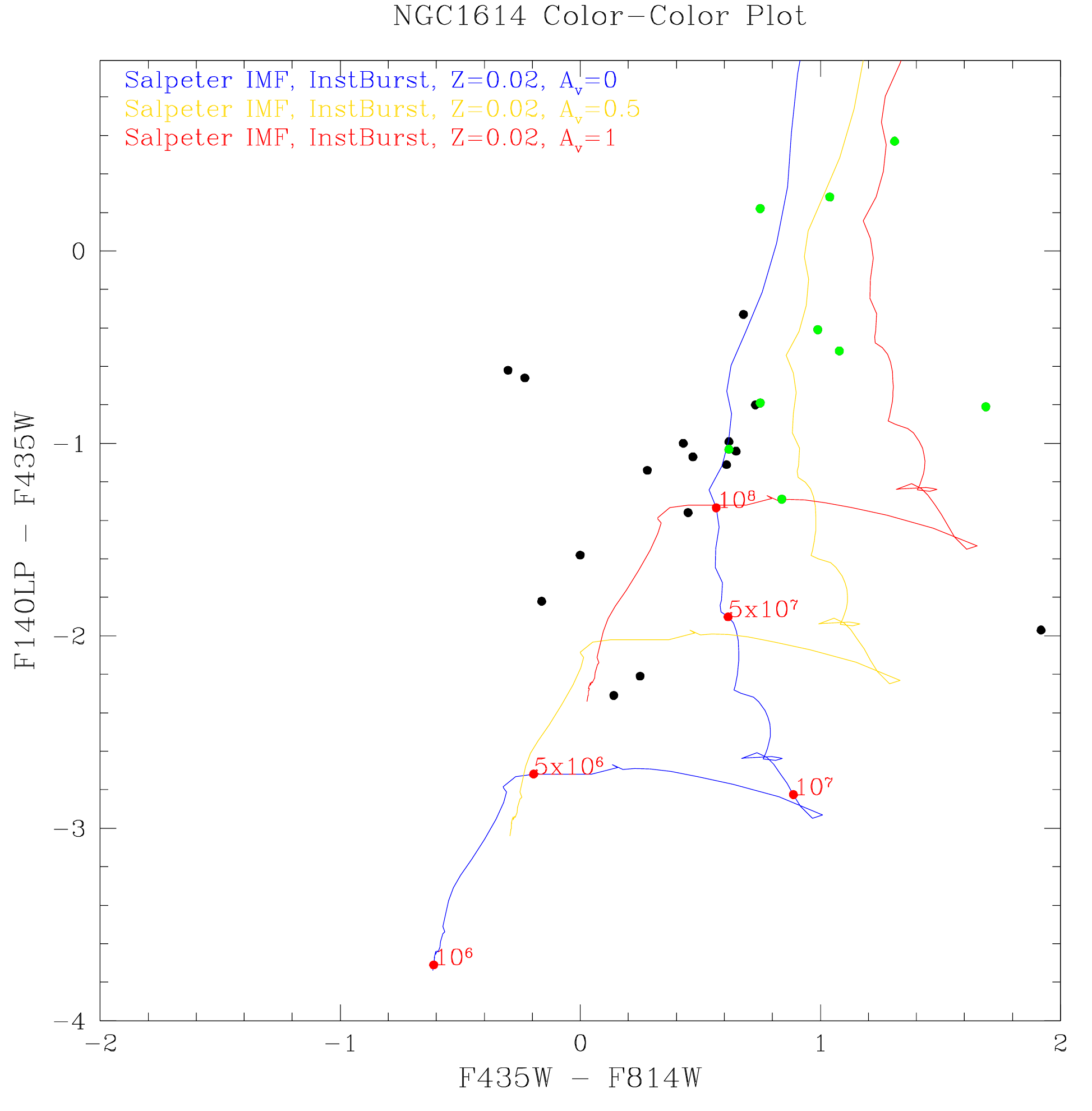}
\caption{Color-Color plot of all star clusters identified in NGC 1614 in F814W, F435W, and F140LP plotted against SSP models with various amount of visual extinction. The green points correspond to the clusters found in dustier regions of the galaxy in Figure 17}
\end{figure*}

\begin{deluxetable*}{ccccccccc}
\tabletypesize{\footnotesize}
\tablewidth{0pt}
\tablecaption {Observed Properties of Star Clusters in NGC 1614}
\tablehead{
\colhead{ID} & \colhead{RA} & \colhead{Dec} & \colhead{$M_{B}$} & \colhead{$\sigma_{B}$} & \colhead{$M_{I}$} & \colhead{$\sigma_{I}$} & \colhead{$M_{FUV}$} & \colhead{$\sigma_{FUV}$}} \\
\startdata
1 & 68.49833746 & -8.5836658 & -11.21 & 0.01 & -11.83 & 0.01 & -12.24 & 0.01 \\
2 & 68.49787214 & -8.579994493 & -11.99 & 0.01 & -12.46 & 0.02 & -13.06 & 0.04 \\
3 & 68.49775125 & -8.5795586 & -12.48 & 0.02 & -13.13 & 0.02 & -13.52 & 0.02 \\
4 & 68.49797702 & -8.579081625 & -13.73 & 0.01 & -14.18 & 0.01 & -15.09 & 0.02 \\
5 & 68.49796061 & -8.578745709 & -12.92 & 0.01 & -13.53 & 0.02 & -14.03 & 0.03 \\
6 & 68.49813488 & -8.578589371 & -11.62 & 0.05 & -11.90 & 0.10 & -12.76 & 0.09 \\
7 & 68.49855752 & -8.57973638 & -11.62 & 0.02 & -12.37 & 0.02 & -12.41 & 0.06 \\
8 & 68.49957784 & -8.583032955 & -10.67 & 0.02 & -10.81 & 0.03 & -12.98 & 0.04 \\
9 & 68.49815445 & -8.578312212 & -13.05 & 0.02 & -13.48 & 0.03 & -14.05 & 0.03 \\
10 & 68.49829417 & -8.578581 & -11.86 & 0.09 & -11.86 & 0.17 & -13.44 & 0.13 \\
11 & 68.49823898 & -8.578206022 & -12.72 & 0.03 & -12.49 & 0.06 & -13.38 & 0.08 \\
12 & 68.49868598 & -8.578484509 & -15.33 & 0.01 & -16.06 & 0.01 & -16.13 & 0.01 \\
13 & 68.49872601 & -8.578589212 & -12.07 & 0.10 & -11.77 & 0.11 & -12.69 & 0.16 \\
14 & 68.49882795 & -8.578206944 & -13.22 & 0.02 & -13.47 & 0.03 & -15.43 & 0.02 \\
15 & 68.4986906 & -8.577771465 & -12.76 & 0.01 & -13.38 & 0.01 & -13.75 & 0.02 \\
16 & 68.49912105 & -8.578165885 & -12.53 & 0.05 & -12.37 & 0.08 & -14.35 & 0.03 \\
17 & 68.49965474 & -8.579516099 & -14.21 & 0.01 & -15.25 & 0.01 & -13.93 & 0.02 \\
18 & 68.49930324 & -8.57824455 & -12.07 & 0.06 & -13.76 & 0.03 & -12.88 & 0.08 \\
19 & 68.49929236 & -8.578158424 & -10.73 & 0.21 & -12.65 & 0.10 & -12.70 & 0.15 \\
20 & 68.49973941 & -8.579291182 & -12.81 & 0.04 & -13.80 & 0.03 & -13.22 & 0.03 \\
21 & 68.49982001 & -8.579497527 & -11.54 & 0.11 & -12.38 & 0.07 & -12.83 & 0.05 \\
22 & 68.49952214 & -8.578509292 & -13.29 & 0.01 & -14.60 & 0.02 & -12.72 & 0.05 \\
23 & 68.49939493 & -8.578038288 & -12.50 & 0.02 & -13.18 & 0.02 & -12.83 & 0.09 \\
24 & 68.49998928 & -8.579944805 & -12.74 & 0.02 & -13.82 & 0.03 & -13.26 & 0.03 \\
25 & 68.50060521 & -8.578602704 & -13.65 & 0.01 & -14.40 & 0.02 & -13.43 & 0.03
\enddata
\end{deluxetable*}

\begin{deluxetable*}{ccccccc}
\tabletypesize{\footnotesize}
\tablewidth{0pt}
\tablecaption {Derived Properties of Star Clusters in NGC 1614}
\tablehead{
\colhead{ID} & \colhead{Log(Age)} & \colhead{$\sigma_{Age}$} & \colhead{Log($M/M_{\odot}$)} & \colhead{$\sigma_{M}$} & \colhead{$A_{V}$} & \colhead{$\sigma_{A_{V}}$}} \\
\startdata  
1 & 6.70 & 0.22 & 5.53 & 0.28 & 1.50 & 0.22 \\
2 & 6.52 & 0.02 & 5.89 & 0.18 & 1.70 & 0.06 \\
3 & 6.66 & 0.73 & 6.10 & 0.66 & 1.60 & 0.70 \\
4 & 6.42 & 0.48 & 6.75 & 0.51 & 1.70 & 0.49 \\
5 & 6.68 & 0.71 & 6.22 & 0.61 & 1.50 & 0.63 \\
6 & 6.52 & 0.32 & 5.68 & 0.39 & 1.60 & 0.35 \\
7 & 6.64 & 0.84 & 5.86 & 0.72 & 1.80 & 0.79 \\
8 & 6.72 & 0.16 & 4.77 & 0.23 & 0.50 & 0.15 \\
9 & 6.52 & 0.01 & 6.31 & 0.17 & 1.70 & 0.04 \\
10 & 6.52 & 0.19 & 5.62 & 0.31 & 1.30 & 0.24 \\
11 & 6.52 & 0.72 & 6.18 & 0.65 & 1.70 & 0.69 \\
12 & 6.64 & 0.81 & 7.34 & 0.76 & 1.80 & 0.85 \\
13 & 6.52 & 0.91 & 5.97 & 0.75 & 1.80 & 0.82 \\
14 & 6.74 & 0.07 & 5.78 & 0.18 & 0.50 & 0.06 \\
15 & 6.36 & 0.10 & 6.59 & 0.22 & 2.00 & 0.13 \\
16 & 6.52 & 0.01 & 5.78 & 0.19 & 1.10 & 0.06 \\
17 & 8.46 & 0.02 & 7.40 & 0.17 & 0.40 & 0.04 \\
18 & 6.82 & 0.01 & 5.86 & 0.20 & 1.40 & 0.08 \\
19 & 7.65 & 0.77 & 5.30 & 0.34 & 0.01 & 0.20 \\
20 & 7.86 & 0.55 & 6.73 & 0.55 & 0.90 & 0.55 \\
21 & 7.59 & 0.40 & 5.91 & 0.42 & 0.60 & 0.37 \\
22 & 6.52 & 0.66 & 7.04 & 0.68 & 2.90 & 0.73 \\
23 & 6.52 & 0.04 & 6.35 & 0.17 & 2.20 & 0.57 \\
24 & 7.72 & 0.35 & 6.68 & 0.31 & 1.00 & 0.25 \\
25 & 8.61 & 0.02 & 7.15 & 0.17 & 0.10 & 0.05
\enddata
\end{deluxetable*}

\subsection{2MASX J06094582-2140234}

2MASX J06094582-2140234 is a mid-stage merger consisting of two face-on galaxies which appear to overlap and have a projected seperation of $\sim 8.4''$ (6.3 kpc). Prominent rings/arms in each galaxy contain the bulk of the visible star clusters. The maximum $A_{V}$ adopted for this galaxy is 1.0 mags of visual extinction (Miralles-Caballero et al. 2012). 

\begin{figure*}
\centering
\includegraphics[scale=0.25]{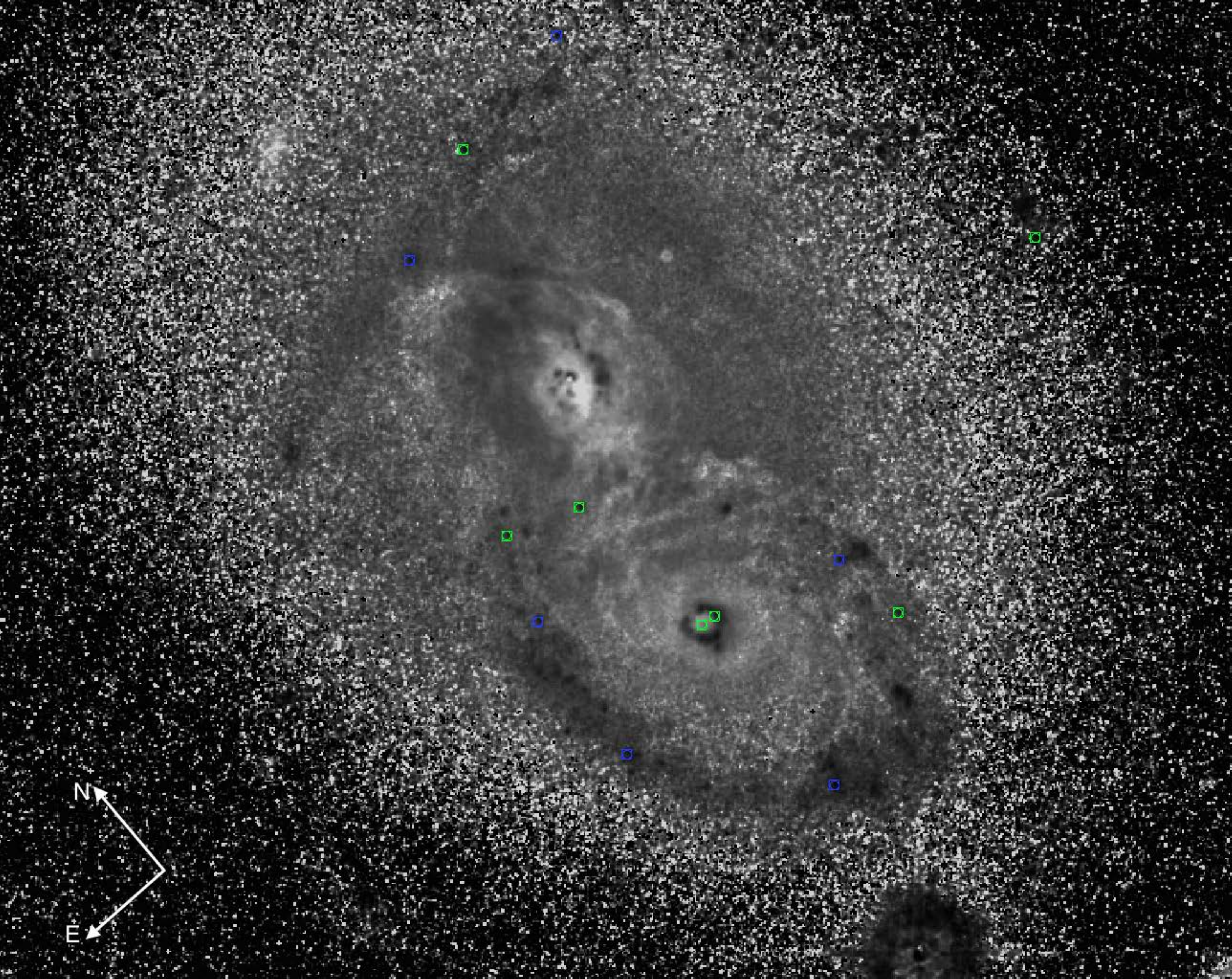}
\caption{Inverted black and white B-I image of 2MASX J06094582-2140234 taken with HST ACS/WFC F814W and F435W. The bright emission corresponds to redder (i.e. dustier) regions of the galaxy. The blue centroids correspond to clusters found in relatively ``dust-free'' regions of these galaxies, whereas the green centroids correspond to clusters found in relatively dustier regions of the galaxy.}
\end{figure*}

\begin{figure*}
\centering
\includegraphics[scale=0.55]{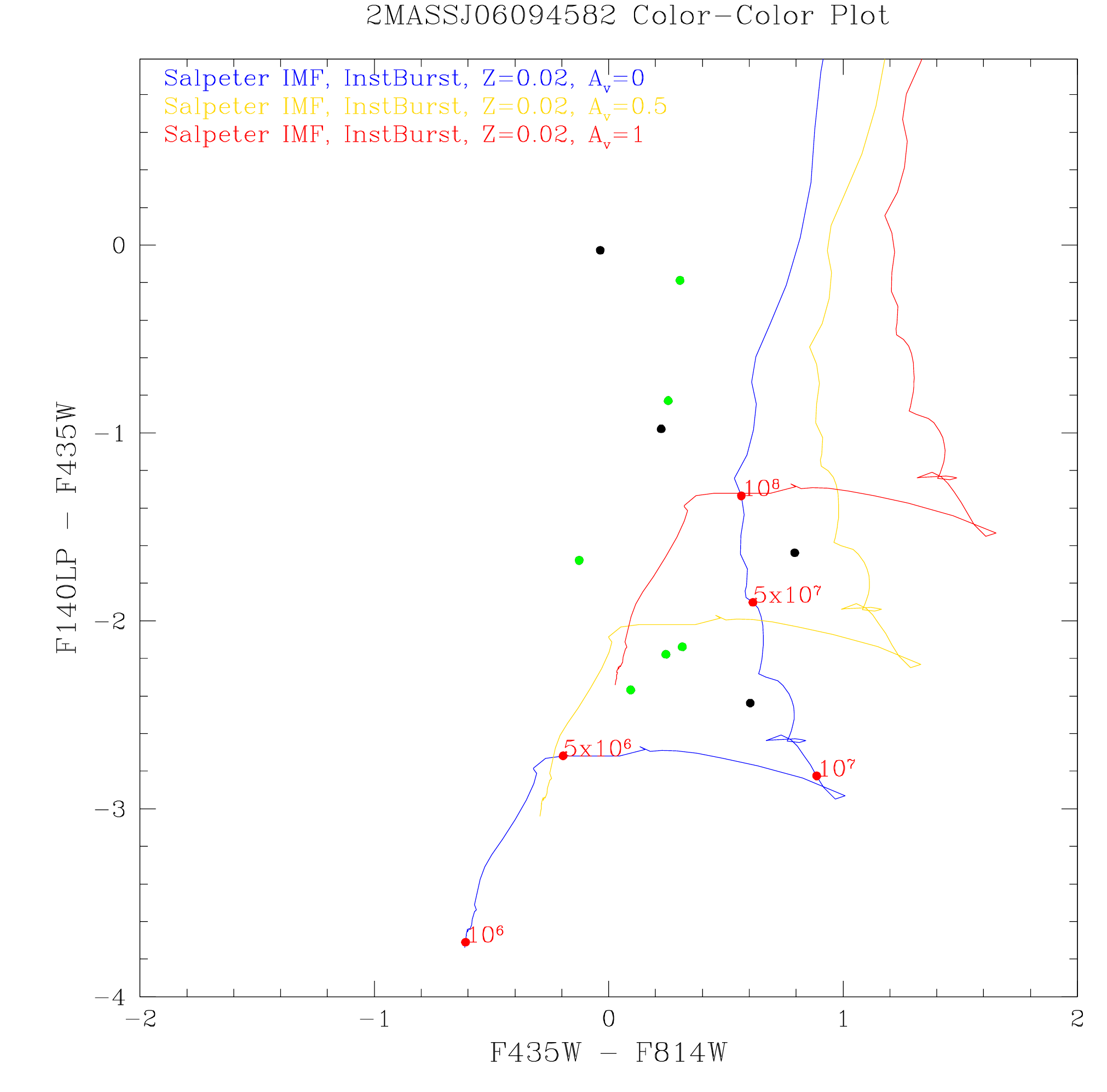}
\caption{Color-Color plot of all star clusters identified in 2MASX J06094582-2140234 in F814W, F435W, and F140LP plotted against SSP models with various amount of visual extinction. The green points correspond to the clusters found in dustier regions of the galaxy in Figure 19}
\end{figure*}

\begin{deluxetable*}{ccccccccc}
\tabletypesize{\footnotesize}
\tablewidth{0pt}
\tablecaption {Observed Properties of Star Clusters in 2MASX J06094582-2140234}
\tablehead{
\colhead{ID} & \colhead{RA} & \colhead{Dec} & \colhead{$M_{B}$} & \colhead{$\sigma_{B}$} & \colhead{$M_{I}$} & \colhead{$\sigma_{I}$} & \colhead{$M_{FUV}$} & \colhead{$\sigma_{FUV}$}} \\
\startdata
1 & 92.438548 & -21.67128502 & -10.07 & 0.09 & -10.29 & 0.14 & -11.05 & 0.04 \\
2 & 92.43987859 & -21.67130163 & -10.68 & 0.06 & -10.77 & 0.10 & -13.05 & 0.01 \\
3 & 92.43715675 & -21.67545341 & -11.44 & 0.02 & -11.68 & 0.04 & -13.62 & 0.02 \\
4 & 92.4409611 & -21.6715528 & -11.60 & 0.03 & -12.20 & 0.05 & -14.04 & 0.05 \\
5 & 92.4416782 & -21.67398257 & -12.05 & 0.03 & -11.92 & 0.12 & -13.73 & 0.02 \\
6 & 92.44229618 & -21.67366993 & -10.76 & 0.07 & -11.07 & 0.15 & -12.90 & 0.01 \\
7 & 92.44270943 & -21.67433539 & -10.88 & 0.06 & -10.84 & 0.18 & -10.91 & 0.03 \\
8 & 92.44164311 & -21.67543932 & -14.49 & 0.03 & -14.79 & 0.09 & -14.68 & 0.05 \\
9 & 92.44054901 & -21.67659996 & -10.76 & 0.05 & -11.01 & 0.10 & -11.59 & 0.04 \\
10 & 92.44310457 & -21.67562528 & -11.02 & 0.07 & -11.81 & 0.07 & -12.66 & 0.01
\enddata
\end{deluxetable*}

\begin{deluxetable*}{ccccccc}
\tabletypesize{\footnotesize}
\tablewidth{0pt}
\tablecaption {Derived Properties of Star Clusters in 2MASX J06094582-2140234}
\tablehead{
\colhead{ID} & \colhead{Log(Age)} & \colhead{$\sigma_{Age}$} & \colhead{Log($M/M_{\odot}$)} & \colhead{$\sigma_{M}$} & \colhead{$A_{V}$} & \colhead{$\sigma_{A_{V}}$}} \\
\startdata  
1 & 6.74 & 0.09 & 4.79 & 0.20 & 1.00 & 0.08 \\
2 & 6.72 & 0.29 & 4.78 & 0.30 & 0.50 & 0.24 \\
3 & 6.74 & 0.03 & 5.07 & 0.19 & 0.50 & 0.09 \\
4 & 6.92 & 0.05 & 5.14 & 0.17 & 0.20 & 0.04 \\
5 & 6.70 & 0.08 & 5.55 & 0.17 & 0.90 & 0.05 \\
6 & 6.74 & 0.36 & 4.85 & 0.35 & 0.60 & 0.30 \\
7 & 7.91 & 0.03 & 6.04 & 0.17 & 1.00 & 0.29 \\
8 & 7.91 & 0.03 & 7.48 & 0.16 & 1.00 & 0.20 \\
9 & 7.16 & 0.09 & 5.52 & 0.18 & 1.00 & 0.07 \\
10 & 7.36 & 0.31 & 5.55 & 0.28 & 0.60 & 0.21
\enddata
\end{deluxetable*}

\subsection{2MASX J08370182-4954302}

2MASX J08370182-4954302 is a mid-stage merger containing two nuclei separated by $\sim 0.66''$ (0.36 kpc). Surrounding the nuclei are multiple bright star clusters in a spiral ridge just northwest and west of the nuclei. The maximum $A_{V}$ adopted for this galaxy is 3.7 mags of visual extinction (Rich et al. 2012). 

\begin{figure*}
\centering
\includegraphics[scale=0.25]{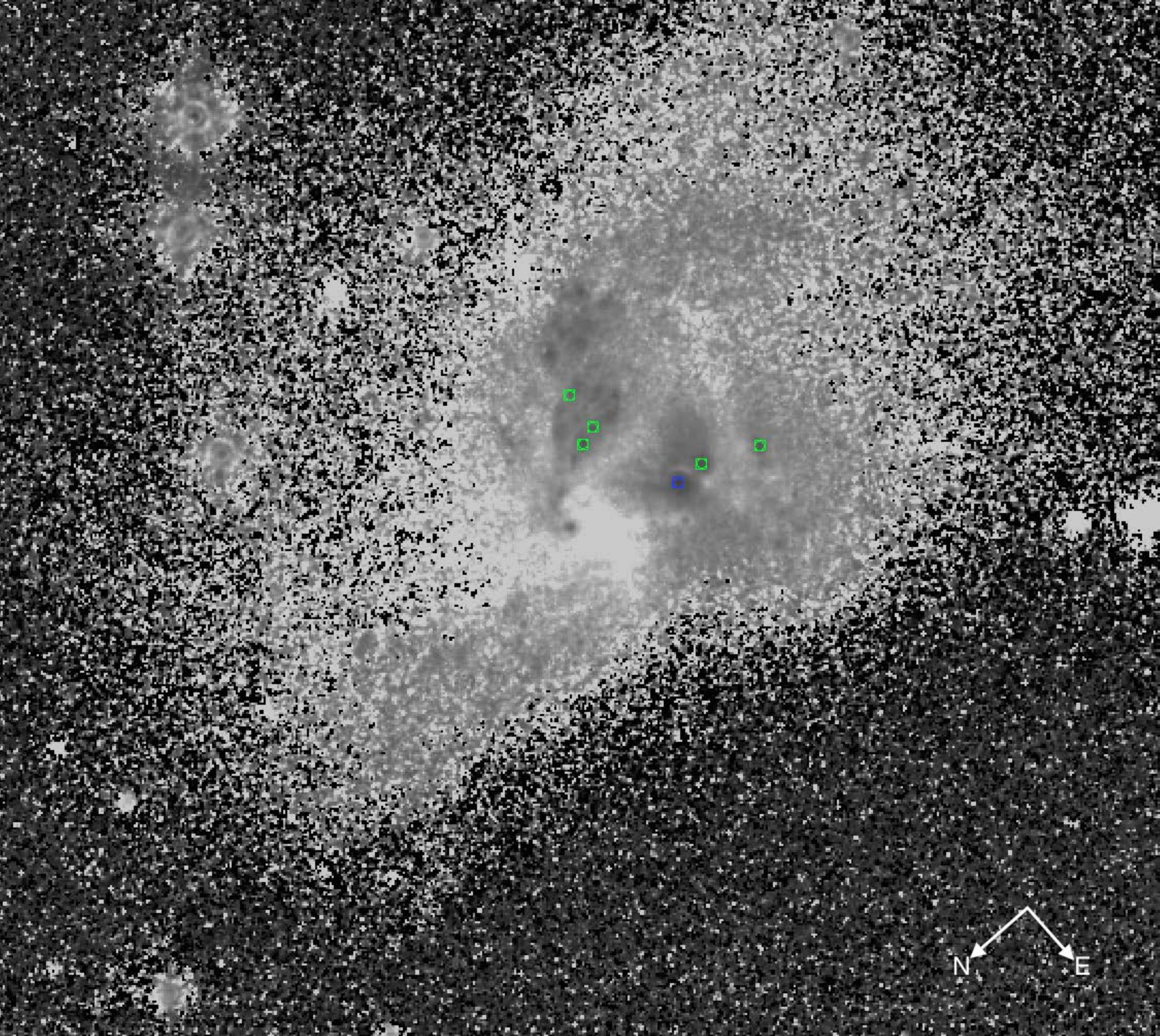}
\caption{Inverted black and white B-I image of 2MASX J08370182-4954302 taken with HST ACS/WFC F814W and F435W. The bright emission corresponds to redder (i.e. dustier) regions of the galaxy. The blue centroids correspond to clusters found in relatively ``dust-free'' regions of these galaxies, whereas the green centroids correspond to clusters found in relatively dustier regions of the galaxy.}
\end{figure*}

\begin{figure*}
\centering
\includegraphics[scale=0.55]{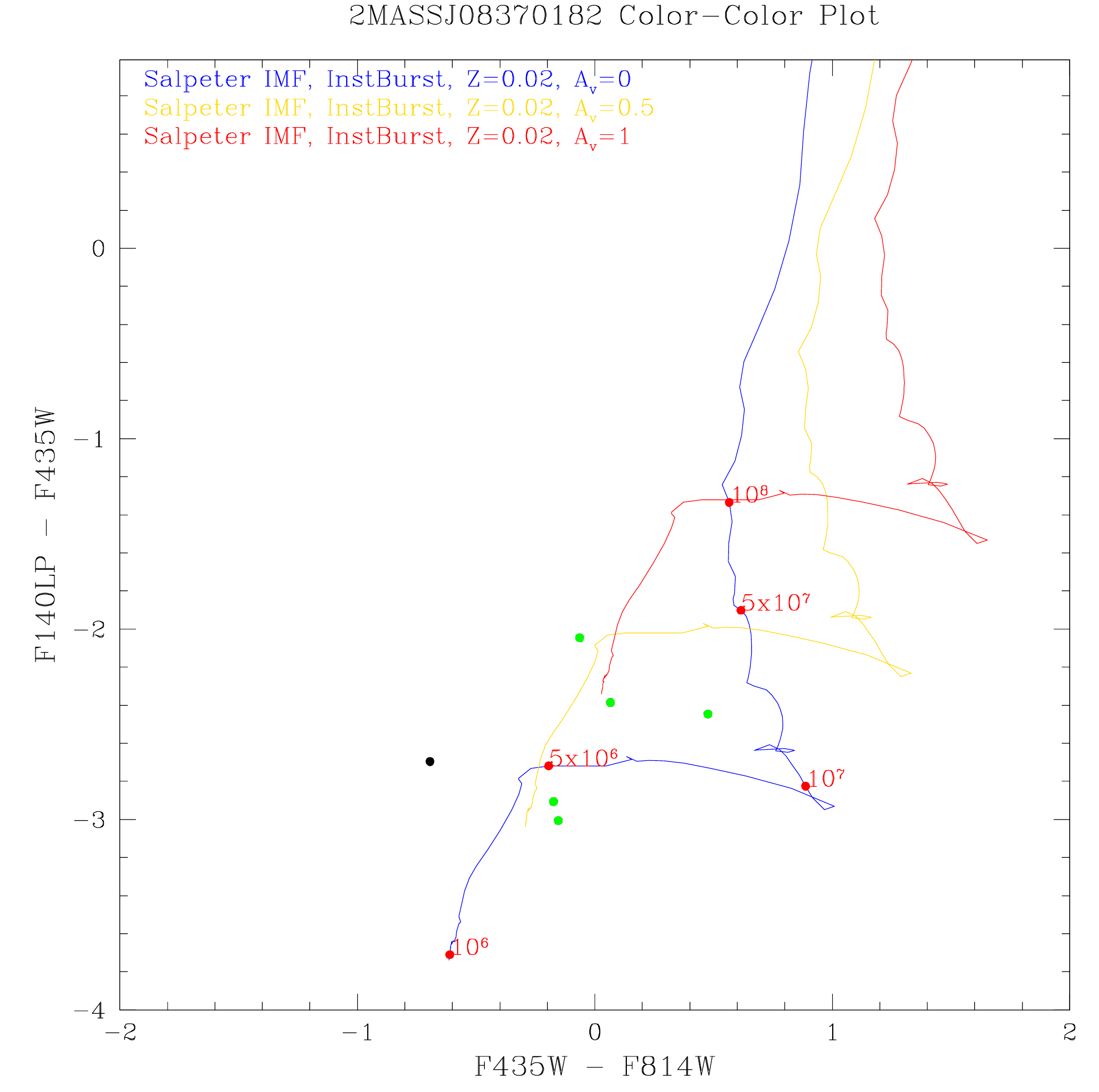}
\caption{Color-Color plot of all star clusters identified in 2MASX J08370182-4954302 in F814W, F435W, and F140LP plotted against SSP models with various amount of visual extinction. The green points correspond to the clusters found in dustier regions of the galaxy in Figure 21}
\end{figure*}

\begin{deluxetable*}{ccccccccc}
\tabletypesize{\footnotesize}
\tablewidth{0pt}
\tablecaption {Observed Properties of Star Clusters in 2MASX J08370182-4954302}
\tablehead{
\colhead{ID} & \colhead{RA} & \colhead{Dec} & \colhead{$M_{B}$} & \colhead{$\sigma_{B}$} & \colhead{$M_{I}$} & \colhead{$\sigma_{I}$} & \colhead{$M_{FUV}$} & \colhead{$\sigma_{FUV}$}} \\
\startdata
1 & 129.2557186 & -49.90846967 & -14.83 & 0.04 & -15.31 & 0.03 & -17.28 & 0.01 \\
2 & 129.2561197 & -49.90840169 & -15.24 & 0.05 & -15.31 & 0.03 & -17.63 & 0.16 \\
3 & 129.2561534 & -49.90828131 & -15.85 & 0.02 & -15.68 & 0.02 & -18.76 & 0.02 \\
4 & 129.2574869 & -49.90904604 & -15.42 & 0.01 & -15.36 & 0.02 & -17.47 & 0.01 \\
5 & 129.2571648 & -49.90869922 & -18.38 & 0.01 & -18.23 & 0.01 & -21.39 & 0.01 \\
6 & 129.2571314 & -49.90851337 & -19.79 & 0.01 & -19.10 & 0.01 & -22.49 & 0.01
\enddata
\end{deluxetable*}

\begin{deluxetable*}{ccccccc}
\tabletypesize{\footnotesize}
\tablewidth{0pt}
\tablecaption {Derived Properties of Star Clusters in 2MASX J08370182-4954302}
\tablehead{
\colhead{ID} & \colhead{Log(Age)} & \colhead{$\sigma_{Age}$} & \colhead{Log($M/M_{\odot}$)} & \colhead{$\sigma_{M}$} & \colhead{$A_{V}$} & \colhead{$\sigma_{A_{V}}$}} \\
\startdata  
1 & 6.86 & 0.02 & 6.38 & 0.18 & 0.20 & 0.05 \\
2 & 5.10 & 0.64 & 7.12 & 0.43 & 1.00 & 0.40 \\
3 & 5.10 & 0.46 & 7.15 & 0.17 & 0.60 & 0.02 \\
4 & 6.66 & 0.03 & 6.59 & 0.16 & 0.50 & 0.04 \\
5 & 5.10 & 0.41 & 8.17 & 0.17 & 0.60 & 0.04 \\
6 & 6.66 & 0.61 & 8.07 & 0.16 & 0.01 & 0.24
\enddata
\end{deluxetable*}

\subsection{NGC 2623}

Evans et al. (2008) discusses the detailed morphology of this galaxy at length. NGC 2623 is a late-stage merger with dust lanes running along its tidal tails into the nucleus. Several bright clusters are distributed throughout the bulge and in a 'pie-wedge' concentration south of the nucleus. The maximum $A_{V}$ adopted for this galaxy is 1.9 mags of visual extinction (Privon et al. 2013). 

\begin{figure*}
\centering
\includegraphics[scale=0.25]{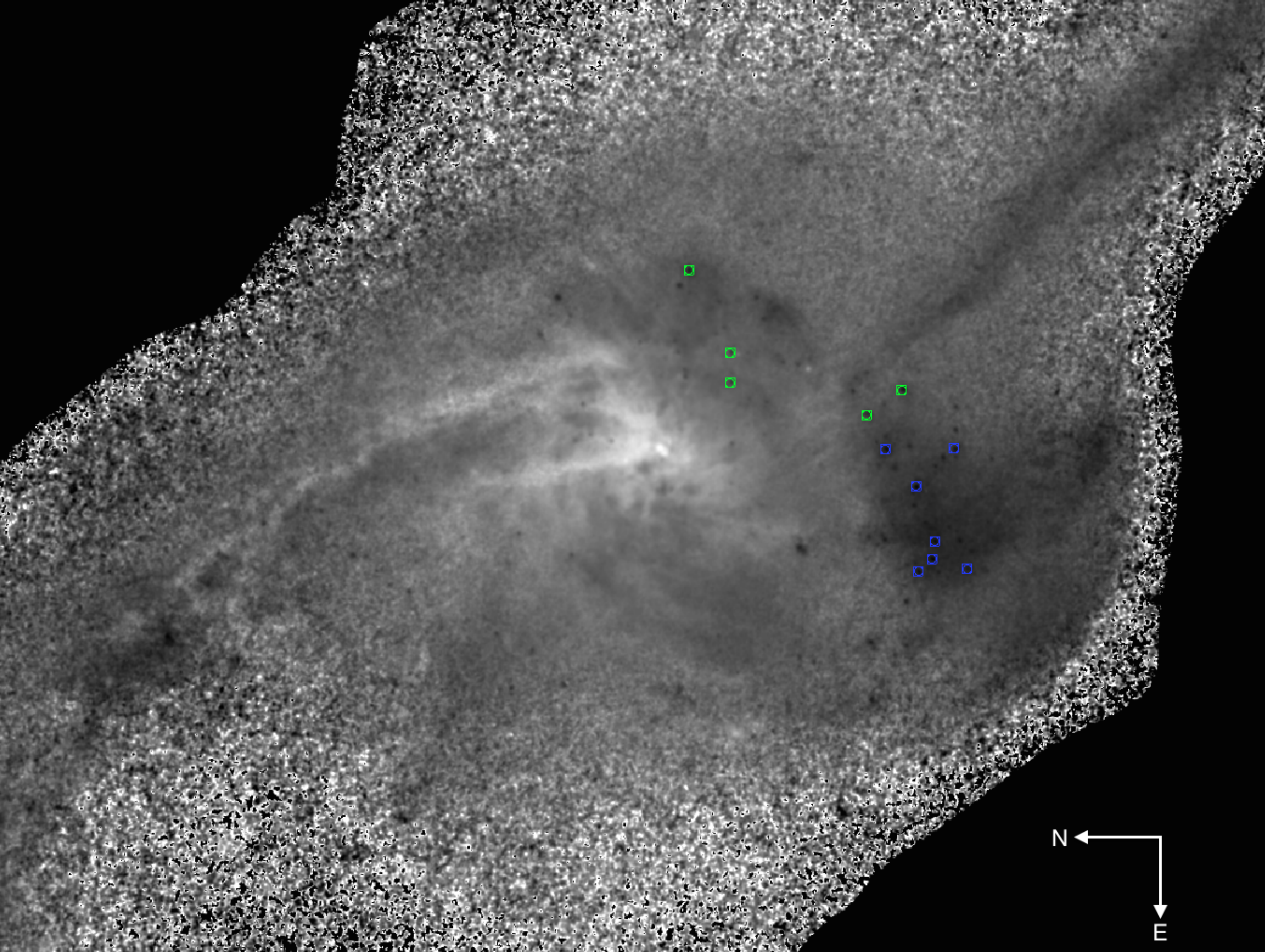}
\caption{Inverted black and white B-I image taken of NGC 2623 with HST ACS/WFC F814W and F435W. The bright emission corresponds to redder (i.e. dustier) regions of the galaxy. The blue centroids correspond to clusters found in relatively ``dust-free'' regions of these galaxies, whereas the green centroids correspond to clusters found in relatively dustier regions of the galaxy.}
\end{figure*}

\begin{figure*}
\centering
\includegraphics[scale=0.55]{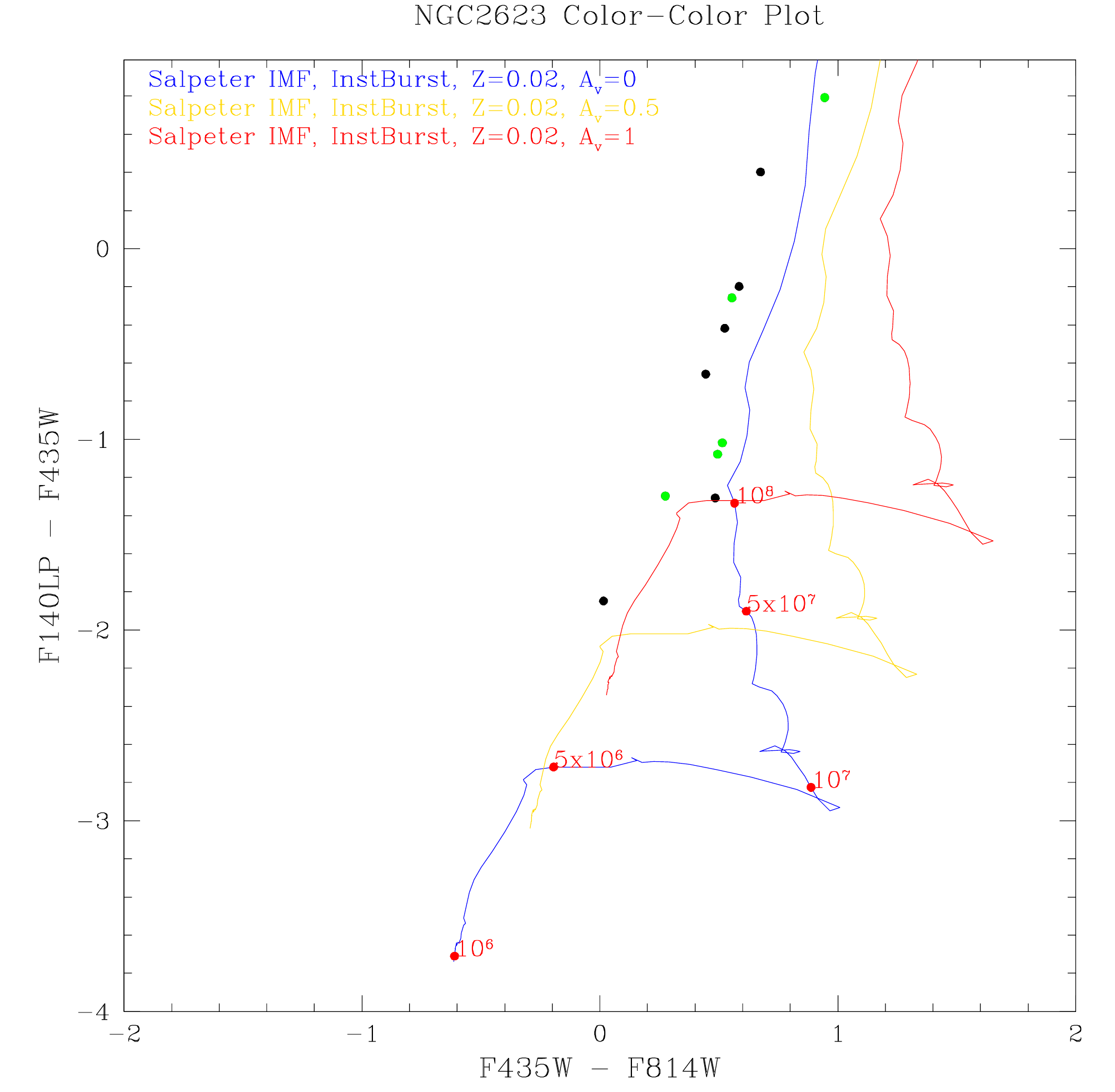}
\caption{Color-Color plot of all star clusters identified in NGC 2623 in F814W, F435W, and F140LP plotted against SSP models with various amount of visual extinction. The green points correspond to the clusters found in dustier regions of the galaxy in Figure 23}
\end{figure*}

\begin{deluxetable*}{ccccccccc}
\tabletypesize{\footnotesize}
\tablewidth{0pt}
\tablecaption {Observed Properties of Star Clusters in NGC 2623}
\tablehead{
\colhead{ID} & \colhead{RA} & \colhead{Dec} & \colhead{$M_{B}$} & \colhead{$\sigma_{B}$} & \colhead{$M_{I}$} & \colhead{$\sigma_{I}$} & \colhead{$M_{FUV}$} & \colhead{$\sigma_{FUV}$}} \\
\startdata
1 & 129.5980304 & 25.75459514 & -14.04 & 0.01 & -14.98 & 0.01 & -13.25 & 0.02 \\
2 & 129.5990031 & 25.7538954 & -11.45 & 0.02 & -11.94 & 0.05 & -12.53 & 0.04 \\
3 & 129.5994079 & 25.75382853 & -11.39 & 0.02 & -11.90 & 0.05 & -12.41 & 0.05 \\
4 & 129.5990874 & 25.751723 & -11.53 & 0.01 & -12.08 & 0.01 & -11.79 & 0.05 \\
5 & 129.5995056 & 25.75209619 & -10.61 & 0.02 & -10.88 & 0.03 & -11.91 & 0.08 \\
6 & 129.599766 & 25.75096143 & -11.27 & 0.01 & -11.85 & 0.01 & -11.47 & 0.03 \\
7 & 129.5999493 & 25.75179613 & -13.15 & 0.01 & -13.82 & 0.01 & -12.75 & 0.04 \\
8 & 129.6003718 & 25.75134324 & -11.08 & 0.02 & -11.09 & 0.05 & -12.93 & 0.03 \\
9 & 129.6010618 & 25.75100371 & -11.58 & 0.01 & -12.06 & 0.02 & -12.89 & 0.03 \\
10 & 129.6013548 & 25.75053983 & -10.88 & 0.03 & -11.40 & 0.03 & -11.30 & 0.02 \\
11 & 129.601522 & 25.75112441 & -11.24 & 0.02 & -11.68 & 0.02 & -11.90 & 0.08
\enddata
\end{deluxetable*}

\begin{deluxetable*}{ccccccc}
\tabletypesize{\footnotesize}
\tablewidth{0pt}
\tablecaption {Derived Properties of Star Clusters in NGC 2623}
\tablehead{
\colhead{ID} & \colhead{Log(Age)} & \colhead{$\sigma_{Age}$} & \colhead{Log($M/M_{\odot}$)} & \colhead{$\sigma_{M}$} & \colhead{$A_{V}$} & \colhead{$\sigma_{A_{V}}$}} \\
\startdata  
1 & 8.56 & 0.18 & 7.33 & 0.16 & 0.10 & 0.02 \\
2 & 6.54 & 0.56 & 5.48 & 0.50 & 1.50 & 0.49 \\
3 & 6.62 & 0.47 & 5.34 & 0.43 & 1.30 & 0.40 \\
4 & 8.36 & 0.64 & 6.16 & 0.16 & 0.10 & 0.03 \\
5 & 6.66 & 0.03 & 4.93 & 0.19 & 1.00 & 0.09 \\
6 & 6.68 & 0.72 & 5.46 & 0.16 & 1.50 & 0.27 \\
7 & 7.86 & 0.67 & 7.25 & 0.16 & 1.50 & 0.29 \\
8 & 6.66 & 0.05 & 4.90 & 0.21 & 0.60 & 0.12 \\
9 & 6.70 & 0.57 & 5.32 & 0.47 & 1.00 & 0.45 \\
10 & 6.68 & 0.02 & 5.30 & 0.17 & 1.50 & 0.03 \\
11 & 6.66 & 0.72 & 5.44 & 0.60 & 1.50 & 0.62
\enddata
\end{deluxetable*}

\subsection{UGC 04881}

UGC 04881 is an early-stage merger containing two nuclei separated by $\sim 11''$ (9 kpc). Spiral dust lanes and strings of star clusters surround the NE nucleus. In the SW nucleus a linear distribution of star clusters and a prominent dust lane are seen. The maximum $A_{V}$ adopted for this galaxy is 1.9 mags of visual extinction (Gonzalez-Martin et al. 2009). 

\begin{figure*}
\centering
\includegraphics[scale=0.25]{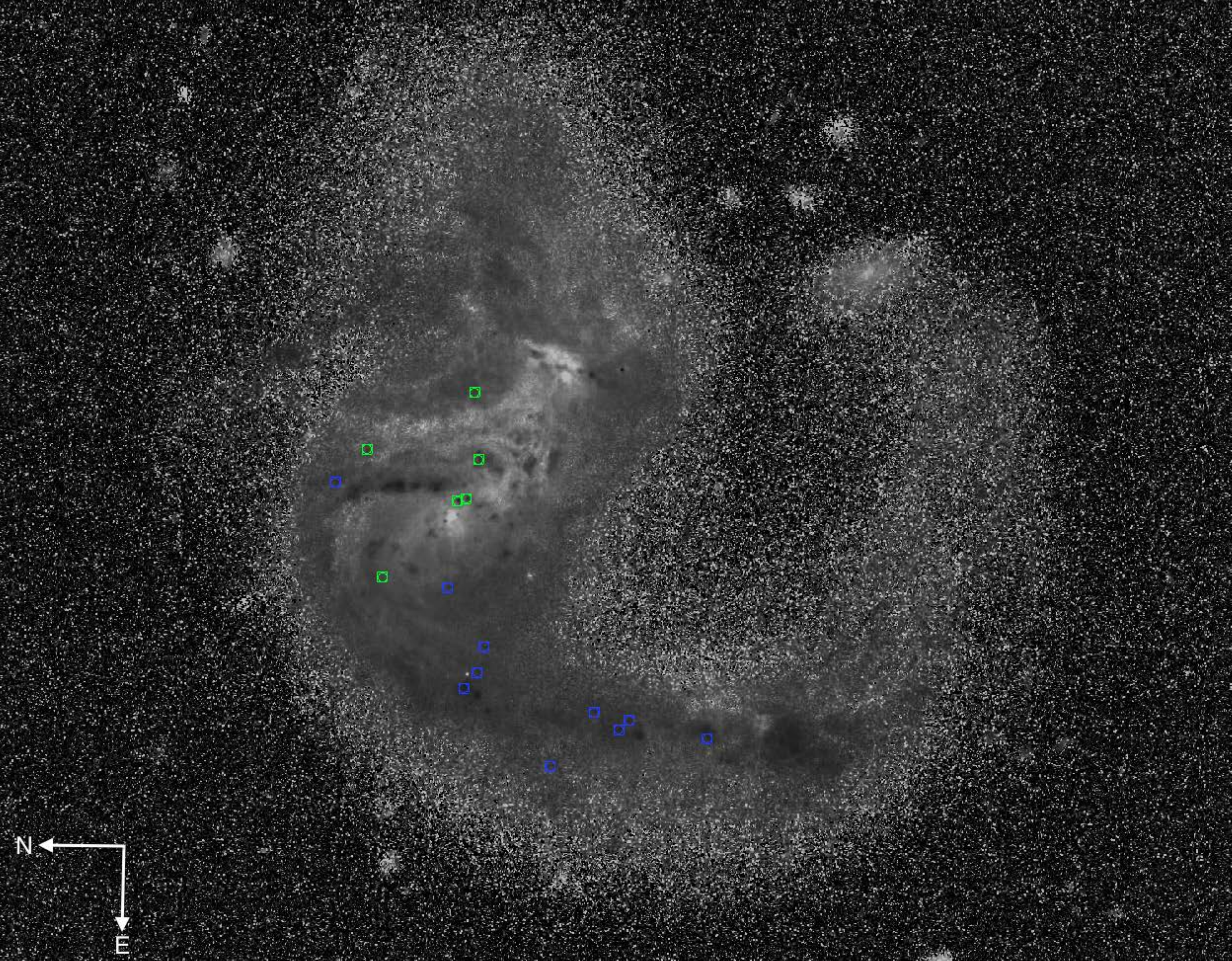}
\caption{Inverted black and white B-I image of UGC 04881 taken with HST ACS/WFC F814W and F435W. The bright emission corresponds to redder (i.e. dustier) regions of the galaxy. The blue centroids correspond to clusters found in relatively ``dust-free'' regions of these galaxies, whereas the green centroids correspond to clusters found in relatively dustier regions of the galaxy.}
\end{figure*}

\begin{figure*}
\centering
\includegraphics[scale=0.55]{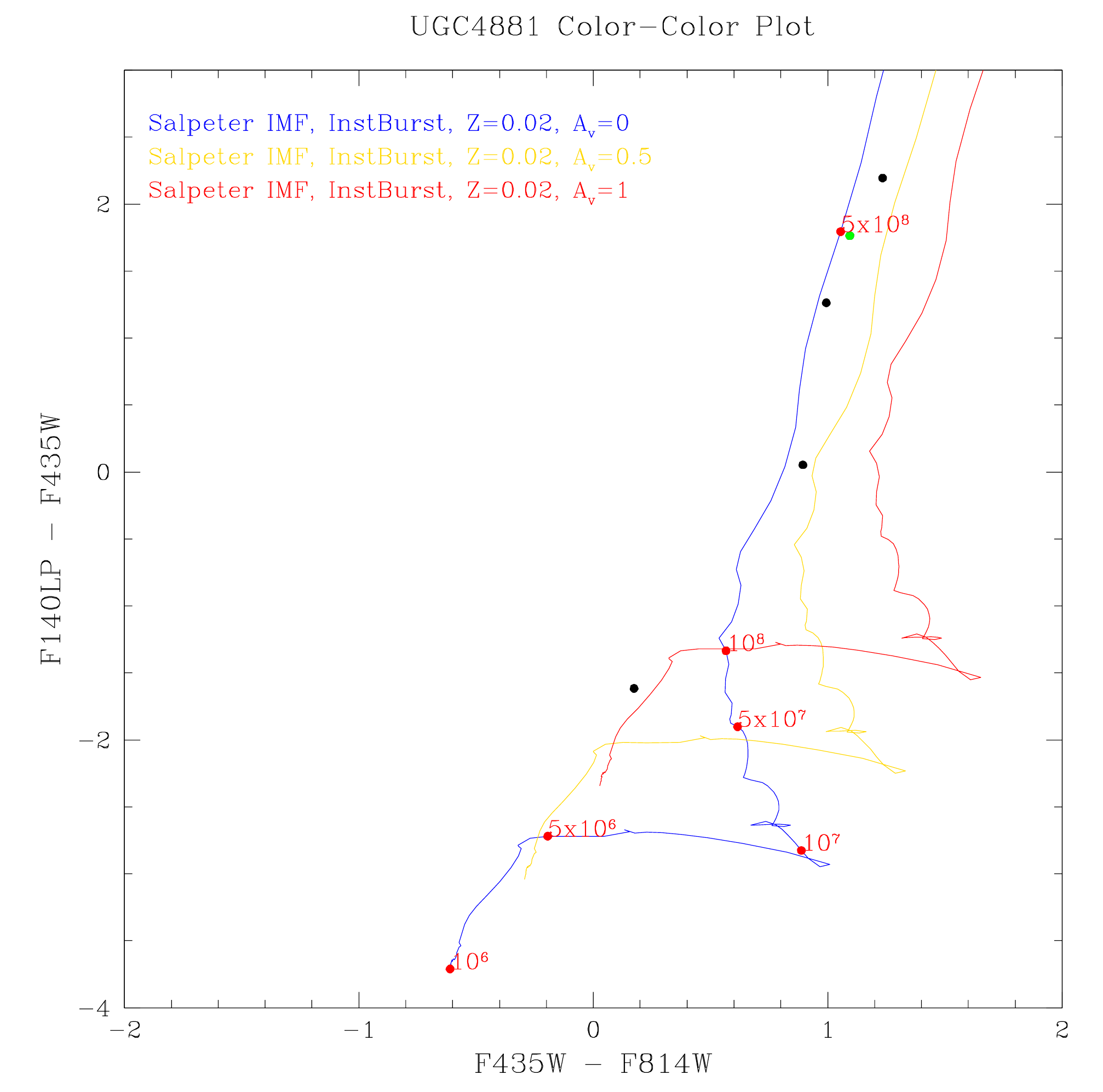}
\caption{Color-Color plot of all star clusters identified in UGC 04881 in F814W, F435W, and F140LP plotted against SSP models with various amount of visual extinction. The green points correspond to the clusters found in dustier regions of the galaxy in Figure 25}
\end{figure*}

\begin{deluxetable*}{ccccccccc}
\tabletypesize{\footnotesize}
\tablewidth{0pt}
\tablecaption {Observed Properties of Star Clusters in UGC 04881}
\tablehead{
\colhead{ID} & \colhead{RA} & \colhead{Dec} & \colhead{$M_{B}$} & \colhead{$\sigma_{B}$} & \colhead{$M_{I}$} & \colhead{$\sigma_{I}$} & \colhead{$M_{FUV}$} & \colhead{$\sigma_{FUV}$}} \\
\startdata
1 & 138.9806254 & 44.33224681 & -13.41 & 0.02 & -14.51 & 0.05 & -11.65 & 0.13 \\
2 & 138.9802968 & 44.33445408 & -12.38 & 0.02 & -13.38 & 0.01 & -11.12 & 0.14 \\
3 & 138.9841104 & 44.33189026 & -13.48 & 0.01 & -14.72 & 0.01 & -11.29 & 0.01 \\
4 & 138.9857806 & 44.32941604 & -11.38 & 0.08 & -12.28 & 0.11 & -11.33 & 0.03 \\
5 & 138.9861709 & 44.32809268 & -10.97 & 0.13 & -11.15 & 0.20 & -12.59 & 0.16
\enddata
\end{deluxetable*}

\begin{deluxetable*}{ccccccc}
\tabletypesize{\footnotesize}
\tablewidth{0pt}
\tablecaption {Derived Properties of Star Clusters in UGC 04881}
\tablehead{
\colhead{ID} & \colhead{Log(Age)} & \colhead{$\sigma_{Age}$} & \colhead{Log($M/M_{\odot}$)} & \colhead{$\sigma_{M}$} & \colhead{$A_{V}$} & \colhead{$\sigma_{A_{V}}$}} \\
\startdata  
1 & 8.71 & 0.03 & 7.17 & 0.23 & 0.01 & 0.16 \\
2 & 8.06 & 0.02 & 7.27 & 0.19 & 1.90 & 0.09 \\
3 & 8.71 & 0.10 & 7.36 & 0.16 & 0.30 & 5.62 \\
4 & 6.68 & 0.44 & 5.71 & 0.53 & 1.90 & 0.52 \\
5 & 6.66 & 0.36 & 4.97 & 0.42 & 0.80 & 0.37
\enddata
\end{deluxetable*}

\subsection{IC 2545}

IC 2545 is a late-stage merger being viewed face-on. Dust lanes and strings of star clusters extend from two unresolved nuclei separated by $\sim 0.8''$ (0.54 kpc) in the center of the galaxy. Multiple star clusters are also visible throughout the tidal tails. The maximum $A_{V}$ adopted for this galaxy is 4.0 mags of visual extinction (van den Broek et al. 1991). 

\begin{figure*}
\centering
\includegraphics[scale=0.25]{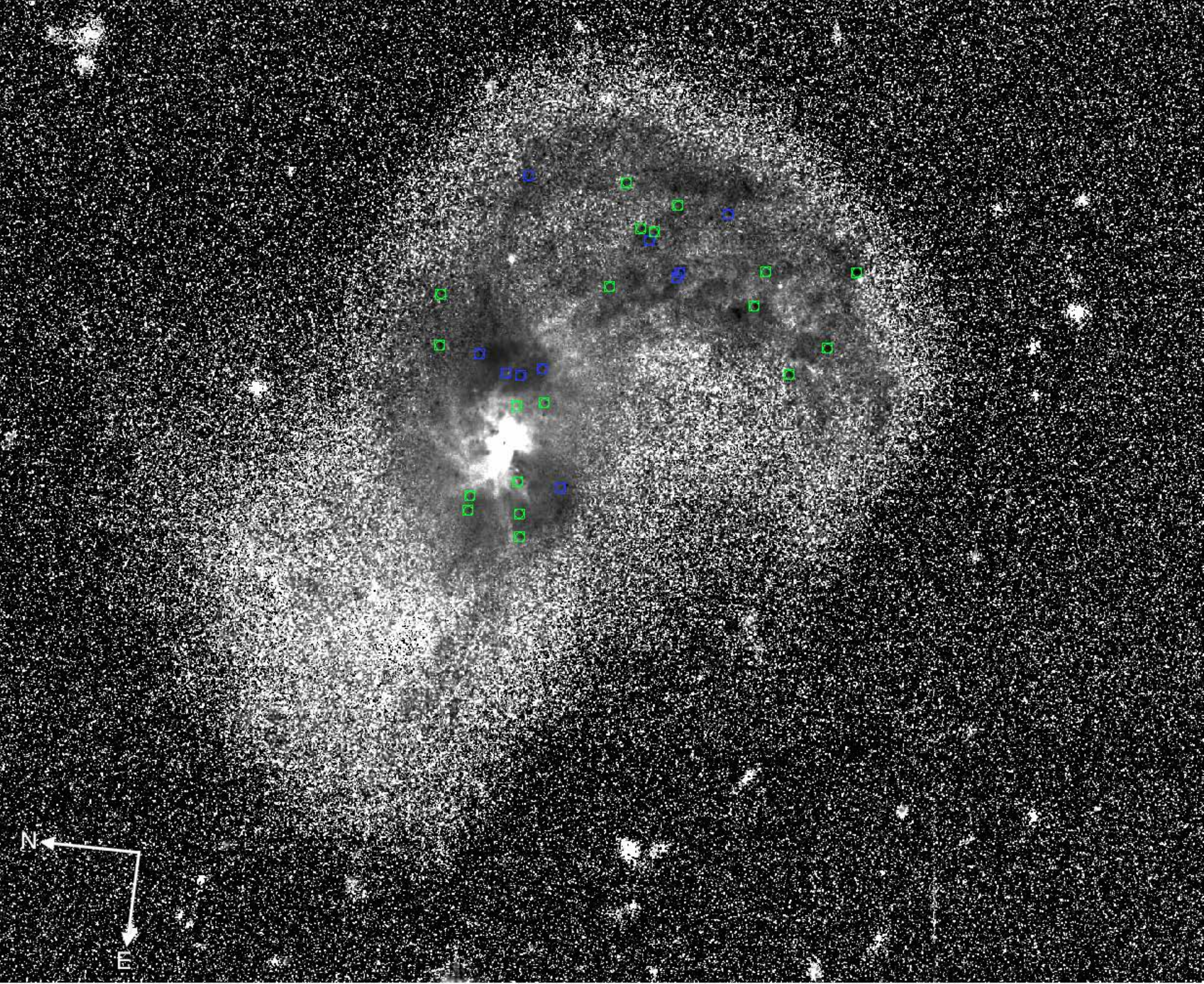}
\caption{Inverted black and white B-I image of IC 2545 taken with HST ACS/WFC F814W and F435W. The bright emission corresponds to redder (i.e. dustier) regions of the galaxy. The blue centroids correspond to clusters found in relatively ``dust-free'' regions of these galaxies, whereas the green centroids correspond to clusters found in relatively dustier regions of the galaxy.}
\end{figure*}

\begin{figure*}
\centering
\includegraphics[scale=0.55]{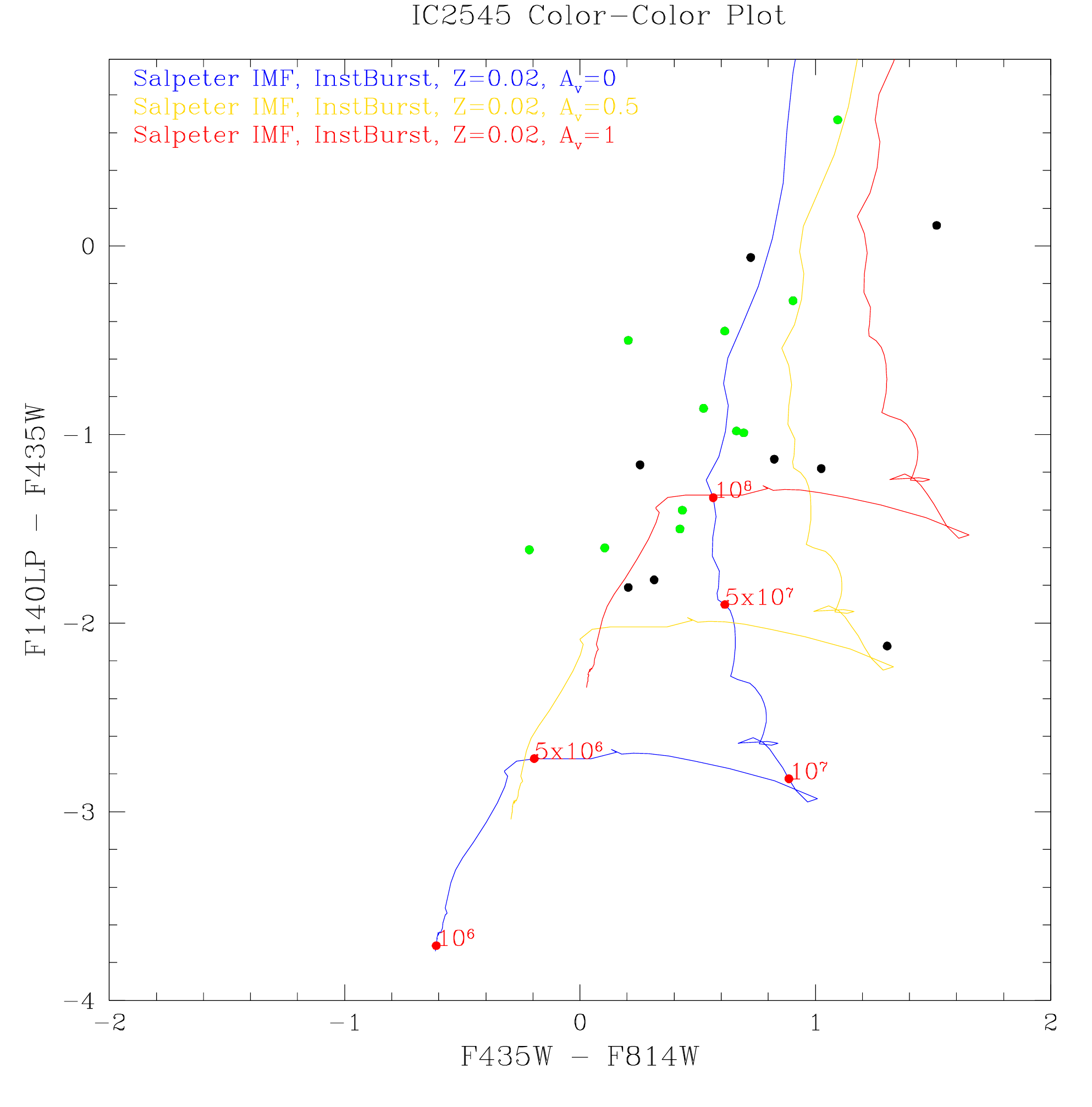}
\caption{Color-Color plot of all star clusters identified in IC 2545 in F814W, F435W, and F140LP plotted against SSP models with various amount of visual extinction. The green points correspond to the clusters found in dustier regions of the galaxy in Figure 27}
\end{figure*}

\begin{deluxetable*}{ccccccccc}
\tabletypesize{\footnotesize}
\tablewidth{0pt}
\tablecaption {Observed Properties of Star Clusters in IC 2545}
\tablehead{
\colhead{ID} & \colhead{RA} & \colhead{Dec} & \colhead{$M_{B}$} & \colhead{$\sigma_{B}$} & \colhead{$M_{I}$} & \colhead{$\sigma_{I}$} & \colhead{$M_{FUV}$} & \colhead{$\sigma_{FUV}$}} \\
\startdata
1 & 151.5136356 & -33.88528106 & -10.23 & 0.08 & -11.06 & 0.08 & -11.36 & 0.11 \\
2 & 151.5136616 & -33.88695096 & -10.55 & 0.07 & -10.99 & 0.08 & -11.95 & 0.11 \\
3 & 151.5140689 & -33.88785616 & -10.73 & 0.07 & -11.43 & 0.06 & -11.72 & 0.13 \\
4 & 151.5146406 & -33.88747231 & -9.72 & 0.13 & -10.15 & 0.15 & -11.22 & 0.26 \\
5 & 151.5148205 & -33.8873956 & -10.48 & 0.09 & -10.74 & 0.13 & -11.64 & 0.17 \\
6 & 151.515442 & -33.88795458 & -11.40 & 0.07 & -11.61 & 0.08 & -13.21 & 0.17 \\
7 & 151.5153282 & -33.88941777 & -11.77 & 0.03 & -12.68 & 0.03 & -12.06 & 0.36 \\
8 & 151.5152205 & -33.89096417 & -11.67 & 0.03 & -11.46 & 0.05 & -13.28 & 0.05 \\
9 & 151.5155505 & -33.88790446 & -10.54 & 0.11 & -10.86 & 0.19 & -12.31 & 0.07 \\
10 & 151.5161603 & -33.88389939 & -10.85 & 0.04 & -11.52 & 0.04 & -11.83 & 0.22 \\
11 & 151.5172199 & -33.88393251 & -11.68 & 0.02 & -12.21 & 0.03 & -12.54 & 0.21 \\
12 & 151.5168118 & -33.8905397 & -10.84 & 0.04 & -10.95 & 0.07 & -12.44 & 0.29 \\
13 & 151.5177074 & -33.88509063 & -14.32 & 0.01 & -15.84 & 0.01 & -14.21 & 0.04 \\
14 & 151.517576 & -33.88570119 & -10.70 & 0.20 & -11.73 & 0.19 & -11.88 & 0.36 \\
15 & 151.5177403 & -33.8853364 & -14.65 & 0.02 & -15.38 & 0.02 & -14.71 & 0.03 \\
16 & 151.5199867 & -33.88613476 & -9.91 & 0.19 & -11.22 & 0.07 & -12.03 & 0.01 \\
17 & 151.5202629 & -33.88460375 & -12.84 & 0.03 & -13.94 & 0.02 & -12.17 & 0.06 \\
18 & 151.5205798 & -33.88546044 & -11.59 & 0.04 & -12.21 & 0.08 & -12.04 & 0.27 \\
19 & 151.5210485 & -33.88549176 & -11.29 & 0.04 & -11.50 & 0.08 & -11.79 & 0.24
\enddata
\end{deluxetable*}

\begin{deluxetable*}{ccccccc}
\tabletypesize{\footnotesize}
\tablewidth{0pt}
\tablecaption {Derived Properties of Star Clusters in IC 2545}
\tablehead{
\colhead{ID} & \colhead{Log(Age)} & \colhead{$\sigma_{Age}$} & \colhead{Log($M/M_{\odot}$)} & \colhead{$\sigma_{M}$} & \colhead{$A_{V}$} & \colhead{$\sigma_{A_{V}}$}} \\
\startdata  
1 & 6.78 & 0.03 & 4.88 & 0.18 & 1.10 & 0.29 \\
2 & 6.34 & 0.40 & 5.47 & 0.18 & 1.60 & 0.46 \\
3 & 8.06 & 0.54 & 5.71 & 0.18 & 0.20 & 0.19 \\
4 & 5.10 & 0.35 & 5.23 & 0.21 & 1.60 & 0.26 \\
5 & 6.66 & 0.21 & 4.93 & 0.19 & 1.10 & 0.10 \\
6 & 6.46 & 0.33 & 5.58 & 0.18 & 1.20 & 0.35 \\
7 & 6.46 & 0.17 & 6.31 & 0.17 & 2.30 & 0.51 \\
8 & 6.66 & 0.10 & 5.19 & 0.17 & 0.70 & 0.07 \\
9 & 6.00 & 0.31 & 5.35 & 0.20 & 1.40 & 0.56 \\
10 & 8.11 & 0.31 & 5.73 & 0.17 & 0.10 & 0.55 \\
11 & 6.64 & 0.57 & 5.54 & 0.17 & 1.40 & 0.41 \\
12 & 6.66 & 0.11 & 4.91 & 0.17 & 0.80 & 0.58 \\
13 & 7.74 & 0.13 & 7.60 & 0.17 & 1.40 & 0.41 \\
14 & 7.63 & 0.07 & 5.66 & 0.27 & 0.60 & 0.08 \\
15 & 6.66 & 0.59 & 7.02 & 0.17 & 1.90 & 0.10 \\
16 & 5.10 & 0.36 & 6.58 & 0.26 & 4.00 & 0.06 \\
17 & 8.41 & 0.38 & 6.99 & 0.17 & 0.60 & 0.28 \\
18 & 8.31 & 0.26 & 6.16 & 0.17 & 0.10 & 0.54 \\
19 & 6.66 & 0.50 & 5.46 & 0.17 & 1.50 & 0.22
\enddata
\end{deluxetable*}

\subsection{NGC 3256}

NGC 3256 is a late-stage merger containing a large number of star clusters along the inner ($\sim 20''$, or 4 kpc) spiral structure of the nuclear region. The spiral dust lanes extending from the nucleus give this galaxy pockets of high and low extinction. The maximum $A_{V}$ adopted for this galaxy is 3.3 mags of visual extinction (Rich et al. 2012).

\begin{figure*}
\centering
\includegraphics[scale=0.25]{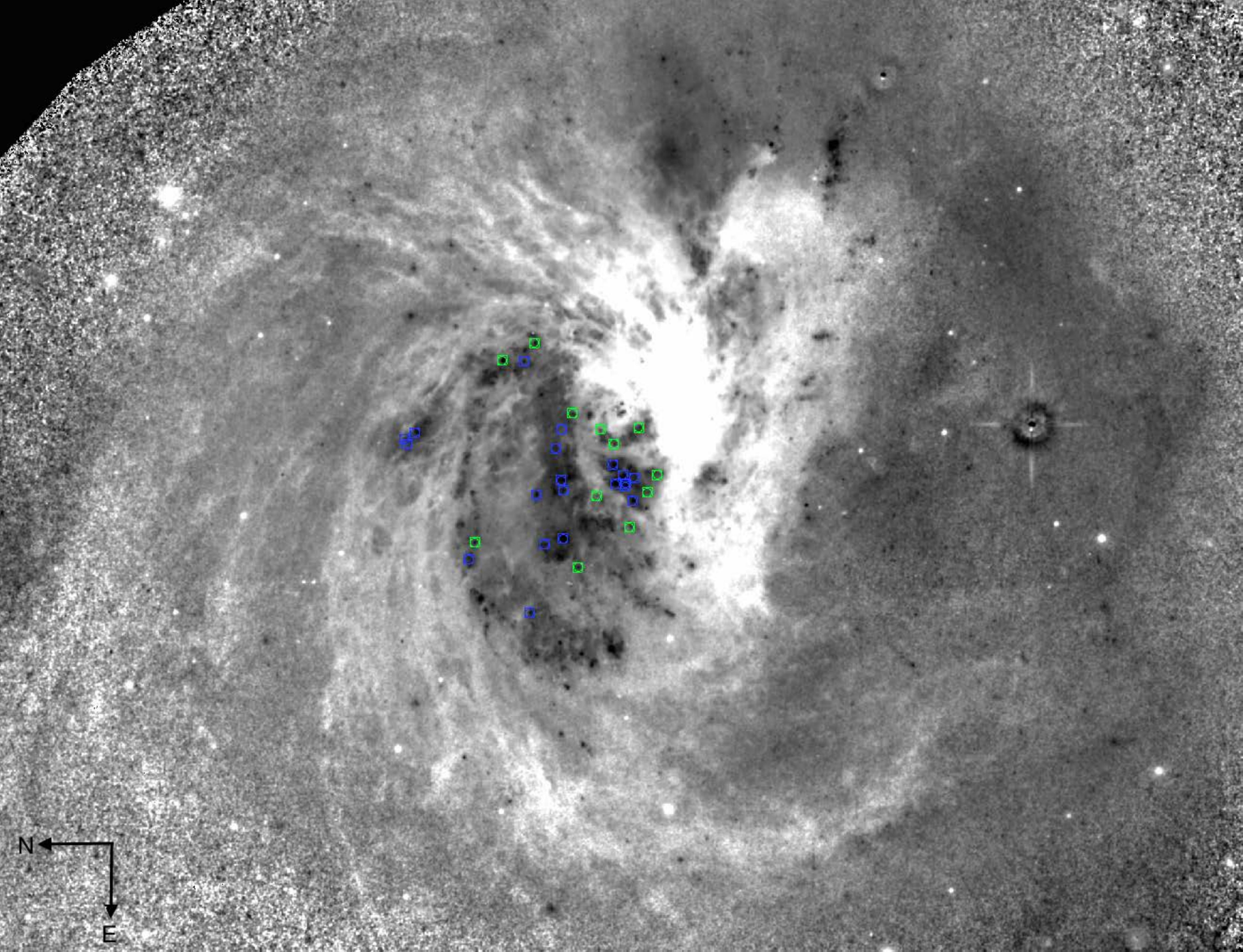}
\caption{Inverted black and white B-I image of NGC 3256 taken with HST ACS/WFC F814W and F435W. The bright emission corresponds to redder (i.e. dustier) regions of the galaxy. The blue centroids correspond to clusters found in relatively ``dust-free'' regions of these galaxies, whereas the green centroids correspond to clusters found in relatively dustier regions of the galaxy.}
\end{figure*}

\begin{figure*}
\centering
\includegraphics[scale=0.55]{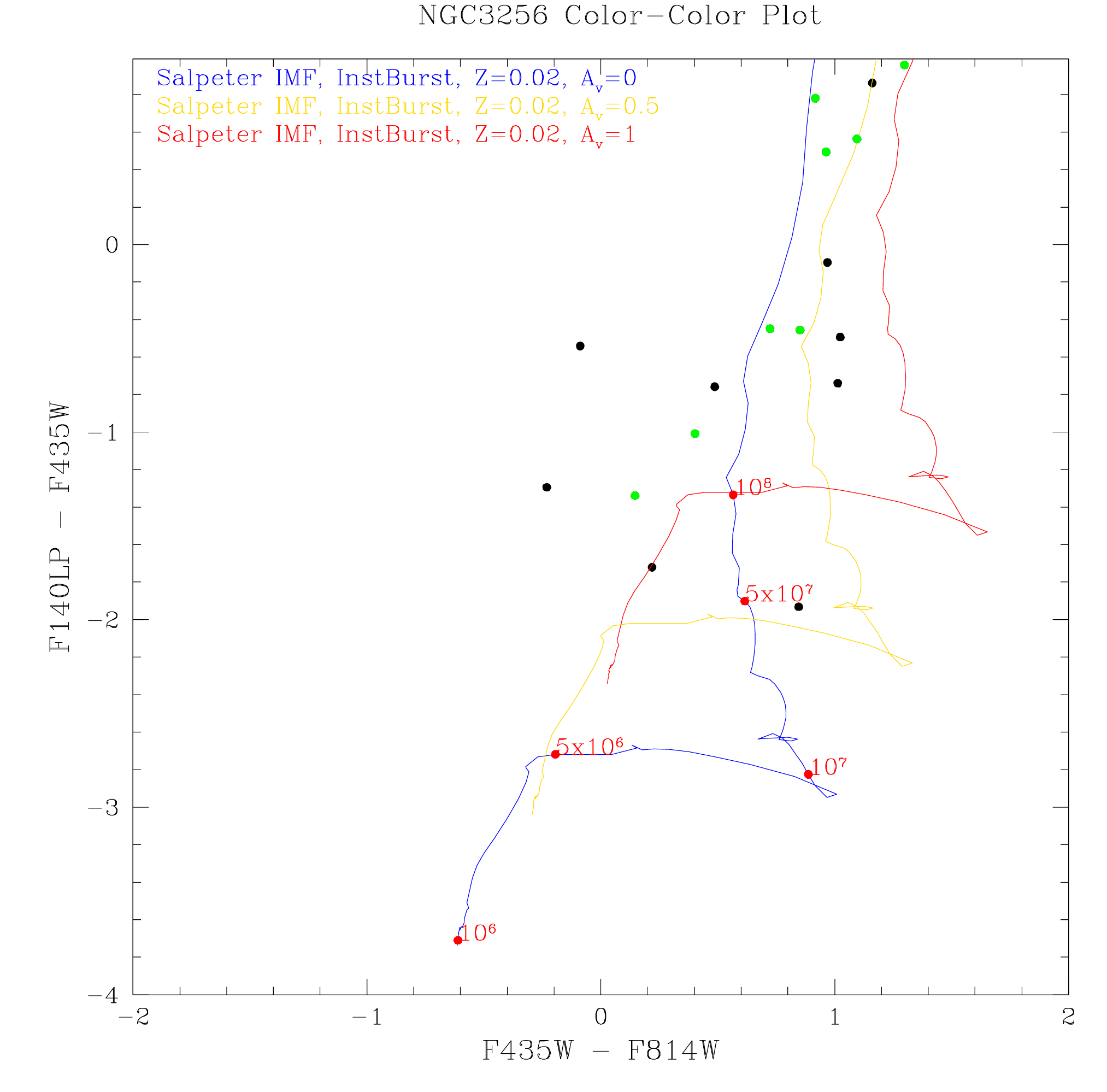}
\caption{Color-Color plot of all star clusters identified in NGC 3256 in F814W, F435W, and F140LP plotted against SSP models with various amount of visual extinction. The green points correspond to the clusters found in dustier regions of the galaxy in Figure 29}
\end{figure*}

\begin{deluxetable*}{ccccccccc}
\tabletypesize{\footnotesize}
\tablewidth{0pt}
\tablecaption {Observed Properties of Star Clusters in NGC 3256}
\tablehead{
\colhead{ID} & \colhead{RA} & \colhead{Dec} & \colhead{$M_{B}$} & \colhead{$\sigma_{B}$} & \colhead{$M_{I}$} & \colhead{$\sigma_{I}$} & \colhead{$M_{FUV}$} & \colhead{$\sigma_{FUV}$}} \\
\startdata
1 & 156.9608449 & -43.9024964 & -12.20 & 0.02 & -12.93 & 0.03 & -12.65 & 0.01 \\
2 & 156.9612737 & -43.90178128 & -12.75 & 0.01 & -12.90 & 0.01 & -14.09 & 0.01 \\
3 & 156.9630369 & -43.90314637 & -12.96 & 0.01 & -13.87 & 0.01 & -12.18 & 0.02 \\
4 & 156.9636042 & -43.9045422 & -14.02 & 0.01 & -14.82 & 0.01 & -12.97 & 0.02 \\
5 & 156.9634535 & -43.90288363 & -12.88 & 0.02 & -14.04 & 0.02 & -12.02 & 0.05 \\
6 & 156.9635947 & -43.90369683 & -13.70 & 0.02 & -15.00 & 0.01 & -12.74 & 0.01 \\
7 & 156.9633212 & -43.89955977 & -12.29 & 0.02 & -13.32 & 0.02 & -12.79 & 0.02 \\
8 & 156.9634618 & -43.89962047 & -11.36 & 0.01 & -12.37 & 0.02 & -12.10 & 0.04 \\
9 & 156.9640714 & -43.90271444 & -13.34 & 0.02 & -14.31 & 0.01 & -13.43 & 0.04 \\
10 & 156.9646123 & -43.90392525 & -13.49 & 0.03 & -13.98 & 0.03 & -14.25 & 0.01 \\
11 & 156.9652121 & -43.90415987 & -13.56 & 0.02 & -13.33 & 0.03 & -14.86 & 0.01 \\
12 & 156.9653068 & -43.90414225 & -13.36 & 0.02 & -13.27 & 0.04 & -13.90 & 0.02 \\
13 & 156.9655228 & -43.9046008 & -12.70 & 0.01 & -13.66 & 0.01 & -12.20 & 0.02 \\
14 & 156.9654982 & -43.90351402 & -13.21 & 0.01 & -14.42 & 0.01 & -11.32 & 0.05 \\
15 & 156.966498 & -43.90415946 & -13.52 & 0.01 & -14.61 & 0.01 & -12.95 & 0.01 \\
16 & 156.9666609 & -43.90272713 & -13.74 & 0.01 & -13.96 & 0.01 & -15.46 & 0.01 \\
17 & 156.9665347 & -43.90083852 & -10.75 & 0.01 & -11.16 & 0.01 & -11.76 & 0.03 \\
18 & 156.9667894 & -43.90232476 & -10.09 & 0.05 & -10.93 & 0.05 & -12.02 & 0.02 \\
19 & 156.9675522 & -43.9029773 & -12.70 & 0.01 & -13.55 & 0.01 & -13.15 & 0.01
\enddata
\end{deluxetable*}

\begin{deluxetable*}{ccccccc}
\tabletypesize{\footnotesize}
\tablewidth{0pt}
\tablecaption {Derived Properties of Star Clusters in NGC 3256}
\tablehead{
\colhead{ID} & \colhead{Log(Age)} & \colhead{$\sigma_{Age}$} & \colhead{Log($M/M_{\odot}$)} & \colhead{$\sigma_{M}$} & \colhead{$A_{V}$} & \colhead{$\sigma_{A_{V}}$}} \\
\startdata  
1 & 6.68 & 0.03 & 5.88 & 0.18 & 1.60 & 0.06 \\
2 & 6.66 & 0.12 & 5.78 & 0.17 & 1.00 & 0.27 \\
3 & 8.36 & 0.19 & 7.10 & 0.17 & 0.80 & 0.07 \\
4 & 8.41 & 0.10 & 7.56 & 0.17 & 0.80 & 0.30 \\
5 & 8.36 & 0.03 & 7.13 & 0.18 & 0.90 & 0.06 \\
6 & 8.36 & 0.03 & 7.51 & 0.20 & 1.00 & 0.11 \\
7 & 7.86 & 0.54 & 6.54 & 0.42 & 0.80 & 0.38 \\
8 & 7.74 & 0.02 & 6.10 & 0.18 & 0.80 & 0.06 \\
9 & 6.54 & 0.13 & 6.61 & 0.27 & 2.20 & 0.20 \\
10 & 6.66 & 0.01 & 6.29 & 0.17 & 1.40 & 0.03 \\
11 & 6.66 & 0.47 & 6.06 & 0.17 & 0.90 & 0.02 \\
12 & 6.66 & 0.22 & 6.24 & 0.17 & 1.40 & 0.03 \\
13 & 8.31 & 0.02 & 6.97 & 0.17 & 0.80 & 0.03 \\
14 & 8.56 & 0.03 & 7.37 & 0.17 & 0.80 & 0.02 \\
15 & 8.31 & 0.10 & 7.30 & 0.17 & 0.80 & 0.08 \\
16 & 6.52 & 0.07 & 6.19 & 0.26 & 1.10 & 0.19 \\
17 & 6.68 & 0.01 & 5.09 & 0.18 & 1.20 & 0.05 \\
18 & 6.86 & 0.03 & 4.80 & 0.17 & 0.80 & 0.37 \\
19 & 6.44 & 0.14 & 6.52 & 0.25 & 2.20 & 0.17
\enddata
\end{deluxetable*}

\subsection{Arp 148}

Arp 148 is an early-stage merger and the only example of a ring galaxy in the sample. This $\sim 23''$ (16 kpc) diameter galaxy is comprised of clumps of star clusters along its perimeter, and throughout much of its interior. The maximum $A_{V}$ adopted for this galaxy is 2.1 mags of visual extinction (Joy \& Harvey 1997). 

\begin{figure*}
\centering
\includegraphics[scale=0.25]{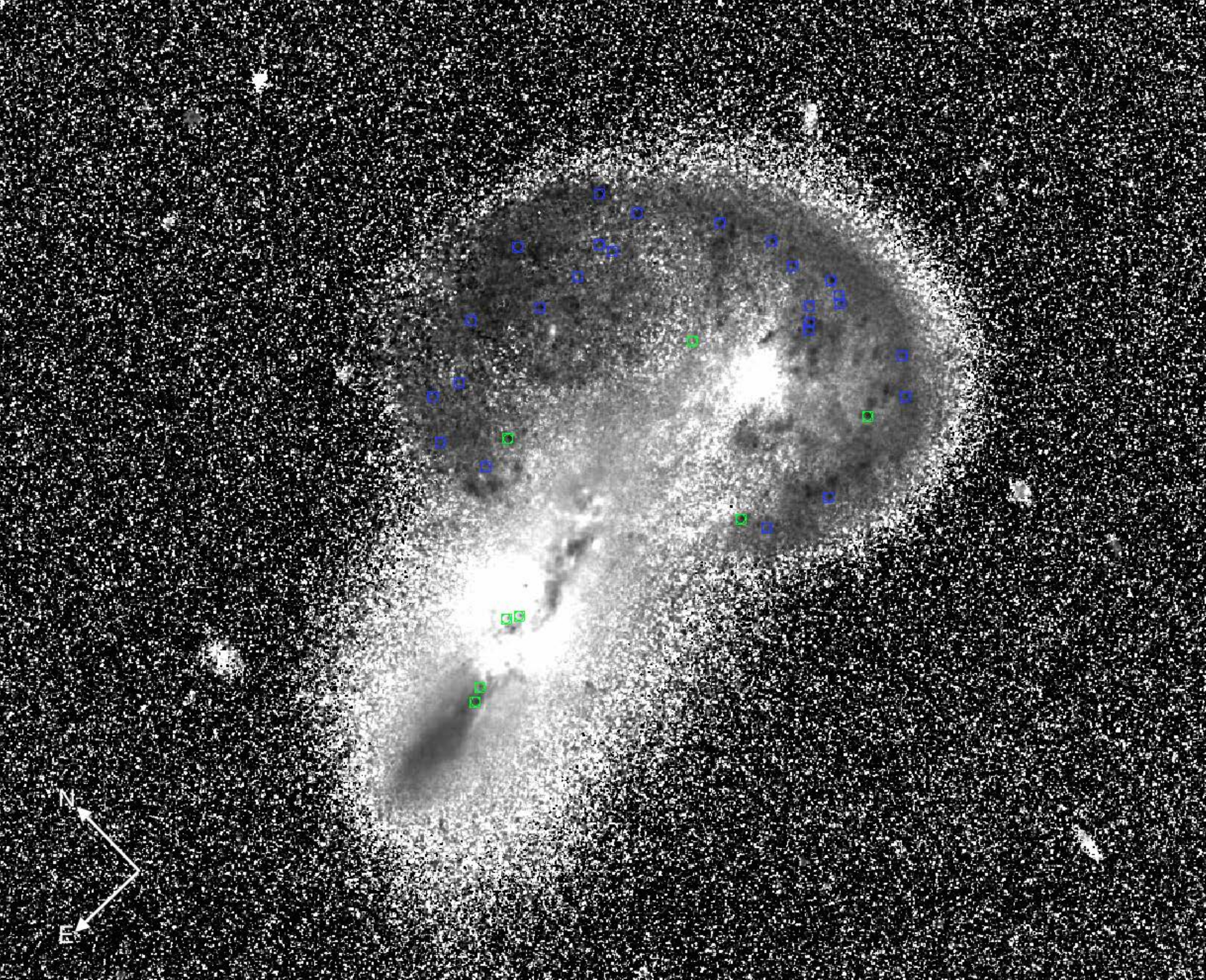}
\caption{Inverted black and white B-I image of Arp 148 taken with HST ACS/WFC F814W and F435W. The bright emission corresponds to redder (i.e. dustier) regions of the galaxy. The blue centroids correspond to clusters found in relatively ``dust-free'' regions of these galaxies, whereas the green centroids correspond to clusters found in relatively dustier regions of the galaxy.}
\end{figure*}

\begin{figure*}
\centering
\includegraphics[scale=0.55]{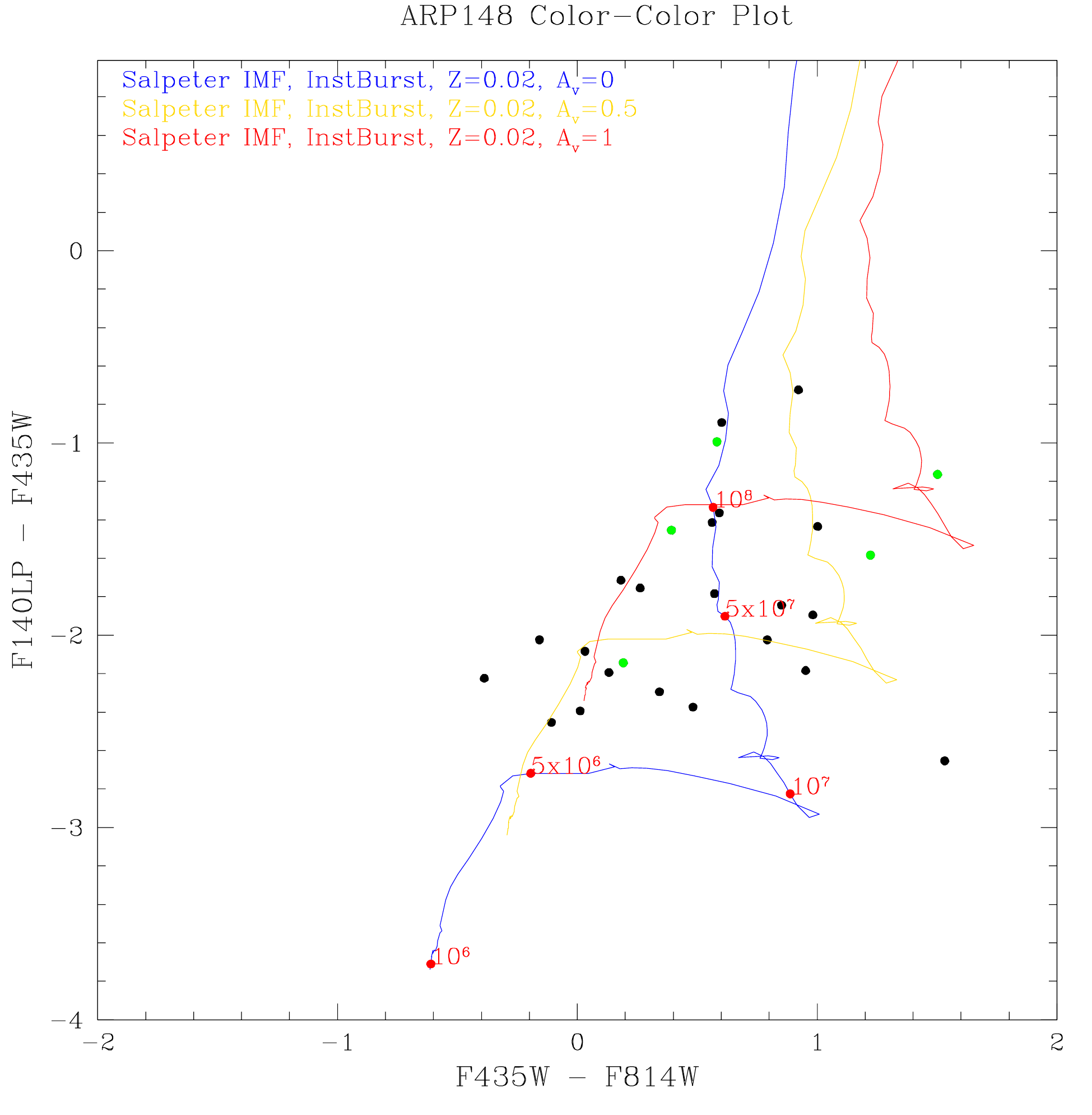}
\caption{Color-Color plot of all star clusters identified in Arp 148 in F814W, F435W, and F140LP plotted against SSP models with various amount of visual extinction. The green points correspond to the clusters found in dustier regions of the galaxy in Figure 31}
\end{figure*}

\begin{deluxetable*}{ccccccccc}
\tabletypesize{\footnotesize}
\tablewidth{0pt}
\tablecaption {Observed Properties of Star Clusters in Arp 148}
\tablehead{
\colhead{ID} & \colhead{RA} & \colhead{Dec} & \colhead{$M_{B}$} & \colhead{$\sigma_{B}$} & \colhead{$M_{I}$} & \colhead{$\sigma_{I}$} & \colhead{$M_{FUV}$} & \colhead{$\sigma_{FUV}$}} \\
\startdata
1 & 165.968436 & 40.8519488 & -12.21 & 0.01 & -12.34 & 0.02 & -14.41 & 0.02 \\
2 & 165.9678663 & 40.8505602 & -11.09 & 0.05 & -11.12 & 0.10 & -13.18 & 0.07 \\
3 & 165.9683844 & 40.85145549 & -11.10 & 0.03 & -11.36 & 0.05 & -12.86 & 0.09 \\
4 & 165.9691046 & 40.85164919 & -11.30 & 0.03 & -11.90 & 0.04 & -12.20 & 0.10 \\
5 & 165.9690976 & 40.85147869 & -11.19 & 0.03 & -12.17 & 0.03 & -13.09 & 0.09 \\
6 & 165.967872 & 40.84958083 & -10.44 & 0.07 & -10.45 & 0.13 & -12.84 & 0.09 \\
7 & 165.9678992 & 40.84893317 & -9.78 & 0.14 & -11.31 & 0.08 & -12.44 & 0.14 \\
8 & 165.9680004 & 40.84887677 & -11.27 & 0.05 & -11.84 & 0.07 & -13.06 & 0.08 \\
9 & 165.9682628 & 40.84916472 & -10.83 & 0.07 & -11.39 & 0.08 & -12.25 & 0.16 \\
10 & 165.9677623 & 40.84910158 & -11.64 & 0.03 & -11.53 & 0.07 & -14.10 & 0.03 \\
11 & 165.9704139 & 40.85187208 & -10.71 & 0.07 & -10.55 & 0.18 & -12.74 & 0.10 \\
12 & 165.968608 & 40.84903015 & -11.16 & 0.06 & -11.64 & 0.09 & -13.54 & 0.05 \\
13 & 165.967709 & 40.84993148 & -11.37 & 0.03 & -11.55 & 0.05 & -13.09 & 0.08 \\
14 & 165.9682107 & 40.84793431 & -10.78 & 0.11 & -11.78 & 0.09 & -12.22 & 0.16 \\
15 & 165.9684927 & 40.8490672 & -11.59 & 0.05 & -11.93 & 0.08 & -13.89 & 0.04 \\
16 & 165.9687209 & 40.84766184 & -12.23 & 0.03 & -12.82 & 0.07 & -13.60 & 0.05 \\
17 & 165.9720438 & 40.85223589 & -10.78 & 0.04 & -11.57 & 0.04 & -12.81 & 0.10 \\
18 & 165.969655 & 40.85013869 & -9.45 & 0.12 & -10.67 & 0.10 & -11.04 & 0.13 \\
19 & 165.9692811 & 40.84792405 & -13.17 & 0.02 & -13.56 & 0.04 & -14.63 & 0.02 \\
20 & 165.9724453 & 40.852421 & -9.47 & 0.14 & -10.32 & 0.14 & -11.32 & 0.01 \\
21 & 165.9724057 & 40.85141491 & -11.51 & 0.02 & -11.70 & 0.04 & -13.66 & 0.07 \\
22 & 165.9716475 & 40.84859071 & -11.98 & 0.02 & -12.56 & 0.03 & -12.98 & 0.09 \\
23 & 165.9715595 & 40.84827508 & -11.99 & 0.03 & -12.91 & 0.04 & -12.72 & 0.11 \\
24 & 165.9706662 & 40.84783769 & -10.84 & 0.04 & -10.45 & 0.16 & -13.07 & 0.08 \\
25 & 165.9729915 & 40.85207186 & -10.18 & 0.12 & -11.13 & 0.11 & -12.37 & 0.14 \\
26 & 165.9748241 & 40.85035121 & -12.83 & 0.04 & -14.33 & 0.09 & -14.00 & 0.03
\enddata
\end{deluxetable*}

\begin{deluxetable*}{ccccccc}
\tabletypesize{\footnotesize}
\tablewidth{0pt}
\tablecaption {Derived Properties of Star Clusters in Arp 148}
\tablehead{
\colhead{ID} & \colhead{Log(Age)} & \colhead{$\sigma_{Age}$} & \colhead{Log($M/M_{\odot}$)} & \colhead{$\sigma_{M}$} & \colhead{$A_{V}$} & \colhead{$\sigma_{A_{V}}$}} \\
\startdata  
1 & 5.10 & 0.67 & 5.96 & 0.39 & 1.10 & 0.35 \\
2 & 6.56 & 0.42 & 4.91 & 0.36 & 0.70 & 0.32 \\
3 & 6.70 & 0.51 & 4.97 & 0.34 & 0.70 & 0.29 \\
4 & 6.60 & 0.56 & 5.35 & 0.40 & 1.40 & 0.36 \\
5 & 7.12 & 0.18 & 5.44 & 0.19 & 0.50 & 0.10 \\
6 & 6.36 & 0.36 & 5.03 & 0.34 & 0.90 & 0.29 \\
7 & 5.10 & 0.01 & 5.52 & 0.27 & 2.10 & 0.15 \\
8 & 6.76 & 0.49 & 5.02 & 0.35 & 0.60 & 0.30 \\
9 & 6.74 & 0.79 & 4.99 & 0.55 & 0.90 & 0.55 \\
10 & 6.60 & 0.57 & 4.90 & 0.36 & 0.30 & 0.32 \\
11 & 6.66 & 0.11 & 4.70 & 0.28 & 0.50 & 0.20 \\
12 & 6.86 & 0.23 & 4.91 & 0.20 & 0.20 & 0.10 \\
13 & 6.52 & 0.18 & 5.25 & 0.31 & 1.10 & 0.26 \\
14 & 7.63 & 0.31 & 5.58 & 0.33 & 0.40 & 0.26 \\
15 & 6.76 & 0.05 & 4.98 & 0.18 & 0.30 & 0.07 \\
16 & 6.72 & 0.70 & 5.59 & 0.58 & 1.00 & 0.60 \\
17 & 7.49 & 0.20 & 5.42 & 0.21 & 0.20 & 0.13 \\
18 & 7.34 & 0.24 & 5.07 & 0.27 & 0.70 & 0.18 \\
19 & 6.42 & 0.22 & 6.39 & 0.33 & 1.50 & 0.28 \\
20 & 7.52 & 0.34 & 4.97 & 0.31 & 0.30 & 0.22 \\
21 & 6.72 & 0.32 & 4.99 & 0.35 & 0.40 & 0.30 \\
22 & 6.46 & 0.46 & 6.13 & 0.50 & 1.80 & 0.49 \\
23 & 6.74 & 0.45 & 5.72 & 0.38 & 1.40 & 0.34 \\
24 & 6.66 & 0.03 & 4.65 & 0.21 & 0.30 & 0.12 \\
25 & 7.30 & 0.23 & 5.13 & 0.27 & 0.30 & 0.17 \\
26 & 7.06 & 0.17 & 6.35 & 0.20 & 1.10 & 0.11
\enddata
\end{deluxetable*}

\subsection{NGC 3690E}

NGC3690 is a mid-stage merger. NGC 3690E contains a multitude of star clusters  and dust lanes from the southeast tip of the galaxy to the northwest. The maximum $A_{V}$ adopted for this galaxy is 3.4 mags of visual extinction (Garcia-Marin et al. 2006). 

\begin{figure*}
\centering
\includegraphics[scale=0.25]{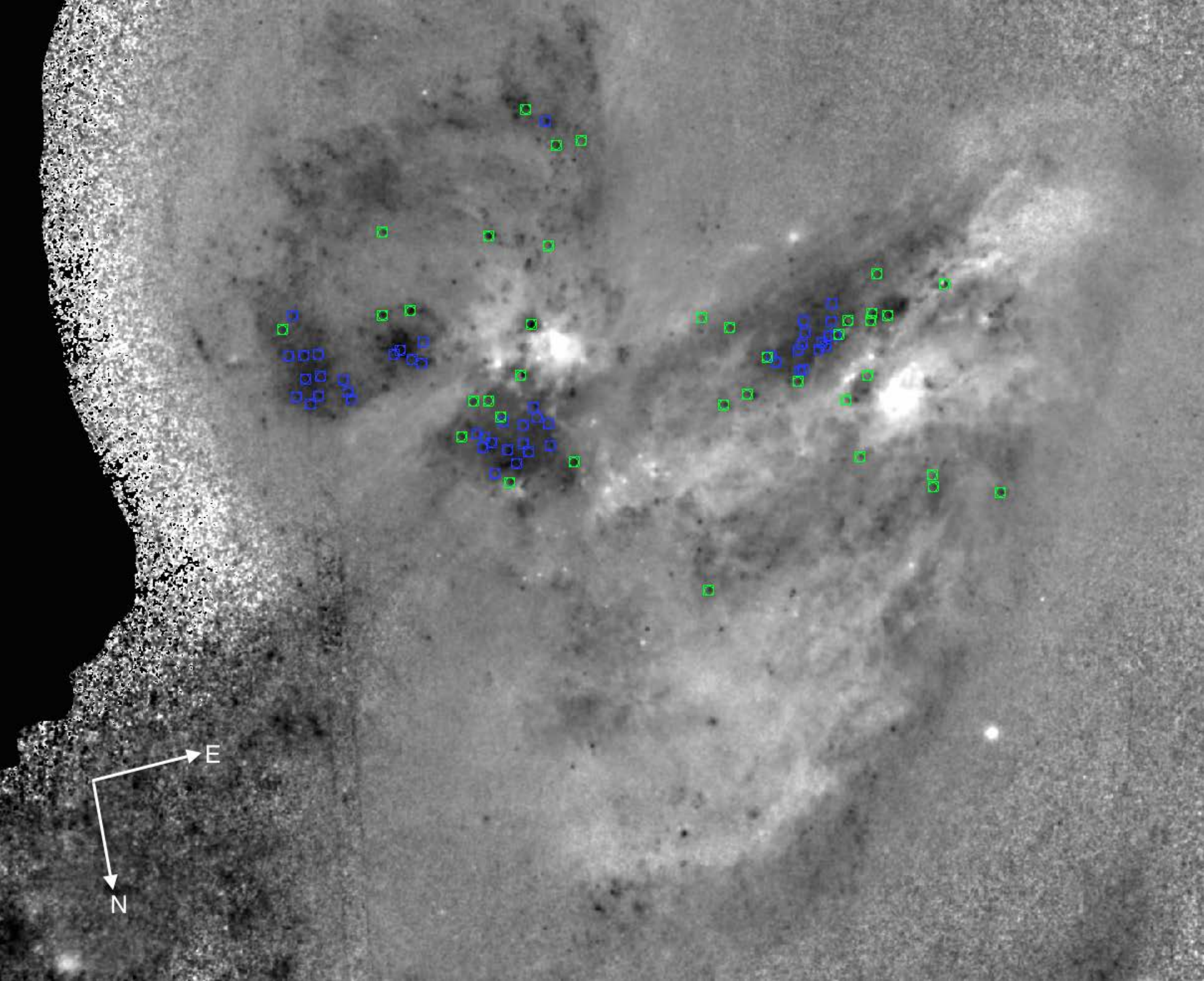}
\caption{Inverted black and white B-I image of NGC 3690E taken with HST ACS/WFC F814W and F435W. The bright emission corresponds to redder (i.e. dustier) regions of the galaxy. The blue centroids correspond to clusters found in relatively ``dust-free'' regions of these galaxies, whereas the green centroids correspond to clusters found in relatively dustier regions of the galaxy.}
\end{figure*}

\begin{figure*}
\centering
\includegraphics[scale=0.55]{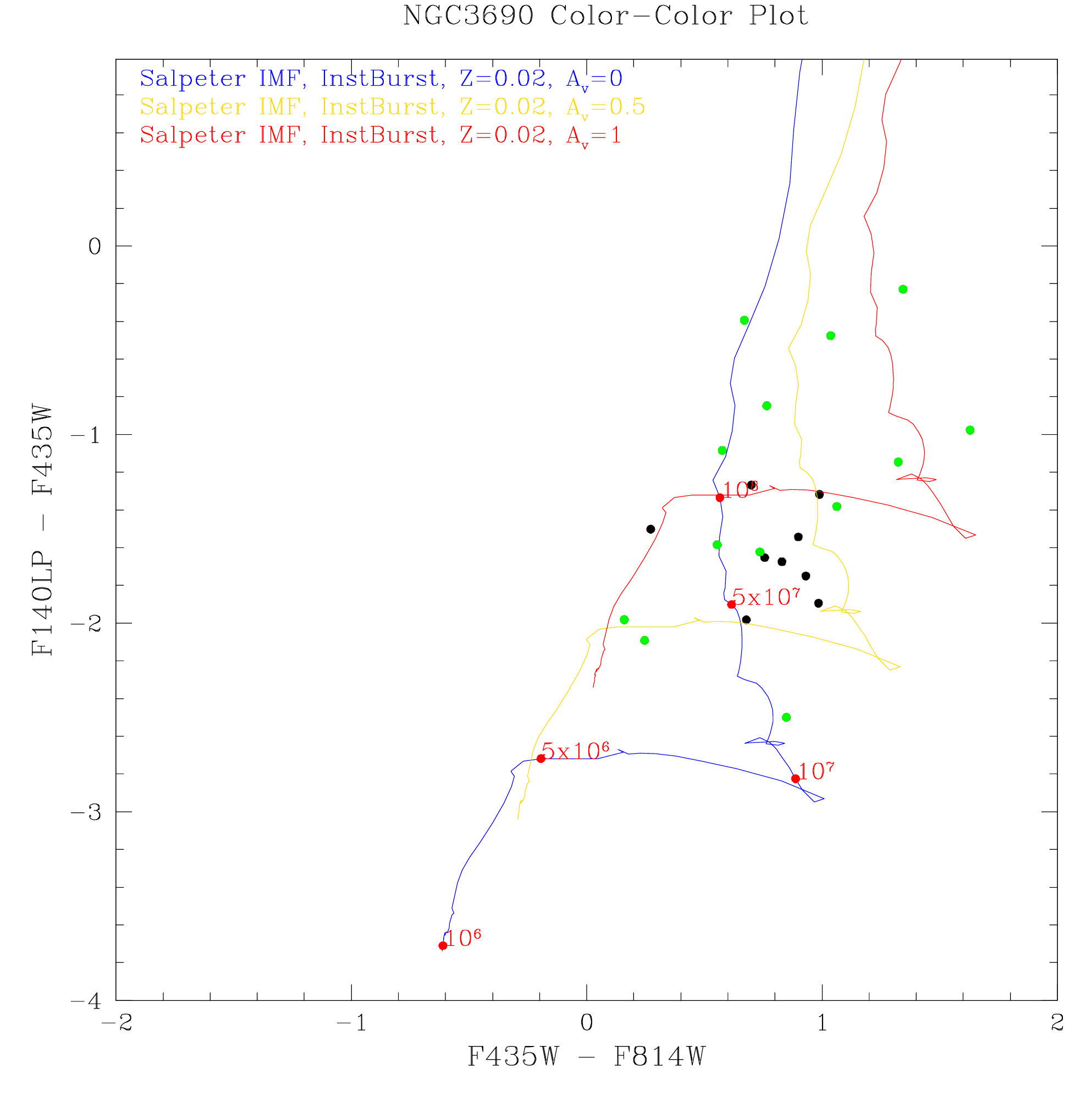}
\caption{Color-Color plot of all star clusters identified in NGC 3690E in F814W, F435W, and F140LP plotted against SSP models with various amount of visual extinction. The green points correspond to the clusters found in dustier regions of the galaxy in Figure 33}
\end{figure*}

\begin{deluxetable*}{ccccccccc}
\tabletypesize{\footnotesize}
\tablewidth{0pt}
\tablecaption {Observed Properties of Star Clusters in NGC 3690E}
\tablehead{
\colhead{ID} & \colhead{RA} & \colhead{Dec} & \colhead{$M_{B}$} & \colhead{$\sigma_{B}$} & \colhead{$M_{I}$} & \colhead{$\sigma_{I}$} & \colhead{$M_{FUV}$} & \colhead{$\sigma_{FUV}$}} \\
\startdata
1 & 172.1336515 & 58.56572969 & -11.56 & 0.01 & -12.23 & 0.01 & -11.95 & 0.02 \\
2 & 172.1424578 & 58.56116578 & -11.76 & 0.01 & -12.83 & 0.02 & -13.14 & 0.01 \\
3 & 172.1388005 & 58.56124188 & -10.20 & 0.05 & -11.19 & 0.04 & -11.52 & 0.05 \\
4 & 172.1400284 & 58.56149757 & -10.81 & 0.04 & -10.97 & 0.14 & -12.79 & 0.02 \\
5 & 172.1345803 & 58.56118873 & -8.96 & 0.04 & -9.81 & 0.05 & -11.46 & 0.03 \\
6 & 172.1392395 & 58.56156972 & -12.52 & 0.02 & -14.15 & 0.01 & -13.50 & 0.02 \\
7 & 172.1399479 & 58.56161791 & -11.73 & 0.02 & -12.28 & 0.02 & -13.31 & 0.02 \\
8 & 172.1386985 & 58.56153798 & -10.07 & 0.08 & -10.97 & 0.10 & -11.61 & 0.10 \\
9 & 172.1378106 & 58.56146996 & -10.01 & 0.04 & -11.00 & 0.04 & -11.91 & 0.04 \\
10 & 172.1387899 & 58.56177443 & -12.85 & 0.02 & -13.61 & 0.03 & -14.51 & 0.02 \\
11 & 172.1385677 & 58.56177224 & -12.15 & 0.03 & -12.98 & 0.03 & -13.82 & 0.02 \\
12 & 172.1376673 & 58.5618607 & -11.11 & 0.03 & -11.81 & 0.03 & -12.38 & 0.04 \\
13 & 172.138187 & 58.56197966 & -11.39 & 0.05 & -12.33 & 0.07 & -13.14 & 0.05 \\
14 & 172.1366287 & 58.56197706 & -10.22 & 0.18 & -12.26 & 0.06 & -11.91 & 0.06 \\
15 & 172.1365065 & 58.56198534 & -10.79 & 0.10 & -12.11 & 0.05 & -11.93 & 0.05 \\
16 & 172.1367658 & 58.56209285 & -10.89 & 0.04 & -11.16 & 0.06 & -12.39 & 0.04 \\
17 & 172.1376122 & 58.56226713 & -10.54 & 0.10 & -11.22 & 0.08 & -12.52 & 0.05 \\
18 & 172.1357033 & 58.56256555 & -9.68 & 0.08 & -9.93 & 0.10 & -11.77 & 0.02 \\
19 & 172.1349188 & 58.56268913 & -11.06 & 0.01 & -11.80 & 0.02 & -12.68 & 0.01 \\
20 & 172.1390305 & 58.56385382 & -11.61 & 0.02 & -12.65 & 0.02 & -12.09 & 0.02 \\
21 & 172.1412556 & 58.56431432 & -11.46 & 0.04 & -12.81 & 0.02 & -11.69 & 0.04 \\
22 & 172.1412392 & 58.56451117 & -11.52 & 0.02 & -12.28 & 0.03 & -12.36 & 0.01 \\
23 & 172.1433592 & 58.56475551 & -11.38 & 0.01 & -11.95 & 0.01 & -12.46 & 0.01
\enddata
\end{deluxetable*}

\begin{deluxetable*}{ccccccc}
\tabletypesize{\footnotesize}
\tablewidth{0pt}
\tablecaption {Derived Properties of Star Clusters in NGC 3690E}
\tablehead{
\colhead{ID} & \colhead{Log(Age)} & \colhead{$\sigma_{Age}$} & \colhead{Log($M/M_{\odot}$)} & \colhead{$\sigma_{M}$} & \colhead{$A_{V}$} & \colhead{$\sigma_{A_{V}}$}} \\
\startdata  
1 & 8.46 & 0.01 & 6.13 & 0.17 & 0.01 & 0.15 \\
2 & 6.94 & 0.05 & 5.66 & 0.17 & 1.00 & 0.35 \\
3 & 7.00 & 0.23 & 5.22 & 0.24 & 1.10 & 0.15 \\
4 & 6.56 & 0.43 & 5.06 & 0.42 & 1.00 & 0.38 \\
5 & 6.82 & 0.01 & 3.99 & 0.19 & 0.20 & 0.06 \\
6 & 6.82 & 0.83 & 5.99 & 0.18 & 1.30 & 0.04 \\
7 & 7.76 & 0.31 & 5.88 & 0.31 & 0.20 & 0.25 \\
8 & 6.98 & 0.30 & 4.99 & 0.27 & 0.90 & 0.19 \\
9 & 6.84 & 0.04 & 4.64 & 0.19 & 0.60 & 0.06 \\
10 & 7.24 & 0.15 & 6.25 & 0.22 & 0.70 & 0.13 \\
11 & 6.98 & 0.20 & 5.77 & 0.20 & 0.80 & 0.10 \\
12 & 6.74 & 0.46 & 5.31 & 0.41 & 1.20 & 0.37 \\
13 & 6.92 & 0.12 & 5.32 & 0.19 & 0.70 & 0.06 \\
14 & 7.65 & 0.24 & 5.21 & 0.30 & 0.20 & 0.15 \\
15 & 6.84 & 0.11 & 5.26 & 0.22 & 1.20 & 0.09 \\
16 & 6.52 & 0.30 & 5.29 & 0.37 & 1.40 & 0.32 \\
17 & 7.42 & 0.30 & 5.23 & 0.30 & 0.30 & 0.21 \\
18 & 6.72 & 0.45 & 4.49 & 0.42 & 0.70 & 0.38 \\
19 & 7.12 & 0.20 & 5.49 & 0.25 & 0.80 & 0.17 \\
20 & 7.76 & 0.27 & 6.26 & 0.29 & 1.00 & 0.22 \\
21 & 7.57 & 0.27 & 6.30 & 0.29 & 1.40 & 0.22 \\
22 & 8.16 & 0.84 & 6.01 & 0.69 & 0.20 & 0.74 \\
23 & 6.42 & 0.42 & 5.91 & 0.47 & 1.90 & 0.44
\enddata
\end{deluxetable*}

\subsection{NGC 3690W}

NGC 3690W has the brightest cluster complexes in the merging system located $\sim 6.8''$ (1.6 kpc) from the resolved nucleus. Cluster-rich spiral arms extend north and westward from the nuclear region out to a maximum projected distance of $\sim 58''$ (14 kpc). The maximum $A_{V}$ adopted for this galaxy is 3.9 mags of visual extinction (Garcia-Marin et al. 2006). 


\begin{figure*}
\centering
\includegraphics[scale=0.55]{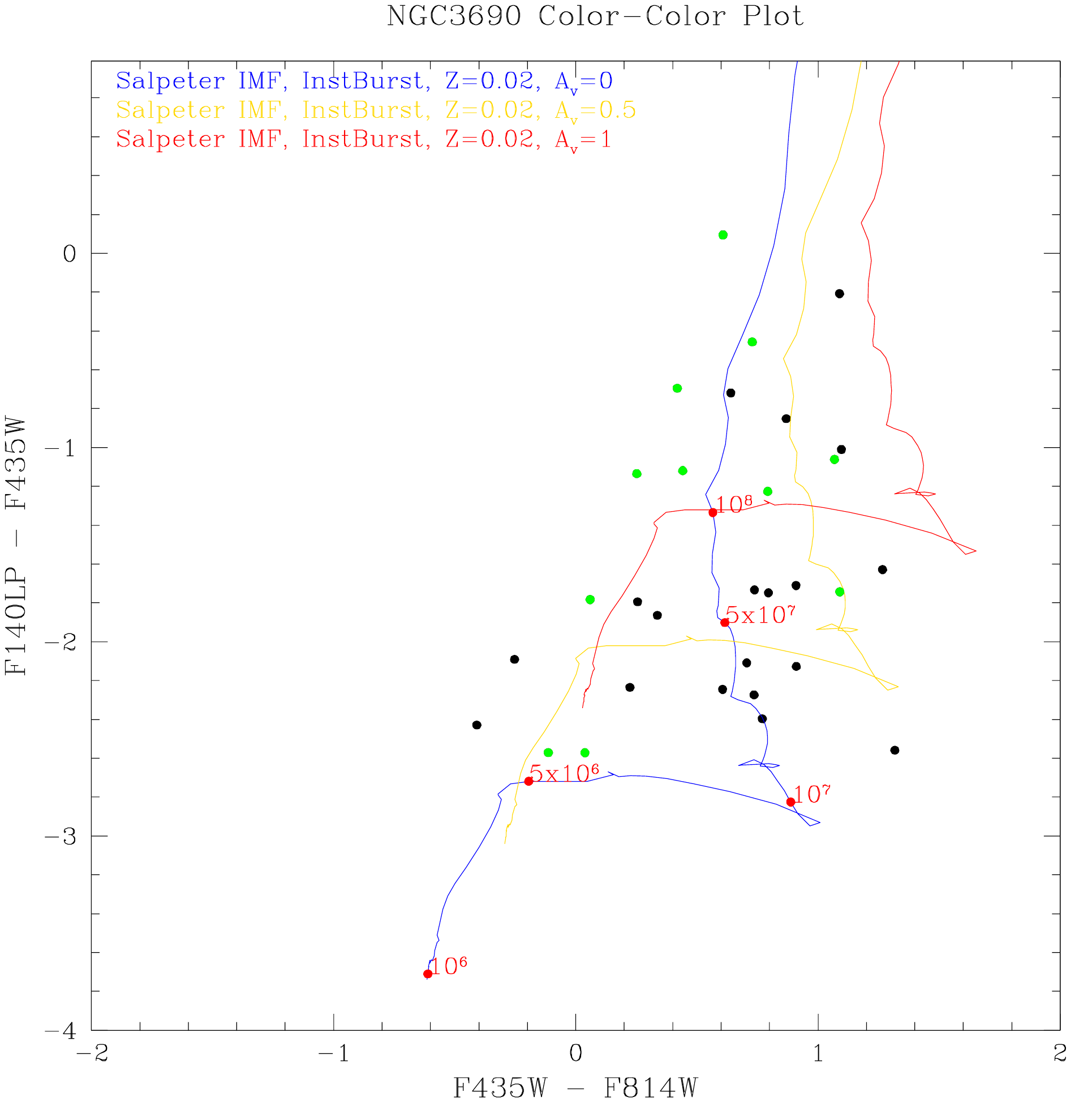}
\caption{Color-Color plot of all star clusters identified in NGC 3690W in F814W, F435W, and F140LP plotted against SSP models with various amount of visual extinction. The green points correspond to the clusters found in dustier regions of the galaxy in Figure 33}
\end{figure*}

\begin{deluxetable*}{ccccccccc}
\tabletypesize{\footnotesize}
\tablewidth{0pt}
\tablecaption {Observed Properties of Star Clusters in NGC 3690W}
\tablehead{
\colhead{ID} & \colhead{RA} & \colhead{Dec} & \colhead{$M_{B}$} & \colhead{$\sigma_{B}$} & \colhead{$M_{I}$} & \colhead{$\sigma_{I}$} & \colhead{$M_{FUV}$} & \colhead{$\sigma_{FUV}$}} \\
\startdata
1 & 172.1298769 & 58.55733473 & -11.32 & 0.01 & -12.11 & 0.01 & -12.55 & 0.02 \\
2 & 172.128158 & 58.55936534 & -10.84 & 0.01 & -10.73 & 0.02 & -13.41 & 0.01 \\
3 & 172.1300076 & 58.5596496 & -11.44 & 0.01 & -12.51 & 0.01 & -12.50 & 0.01 \\
4 & 172.1253481 & 58.56042622 & -11.36 & 0.01 & -11.42 & 0.01 & -13.14 & 0.02 \\
5 & 172.1244337 & 58.56044047 & -10.15 & 0.03 & -10.19 & 0.04 & -12.73 & 0.02 \\
6 & 172.1291326 & 58.56092032 & -14.92 & 0.01 & -15.17 & 0.02 & -16.05 & 0.02 \\
7 & 172.1248588 & 58.5610534 & -14.22 & 0.01 & -14.83 & 0.01 & -16.47 & 0.01 \\
8 & 172.1246296 & 58.56111739 & -11.76 & 0.03 & -12.53 & 0.01 & -14.16 & 0.02 \\
9 & 172.1222293 & 58.56095288 & -10.70 & 0.02 & -11.44 & 0.03 & -12.43 & 0.03 \\
10 & 172.1251824 & 58.56124124 & -10.38 & 0.14 & -11.70 & 0.06 & -12.94 & 0.04 \\
11 & 172.1254777 & 58.56133083 & -10.78 & 0.04 & -11.00 & 0.06 & -13.01 & 0.03 \\
12 & 172.1222096 & 58.56131918 & -11.41 & 0.03 & -12.32 & 0.03 & -13.12 & 0.03 \\
13 & 172.1230124 & 58.56163962 & -12.19 & 0.01 & -12.99 & 0.01 & -13.94 & 0.01 \\
14 & 172.122073 & 58.5616329 & -10.06 & 0.04 & -9.65 & 0.14 & -12.49 & 0.02 \\
15 & 172.1231006 & 58.56176699 & -10.62 & 0.02 & -11.36 & 0.02 & -12.90 & 0.02 \\
16 & 172.1274401 & 58.56209845 & -11.04 & 0.07 & -12.13 & 0.05 & -12.79 & 0.03 \\
17 & 172.126965 & 58.5620665 & -13.05 & 0.01 & -13.66 & 0.02 & -12.96 & 0.04 \\
18 & 172.1217838 & 58.56176205 & -11.84 & 0.01 & -12.55 & 0.01 & -13.95 & 0.01 \\
19 & 172.128929 & 58.56247605 & -10.89 & 0.09 & -12.16 & 0.07 & -12.52 & 0.07 \\
20 & 172.1292452 & 58.56260576 & -12.98 & 0.01 & -14.07 & 0.01 & -13.19 & 0.02 \\
21 & 172.1284475 & 58.56257873 & -10.60 & 0.06 & -11.51 & 0.07 & -12.73 & 0.03 \\
22 & 172.1269226 & 58.56262585 & -12.87 & 0.02 & -13.74 & 0.02 & -13.72 & 0.02 \\
23 & 172.1264357 & 58.56263441 & -13.24 & 0.02 & -13.66 & 0.03 & -13.93 & 0.02 \\
24 & 172.1271745 & 58.56269741 & -14.01 & 0.01 & -15.11 & 0.01 & -15.02 & 0.01 \\
25 & 172.1292261 & 58.56297394 & -12.57 & 0.01 & -13.21 & 0.01 & -13.29 & 0.02 \\
26 & 172.1278642 & 58.56295651 & -12.73 & 0.03 & -12.98 & 0.04 & -14.52 & 0.02 \\
27 & 172.1285202 & 58.56303529 & -11.18 & 0.04 & -10.93 & 0.13 & -13.27 & 0.03 \\
28 & 172.1299061 & 58.56329793 & -12.44 & 0.01 & -12.88 & 0.02 & -13.56 & 0.01 \\
29 & 172.1280723 & 58.56319834 & -12.29 & 0.05 & -12.62 & 0.05 & -14.15 & 0.02 \\
30 & 172.1277823 & 58.56349895 & -14.73 & 0.01 & -15.46 & 0.02 & -15.18 & 0.02
\enddata
\end{deluxetable*}

\begin{deluxetable*}{ccccccc}
\tabletypesize{\footnotesize}
\tablewidth{0pt}
\tablecaption {Derived Properties of Star Clusters in NGC 3690W}
\tablehead{
\colhead{ID} & \colhead{Log(Age)} & \colhead{$\sigma_{Age}$} & \colhead{Log($M/M_{\odot}$)} & \colhead{$\sigma_{M}$} & \colhead{$A_{V}$} & \colhead{$\sigma_{A_{V}}$}} \\
\startdata  
1 & 6.78 & 0.02 & 5.27 & 0.18 & 1.00 & 0.05 \\
2 & 6.42 & 0.16 & 5.03 & 0.19 & 0.70 & 0.06 \\
3 & 7.63 & 0.34 & 6.00 & 0.30 & 0.70 & 0.23 \\
4 & 6.66 & 0.10 & 5.07 & 0.17 & 0.70 & 1.12 \\
5 & 5.10 & 0.40 & 5.03 & 0.18 & 0.90 & 0.04 \\
6 & 6.66 & 8.74 & 6.70 & 0.18 & 1.10 & 0.03 \\
7 & 6.84 & 0.01 & 6.26 & 0.17 & 0.50 & 0.30 \\
8 & 6.88 & 0.01 & 5.35 & 0.17 & 0.50 & 0.17 \\
9 & 6.84 & 0.01 & 4.95 & 0.18 & 0.70 & 0.04 \\
10 & 5.10 & 0.01 & 6.71 & 0.23 & 3.90 & 0.02 \\
11 & 5.10 & 0.67 & 5.39 & 0.29 & 1.10 & 0.22 \\
12 & 6.88 & 0.04 & 5.32 & 0.18 & 0.70 & 0.05 \\
13 & 6.86 & 0.10 & 5.59 & 0.17 & 0.70 & 1.12 \\
14 & 6.52 & 0.02 & 4.40 & 0.19 & 0.50 & 0.06 \\
15 & 6.88 & 0.01 & 4.90 & 0.17 & 0.50 & 0.20 \\
16 & 7.42 & 0.20 & 5.65 & 0.21 & 0.50 & 0.09 \\
17 & 8.31 & 0.35 & 6.95 & 0.35 & 0.50 & 0.30 \\
18 & 6.86 & 0.01 & 5.34 & 0.17 & 0.50 & 0.99 \\
19 & 5.10 & 0.19 & 6.92 & 0.24 & 3.90 & 0.12 \\
20 & 8.01 & 0.48 & 6.89 & 0.41 & 0.80 & 0.37 \\
21 & 6.90 & 0.10 & 4.94 & 0.19 & 0.50 & 0.42 \\
22 & 6.74 & 0.58 & 6.02 & 0.44 & 1.30 & 0.41 \\
23 & 6.66 & 6.61 & 6.19 & 0.18 & 1.40 & 0.04 \\
24 & 7.65 & 0.29 & 7.05 & 0.25 & 0.70 & 0.18 \\
25 & 6.62 & 0.06 & 5.92 & 0.23 & 1.50 & 0.14 \\
26 & 6.38 & 0.49 & 6.18 & 0.36 & 1.30 & 0.30 \\
27 & 6.66 & 0.08 & 4.89 & 0.22 & 0.50 & 0.12 \\
28 & 6.64 & 0.06 & 5.73 & 0.23 & 1.20 & 0.15 \\
29 & 6.72 & 0.43 & 5.40 & 0.35 & 0.60 & 0.30 \\
30 & 6.68 & 0.03 & 6.89 & 0.18 & 1.60 & 0.05
\enddata
\end{deluxetable*}

\subsection{NGC 5257E}

NGC5257/8 is an early-stage merger system with the E and W nuclei separated by $\sim 80''$ (40 kpc). Star clusters and dust lanes make up the prominent spiral arms seen in the eastern galaxy. The maximum $A_{V}$ adopted for this galaxy is 2.6 mags of visual extinction (Smith et al. 2014). 

\begin{figure*}
\centering
\includegraphics[scale=0.25]{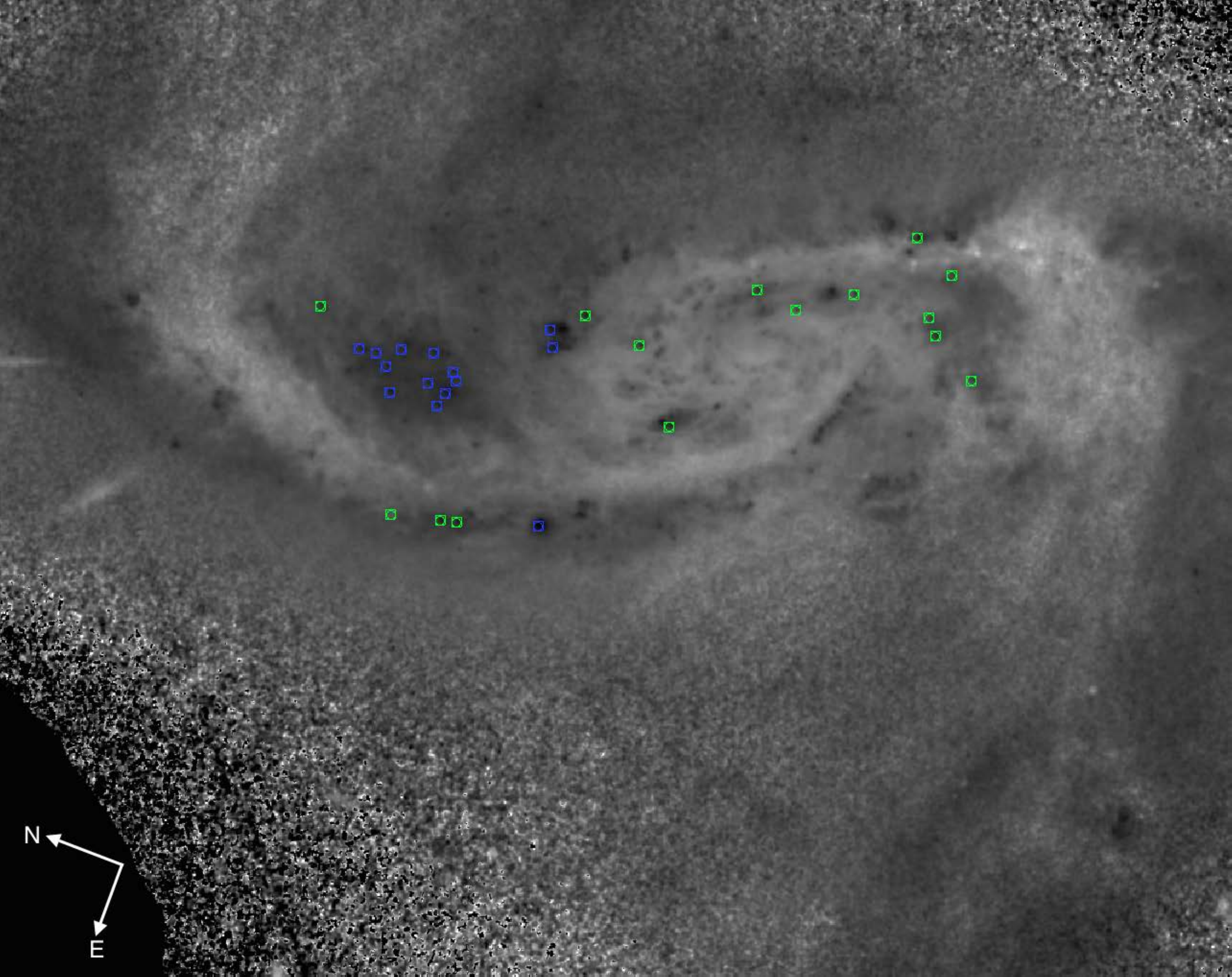}
\caption{Inverted black and white B-I image of NGC 5257E taken with HST ACS/WFC F814W and F435W. The bright emission corresponds to redder (i.e. dustier) regions of the galaxy. The blue centroids correspond to clusters found in relatively ``dust-free'' regions of these galaxies, whereas the green centroids correspond to clusters found in relatively dustier regions of the galaxy.}
\end{figure*}

\begin{figure*}
\centering
\includegraphics[scale=0.55]{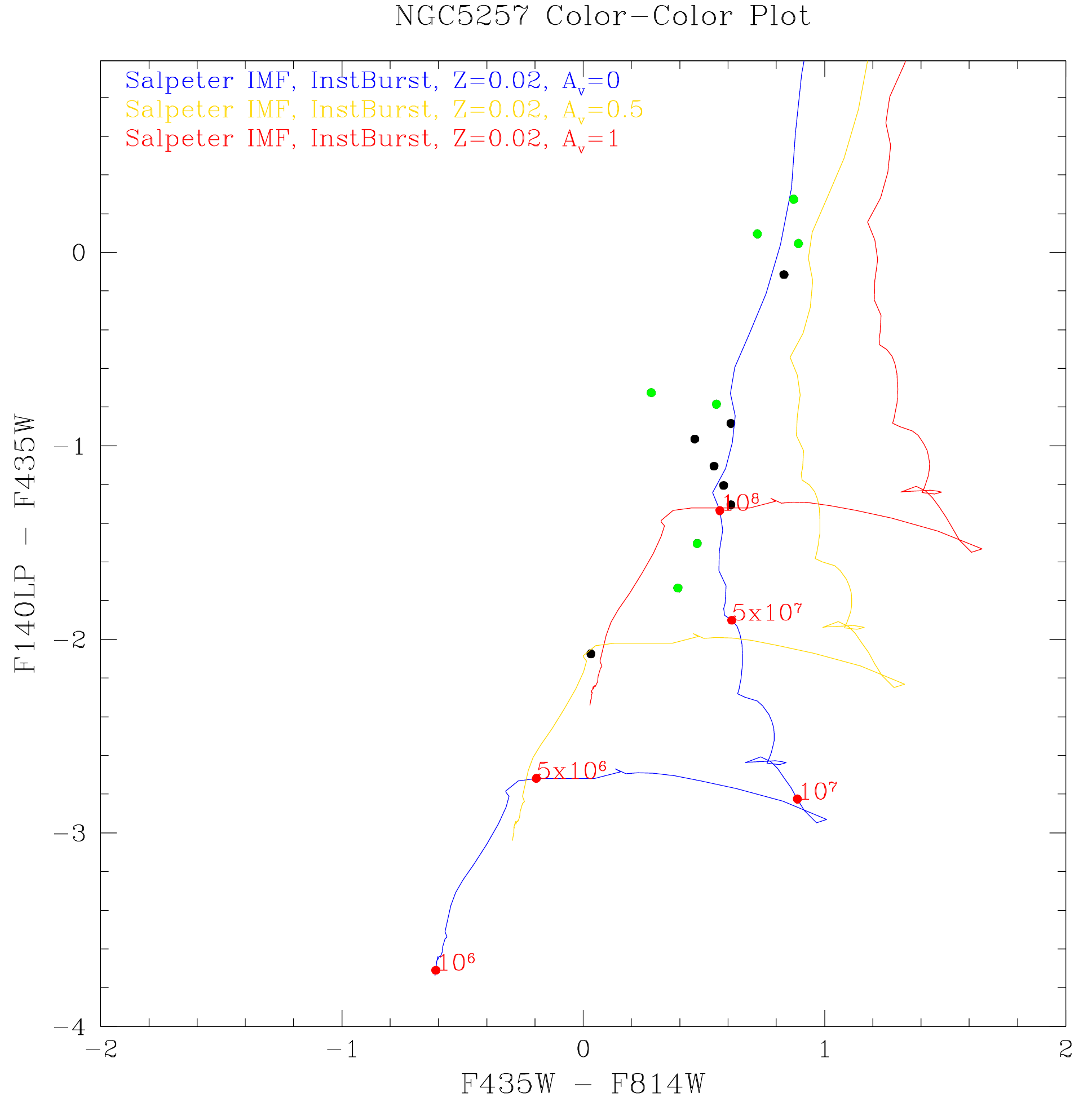}
\caption{Color-Color plot of all star clusters identified in NGC 5257E F814W, F435W, and F140LP plotted against SSP models with various amount of visual extinction. The green points correspond to the clusters found in dustier regions of the galaxy in Figure 36}
\end{figure*}

\begin{deluxetable*}{ccccccccc}
\tabletypesize{\footnotesize}
\tablewidth{0pt}
\tablecaption {Observed Properties of Star Clusters in NGC 5257E}
\tablehead{
\colhead{ID} & \colhead{RA} & \colhead{Dec} & \colhead{$M_{B}$} & \colhead{$\sigma_{B}$} & \colhead{$M_{I}$} & \colhead{$\sigma_{I}$} & \colhead{$M_{FUV}$} & \colhead{$\sigma_{FUV}$}} \\
\startdata
1 & 204.988389 & 0.82941763 & -15.90 & 0.01 & -16.62 & 0.02 & -15.81 & 0.01 \\
2 & 204.9886135 & 0.828844213 & -13.70 & 0.01 & -14.98 & 0.01 & -12.25 & 0.07 \\
3 & 204.9893355 & 0.829802618 & -12.13 & 0.01 & -12.41 & 0.04 & -12.86 & 0.04 \\
4 & 204.99099 & 0.832602833 & -12.33 & 0.01 & -12.80 & 0.02 & -13.84 & 0.02 \\
5 & 204.9913336 & 0.832908072 & -12.27 & 0.05 & -12.88 & 0.03 & -13.58 & 0.04 \\
6 & 204.9893535 & 0.828699614 & -12.67 & 0.02 & -13.22 & 0.03 & -13.46 & 0.02 \\
7 & 204.9910317 & 0.83185232 & -13.36 & 0.01 & -14.25 & 0.02 & -13.32 & 0.03 \\
8 & 204.9915112 & 0.832786319 & -12.80 & 0.06 & -13.26 & 0.05 & -13.77 & 0.05 \\
9 & 204.992339 & 0.834417957 & -11.11 & 0.10 & -11.94 & 0.07 & -11.23 & 0.18 \\
10 & 204.9921984 & 0.834047212 & -11.78 & 0.04 & -12.36 & 0.05 & -12.99 & 0.04 \\
11 & 204.9928594 & 0.834311205 & -13.27 & 0.01 & -13.81 & 0.01 & -14.38 & 0.01 \\
12 & 204.9927539 & 0.833729231 & -12.57 & 0.02 & -13.18 & 0.02 & -13.46 & 0.02 \\
13 & 204.9917487 & 0.831103146 & -14.21 & 0.02 & -15.08 & 0.02 & -13.94 & 0.01 \\
14 & 204.9941781 & 0.833654938 & -13.58 & 0.01 & -15.28 & 0.01 & -12.41 & 0.06 \\
15 & 204.9935202 & 0.831994291 & -15.04 & 0.01 & -15.07 & 0.01 & -17.12 & 0.01 \\
16 & 204.993914 & 0.832897563 & -12.26 & 0.04 & -12.65 & 0.05 & -14.00 & 0.01
\enddata
\end{deluxetable*}

\begin{deluxetable*}{ccccccc}
\tabletypesize{\footnotesize}
\tablewidth{0pt}
\tablecaption {Derived Properties of Star Clusters in NGC 5257E}
\tablehead{
\colhead{ID} & \colhead{Log(Age)} & \colhead{$\sigma_{Age}$} & \colhead{Log($M/M_{\odot}$)} & \colhead{$\sigma_{M}$} & \colhead{$A_{V}$} & \colhead{$\sigma_{A_{V}}$}} \\
\startdata  
1 & 8.31 & 0.01 & 8.04 & 0.17 & 0.40 & 0.23 \\
2 & 8.56 & 0.03 & 7.46 & 0.18 & 0.60 & 0.07 \\
3 & 6.66 & 0.39 & 5.75 & 0.17 & 1.40 & 0.04 \\
4 & 6.72 & 0.67 & 5.58 & 0.39 & 0.90 & 0.34 \\
5 & 6.72 & 0.58 & 5.61 & 0.36 & 1.00 & 0.30 \\
6 & 6.66 & 0.03 & 5.96 & 0.19 & 1.40 & 0.07 \\
7 & 8.31 & 0.49 & 7.02 & 0.48 & 0.40 & 0.46 \\
8 & 6.66 & 0.40 & 5.96 & 0.29 & 1.30 & 0.22 \\
9 & 6.66 & 0.81 & 5.60 & 0.66 & 1.90 & 0.70 \\
10 & 6.00 & 0.44 & 6.06 & 0.37 & 1.80 & 0.32 \\
11 & 6.30 & 0.09 & 6.68 & 0.22 & 1.80 & 0.13 \\
12 & 6.52 & 0.39 & 6.04 & 0.31 & 1.70 & 0.24 \\
13 & 8.36 & 0.02 & 7.45 & 0.18 & 0.50 & 0.06 \\
14 & 8.01 & 0.05 & 7.66 & 0.18 & 1.80 & 0.04 \\
15 & 6.56 & 0.05 & 6.49 & 0.21 & 0.70 & 0.11 \\
16 & 6.72 & 0.43 & 5.45 & 0.36 & 0.70 & 0.31
\enddata
\end{deluxetable*}

\subsection{NGC 5257W}

NGC 5257W contains a prominent group of bright clusters located $\sim 10''$ (5 kpc) south from the nucleus in a $\sim 17''$ (4.2 kpc) long spiral arm. The maximum $A_{V}$ adopted for this galaxy is 1.8 mags of visual extinction (Smith et al. 2014). 

\begin{figure*}
\centering
\includegraphics[scale=0.3]{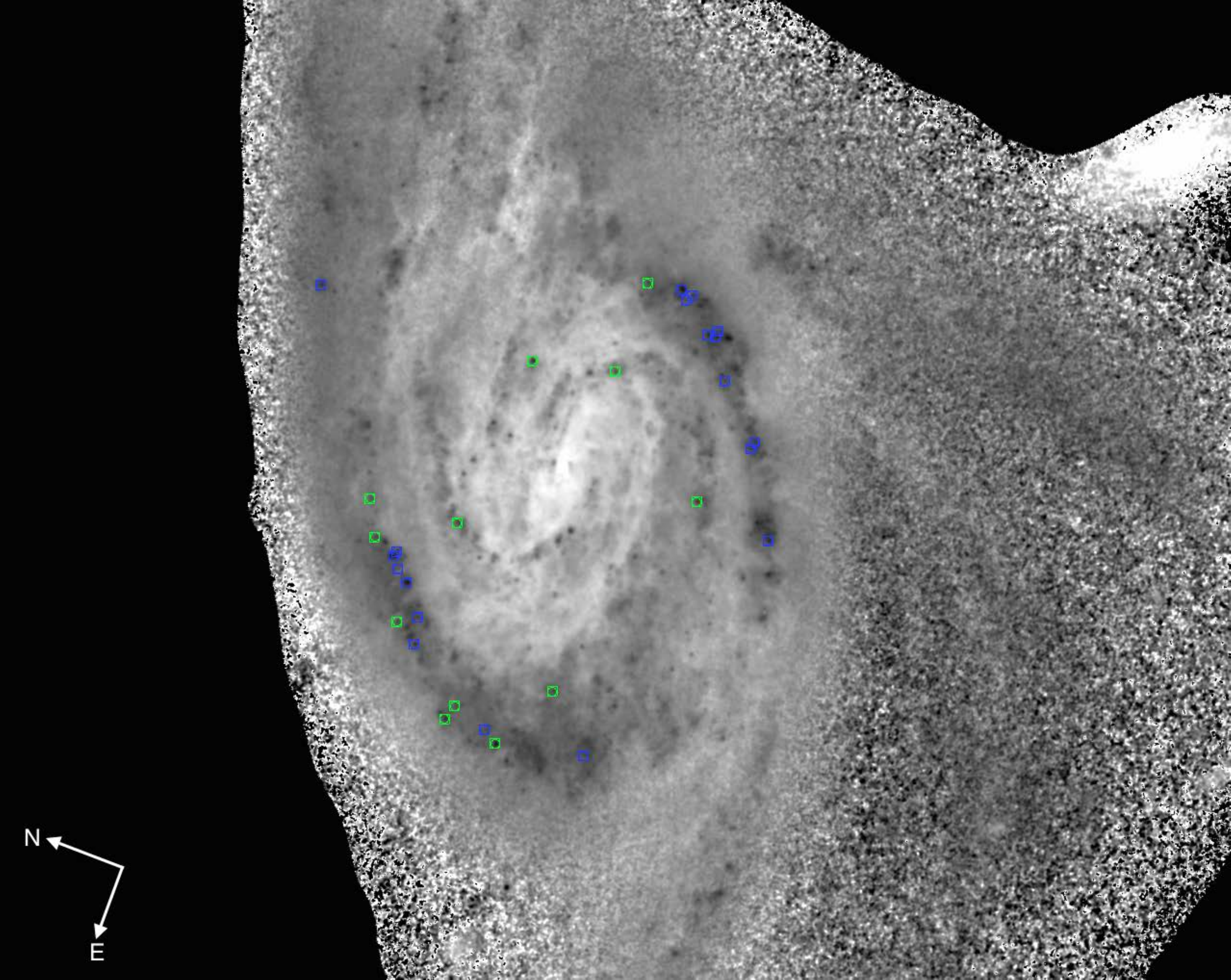}
\caption{Inverted black and white B-I image of NGC 5257W taken with HST ACS/WFC F814W and F435W. The bright emission corresponds to redder (i.e. dustier) regions of the galaxy. The blue centroids correspond to clusters found in relatively ``dust-free'' regions of these galaxies, whereas the green centroids correspond to clusters found in relatively dustier regions of the galaxy.}
\end{figure*}

\begin{figure*}
\centering
\includegraphics[scale=0.55]{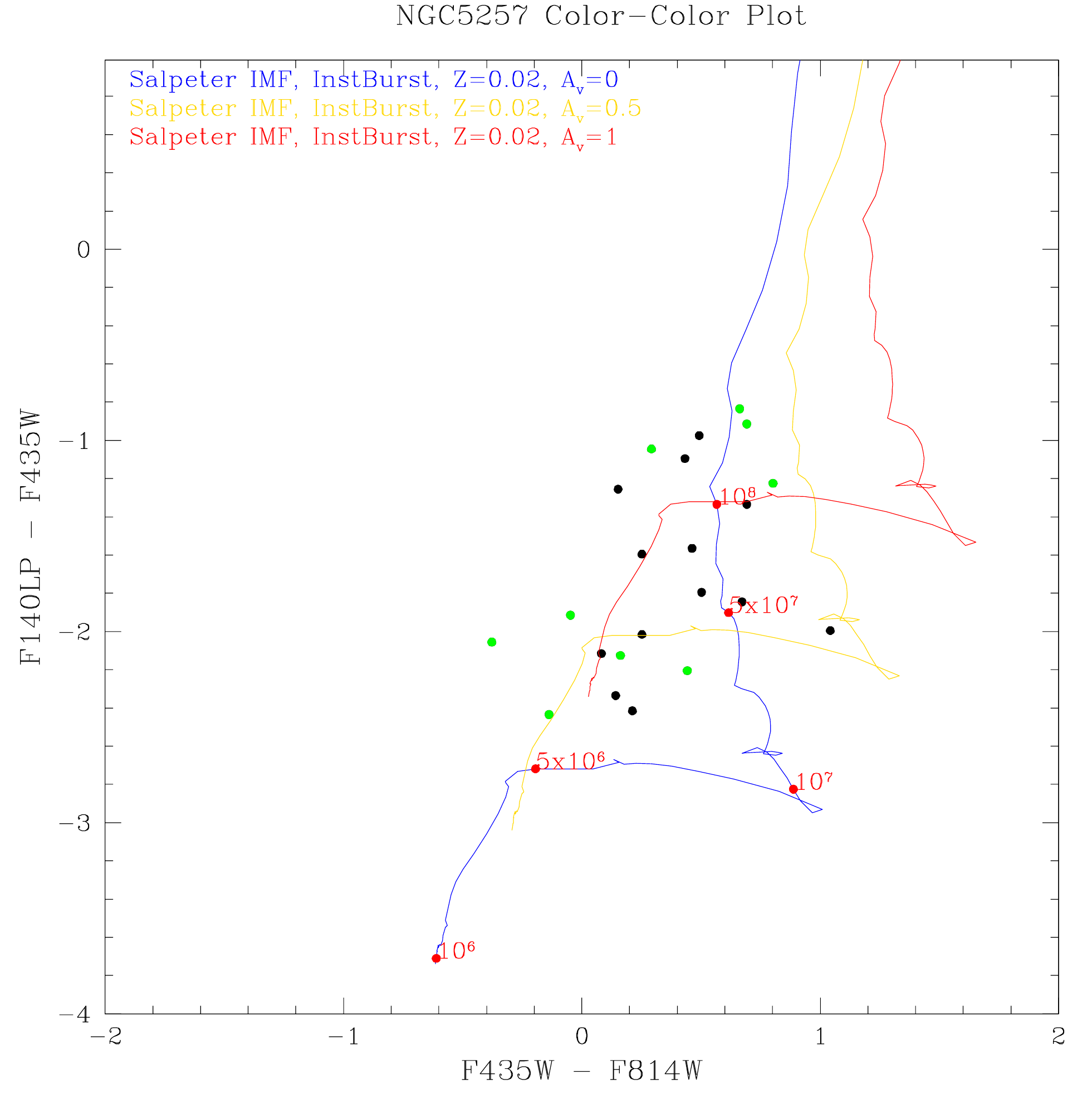}
\caption{Color-Color plot of all star clusters identified in NGC 5257W in F814W, F435W, and F140LP plotted against SSP models with various amount of visual extinction. The green points correspond to the clusters found in dustier regions of the galaxy in Figure 38}
\end{figure*}

\begin{deluxetable*}{ccccccccc}
\tabletypesize{\footnotesize}
\tablewidth{0pt}
\tablecaption {Observed Properties of Star Clusters in NGC 5257W}
\tablehead{
\colhead{ID} & \colhead{RA} & \colhead{Dec} & \colhead{$M_{B}$} & \colhead{$\sigma_{B}$} & \colhead{$M_{I}$} & \colhead{$\sigma_{I}$} & \colhead{$M_{FUV}$} & \colhead{$\sigma_{FUV}$}} \\
\startdata
1 & 204.9677325 & 0.839888315 & -12.60 & 0.04 & -13.26 & 0.04 & -13.43 & 0.07 \\
2 & 204.9676008 & 0.839405141 & -14.32 & 0.02 & -14.40 & 0.02 & -16.43 & 0.01 \\
3 & 204.9676117 & 0.839232784 & -13.47 & 0.04 & -13.90 & 0.05 & -14.56 & 0.02 \\
4 & 204.9679115 & 0.838666338 & -12.25 & 0.06 & -13.29 & 0.02 & -14.24 & 0.06 \\
5 & 204.9680284 & 0.838781875 & -13.85 & 0.02 & -14.10 & 0.03 & -15.44 & 0.01 \\
6 & 204.9679985 & 0.838669059 & -13.07 & 0.06 & -13.32 & 0.06 & -15.08 & 0.04 \\
7 & 204.969473 & 0.840889786 & -13.17 & 0.01 & -13.46 & 0.01 & -14.21 & 0.01 \\
8 & 204.9690871 & 0.839750205 & -12.72 & 0.05 & -13.41 & 0.07 & -13.63 & 0.02 \\
9 & 204.96912 & 0.837489752 & -13.63 & 0.05 & -14.12 & 0.08 & -14.60 & 0.06 \\
10 & 204.9692242 & 0.837503427 & -15.05 & 0.02 & -15.74 & 0.02 & -16.38 & 0.01 \\
11 & 204.970253 & 0.837864699 & -12.01 & 0.02 & -12.17 & 0.04 & -14.13 & 0.01 \\
12 & 204.9727531 & 0.841813782 & -11.72 & 0.06 & -11.67 & 0.09 & -13.63 & 0.04 \\
13 & 204.9703117 & 0.836690288 & -13.97 & 0.02 & -14.12 & 0.03 & -15.22 & 0.03 \\
14 & 204.9728795 & 0.841450503 & -13.65 & 0.02 & -13.86 & 0.03 & -16.06 & 0.01 \\
15 & 204.9730222 & 0.841315721 & -12.87 & 0.04 & -13.54 & 0.03 & -14.71 & 0.02 \\
16 & 204.9731456 & 0.841116419 & -14.47 & 0.01 & -14.61 & 0.02 & -16.80 & 0.01 \\
17 & 204.9735238 & 0.840747 & -12.24 & 0.06 & -12.70 & 0.07 & -13.80 & 0.08 \\
18 & 204.9737106 & 0.840991159 & -12.21 & 0.04 & -12.65 & 0.07 & -14.41 & 0.02 \\
19 & 204.9738917 & 0.840626292 & -12.65 & 0.04 & -13.15 & 0.03 & -14.44 & 0.02 \\
20 & 204.9736297 & 0.838529841 & -13.22 & 0.03 & -14.02 & 0.02 & -14.44 & 0.02 \\
21 & 204.974667 & 0.839754239 & -12.11 & 0.03 & -11.73 & 0.09 & -14.16 & 0.02 \\
22 & 204.9746698 & 0.838945129 & -12.42 & 0.02 & -12.28 & 0.05 & -14.85 & 0.01
\enddata
\end{deluxetable*}

\begin{deluxetable*}{ccccccc}
\tabletypesize{\footnotesize}
\tablewidth{0pt}
\tablecaption {Derived Properties of Star Clusters in NGC 5257W}
\tablehead{
\colhead{ID} & \colhead{Log(Age)} & \colhead{$\sigma_{Age}$} & \colhead{Log($M/M_{\odot}$)} & \colhead{$\sigma_{M}$} & \colhead{$A_{V}$} & \colhead{$\sigma_{A_{V}}$}} \\
\startdata  
1 & 6.56 & 0.64 & 6.19 & 0.59 & 1.80 & 0.60 \\
2 & 6.48 & 0.09 & 6.57 & 0.22 & 1.10 & 0.13 \\
3 & 6.48 & 0.04 & 6.60 & 0.20 & 1.80 & 0.09 \\
4 & 6.24 & 0.02 & 6.35 & 0.20 & 1.80 & 0.07 \\
5 & 6.54 & 0.03 & 6.42 & 0.19 & 1.30 & 0.07 \\
6 & 6.72 & 0.34 & 5.84 & 0.32 & 0.70 & 0.26 \\
7 & 6.54 & 0.01 & 6.30 & 0.18 & 1.60 & 0.05 \\
8 & 6.64 & 0.59 & 6.25 & 0.54 & 1.70 & 0.53 \\
9 & 6.54 & 0.29 & 6.54 & 0.35 & 1.70 & 0.29 \\
10 & 6.74 & 0.32 & 6.83 & 0.29 & 1.10 & 0.23 \\
11 & 6.58 & 0.30 & 5.50 & 0.29 & 0.90 & 0.22 \\
12 & 6.54 & 0.03 & 5.40 & 0.21 & 1.00 & 0.09 \\
13 & 6.52 & 0.01 & 6.57 & 0.18 & 1.50 & 0.03 \\
14 & 6.74 & 0.01 & 5.90 & 0.17 & 0.40 & 0.02 \\
15 & 7.30 & 0.19 & 6.19 & 0.21 & 0.50 & 0.10 \\
16 & 6.72 & 0.01 & 6.29 & 0.18 & 0.50 & 0.05 \\
17 & 6.58 & 0.24 & 5.81 & 0.31 & 1.30 & 0.24 \\
18 & 6.76 & 0.16 & 5.37 & 0.19 & 0.50 & 0.07 \\
19 & 6.74 & 0.17 & 5.71 & 0.19 & 0.80 & 0.07 \\
20 & 7.63 & 0.27 & 6.60 & 0.26 & 0.60 & 0.19 \\
21 & 6.52 & 0.01 & 5.51 & 0.18 & 0.90 & 0.05 \\
22 & 6.52 & 0.05 & 5.53 & 0.20 & 0.70 & 0.10
\enddata
\end{deluxetable*}

\clearpage

\subsection{NGC 5331S}

NGC 5331 is a mid-stage mergering system. NGC 5331S and N have a projected nuclear separation of $\sim 27''$ (19 kpc). Large dust lanes are visible along the near edge of the galaxy, and only a small number of star clusters are visible. The maximum $A_{V}$ adopted for this galaxy is 3.6 mags of visual extinction (Lutz 1992). 

\begin{figure*}
\centering
\includegraphics[scale=0.3]{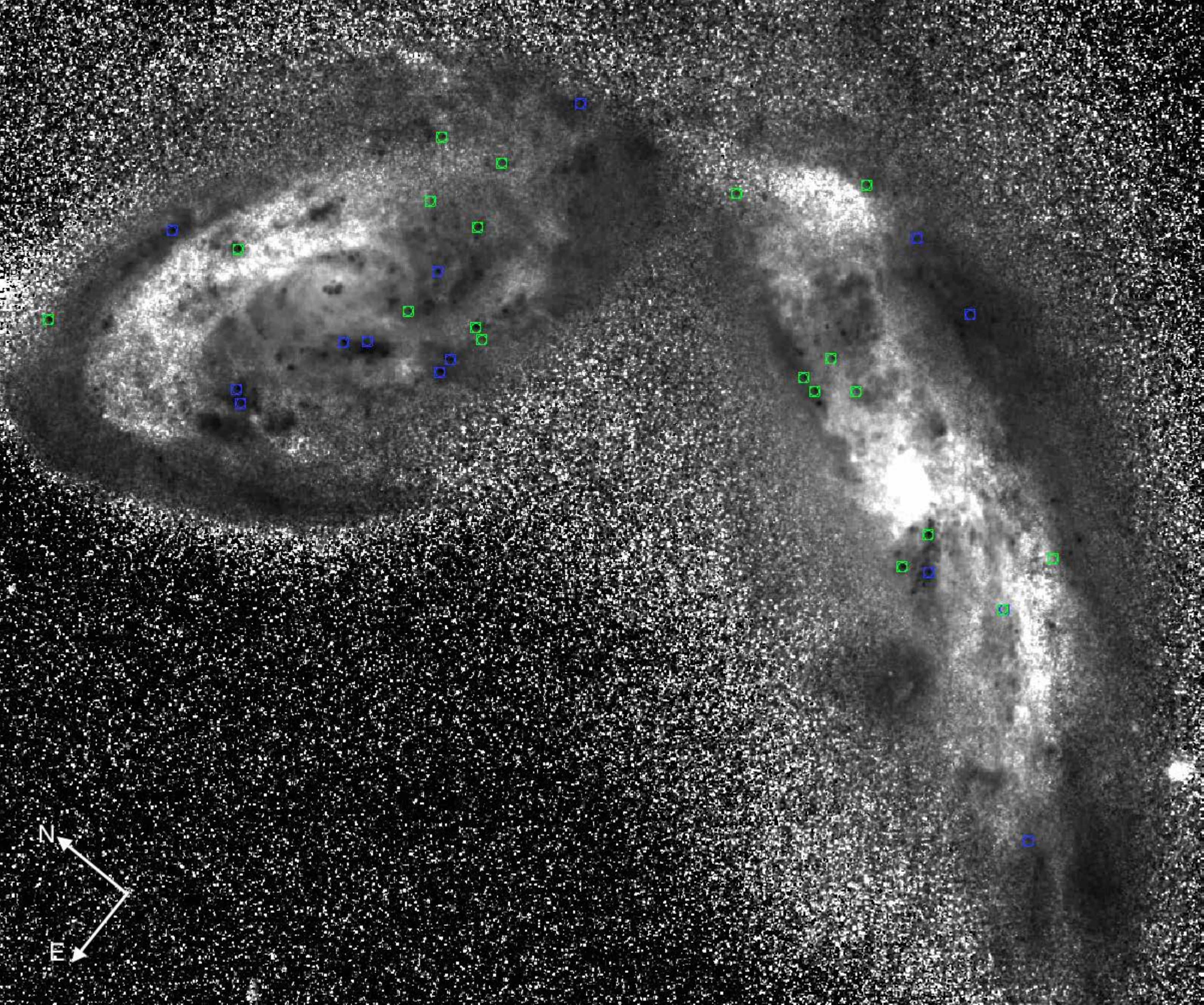}
\caption{Inverted black and white B-I image of NGC 5331S taken with HST ACS/WFC F814W and F435W. The bright emission corresponds to redder (i.e. dustier) regions of the galaxy. The blue centroids correspond to clusters found in relatively ``dust-free'' regions of these galaxies, whereas the green centroids correspond to clusters found in relatively dustier regions of the galaxy.}
\end{figure*}

\begin{figure*}
\centering
\includegraphics[scale=0.55]{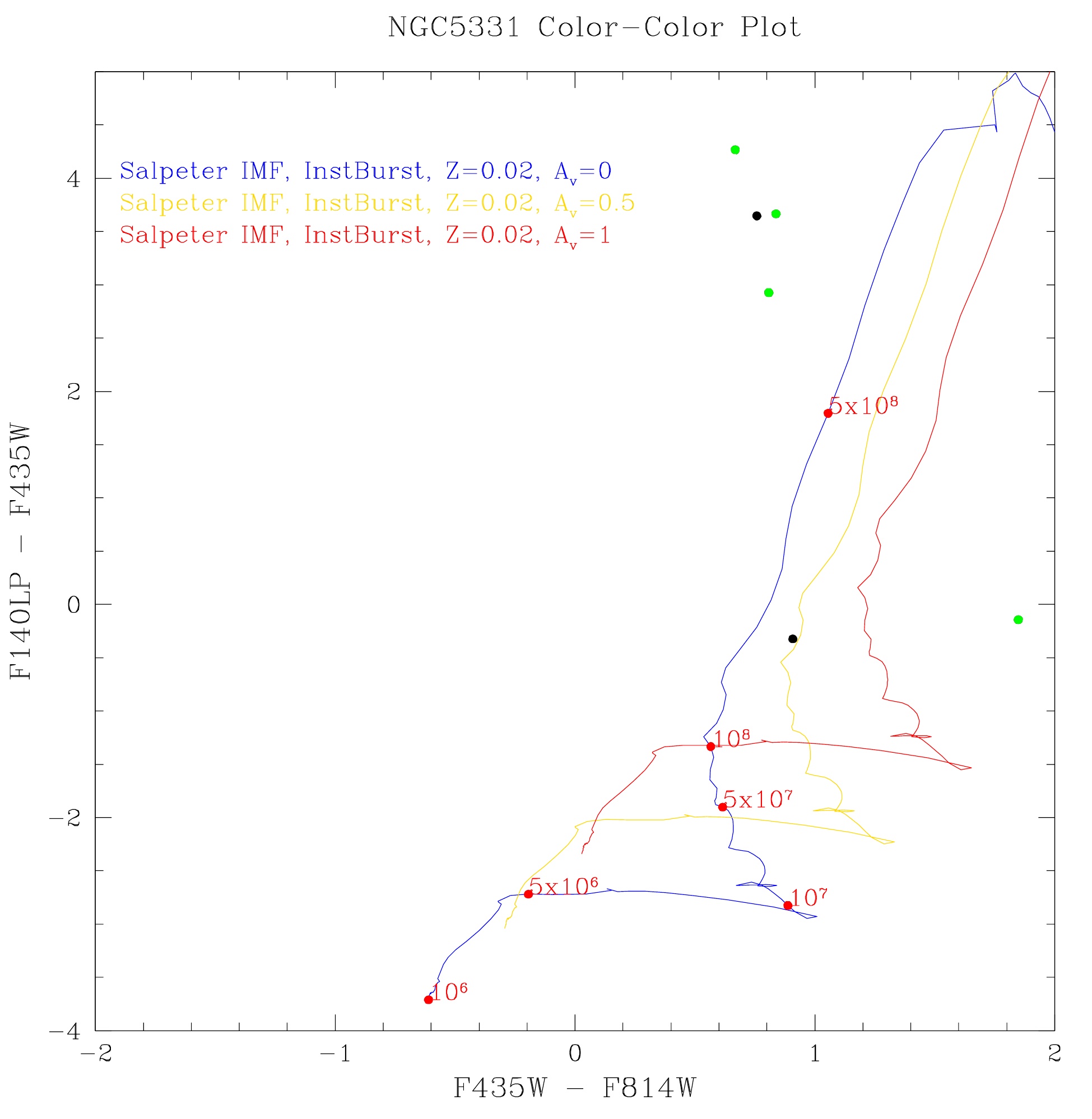}
\caption{Color-Color plot of all star clusters identified in NGC 5331S in F814W, F435W, and F140LP plotted against SSP models with various amount of visual extinction. The green points correspond to the clusters found in dustier regions of the galaxy in Figure 40}
\end{figure*}

\begin{deluxetable*}{ccccccccc}
\tabletypesize{\footnotesize}
\tablewidth{0pt}
\tablecaption {Observed Properties of Star Clusters in NGC 5331S}
\tablehead{
\colhead{ID} & \colhead{RA} & \colhead{Dec} & \colhead{$M_{B}$} & \colhead{$\sigma_{B}$} & \colhead{$M_{I}$} & \colhead{$\sigma_{I}$} & \colhead{$M_{FUV}$} & \colhead{$\sigma_{FUV}$}} \\
\startdata
1 & 208.0703109 & 2.098268967 & -9.51 & 0.10 & -10.42 & 0.10 & -9.83 & 0.11 \\
2 & 208.0667504 & 2.102795958 & -12.63 & 0.03 & -13.47 & 0.03 & -8.96 & 0.07 \\
3 & 208.0676478 & 2.100879121 & -15.03 & 0.01 & -15.70 & 0.01 & -10.76 & 0.27 \\
4 & 208.0681184 & 2.100981506 & -12.59 & 0.02 & -13.40 & 0.03 & -9.66 & 0.19 \\
5 & 208.0672543 & 2.099466214 & -9.90 & 0.07 & -12.06 & 0.03 & -9.13 & 0.18 \\
6 & 208.0680338 & 2.099718103 & -9.64 & 0.10 & -11.49 & 0.10 & -9.78 & 0.26 \\
7 & 208.068045 & 2.100690283 & -13.34 & 0.03 & -14.10 & 0.02 & -9.69 & 0.14
\enddata
\end{deluxetable*}

\begin{deluxetable*}{ccccccc}
\tabletypesize{\footnotesize}
\tablewidth{0pt}
\tablecaption {Derived Properties of Star Clusters in NGC 5331S}
\tablehead{
\colhead{ID} & \colhead{Log(Age)} & \colhead{$\sigma_{Age}$} & \colhead{Log($M/M_{\odot}$)} & \colhead{$\sigma_{M}$} & \colhead{$A_{V}$} & \colhead{$\sigma_{A_{V}}$}} \\
\startdata  
1 & 6.50 & 0.38 & 5.53 & 0.20 & 2.30 & 0.37 \\
2 & 8.76 & 0.58 & 8.08 & 0.17 & 1.80 & 0.55 \\
3 & 8.86 & 0.44 & 9.13 & 0.17 & 1.80 & 0.31 \\
4 & 8.61 & 0.43 & 7.93 & 0.17 & 1.80 & 0.28 \\
5 & 6.80 & 0.35 & 5.85 & 0.19 & 2.50 & 0.92 \\
6 & 6.82 & 0.51 & 5.46 & 0.20 & 1.90 & 0.70 \\
7 & 7.91 & 0.13 & 8.70 & 0.18 & 3.60 & 0.05
\enddata
\end{deluxetable*}

\subsection{NGC 5331N}

NGC 5331N has a nucleus and two distinct spiral arms, with a small number of star clusters visible throughout the galaxy. The maximum $A_{V}$ adopted for this galaxy is 1.8 mags of visual extinction (Lutz 1992). 


\begin{figure*}
\centering
\includegraphics[scale=0.55]{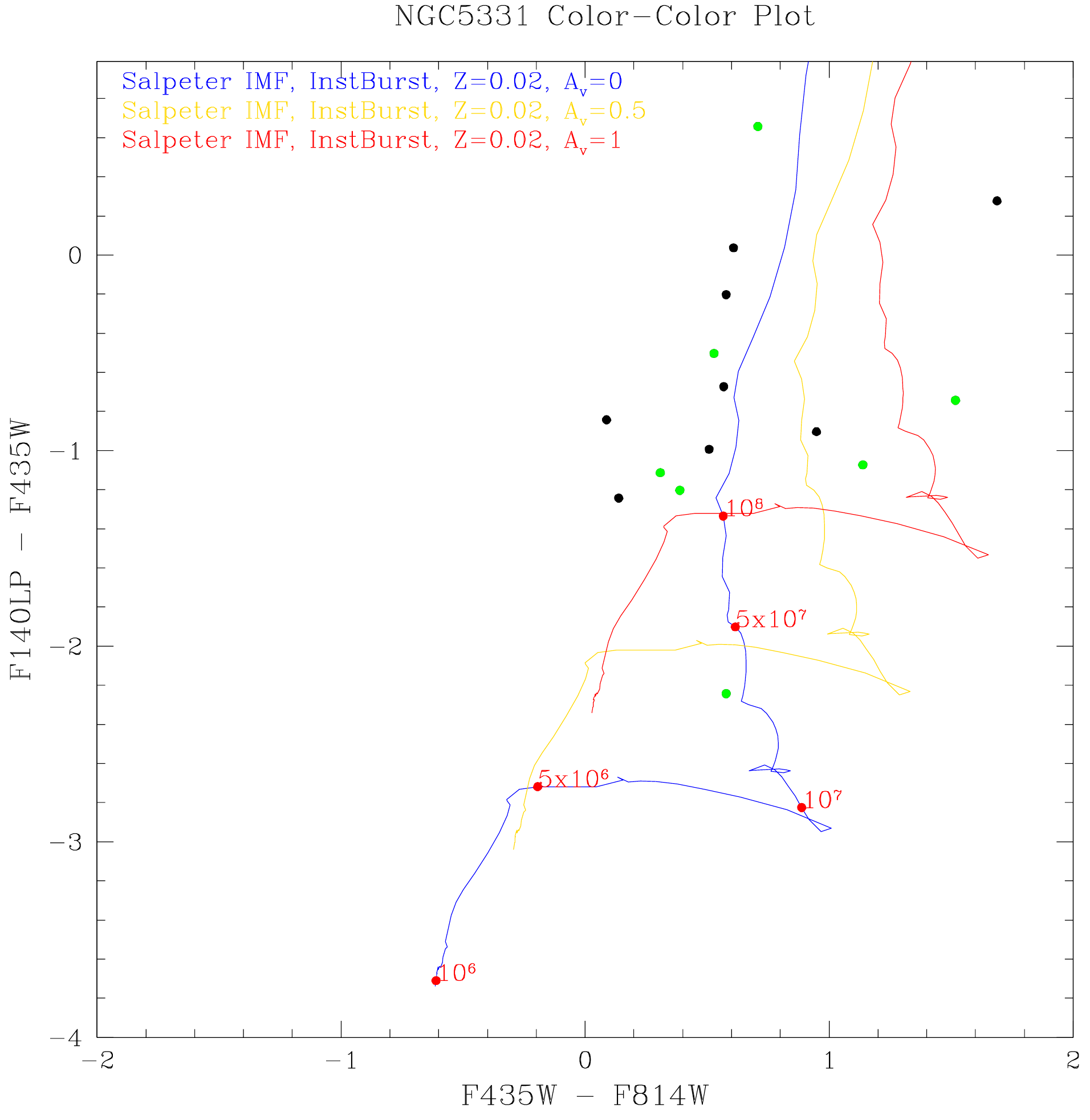}
\caption{Color-Color plot of all star clusters identified in NGC 5331N in F814W, F435W, and F140LP plotted against SSP models with various amount of visual extinction. The green points correspond to the clusters found in dustier regions of the galaxy in Figure 40}
\end{figure*}

\begin{deluxetable*}{ccccccccc}
\tabletypesize{\footnotesize}
\tablewidth{0pt}
\tablecaption {Observed Properties of Star Clusters in NGC 5331N}
\tablehead{
\colhead{ID} & \colhead{RA} & \colhead{Dec} & \colhead{$M_{B}$} & \colhead{$\sigma_{B}$} & \colhead{$M_{I}$} & \colhead{$\sigma_{I}$} & \colhead{$M_{FUV}$} & \colhead{$\sigma_{FUV}$}} \\
\startdata
1 & 208.0660418 & 2.10796331 & -9.62618 & 0.07 & -10.20 & 0.10 & -11.87 & 0.10 \\
2 & 208.0660027 & 2.107205416 & -9.75618 & 0.07 & -10.89 & 0.07 & -10.83 & 0.07 \\
3 & 208.0667669 & 2.107753228 & -9.37618 & 0.09 & -10.89 & 0.08 & -10.12 & 0.11 \\
4 & 208.068404 & 2.11027354 & -12.4562 & 0.02 & -13.02 & 0.02 & -13.13 & 0.03 \\
5 & 208.0682536 & 2.109496656 & -11.5962 & 0.02 & -11.98 & 0.05 & -12.80 & 0.04 \\
6 & 208.0667871 & 2.107126587 & -11.0962 & 0.03 & -11.62 & 0.05 & -11.60 & 0.14 \\
7 & 208.0674622 & 2.107317376 & -12.2362 & 0.03 & -13.18 & 0.04 & -13.14 & 0.03 \\
8 & 208.0684892 & 2.106770333 & -12.9062 & 0.01 & -13.48 & 0.01 & -13.11 & 0.03 \\
9 & 208.0678413 & 2.106630757 & -11.5662 & 0.03 & -11.87 & 0.05 & -12.68 & 0.05 \\
10 & 208.0683038 & 2.10672568 & -11.4162 & 0.03 & -11.50 & 0.08 & -12.26 & 0.07 \\
11 & 208.0686796 & 2.107921587 & -13.8362 & 0.01 & -14.44 & 0.01 & -13.80 & 0.02 \\
12 & 208.0666683 & 2.102976709 & -12.5562 & 0.02 & -13.41 & 0.02 & -11.52 & 0.01 \\
13 & 208.0698361 & 2.108671969 & -10.3762 & 0.09 & -10.51 & 0.14 & -11.62 & 0.12 \\
14 & 208.0685396 & 2.107683102 & -11.6562 & 0.07 & -13.34 & 0.02 & -11.38 & 0.16 \\
15 & 208.0699658 & 2.111094004 & -10.4662 & 0.03 & -11.17 & 0.03 & -9.81 & 0.13 \\
16 & 208.0697116 & 2.108783852 & -11.8962 & 0.02 & -12.40 & 0.02 & -12.89 & 0.04
\enddata
\end{deluxetable*}

\begin{deluxetable*}{ccccccc}
\tabletypesize{\footnotesize}
\tablewidth{0pt}
\tablecaption {Derived Properties of Star Clusters in NGC 5331N}
\tablehead{
\colhead{ID} & \colhead{Log(Age)} & \colhead{$\sigma_{Age}$} & \colhead{Log($M/M_{\odot}$)} & \colhead{$\sigma_{M}$} & \colhead{$A_{V}$} & \colhead{$\sigma_{A_{V}}$}} \\
\startdata  
1 & 6.86 & 0.33 & 4.64 & 0.25 & 0.30 & 0.15 \\
2 & 7.54 & 0.34 & 5.65 & 0.30 & 0.80 & 0.23 \\
3 & 7.42 & 0.24 & 5.65 & 0.27 & 1.20 & 0.17 \\
4 & 6.66 & 0.35 & 6.23 & 0.37 & 1.50 & 0.32 \\
5 & 6.66 & 0.11 & 5.67 & 0.27 & 1.10 & 0.20 \\
6 & 8.31 & 0.01 & 6.20 & 0.18 & 0.01 & 0.04 \\
7 & 6.78 & 0.44 & 6.09 & 0.36 & 1.30 & 0.30 \\
8 & 6.66 & 0.02 & 6.56 & 0.18 & 1.80 & 0.05 \\
9 & 6.66 & 0.04 & 5.71 & 0.21 & 1.20 & 0.10 \\
10 & 6.66 & 0.04 & 5.70 & 0.19 & 1.30 & 0.07 \\
11 & 6.68 & 0.30 & 6.94 & 0.17 & 1.80 & 0.08 \\
12 & 8.61 & 0.15 & 7.08 & 0.18 & 0.10 & 0.05 \\
13 & 6.66 & 0.35 & 5.13 & 0.28 & 1.00 & 0.19 \\
14 & 7.65 & 0.11 & 6.88 & 0.22 & 1.60 & 0.11 \\
15 & 8.56 & 0.02 & 6.14 & 0.17 & 0.01 & 0.13 \\
16 & 6.64 & 0.08 & 5.86 & 0.26 & 1.30 & 0.18
\enddata
\end{deluxetable*}

\subsection{UGC 09618NED02}

Armus et al. (2009) discusses the detailed morphology of this galaxy at length. UGC 09618NED02 is an early-stage merger with the two nuclei separated by $\sim 40''$ (30 kpc). Multiple star clusters are visible along the spiral arms in the face-on galaxy (VV430A). The maximum $A_{V}$ adopted for this galaxy is 2.4 mags of visual extinction (Leech et al. 1989). 

\begin{figure*}
\centering
\includegraphics[scale=0.25]{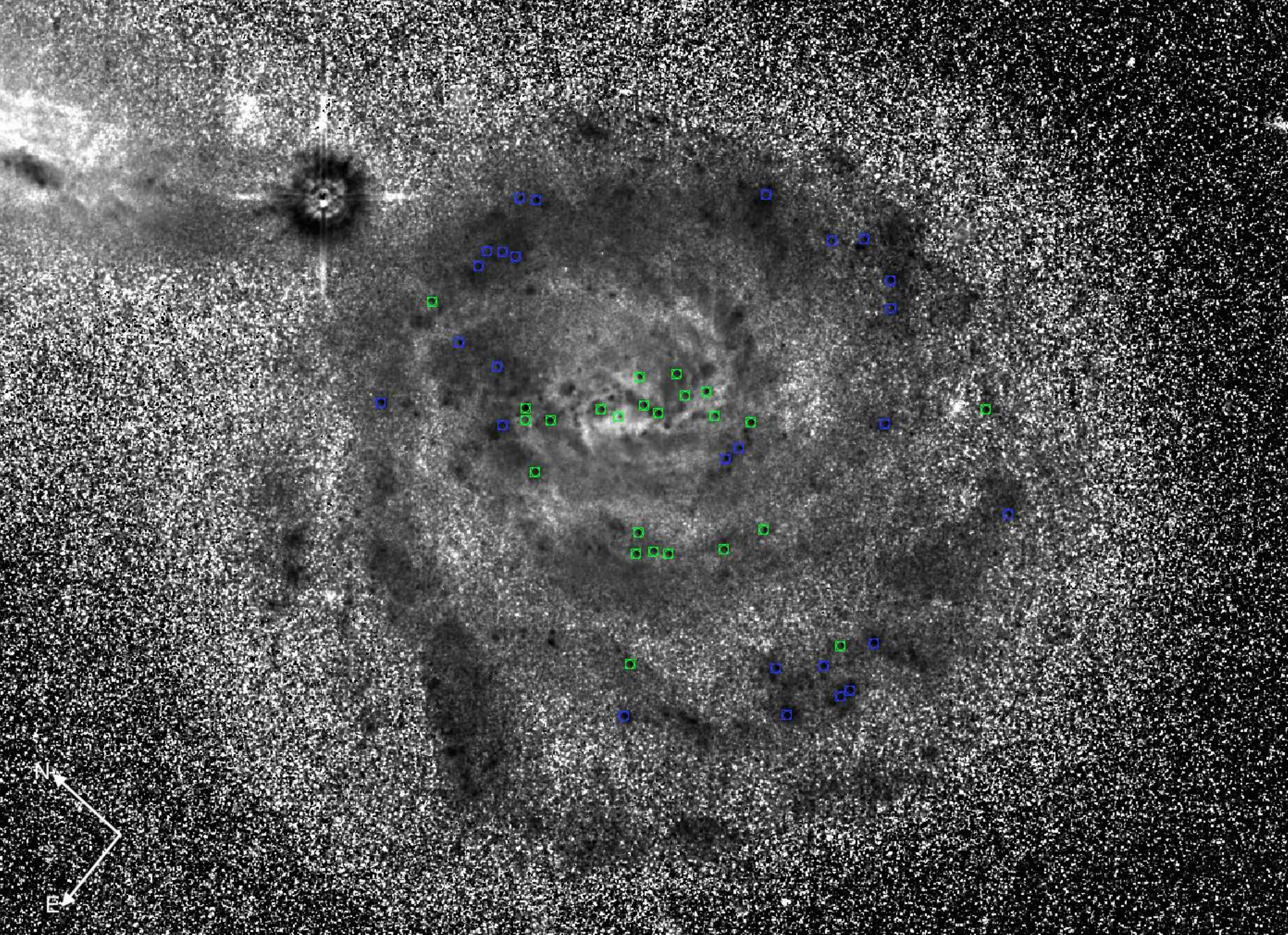}
\caption{Inverted black and white B-I image of UGC 09618NED02 taken with HST ACS/WFC F814W and F435W. The bright emission corresponds to redder (i.e. dustier) regions of the galaxy. The blue centroids correspond to clusters found in relatively ``dust-free'' regions of these galaxies, whereas the green centroids correspond to clusters found in relatively dustier regions of the galaxy.}
\end{figure*}

\begin{figure*}
\centering
\includegraphics[scale=0.55]{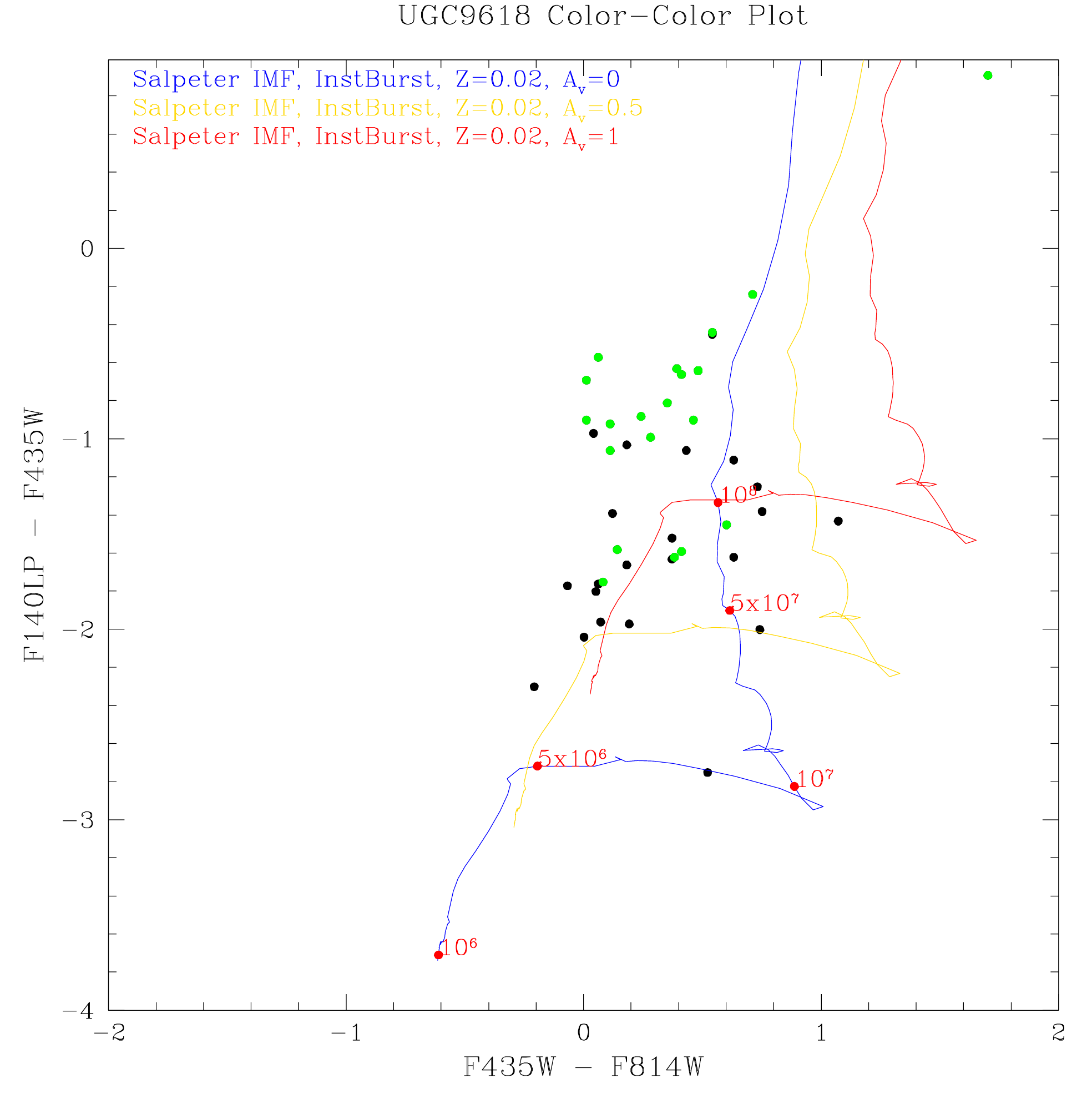}
\caption{Color-Color plot of all star clusters identified in UGC 09618NED02 in F814W, F435W, and F140LP plotted against SSP models with various amount of visual extinction. The green points correspond to the clusters found in dustier regions of the galaxy in Figure 43}
\end{figure*}

\begin{deluxetable*}{ccccccccc}
\tabletypesize{\footnotesize}
\tablewidth{0pt}
\tablecaption {Observed Properties of Star Clusters in UGC 09618NED02}
\tablehead{
\colhead{ID} & \colhead{RA} & \colhead{Dec} & \colhead{$M_{B}$} & \colhead{$\sigma_{B}$} & \colhead{$M_{I}$} & \colhead{$\sigma_{I}$} & \colhead{$M_{FUV}$} & \colhead{$\sigma_{FUV}$}} \\
\startdata
1 & 224.2502774 & 24.60907674 & -10.22 & 0.19 & -10.97 & 0.15 & -12.23 & 0.18 \\
2 & 224.2495249 & 24.60924743 & -11.10 & 0.03 & -11.63 & 0.04 & -13.86 & 0.18 \\
3 & 224.2494326 & 24.60905594 & -11.83 & 0.03 & -11.84 & 0.04 & -13.88 & 0.04 \\
4 & 224.247693 & 24.60664702 & -11.07 & 0.08 & -12.15 & 0.05 & -12.51 & 0.14 \\
5 & 224.2474984 & 24.6053126 & -11.00 & 0.04 & -11.08 & 0.08 & -12.97 & 0.08 \\
6 & 224.2477925 & 24.60476143 & -11.71 & 0.03 & -11.78 & 0.05 & -13.48 & 0.06 \\
7 & 224.2513684 & 24.6094958 & -12.99 & 0.02 & -13.71 & 0.02 & -13.24 & 0.07 \\
8 & 224.2481156 & 24.6045755 & -11.08 & 0.05 & -11.84 & 0.05 & -12.47 & 0.18 \\
9 & 224.2516575 & 24.60893269 & -13.11 & 0.01 & -13.66 & 0.01 & -13.57 & 0.02 \\
10 & 224.2516642 & 24.60837604 & -11.69 & 0.07 & -12.07 & 0.08 & -13.22 & 0.07 \\
11 & 224.2504346 & 24.60641579 & -12.52 & 0.05 & -12.88 & 0.06 & -13.34 & 0.06 \\
12 & 224.2504328 & 24.60598905 & -12.40 & 0.02 & -12.52 & 0.06 & -13.33 & 0.17 \\
13 & 224.25063 & 24.60618376 & -11.50 & 0.07 & -11.92 & 0.11 & -12.17 & 0.05 \\
14 & 224.2519383 & 24.60779872 & -14.50 & 0.01 & -14.62 & 0.01 & -15.57 & 0.01 \\
15 & 224.2510431 & 24.60655958 & -13.33 & 0.04 & -13.35 & 0.13 & -14.03 & 0.03 \\
16 & 224.2514028 & 24.6069871 & -13.30 & 0.07 & -13.37 & 0.14 & -13.88 & 0.04 \\
17 & 224.2510282 & 24.6063554 & -13.65 & 0.02 & -13.90 & 0.03 & -14.54 & 0.02 \\
18 & 224.2506515 & 24.60573423 & -11.61 & 0.05 & -11.63 & 0.08 & -12.52 & 0.14 \\
19 & 224.2513638 & 24.60674976 & -15.35 & 0.01 & -17.06 & 0.01 & -14.45 & 0.02 \\
20 & 224.2518985 & 24.60745116 & -13.16 & 0.02 & -13.71 & 0.02 & -13.61 & 0.05 \\
21 & 224.2520755 & 24.60771662 & -11.17 & 0.14 & -11.78 & 0.13 & -12.63 & 0.12 \\
22 & 224.2523005 & 24.60793133 & -12.08 & 0.06 & -12.72 & 0.06 & -13.20 & 0.07 \\
23 & 224.2504648 & 24.6053105 & -12.88 & 0.01 & -13.37 & 0.02 & -13.53 & 0.01 \\
24 & 224.2495097 & 24.60387414 & -11.68 & 0.03 & -12.32 & 0.05 & -13.31 & 0.14 \\
25 & 224.2508571 & 24.60526342 & -11.73 & 0.04 & -12.17 & 0.06 & -12.80 & 0.11 \\
26 & 224.2510791 & 24.6053372 & -12.61 & 0.02 & -12.80 & 0.04 & -13.65 & 0.05 \\
27 & 224.2526081 & 24.60727308 & -12.41 & 0.04 & -12.81 & 0.04 & -13.05 & 0.08 \\
28 & 224.2496684 & 24.60197058 & -11.11 & 0.04 & -11.31 & 0.06 & -13.09 & 0.08 \\
29 & 224.2516261 & 24.6044602 & -11.22 & 0.04 & -11.31 & 0.07 & -12.98 & 0.03 \\
30 & 224.2525706 & 24.60577791 & -11.27 & 0.03 & -11.56 & 0.06 & -12.27 & 0.17 \\
31 & 224.2521417 & 24.60475683 & -11.61 & 0.04 & -12.03 & 0.03 & -13.21 & 0.03 \\
32 & 224.2525986 & 24.60531766 & -11.85 & 0.02 & -12.32 & 0.03 & -12.76 & 0.18 \\
33 & 224.2528334 & 24.60566186 & -11.89 & 0.03 & -12.38 & 0.03 & -12.54 & 0.15 \\
34 & 224.2521531 & 24.60254189 & -12.79 & 0.01 & -12.85 & 0.02 & -14.60 & 0.02 \\
35 & 224.2524271 & 24.6028813 & -10.56 & 0.07 & -10.95 & 0.07 & -12.19 & 0.11 \\
36 & 224.2531559 & 24.60341951 & -11.72 & 0.03 & -11.91 & 0.04 & -13.39 & 0.06 \\
37 & 224.2527859 & 24.60292948 & -11.65 & 0.02 & -11.59 & 0.04 & -13.43 & 0.02 \\
38 & 224.2541651 & 24.60499864 & -11.06 & 0.03 & -11.21 & 0.06 & -12.65 & 0.02 \\
39 & 224.2528804 & 24.6024864 & -12.38 & 0.02 & -12.51 & 0.03 & -13.78 & 0.04 \\
40 & 224.2530173 & 24.60254644 & -11.29 & 0.06 & -11.34 & 0.07 & -12.27 & 0.18 \\
41 & 224.2536233 & 24.60299199 & -13.79 & 0.01 & -13.59 & 0.01 & -16.10 & 0.01 \\
42 & 224.2548158 & 24.60471647 & -11.68 & 0.02 & -12.06 & 0.03 & -13.32 & 0.12 \\
43 & 224.2506179 & 24.60923582 & -12.07 & 0.03 & -12.81 & 0.04 & -13.33 & 0.06
\enddata
\end{deluxetable*}

\begin{deluxetable*}{ccccccc}
\tabletypesize{\footnotesize}
\tablewidth{0pt}
\tablecaption {Derived Properties of Star Clusters in UGC 09618NED02}
\tablehead{
\colhead{ID} & \colhead{Log(Age)} & \colhead{$\sigma_{Age}$} & \colhead{Log($M/M_{\odot}$)} & \colhead{$\sigma_{M}$} & \colhead{$A_{V}$} & \colhead{$\sigma_{A_{V}}$}} \\
1 & 7.57 & 0.39 & 5.17 & 0.41 & 0.10 & 0.32 \\
2 & 6.88 & 0.05 & 4.82 & 0.16 & 0.01 & 0.45 \\
3 & 6.66 & 0.09 & 5.15 & 0.25 & 0.50 & 0.18 \\
4 & 7.63 & 0.28 & 5.70 & 0.28 & 0.40 & 0.21 \\
5 & 6.64 & 0.20 & 4.84 & 0.34 & 0.60 & 0.29 \\
6 & 6.66 & 0.03 & 5.21 & 0.19 & 0.70 & 0.09 \\
7 & 8.41 & 0.03 & 6.73 & 0.17 & 0.01 & 0.05 \\
8 & 6.76 & 0.52 & 5.10 & 0.38 & 0.90 & 0.34 \\
9 & 6.66 & 0.02 & 6.25 & 0.16 & 1.60 & 0.04 \\
10 & 6.38 & 0.65 & 5.87 & 0.53 & 1.50 & 0.52 \\
11 & 6.66 & 0.57 & 5.85 & 0.49 & 1.30 & 0.47 \\
12 & 6.66 & 0.53 & 5.75 & 0.16 & 1.20 & 0.03 \\
13 & 6.66 & 0.75 & 5.55 & 0.63 & 1.50 & 0.66 \\
14 & 6.66 & 0.95 & 6.54 & 0.16 & 1.10 & 0.03 \\
15 & 6.66 & 0.57 & 6.18 & 0.48 & 1.30 & 0.46 \\
16 & 6.66 & 0.77 & 6.22 & 0.61 & 1.40 & 0.63 \\
17 & 6.66 & 0.67 & 6.30 & 0.16 & 1.30 & 0.04 \\
18 & 6.66 & 0.16 & 5.43 & 0.18 & 1.20 & 0.05 \\
19 & 7.86 & 0.04 & 8.29 & 0.17 & 1.80 & 0.05 \\
20 & 8.36 & 0.03 & 6.76 & 0.16 & 0.01 & 0.03 \\
21 & 6.74 & 0.75 & 5.13 & 0.53 & 0.90 & 0.51 \\
22 & 5.10 & 0.97 & 6.34 & 0.66 & 1.90 & 0.70 \\
23 & 6.66 & 0.03 & 6.10 & 0.16 & 1.50 & 0.02 \\
24 & 6.76 & 0.47 & 5.29 & 0.37 & 0.80 & 0.33 \\
25 & 6.66 & 0.26 & 5.48 & 0.35 & 1.20 & 0.31 \\
26 & 6.66 & 0.51 & 5.83 & 0.17 & 1.20 & 0.05 \\
27 & 6.66 & 0.75 & 5.91 & 0.61 & 1.50 & 0.63 \\
28 & 6.36 & 0.46 & 5.45 & 0.36 & 1.20 & 0.31 \\
29 & 6.66 & 0.09 & 5.01 & 0.24 & 0.70 & 0.17 \\
30 & 6.66 & 0.43 & 5.30 & 0.17 & 1.20 & 0.04 \\
31 & 6.72 & 0.64 & 5.24 & 0.47 & 0.80 & 0.45 \\
32 & 6.66 & 0.01 & 5.58 & 0.16 & 1.30 & 0.02 \\
33 & 6.66 & 0.03 & 5.71 & 0.16 & 1.50 & 0.04 \\
34 & 6.66 & 0.01 & 5.64 & 0.16 & 0.70 & 0.03 \\
35 & 6.00 & 0.65 & 5.42 & 0.50 & 1.50 & 0.48 \\
36 & 6.66 & 0.25 & 5.27 & 0.25 & 0.80 & 0.18 \\
37 & 6.66 & 0.86 & 5.19 & 0.16 & 0.70 & 0.04 \\
38 & 6.66 & 0.04 & 5.00 & 0.20 & 0.80 & 0.11 \\
39 & 6.66 & 0.07 & 5.58 & 0.17 & 0.90 & 0.05 \\
40 & 6.66 & 0.35 & 5.31 & 0.34 & 1.20 & 0.28 \\
41 & 6.66 & 0.36 & 5.83 & 0.16 & 0.30 & 0.06 \\
42 & 5.10 & 0.67 & 5.96 & 0.38 & 1.50 & 0.34 \\
43 & 6.78 & 0.55 & 5.57 & 0.43 & 1.00 & 0.40
\enddata
\end{deluxetable*}

\subsection{IC 4687N}

IC 4687 is an early-stage merging system. The two primary galaxies (IC 4686 and IC 4687) have a nuclear separation of $\sim 84''$ (31 kpc). The northern galaxy contains several bright clusters in a nuclear arm stretching north and westward. The maximum $A_{V}$ adopted for this galaxy is 2.8 mags of visual extinction (Rich et al. 2012). 

\begin{figure*}
\centering
\includegraphics[scale=0.25]{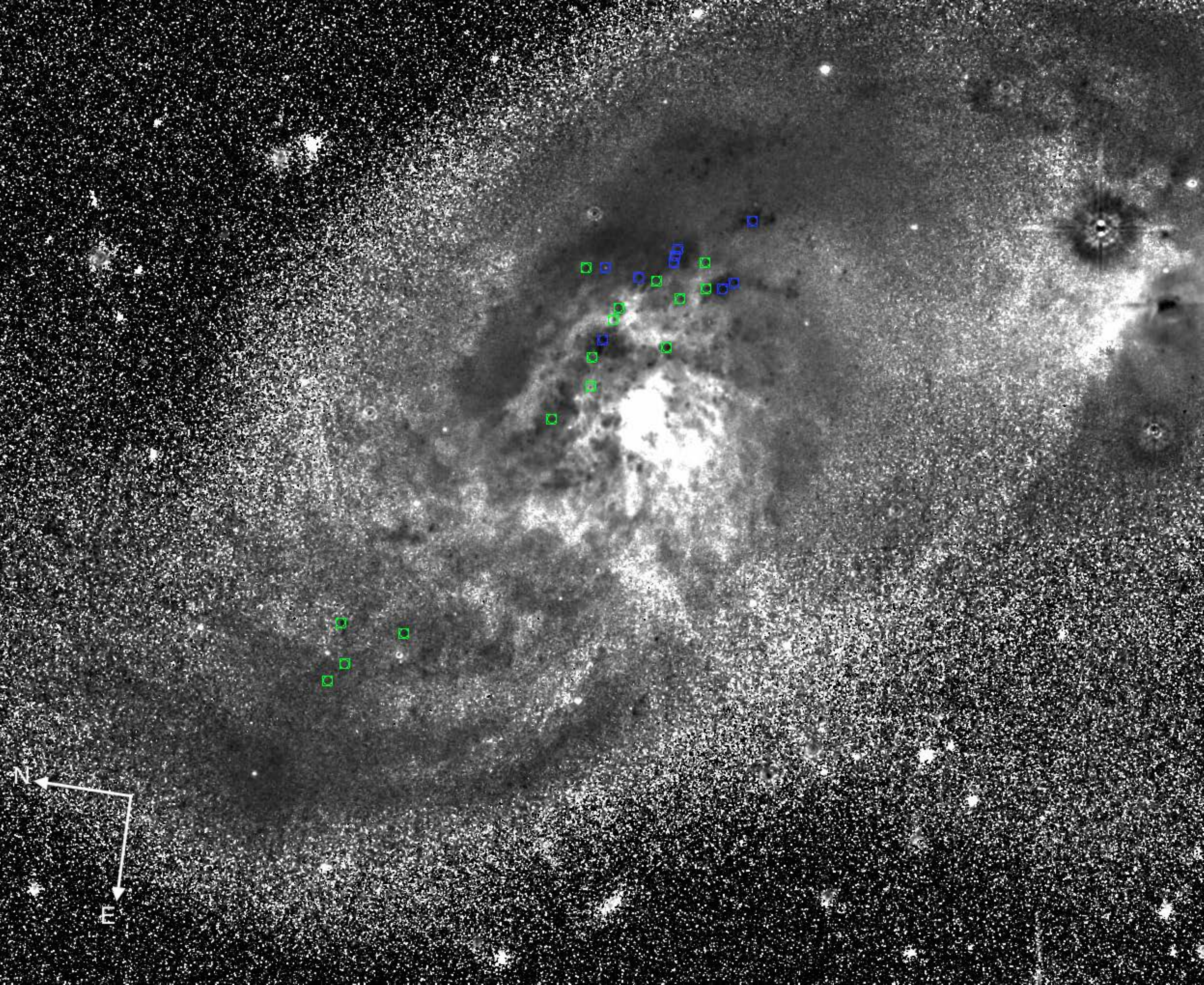}
\caption{Inverted black and white B-I image of IC 4687N taken with HST ACS/WFC F814W and F435W. The bright emission corresponds to redder (i.e. dustier) regions of the galaxy. The blue centroids correspond to clusters found in relatively ``dust-free'' regions of these galaxies, whereas the green centroids correspond to clusters found in relatively dustier regions of the galaxy.}
\end{figure*}

\begin{figure*}
\centering
\includegraphics[scale=0.55]{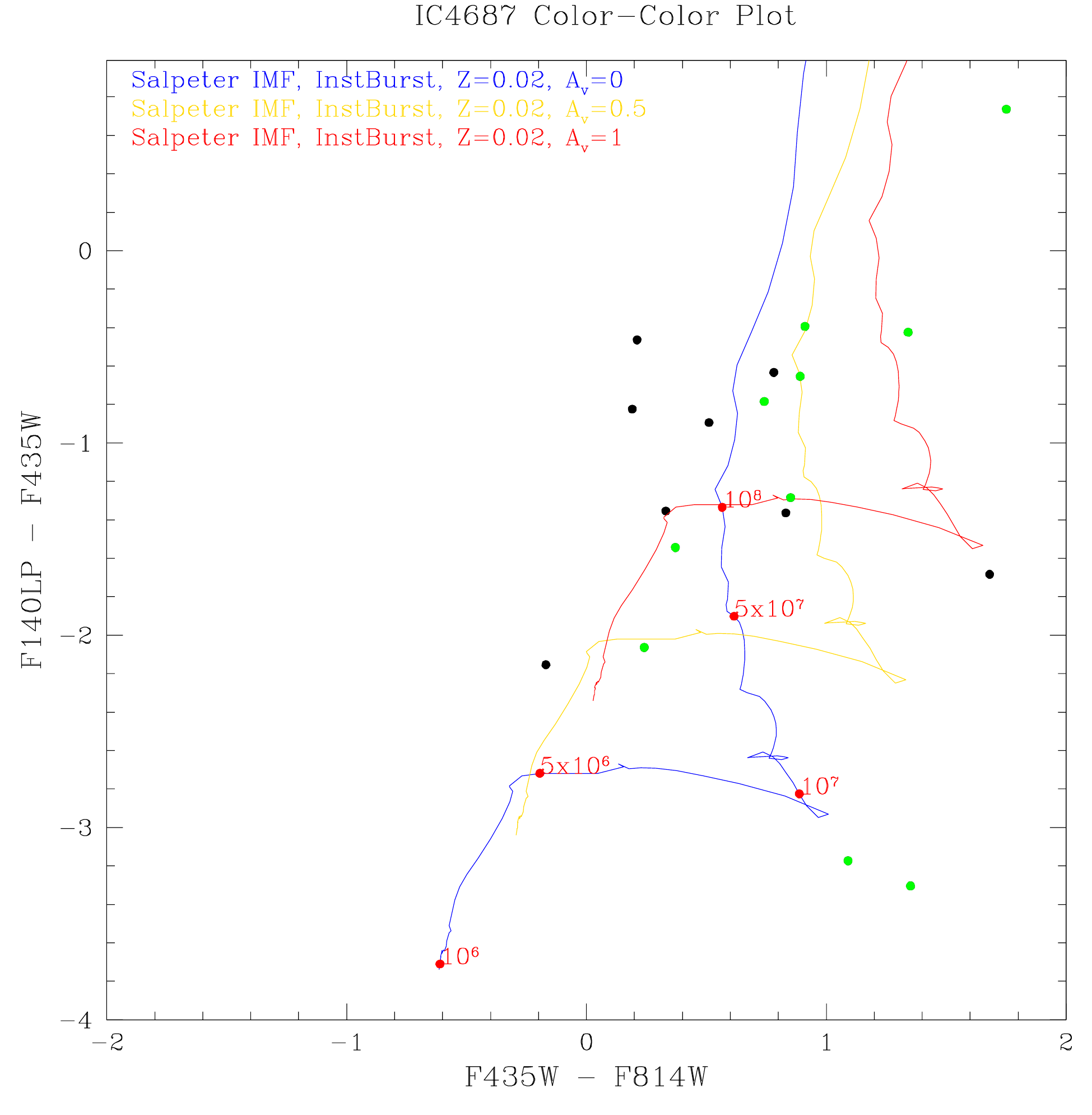}
\caption{Color-Color plot of all star clusters identified in IC 4687N in F814W, F435W, and F140LP plotted against SSP models with various amount of visual extinction. The green points correspond to the clusters found in dustier regions of the galaxy in Figure 45}
\end{figure*}

\begin{deluxetable*}{ccccccccc}
\tabletypesize{\footnotesize}
\tablewidth{0pt}
\tablecaption {Observed Properties of Star Clusters in IC 4687N}
\tablehead{
\colhead{ID} & \colhead{RA} & \colhead{Dec} & \colhead{$M_{B}$} & \colhead{$\sigma_{B}$} & \colhead{$M_{I}$} & \colhead{$\sigma_{I}$} & \colhead{$M_{FUV}$} & \colhead{$\sigma_{FUV}$}} \\
\startdata
1 & 273.4245591 & -57.72132281 & -8.50 & 0.12 & -9.60 & 0.08 & -11.68 & 0.07 \\
2 & 273.4139193 & -57.72470603 & -12.61 & 0.02 & -14.37 & 0.02 & -11.88 & 0.10 \\
3 & 273.4144883 & -57.72459737 & -14.21 & 0.03 & -14.55 & 0.03 & -15.57 & 0.02 \\
4 & 273.4149949 & -57.72446495 & -13.59 & 0.03 & -14.94 & 0.03 & -14.02 & 0.07 \\
5 & 273.4167849 & -57.72402211 & -11.55 & 0.05 & -12.45 & 0.04 & -12.21 & 0.07 \\
6 & 273.4229748 & -57.72139585 & -10.56 & 0.03 & -10.94 & 0.04 & -12.11 & 0.08 \\
7 & 273.4230317 & -57.72231699 & -9.93 & 0.04 & -10.18 & 0.08 & -12.00 & 0.02 \\
8 & 273.4240561 & -57.72152951 & -8.93 & 0.09 & -10.29 & 0.05 & -12.24 & 0.09 \\
9 & 273.4126981 & -57.72621106 & -12.88 & 0.05 & -13.10 & 0.03 & -13.35 & 0.03 \\
10 & 273.4127251 & -57.72525032 & -11.08 & 0.09 & -11.83 & 0.12 & -11.87 & 0.10 \\
11 & 273.4124913 & -57.72636619 & -11.88 & 0.02 & -12.08 & 0.04 & -12.71 & 0.05 \\
12 & 273.4126885 & -57.72499269 & -12.73 & 0.05 & -13.52 & 0.03 & -13.37 & 0.02 \\
13 & 273.4126222 & -57.72421981 & -10.61 & 0.05 & -11.47 & 0.03 & -11.90 & 0.10 \\
14 & 273.4126044 & -57.72447105 & -10.59 & 0.06 & -12.28 & 0.02 & -12.28 & 0.07 \\
15 & 273.4122004 & -57.72543793 & -11.09 & 0.13 & -11.61 & 0.11 & -11.99 & 0.09 \\
16 & 273.4120619 & -57.72591473 & -12.36 & 0.02 & -13.28 & 0.02 & -12.76 & 0.04 \\
17 & 273.4118376 & -57.72547113 & -11.04 & 0.15 & -11.88 & 0.09 & -12.41 & 0.06 \\
18 & 273.4107751 & -57.72651001 & -11.29 & 0.05 & -11.13 & 0.09 & -13.45 & 0.05
\enddata
\end{deluxetable*}

\begin{deluxetable*}{ccccccc}
\tabletypesize{\footnotesize}
\tablewidth{0pt}
\tablecaption {Derived Properties of Star Clusters in IC 4687N}
\tablehead{
\colhead{ID} & \colhead{Log(Age)} & \colhead{$\sigma_{Age}$} & \colhead{Log($M/M_{\odot}$)} & \colhead{$\sigma_{M}$} & \colhead{$A_{V}$} & \colhead{$\sigma_{A_{V}}$}} \\
\startdata  
1 & 6.94 & 0.01 & 3.98 & 0.22 & 0.01 & 0.03 \\
2 & 7.76 & 0.43 & 7.13 & 0.35 & 1.80 & 0.30 \\
3 & 6.66 & 0.09 & 6.37 & 0.27 & 1.00 & 0.19 \\
4 & 7.65 & 0.37 & 7.09 & 0.31 & 1.10 & 0.25 \\
5 & 8.01 & 0.75 & 6.16 & 0.58 & 0.50 & 0.59 \\
6 & 6.36 & 0.35 & 5.39 & 0.37 & 1.50 & 0.32 \\
7 & 6.72 & 0.45 & 4.41 & 0.37 & 0.50 & 0.32 \\
8 & 6.94 & 0.01 & 4.15 & 0.20 & 0.01 & 0.01 \\
9 & 6.66 & 0.01 & 6.10 & 0.18 & 1.50 & 0.53 \\
10 & 6.00 & 0.85 & 5.94 & 0.66 & 2.10 & 0.69 \\
11 & 6.66 & 0.31 & 5.60 & 0.18 & 1.30 & 0.05 \\
12 & 6.70 & 0.97 & 6.05 & 0.70 & 1.50 & 0.75 \\
13 & 7.74 & 0.38 & 5.58 & 0.33 & 0.40 & 0.27 \\
14 & 7.65 & 0.01 & 5.36 & 0.20 & 0.10 & 0.08 \\
15 & 6.68 & 0.66 & 5.28 & 0.62 & 1.30 & 0.63 \\
16 & 8.06 & 0.86 & 6.57 & 0.70 & 0.60 & 0.76 \\
17 & 6.80 & 0.48 & 5.20 & 0.45 & 1.00 & 0.39 \\
18 & 6.66 & 0.09 & 4.88 & 0.24 & 0.40 & 0.15
\enddata
\end{deluxetable*}

\subsection{IC 4687S}

The southern galaxy contains several bright clusters in the  nuclear region as well as a series of dust lanes. Their is little evidence of extended tidal structures which contain star clusters. The maximum $A_{V}$ adopted for this galaxy is 3.7 mags of visual extinction (Rich et al. 2012). 

\begin{figure*}
\centering
\includegraphics[scale=0.25]{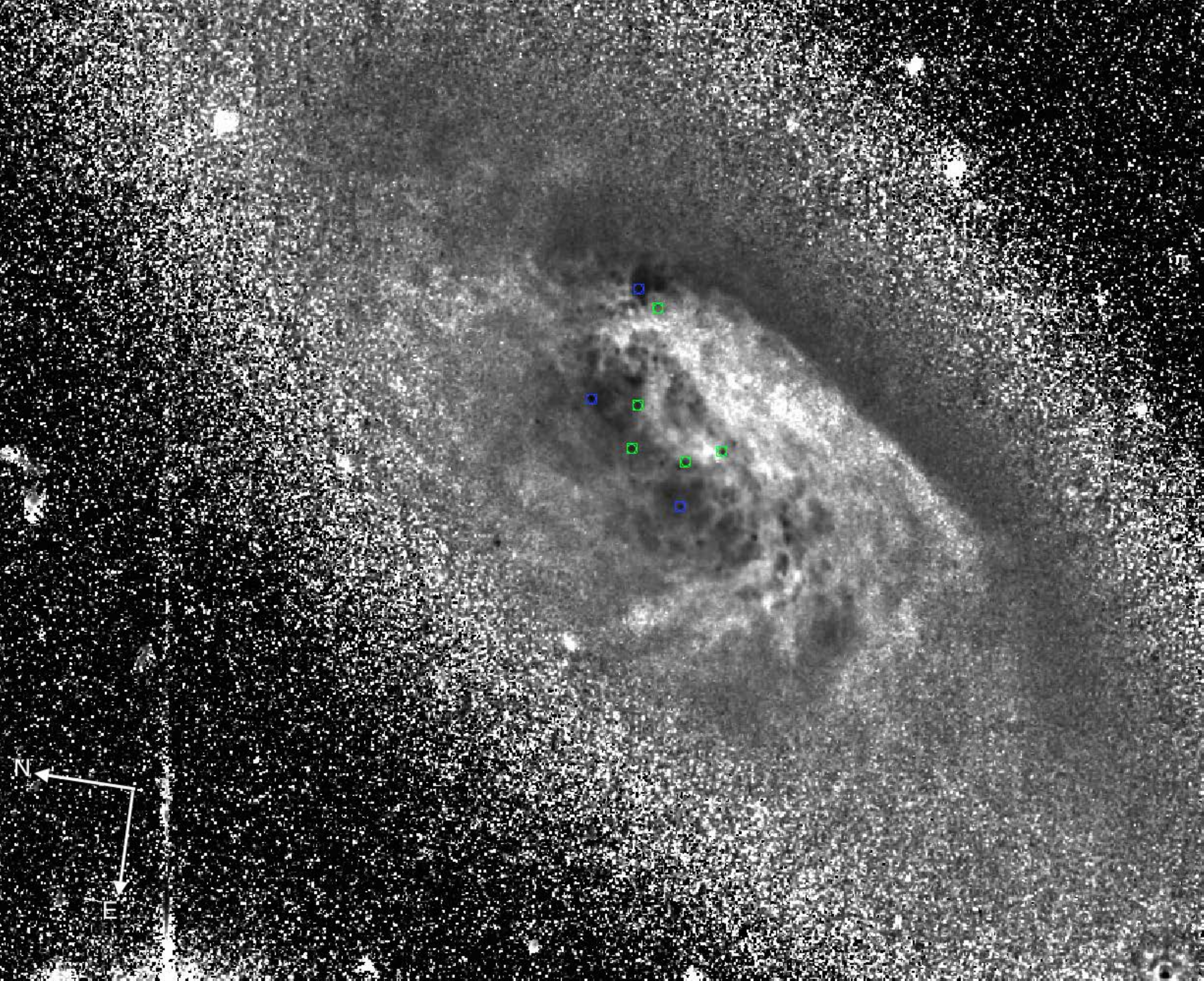}
\caption{Inverted black and white B-I image of IC 4687S taken with HST ACS/WFC F814W and F435W. The bright emission corresponds to redder (i.e. dustier) regions of the galaxy. The blue centroids correspond to clusters found in relatively ``dust-free'' regions of these galaxies, whereas the green centroids correspond to clusters found in relatively dustier regions of the galaxy.}
\end{figure*}

\begin{figure*}
\centering
\includegraphics[scale=0.55]{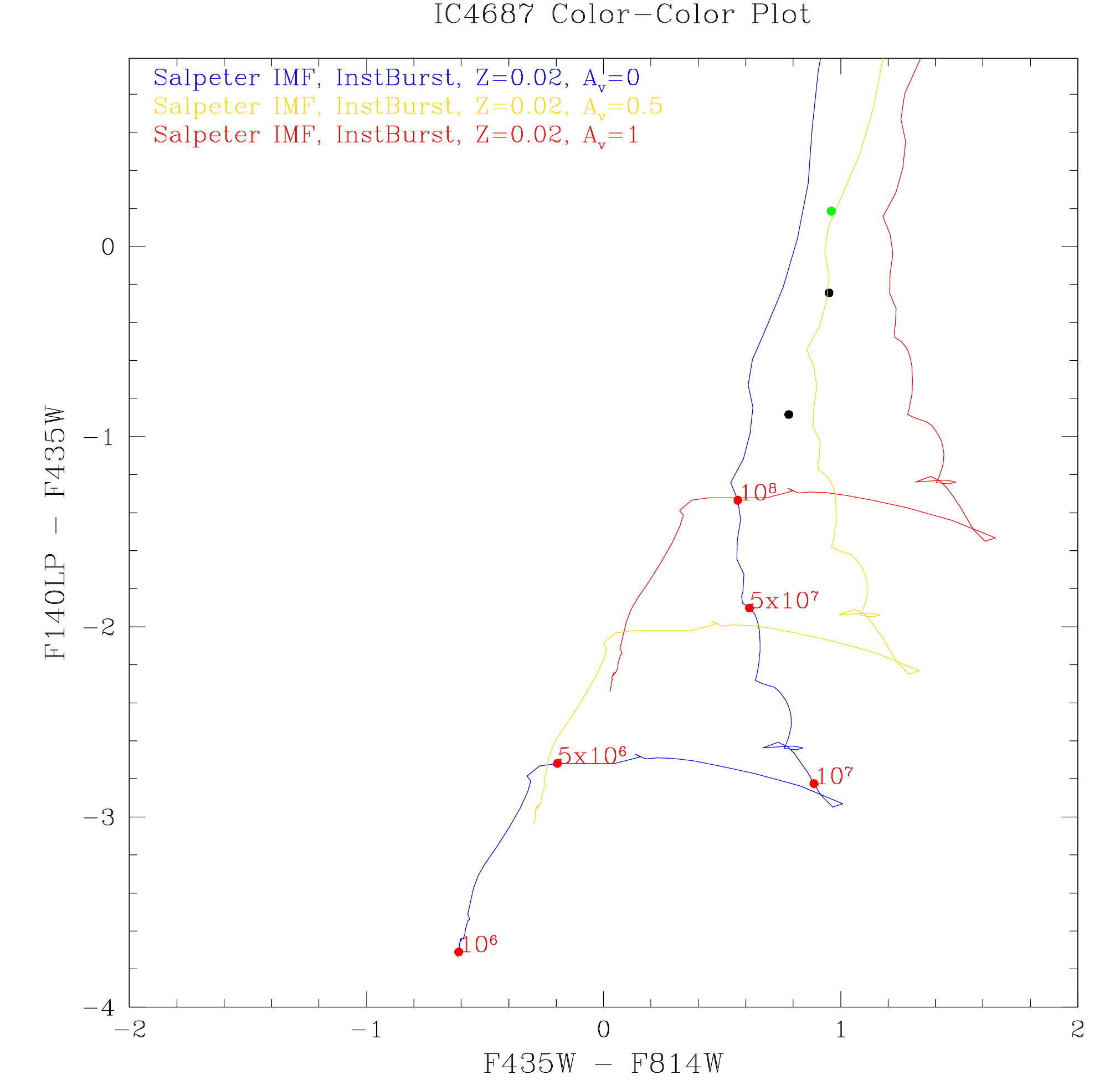}
\caption{Color-Color plot of all star clusters identified in IC 4687S in F814W, F435W, and F140LP plotted against SSP models with various amount of visual extinction. The green points correspond to the clusters found in dustier regions of the galaxy in Figure 47}
\end{figure*}

\begin{deluxetable*}{ccccccccc}
\tabletypesize{\footnotesize}
\tablewidth{0pt}
\tablecaption {Observed Properties of Star Clusters in IC 4687S}
\tablehead{
\colhead{ID} & \colhead{RA} & \colhead{Dec} & \colhead{$M_{B}$} & \colhead{$\sigma_{B}$} & \colhead{$M_{I}$} & \colhead{$\sigma_{I}$} & \colhead{$M_{FUV}$} & \colhead{$\sigma_{FUV}$}} \\
\startdata
1 & 273.4156651 & -57.7476378 & -9.47 & 0.16 & -11.63 & 0.09 & -10.47 & 0.36 \\
2 & 273.4174919 & -57.74709814 & -12.71 & 0.02 & -13.50 & 0.01 & -13.60 & 0.02 \\
3 & 273.4181417 & -57.74844252 & -13.11 & 0.02 & -14.08 & 0.04 & -12.93 & 0.04 \\
4 & 273.4184439 & -57.74809628 & -12.50 & 0.05 & -13.27 & 0.07 & -11.14 & 0.18 \\
5 & 273.4192425 & -57.74811065 & -11.58 & 0.05 & -12.54 & 0.03 & -11.83 & 0.10
\enddata
\end{deluxetable*}

\begin{deluxetable*}{ccccccc}
\tabletypesize{\footnotesize}
\tablewidth{0pt}
\tablecaption {Derived Properties of Star Clusters in IC 4687S}
\tablehead{
\colhead{ID} & \colhead{Log(Age)} & \colhead{$\sigma_{Age}$} & \colhead{Log($M/M_{\odot}$)} & \colhead{$\sigma_{M}$} & \colhead{$A_{V}$} & \colhead{$\sigma_{A_{V}}$}} \\
\startdata  
1 & 6.94 & 0.11 & 5.16 & 0.37 & 1.50 & 0.28 \\
2 & 6.72 & 0.34 & 5.94 & 0.32 & 1.30 & 0.26 \\
3 & 8.41 & 0.66 & 6.94 & 0.66 & 0.30 & 0.70 \\
4 & 8.66 & 0.03 & 6.75 & 0.19 & 0.01 & 0.05 \\
5 & 6.40 & 0.76 & 6.19 & 0.73 & 2.40 & 0.79
\enddata
\end{deluxetable*}

\subsection{NGC 6786}

NGC 6786 is an early-stage merger consisting of a pair of face-on galaxies with a nuclear separation of $\sim 72''$ (37 kpc). Faint star clusters are seen along the inner spiral structure. The brightest clusters sit at roughly $5''$ (3 kpc) from the southwest nucleus in the large arm/tail extending between the two galaxies. The maximum $A_{V}$ adopted for this galaxy is 2.0 mags of visual extinction (Martin et al. 1991). 

\begin{figure*}
\centering
\includegraphics[scale=0.25]{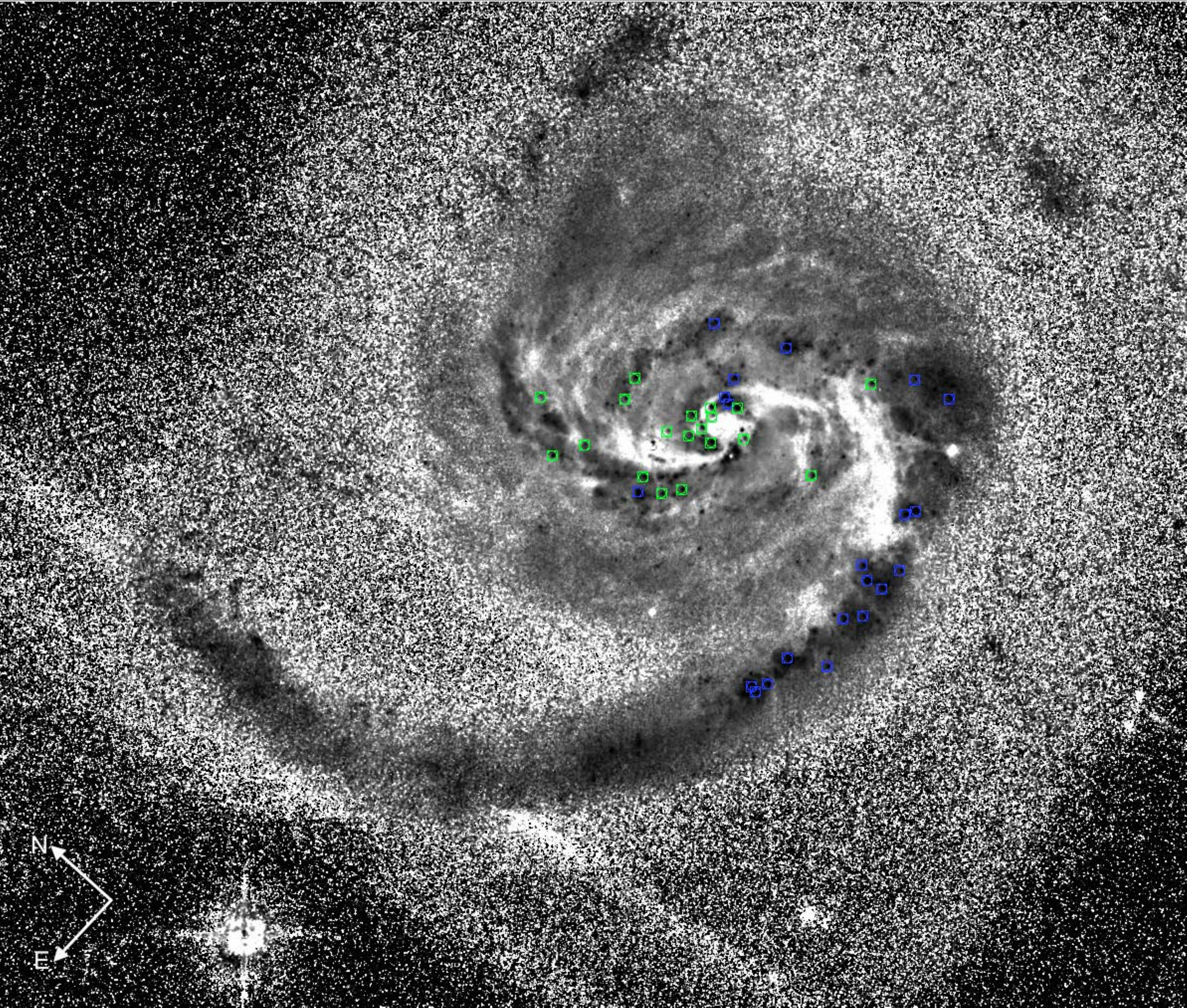}
\caption{Inverted black and white B-I image of NGC 6786 taken with HST ACS/WFC F814W and F435W. The bright emission corresponds to redder (i.e. dustier) regions of the galaxy. The blue centroids correspond to clusters found in relatively ``dust-free'' regions of these galaxies, whereas the green centroids correspond to clusters found in relatively dustier regions of the galaxy.}
\end{figure*}

\begin{figure*}
\centering
\includegraphics[scale=0.55]{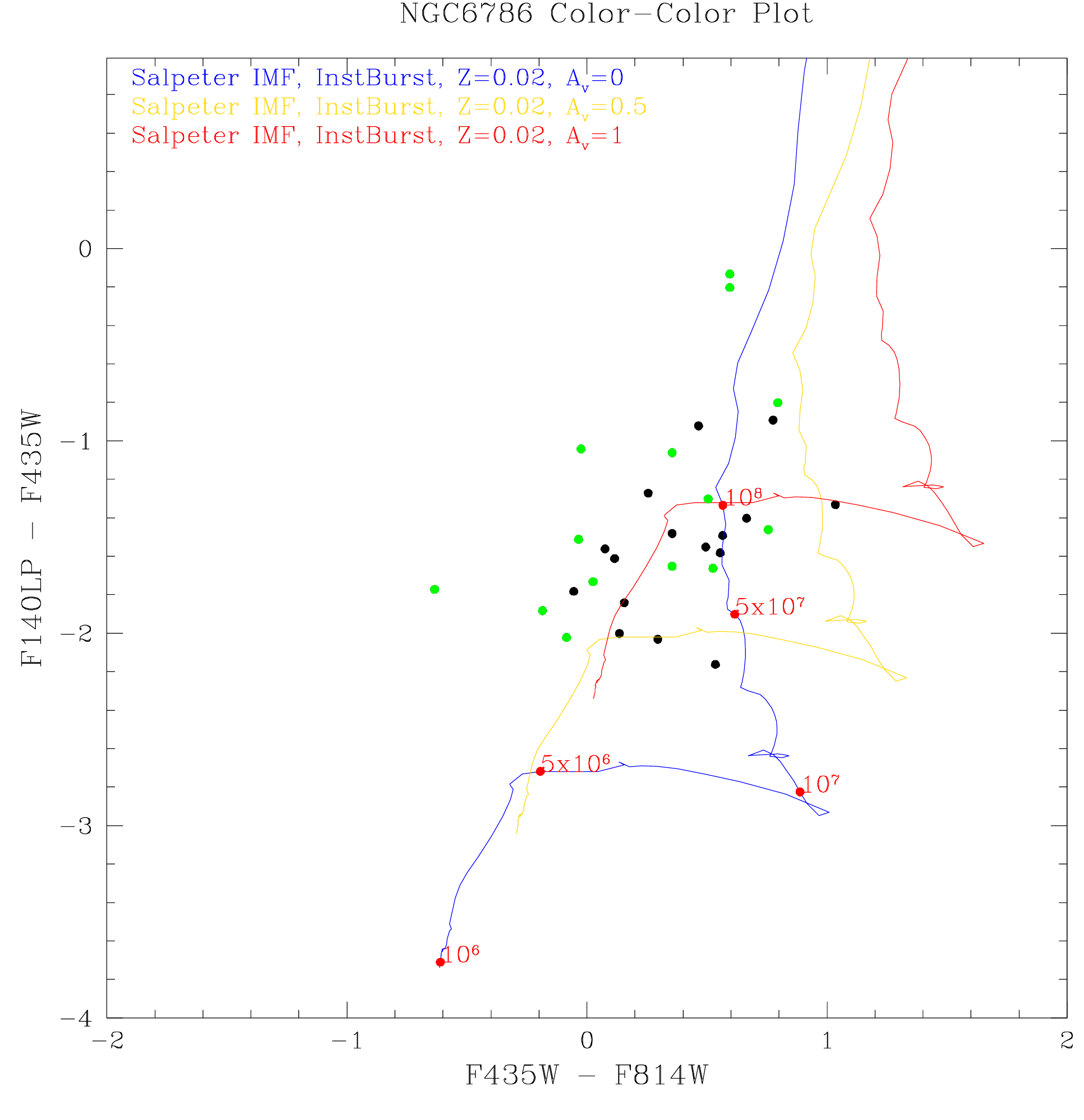}
\caption{Color-Color plot of all star clusters identified in NGC 6786 in F814W, F435W, and F140LP plotted against SSP models with various amount of visual extinction. The green points correspond to the clusters found in dustier regions of the galaxy in Figure 49}
\end{figure*}

\begin{deluxetable*}{ccccccccc}
\tabletypesize{\footnotesize}
\tablewidth{0pt}
\tablecaption {Observed Properties of Star Clusters in NGC 6786}
\tablehead{
\colhead{ID} & \colhead{RA} & \colhead{Dec} & \colhead{$M_{B}$} & \colhead{$\sigma_{B}$} & \colhead{$M_{I}$} & \colhead{$\sigma_{I}$} & \colhead{$M_{FUV}$} & \colhead{$\sigma_{FUV}$}} \\
\startdata
1  & 287.7197193 & 73.41100197 & -10.99 & 0.02 & -11.66 & 0.03 & -12.39 & 0.05 \\
2  & 287.719232  & 73.40994657 & -12.43 & 0.02 & -12.93 & 0.04 & -13.98 & 0.03 \\
3  & 287.7238745 & 73.41165138 & -10.79 & 0.04 & -11.30 & 0.04 & -12.09 & 0.07 \\
4  & 287.7178111 & 73.40815357 & -9.46  & 0.06 & -10.03 & 0.06 & -10.95 & 0.22 \\
5  & 287.7238495 & 73.41044623 & -11.77 & 0.07 & -12.37 & 0.14 & -11.97 & 0.08 \\
6  & 287.7234966 & 73.41051976 & -13.43 & 0.02 & -13.41 & 0.06 & -14.47 & 0.02 \\
7  & 287.7229265 & 73.41018561 & -15.39 & 0.01 & -15.31 & 0.01 & -17.41 & 0.01 \\
8  & 287.7227232 & 73.41040649 & -15.05 & 0.01 & -15.31 & 0.01 & -16.32 & 0.01 \\
9  & 287.7229402 & 73.41032417 & -14.37 & 0.02 & -14.51 & 0.03 & -16.37 & 0.01 \\
10 & 287.7267694 & 73.41270096 & -9.77  & 0.18 & -11.79 & 0.04 & -10.19 & 0.11 \\
11 & 287.7250088 & 73.41164334 & -10.13 & 0.06 & -10.10 & 0.11 & -11.64 & 0.11 \\
12 & 287.7242707 & 73.4107057  & -12.76 & 0.03 & -13.56 & 0.04 & -13.56 & 0.02 \\
13 & 287.7250077 & 73.41029358 & -14.21 & 0.03 & -13.58 & 0.10 & -15.98 & 0.01 \\
14 & 287.7251949 & 73.41061029 & -12.67 & 0.06 & -13.27 & 0.10 & -12.80 & 0.04 \\
15 & 287.7254972 & 73.41090859 & -10.45 & 0.06 & -10.27 & 0.20 & -12.33 & 0.02 \\
16 & 287.7241289 & 73.40990629 & -12.92 & 0.02 & -14.11 & 0.05 & -11.81 & 0.09 \\
17 & 287.7290544 & 73.4121962  & -11.94 & 0.03 & -12.30 & 0.03 & -13.00 & 0.03 \\
18 & 287.7242387 & 73.40883366 & -10.24 & 0.04 & -10.27 & 0.08 & -11.97 & 0.08 \\
19 & 287.7279844 & 73.41092482 & -11.98 & 0.02 & -12.74 & 0.03 & -13.44 & 0.02 \\
20 & 287.7277016 & 73.41036998 & -12.38 & 0.02 & -12.91 & 0.02 & -14.04 & 0.01 \\
21 & 287.7287715 & 73.41090241 & -13.47 & 0.01 & -13.63 & 0.01 & -15.31 & 0.01 \\
22 & 287.7283128 & 73.41059111 & -11.42 & 0.03 & -11.78 & 0.04 & -13.07 & 0.03 \\
23 & 287.7235043 & 73.40730799 & -11.63 & 0.05 & -12.67 & 0.04 & -12.96 & 0.03 \\
24 & 287.7270499 & 73.40763824 & -11.77 & 0.01 & -12.55 & 0.02 & -12.66 & 0.04 \\
25 & 287.7264275 & 73.40713593 & -10.01 & 0.07 & -10.54 & 0.10 & -12.16 & 0.07 \\
26 & 287.7276317 & 73.40724684 & -11.46 & 0.04 & -11.93 & 0.03 & -12.38 & 0.06 \\
27 & 287.727598  & 73.40748488 & -12.61 & 0.01 & -12.56 & 0.02 & -14.39 & 0.01 \\
28 & 287.7297973 & 73.40754253 & -10.67 & 0.03 & -11.23 & 0.03 & -12.25 & 0.06 \\
29 & 287.7327209 & 73.40798896 & -13.58 & 0.01 & -13.70 & 0.01 & -15.19 & 0.01 \\
30 & 287.7322458 & 73.40744712 & -9.67  & 0.04 & -9.97  & 0.07 & -11.70 & 0.20 \\
31 & 287.7343067 & 73.40807688 & -11.58 & 0.02 & -11.66 & 0.04 & -13.14 & 0.03 \\
32 & 287.7349114 & 73.40817718 & -11.63 & 0.04 & -11.99 & 0.05 & -13.11 & 0.06
\enddata
\end{deluxetable*}

\begin{deluxetable*}{ccccccc}
\tabletypesize{\footnotesize}
\tablewidth{0pt}
\tablecaption {Derived Properties of Star Clusters in NGC 6786}
\tablehead{
\colhead{ID} & \colhead{Log(Age)} & \colhead{$\sigma_{Age}$} & \colhead{Log($M/M_{\odot}$)} & \colhead{$\sigma_{M}$} & \colhead{$A_{V}$} & \colhead{$\sigma_{A_{V}}$}} \\
\startdata  
1 & 6.78 & 0.49 & 5.48 & 0.39 & 0.90 & 0.35 \\
2 & 6.74 & 0.56 & 5.97 & 0.43 & 0.80 & 0.40 \\
3 & 6.22 & 0.67 & 5.41 & 0.50 & 1.70 & 0.48 \\
4 & 6.74 & 0.55 & 4.84 & 0.38 & 0.90 & 0.33 \\
5 & 8.41 & 0.07 & 6.64 & 0.23 & 0.01 & 0.13 \\
6 & 6.66 & 0.01 & 6.45 & 0.17 & 1.10 & 0.46 \\
7 & 6.66 & 0.34 & 6.97 & 0.17 & 0.50 & 0.01 \\
8 & 6.66 & 0.01 & 7.10 & 0.17 & 1.00 & 0.11 \\
9 & 6.44 & 0.06 & 6.57 & 0.21 & 1.10 & 0.11 \\
10 & 7.00 & 0.17 & 5.91 & 0.26 & 1.70 & 0.16 \\
11 & 6.66 & 0.01 & 5.03 & 0.20 & 0.80 & 0.07 \\
12 & 8.01 & 0.08 & 6.99 & 0.22 & 0.40 & 0.13 \\
13 & 6.66 & 0.04 & 6.50 & 0.18 & 0.50 & 0.05 \\
14 & 8.41 & 0.04 & 7.00 & 0.20 & 0.01 & 0.07 \\
15 & 6.66 & 0.04 & 5.05 & 0.23 & 0.60 & 0.13 \\
16 & 8.51 & 0.05 & 7.44 & 0.23 & 0.50 & 0.15 \\
17 & 6.66 & 0.57 & 5.86 & 0.43 & 1.20 & 0.40 \\
18 & 6.66 & 0.03 & 5.02 & 0.21 & 0.70 & 0.10 \\
19 & 7.72 & 0.44 & 6.47 & 0.35 & 0.30 & 0.30 \\
20 & 7.86 & 0.55 & 6.55 & 0.40 & 0.01 & 0.36 \\
21 & 6.48 & 0.04 & 6.43 & 0.22 & 1.10 & 0.12 \\
22 & 5.10 & 0.04 & 5.55 & 0.19 & 1.50 & 0.07 \\
23 & 7.63 & 0.24 & 6.37 & 0.25 & 0.50 & 0.16 \\
24 & 6.72 & 0.05 & 5.84 & 0.18 & 1.30 & 0.06 \\
25 & 6.84 & 0.35 & 4.91 & 0.27 & 0.40 & 0.19 \\
26 & 6.66 & 0.05 & 5.67 & 0.18 & 1.30 & 0.02 \\
27 & 6.66 & 0.35 & 5.97 & 0.18 & 0.70 & 0.03 \\
28 & 6.74 & 0.57 & 5.27 & 0.43 & 0.80 & 0.40 \\
29 & 6.66 & 0.29 & 6.41 & 0.17 & 0.80 & 0.04 \\
30 & 6.72 & 0.05 & 4.70 & 0.20 & 0.50 & 0.09 \\
31 & 6.66 & 0.58 & 5.61 & 0.18 & 0.80 & 0.04 \\
32 & 6.40 & 0.05 & 5.67 & 0.20 & 1.50 & 0.08
\enddata
\end{deluxetable*}

\subsection{IRAS 20351+2521}

IRAS 20351+2521 is an early-stage merger containing multiple star clusters in the northern region where the spiral arms diffuse into multiple components beyond the inner $\sim 5''$ (4 kpc). The maximum $A_{V}$ adopted for this galaxy is 4.7 mags of visual extinction, which is lower than the cited value of 9.4 mags, but prevents our model from predicting masses unrealistically high for even the most massive YSCs found in the sample. (Stierwalt et al. 2013). 

\begin{figure*}
\centering
\includegraphics[scale=0.25]{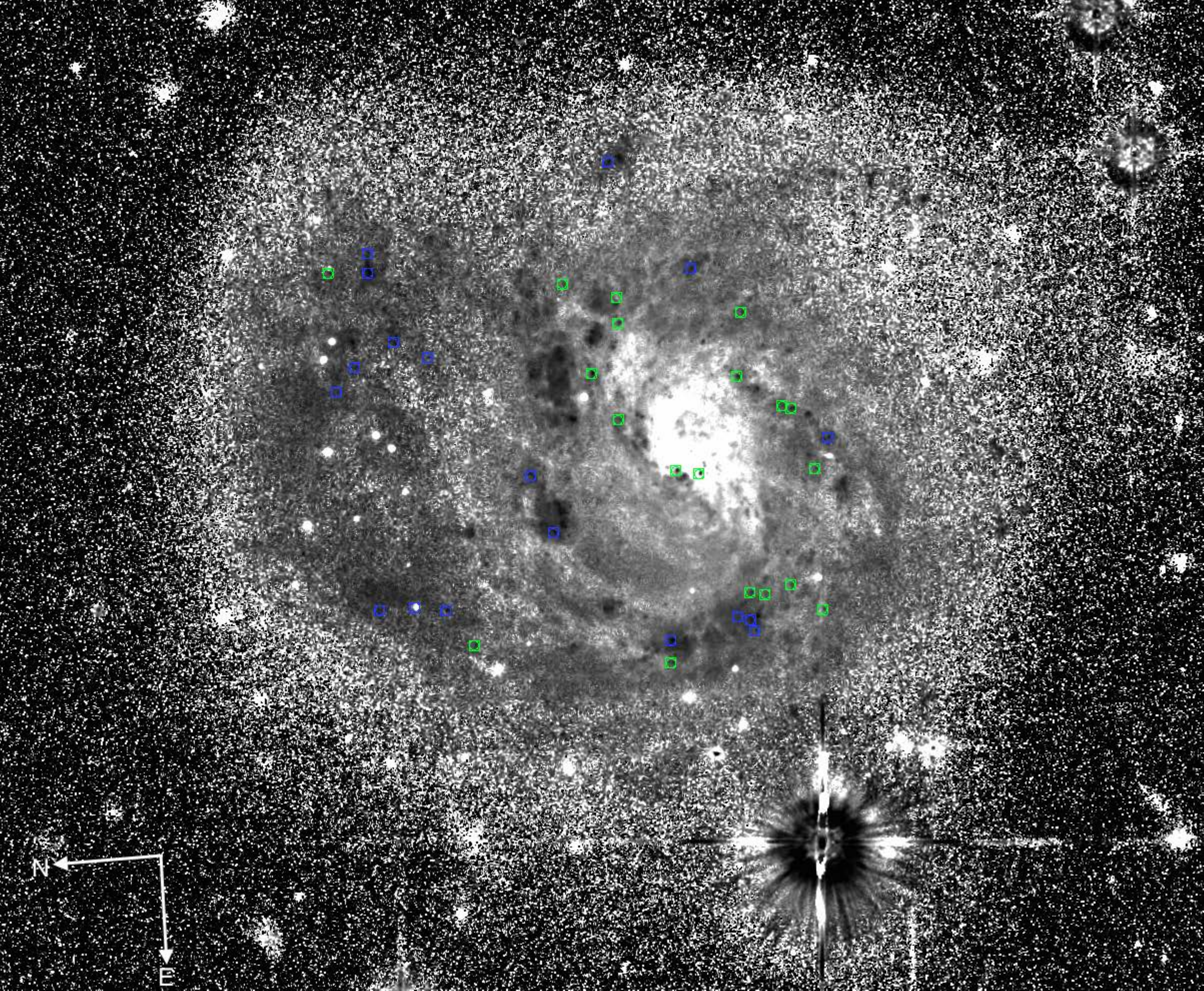}
\caption{Inverted black and white B-I image of IRAS 20351+2521 taken with HST ACS/WFC F814W and F435W. The bright emission corresponds to redder (i.e. dustier) regions of the galaxy. The blue centroids correspond to clusters found in relatively ``dust-free'' regions of these galaxies, whereas the green centroids correspond to clusters found in relatively dustier regions of the galaxy.}
\end{figure*}

\begin{figure*}
\centering
\includegraphics[scale=0.55]{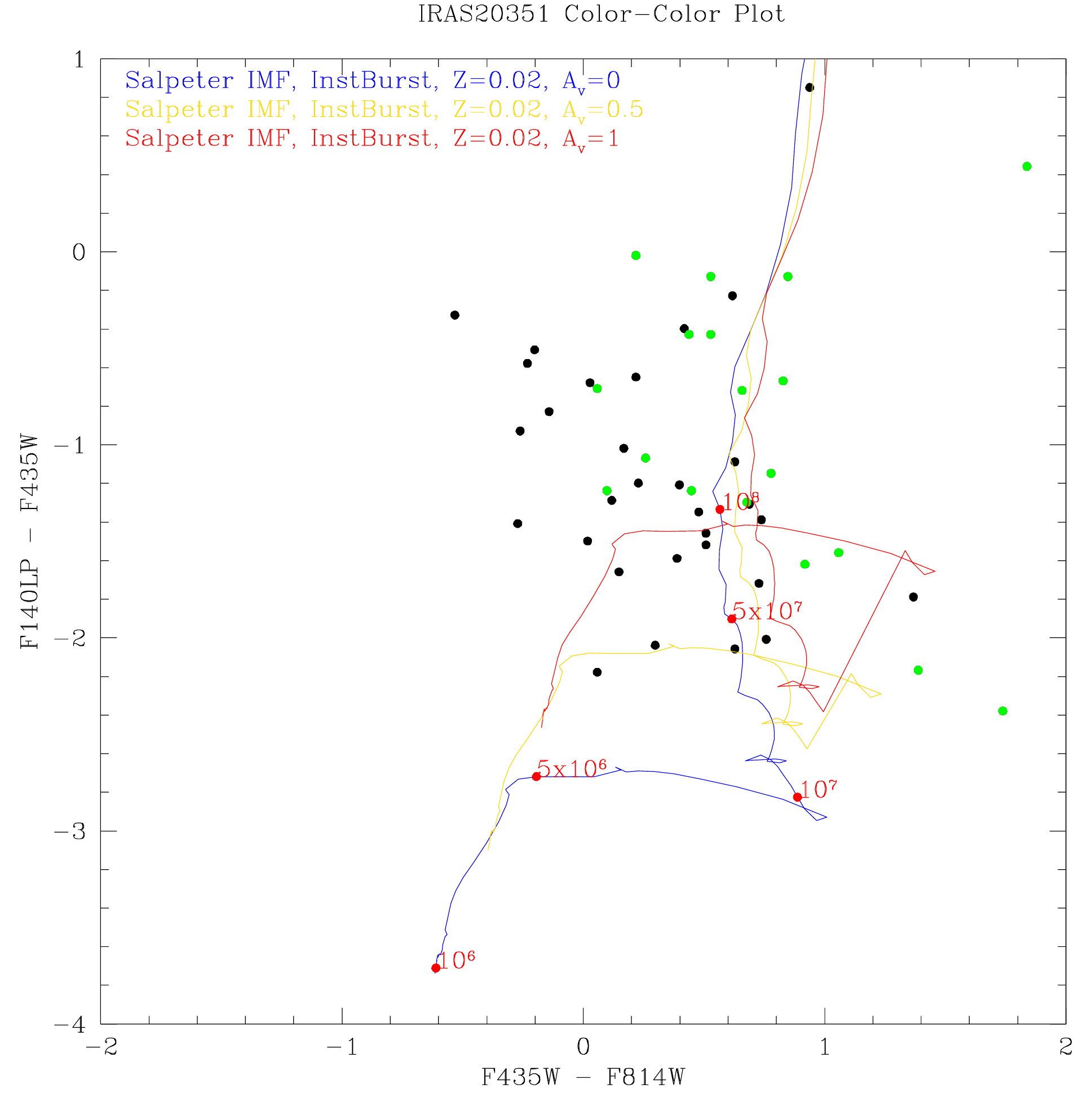}
\caption{Color-Color plot of all star clusters identified in IRAS 20351+2521 in F814W, F435W, and F140LP plotted against SSP models with various amount of visual extinction. The green points correspond to the clusters found in dustier regions of the galaxy in Figure 51}
\end{figure*}

\begin{deluxetable*}{ccccccccc}
\tabletypesize{\footnotesize}
\tablewidth{0pt}
\tablecaption {Observed Properties of Star Clusters in IRAS 20351+2521}
\tablehead{
\colhead{ID} & \colhead{RA} & \colhead{Dec} & \colhead{$M_{B}$} & \colhead{$\sigma_{B}$} & \colhead{$M_{I}$} & \colhead{$\sigma_{I}$} & \colhead{$M_{FUV}$} & \colhead{$\sigma_{FUV}$}} \\
\startdata
1 & 309.3189111 & 25.52810092 & -11.77 & 0.04 & -11.99 & 0.06 & -12.96 & 0.25 \\
2 & 309.3201669 & 25.53148902 & -12.10 & 0.03 & -12.78 & 0.03 & -13.40 & 0.14 \\
3 & 309.3204655 & 25.53148786 & -14.20 & 0.01 & -14.34 & 0.02 & -15.85 & 0.04 \\
4 & 309.3207482 & 25.52879944 & -11.57 & 0.05 & -12.24 & 0.05 & -12.86 & 0.05 \\
5 & 309.3205846 & 25.52701311 & -12.18 & 0.08 & -11.64 & 0.16 & -12.50 & 0.26 \\
6 & 309.3209895 & 25.52805392 & -12.80 & 0.04 & -14.63 & 0.01 & -12.35 & 0.24 \\
7 & 309.3213802 & 25.52804452 & -12.92 & 0.03 & -13.44 & 0.03 & -13.04 & 0.17 \\
8 & 309.3212834 & 25.52633971 & -11.39 & 0.06 & -12.12 & 0.06 & -12.77 & 0.22 \\
9 & 309.3215513 & 25.53117662 & -11.92 & 0.07 & -11.68 & 0.12 & -12.49 & 0.16 \\
10 & 309.3227499 & 25.52581676 & -12.05 & 0.06 & -12.70 & 0.07 & -12.76 & 0.16 \\
11 & 309.3227896 & 25.52569385 & -12.91 & 0.02 & -12.96 & 0.05 & -13.61 & 0.07 \\
12 & 309.3232605 & 25.52520015 & -12.72 & 0.05 & -13.34 & 0.06 & -13.80 & 0.12 \\
13 & 309.3237357 & 25.52539922 & -11.60 & 0.05 & -12.42 & 0.03 & -12.26 & 0.22 \\
14 & 309.3236793 & 25.52933194 & -11.99 & 0.06 & -12.00 & 0.07 & -13.48 & 0.12 \\
15 & 309.3256351 & 25.52614986 & -11.60 & 0.08 & -12.04 & 0.10 & -12.83 & 0.04 \\
16 & 309.3255973 & 25.526354 & -12.15 & 0.05 & -12.24 & 0.08 & -13.38 & 0.06 \\
17 & 309.3256314 & 25.53101137 & -11.60 & 0.05 & -14.92 & 0.01 & -12.71 & 0.01 \\
18 & 309.3256579 & 25.53150391 & -11.02 & 0.08 & -11.77 & 0.06 & -13.02 & 0.21 \\
19 & 309.3259664 & 25.52654554 & -11.17 & 0.15 & -12.53 & 0.06 & -12.95 & 0.27 \\
20 & 309.3261938 & 25.52631453 & -13.23 & 0.02 & -13.84 & 0.02 & -13.45 & 0.16 \\
21 & 309.3262881 & 25.52747562 & -13.98 & 0.02 & -14.03 & 0.02 & -16.15 & 0.01 \\
22 & 309.3262506 & 25.53020748 & -11.03 & 0.05 & -11.65 & 0.05 & -13.08 & 0.03
\enddata
\end{deluxetable*}

\begin{deluxetable*}{ccccccc}
\tabletypesize{\footnotesize}
\tablewidth{0pt}
\tablecaption {Derived Properties of Star Clusters in IRAS 20351+2521}
\tablehead{
\colhead{ID} & \colhead{Log(Age)} & \colhead{$\sigma_{Age}$} & \colhead{Log($M/M_{\odot}$)} & \colhead{$\sigma_{M}$} & \colhead{$A_{V}$} & \colhead{$\sigma_{A_{V}}$}} \\
\startdata  
1 & 6.54 & 0.33 & 5.69 & 0.16 & 1.50 & 0.02 \\
2 & 6.74 & 0.19 & 5.65 & 0.16 & 1.10 & 0.31 \\
3 & 6.54 & 0.32 & 6.50 & 0.15 & 1.20 & 0.23 \\
4 & 7.76 & 0.42 & 5.92 & 0.16 & 0.40 & 0.20 \\
5 & 6.52 & 0.40 & 6.07 & 0.18 & 1.90 & 0.45 \\
6 & 6.78 & 0.05 & 6.57 & 0.16 & 2.30 & 0.59 \\
7 & 8.51 & 0.11 & 6.72 & 0.16 & 0.01 & 0.53 \\
8 & 7.70 & 0.12 & 5.81 & 0.17 & 0.40 & 0.16 \\
9 & 6.52 & 0.23 & 5.91 & 0.17 & 1.80 & 0.46 \\
10 & 8.36 & 0.53 & 6.26 & 0.17 & 0.01 & 0.19 \\
11 & 6.52 & 0.08 & 6.31 & 0.16 & 1.80 & 0.05 \\
12 & 6.64 & 0.37 & 6.19 & 0.16 & 1.60 & 0.54 \\
13 & 7.96 & 0.12 & 6.13 & 0.16 & 0.60 & 0.08 \\
14 & 6.52 & 0.18 & 5.67 & 0.17 & 1.30 & 0.12 \\
15 & 6.52 & 0.29 & 5.68 & 0.18 & 1.60 & 0.24 \\
16 & 6.52 & 0.32 & 5.84 & 0.16 & 1.50 & 0.24 \\
17 & 8.11 & 0.39 & 6.22 & 0.16 & 0.60 & 0.23 \\
18 & 6.94 & 0.52 & 5.09 & 0.18 & 0.50 & 0.59 \\
19 & 7.65 & 0.53 & 5.53 & 0.22 & 0.10 & 0.36 \\
20 & 6.52 & 0.07 & 6.65 & 0.16 & 2.20 & 0.41 \\
21 & 6.44 & 0.19 & 6.49 & 0.16 & 1.10 & 0.19 \\
22 & 7.36 & 0.24 & 5.39 & 0.16 & 0.30 & 0.58
\enddata
\end{deluxetable*}

\subsection{II ZW 096}

Inami et al. (2010) discusses the detailed morphology of this galaxy at length. II Zw 096 is a mid-stage merging system. The western component is a roughly face-on spiral galaxy with star clusters along the spiral arms. The southeast end of the spiral, approximately $11.6''$ (8.4 kpc) from the nucleus, contains a distinct cluster-rich region. The maximum $A_{V}$ adopted for this galaxy is 3.0 mags of visual extinction (Inami et al. 2010). 

\begin{figure*}
\centering
\includegraphics[scale=0.25]{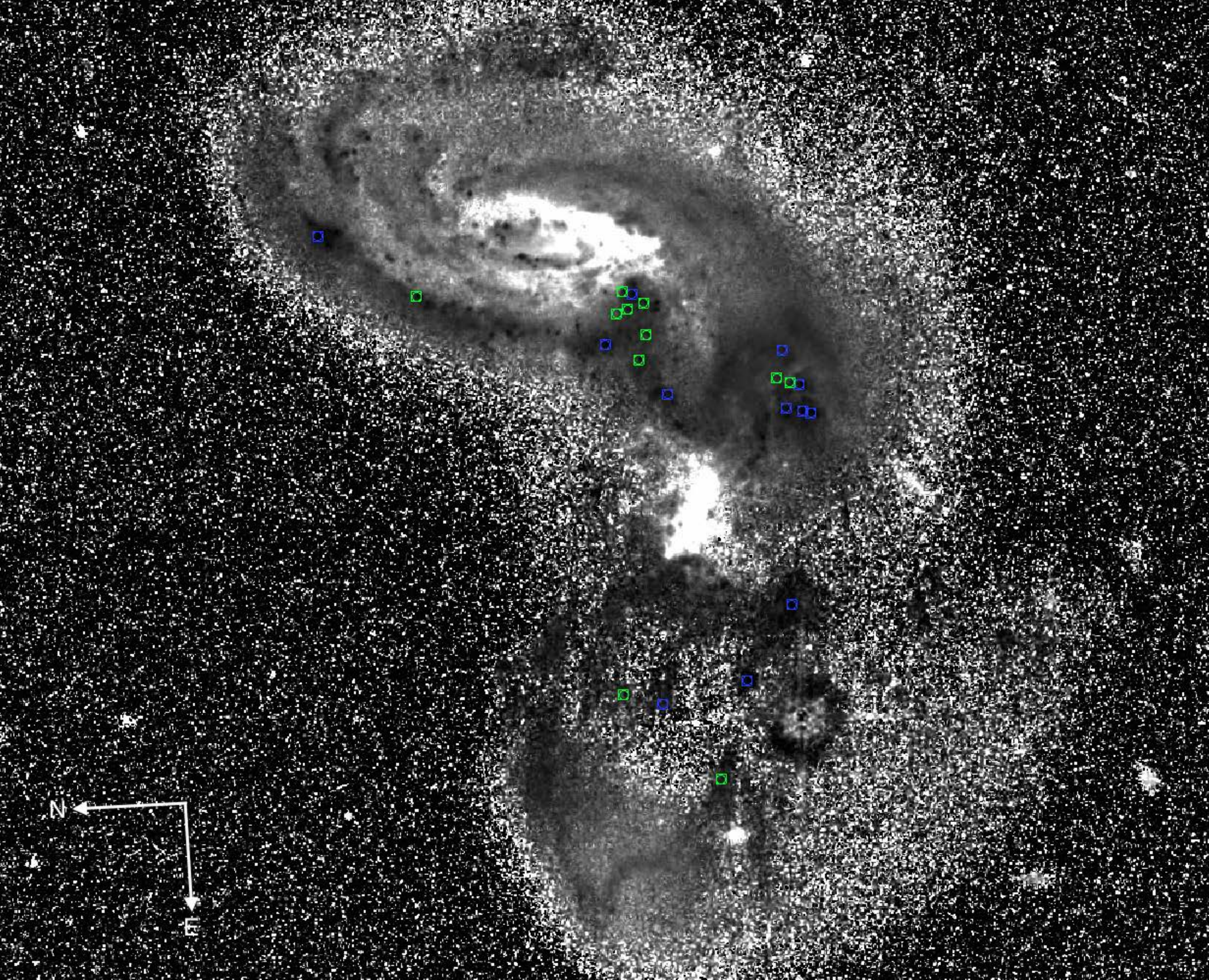}
\caption{Inverted black and white B-I image of II ZW 096 taken with HST ACS/WFC F814W and F435W. The bright emission corresponds to redder (i.e. dustier) regions of the galaxy. The blue centroids correspond to clusters found in relatively ``dust-free'' regions of these galaxies, whereas the green centroids correspond to clusters found in relatively dustier regions of the galaxy.}
\end{figure*}

\begin{figure*}
\centering
\includegraphics[scale=0.55]{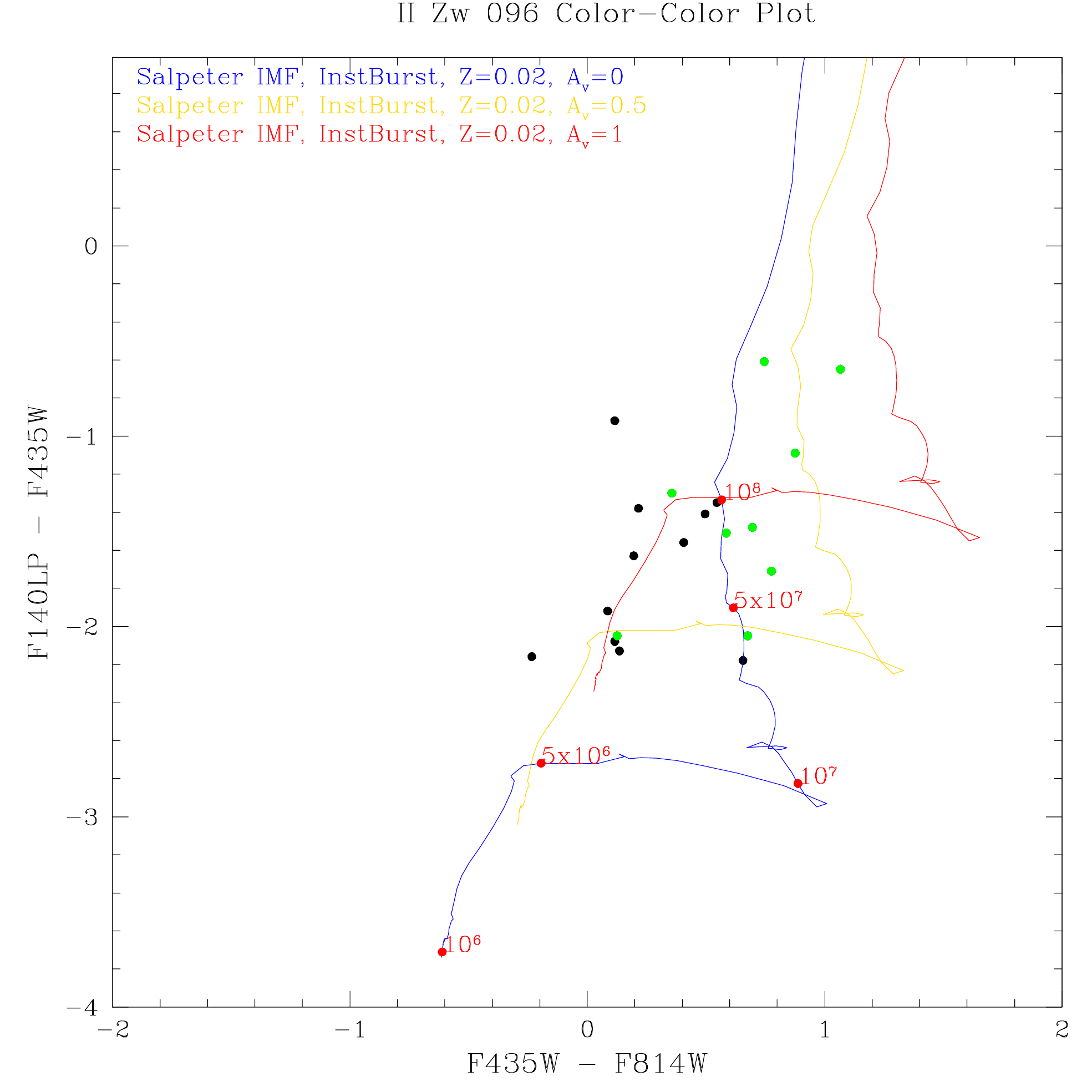}
\caption{Color-Color plot of all star clusters identified in II ZW 096 in F814W, F435W, and F140LP plotted against SSP models with various amount of visual extinction. The green points correspond to the clusters found in dustier regions of the galaxy in Figure 53}
\end{figure*}

\begin{deluxetable*}{ccccccccc}
\tabletypesize{\footnotesize}
\tablewidth{0pt}
\tablecaption {Observed Properties of Star Clusters in II ZW 096}
\tablehead{
\colhead{ID} & \colhead{RA} & \colhead{Dec} & \colhead{$M_{B}$} & \colhead{$\sigma_{B}$} & \colhead{$M_{I}$} & \colhead{$\sigma_{I}$} & \colhead{$M_{FUV}$} & \colhead{$\sigma_{FUV}$}} \\
\startdata
1 & 314.347937 & 17.13209914 & -14.40 & 0.02 & -14.52 & 0.01 & -16.48 & 0.01 \\
2 & 314.3488605 & 17.12855532 & -13.82 & 0.03 & -14.60 & 0.03 & -15.53 & 0.05 \\
3 & 314.3490196 & 17.12831079 & -13.91 & 0.02 & -14.50 & 0.05 & -15.42 & 0.02 \\
4 & 314.3489006 & 17.12844657 & -14.02 & 0.05 & -14.14 & 0.07 & -14.94 & 0.12 \\
5 & 314.3487591 & 17.13097999 & -13.81 & 0.02 & -13.94 & 0.02 & -15.86 & 0.01 \\
6 & 314.3491325 & 17.12863715 & -13.77 & 0.02 & -14.84 & 0.01 & -14.42 & 0.03 \\
7 & 314.349413 & 17.12830611 & -13.18 & 0.02 & -14.06 & 0.02 & -14.27 & 0.04 \\
8 & 314.3494999 & 17.12879568 & -14.29 & 0.01 & -14.38 & 0.02 & -16.21 & 0.01 \\
9 & 314.3497151 & 17.12841414 & -16.14 & 0.01 & -16.84 & 0.01 & -17.62 & 0.01 \\
10 & 314.3501466 & 17.12654531 & -17.92 & 0.01 & -18.33 & 0.01 & -19.48 & 0.01 \\
11 & 314.3501115 & 17.12664894 & -15.51 & 0.05 & -16.26 & 0.05 & -16.12 & 0.03 \\
12 & 314.3500502 & 17.12680274 & -13.52 & 0.11 & -13.88 & 0.19 & -14.82 & 0.07 \\
13 & 314.3501617 & 17.12810109 & -14.29 & 0.02 & -14.84 & 0.02 & -15.64 & 0.02 \\
14 & 314.3504248 & 17.12671799 & -14.41 & 0.10 & -14.91 & 0.09 & -15.82 & 0.05 \\
15 & 314.3504765 & 17.12652561 & -16.42 & 0.01 & -16.64 & 0.02 & -17.80 & 0.02 \\
16 & 314.350505 & 17.12642631 & -16.31 & 0.01 & -16.51 & 0.02 & -17.94 & 0.01 \\
17 & 314.352852 & 17.12679784 & -14.29 & 0.01 & -14.06 & 0.03 & -16.45 & 0.01 \\
18 & 314.353755 & 17.1273767 & -13.74 & 0.01 & -13.88 & 0.02 & -15.87 & 0.01 \\
19 & 314.3538332 & 17.12884574 & -13.71 & 0.01 & -14.39 & 0.01 & -15.76 & 0.01 \\
20 & 314.3539789 & 17.12839369 & -12.26 & 0.02 & -12.92 & 0.02 & -14.44 & 0.03
\enddata
\end{deluxetable*}

\begin{deluxetable*}{ccccccc}
\tabletypesize{\footnotesize}
\tablewidth{0pt}
\tablecaption {Derived Properties of Star Clusters in II ZW 096}
\tablehead{
\colhead{ID} & \colhead{Log(Age)} & \colhead{$\sigma_{Age}$} & \colhead{Log($M/M_{\odot}$)} & \colhead{$\sigma_{M}$} & \colhead{$A_{V}$} & \colhead{$\sigma_{A_{V}}$}} \\
\startdata  
1 & 6.40 & 0.08 & 6.76 & 0.22 & 1.20 & 0.12 \\
2 & 7.40 & 0.25 & 6.64 & 0.24 & 0.50 & 0.16 \\
3 & 6.74 & 0.45 & 6.32 & 0.39 & 1.00 & 0.35 \\
4 & 6.52 & 0.53 & 6.70 & 0.52 & 1.70 & 0.50 \\
5 & 6.42 & 0.12 & 6.52 & 0.24 & 1.20 & 0.16 \\
6 & 7.63 & 0.27 & 7.04 & 0.26 & 1.00 & 0.19 \\
7 & 7.63 & 0.26 & 6.64 & 0.26 & 0.70 & 0.18 \\
8 & 6.52 & 0.02 & 6.49 & 0.18 & 1.10 & 0.06 \\
9 & 7.65 & 0.01 & 7.68 & 0.17 & 0.40 & 0.13 \\
10 & 5.70 & 0.62 & 8.60 & 0.44 & 1.60 & 0.41 \\
11 & 6.54 & 0.85 & 7.45 & 0.81 & 2.00 & 0.93 \\
12 & 6.54 & 0.55 & 6.39 & 0.53 & 1.50 & 0.52 \\
13 & 6.72 & 0.61 & 6.60 & 0.54 & 1.20 & 0.54 \\
14 & 6.58 & 0.62 & 6.73 & 0.55 & 1.40 & 0.55 \\
15 & 6.54 & 0.01 & 7.50 & 0.17 & 1.40 & 0.03 \\
16 & 6.52 & 0.01 & 7.40 & 0.18 & 1.30 & 0.04 \\
17 & 6.54 & 0.01 & 6.33 & 0.18 & 0.80 & 0.05 \\
18 & 5.70 & 0.41 & 6.71 & 0.27 & 1.20 & 0.20 \\
19 & 6.78 & 0.07 & 5.98 & 0.17 & 0.50 & 0.02 \\
20 & 6.96 & 0.15 & 5.57 & 0.18 & 0.40 & 0.06
\enddata
\end{deluxetable*}

\subsection{ESO 148-IG002}

ESO 148-IG002 is a late-stage merger with a projected nuclear separation of $\sim 4.7''$ (4.2 kpc). The galaxy has a series of bright clusters which lie along a north-south ridge to the east of the bulge. The maximum $A_{V}$ adopted for this galaxy is 2.5 mags of visual extinction (Johansson \& Bergvall 1988). 

\begin{figure*}
\centering
\includegraphics[scale=0.25]{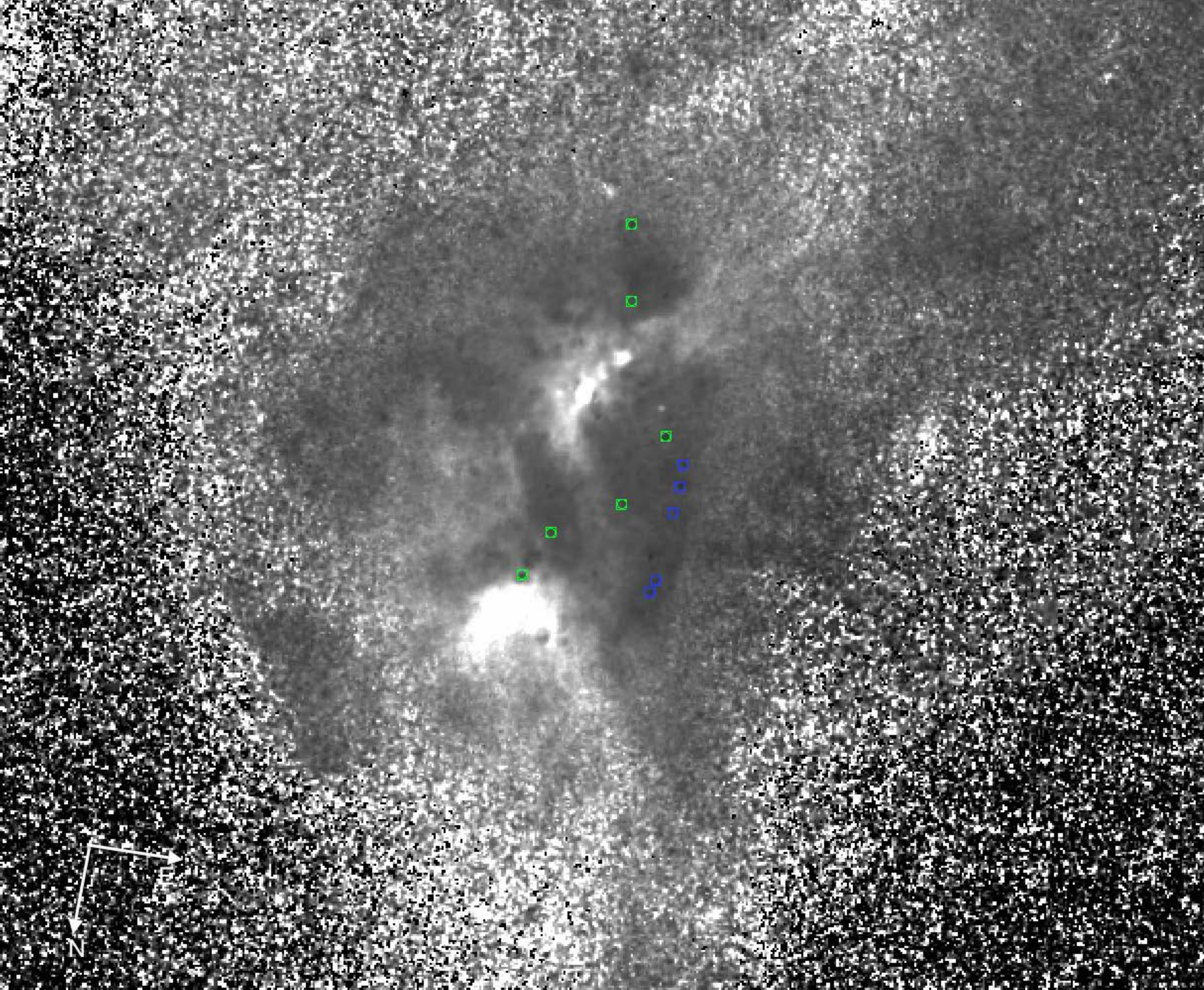}
\caption{Inverted black and white B-I image of ESO 148-IG002 taken with HST ACS/WFC F814W and F435W. The bright emission corresponds to redder (i.e. dustier) regions of the galaxy. The blue centroids correspond to clusters found in relatively ``dust-free'' regions of these galaxies, whereas the green centroids correspond to clusters found in relatively dustier regions of the galaxy.}
\end{figure*}

\begin{figure*}
\centering
\includegraphics[scale=0.55]{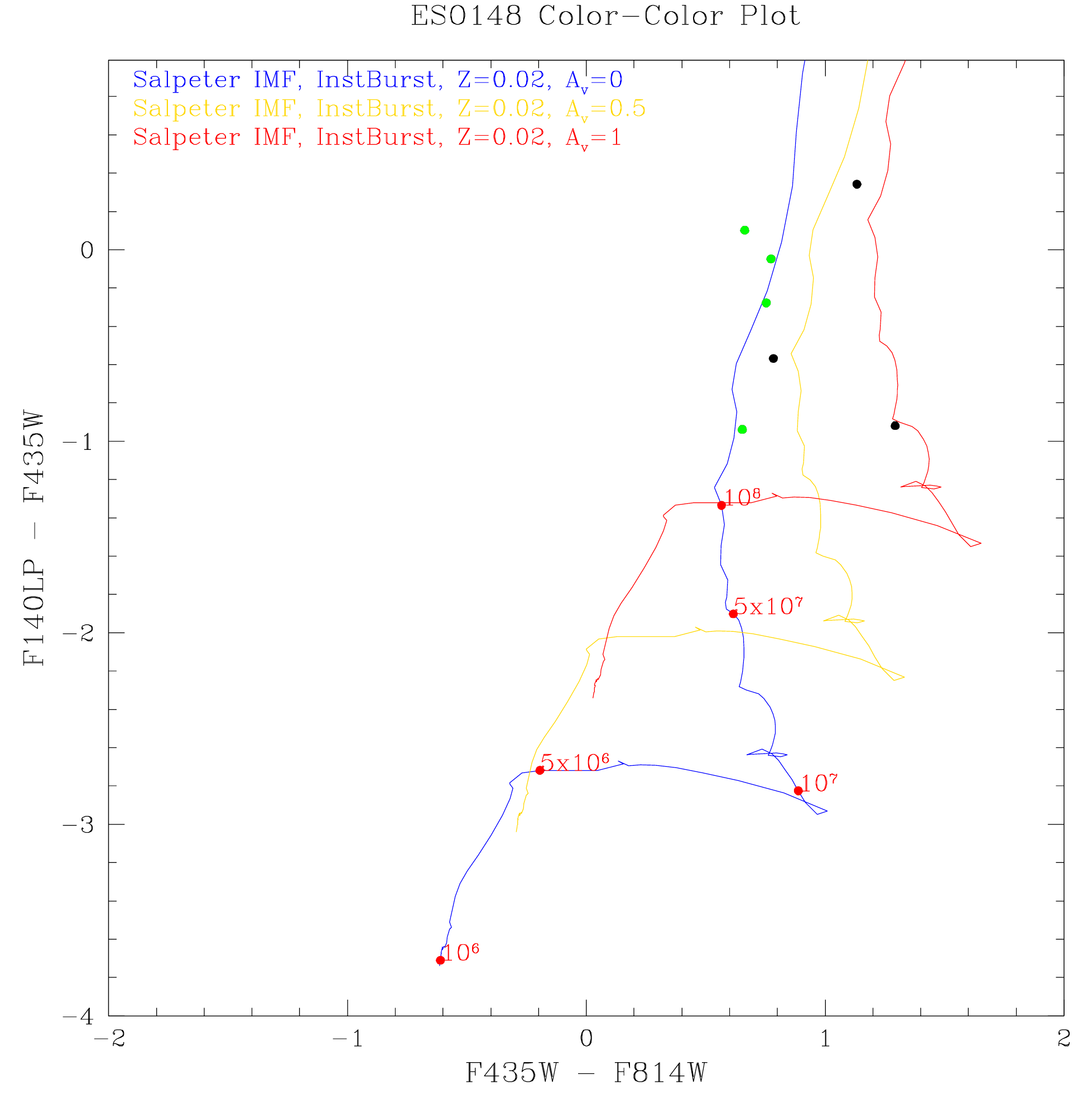}
\caption{Color-Color plot of all star clusters identified in ESO 148-IG002 in F814W, F435W, and F140LP plotted against SSP models with various amount of visual extinction. The green points correspond to the clusters found in dustier regions of the galaxy in Figure 55}
\end{figure*}

\begin{deluxetable*}{ccccccccc}
\tabletypesize{\footnotesize}
\tablewidth{0pt}
\tablecaption {Observed Properties of Star Clusters in ESO 148-IG002}
\tablehead{
\colhead{ID} & \colhead{RA} & \colhead{Dec} & \colhead{$M_{B}$} & \colhead{$\sigma_{B}$} & \colhead{$M_{I}$} & \colhead{$\sigma_{I}$} & \colhead{$M_{FUV}$} & \colhead{$\sigma_{FUV}$}} \\
\startdata
1 & 348.9465405 & -59.05411541 & -14.15 & 0.01 & -14.90 & 0.01 & -14.43 & 0.03 \\
2 & 348.9468576 & -59.0539657 & -13.60 & 0.02 & -14.73 & 0.02 & -13.26 & 0.10 \\
3 & 348.9462256 & -59.05361789 & -12.73 & 0.03 & -13.39 & 0.04 & -12.63 & 0.18 \\
4 & 348.9469037 & -59.05365887 & -11.95 & 0.09 & -13.24 & 0.08 & -12.87 & 0.14 \\
5 & 348.9454074 & -59.05331294 & -13.54 & 0.15 & -14.19 & 0.09 & -14.48 & 0.06 \\
6 & 348.9468404 & -59.05307144 & -14.27 & 0.02 & -15.05 & 0.01 & -14.84 & 0.02 \\
7 & 348.9456965 & -59.05497655 & -14.45 & 0.04 & -15.22 & 0.03 & -14.50 & 0.06
\enddata
\end{deluxetable*}

\begin{deluxetable*}{ccccccc}
\tabletypesize{\footnotesize}
\tablewidth{0pt}
\tablecaption {Derived Properties of Star Clusters in ESO 148-IG002}
\tablehead{
\colhead{ID} & \colhead{Log(Age)} & \colhead{$\sigma_{Age}$} & \colhead{Log($M/M_{\odot}$)} & \colhead{$\sigma_{M}$} & \colhead{$A_{V}$} & \colhead{$\sigma_{A_{V}}$}} \\
\startdata  
1 & 8.46 & 0.03 & 7.17 & 0.16 & 0.01 & 0.78 \\
2 & 8.41 & 0.40 & 7.23 & 0.45 & 0.60 & 0.43 \\
3 & 8.61 & 0.06 & 6.73 & 0.17 & 0.01 & 0.02 \\
4 & 7.65 & 0.25 & 6.21 & 0.24 & 0.80 & 0.14 \\
5 & 8.26 & 0.79 & 6.80 & 0.69 & 0.01 & 0.73 \\
6 & 6.48 & 0.90 & 7.13 & 0.84 & 2.20 & 0.97 \\
7 & 8.56 & 0.04 & 7.38 & 0.17 & 0.01 & 0.04
\enddata
\end{deluxetable*}

\clearpage

\subsection{NGC 7674}

NGC 7674 an early-stage merger with a face-on spiral galaxy and companions to the northeast and southeast. Star clusters are visible along the prominent spiral arms throughout the galaxy. The maximum $A_{V}$ adopted for this galaxy is 2.0 mags of visual extinction (Momjian et al. 2003). 

\begin{figure*}
\centering
\includegraphics[scale=0.25]{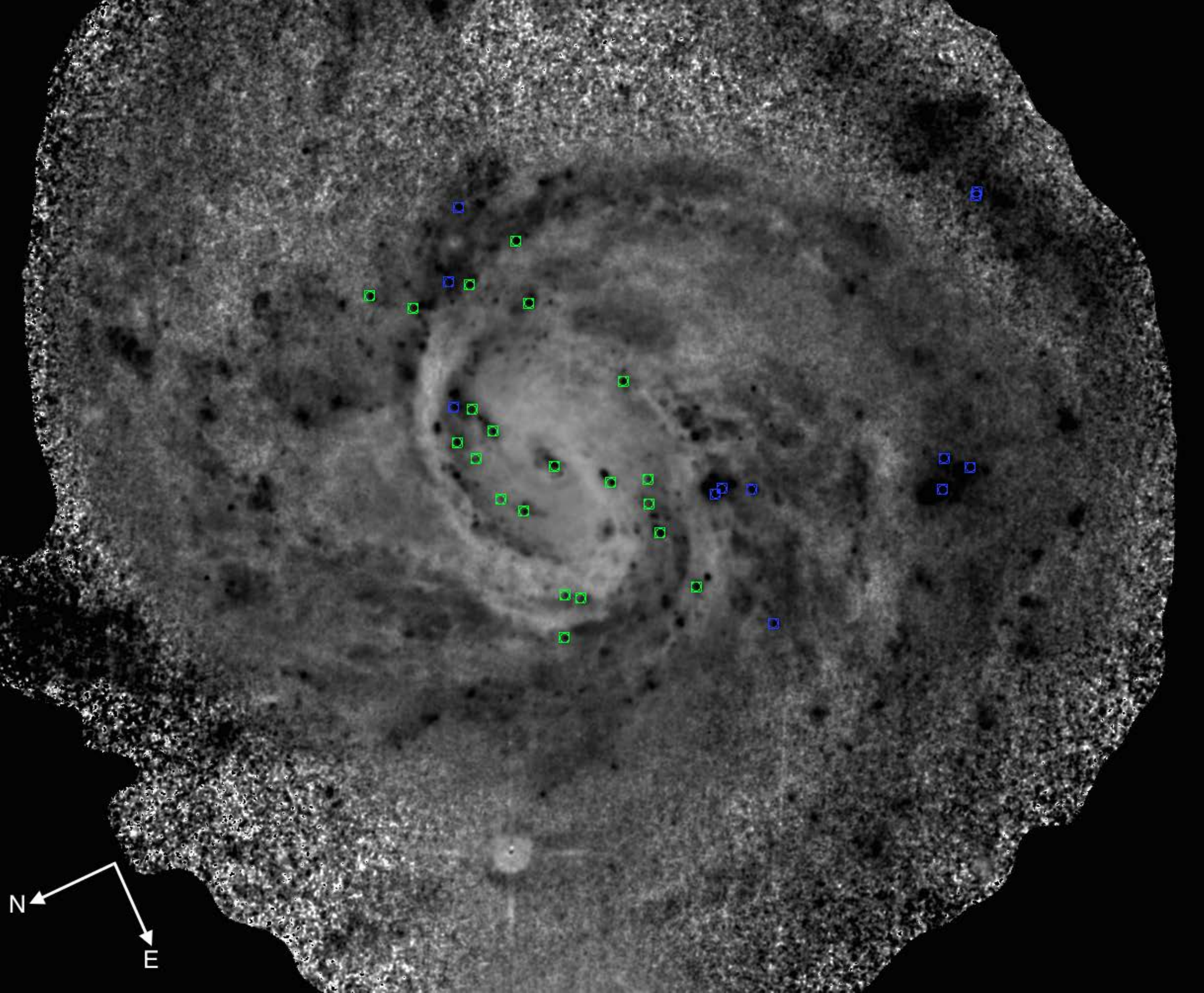}
\caption{Inverted black and white B-I image of NGC 7674 taken with HST ACS/WFC F814W and F435W. The bright emission corresponds to redder (i.e. dustier) regions of the galaxy. The blue centroids correspond to clusters found in relatively ``dust-free'' regions of these galaxies, whereas the green centroids correspond to clusters found in relatively dustier regions of the galaxy.}
\end{figure*}

\begin{figure*}
\centering
\includegraphics[scale=0.55]{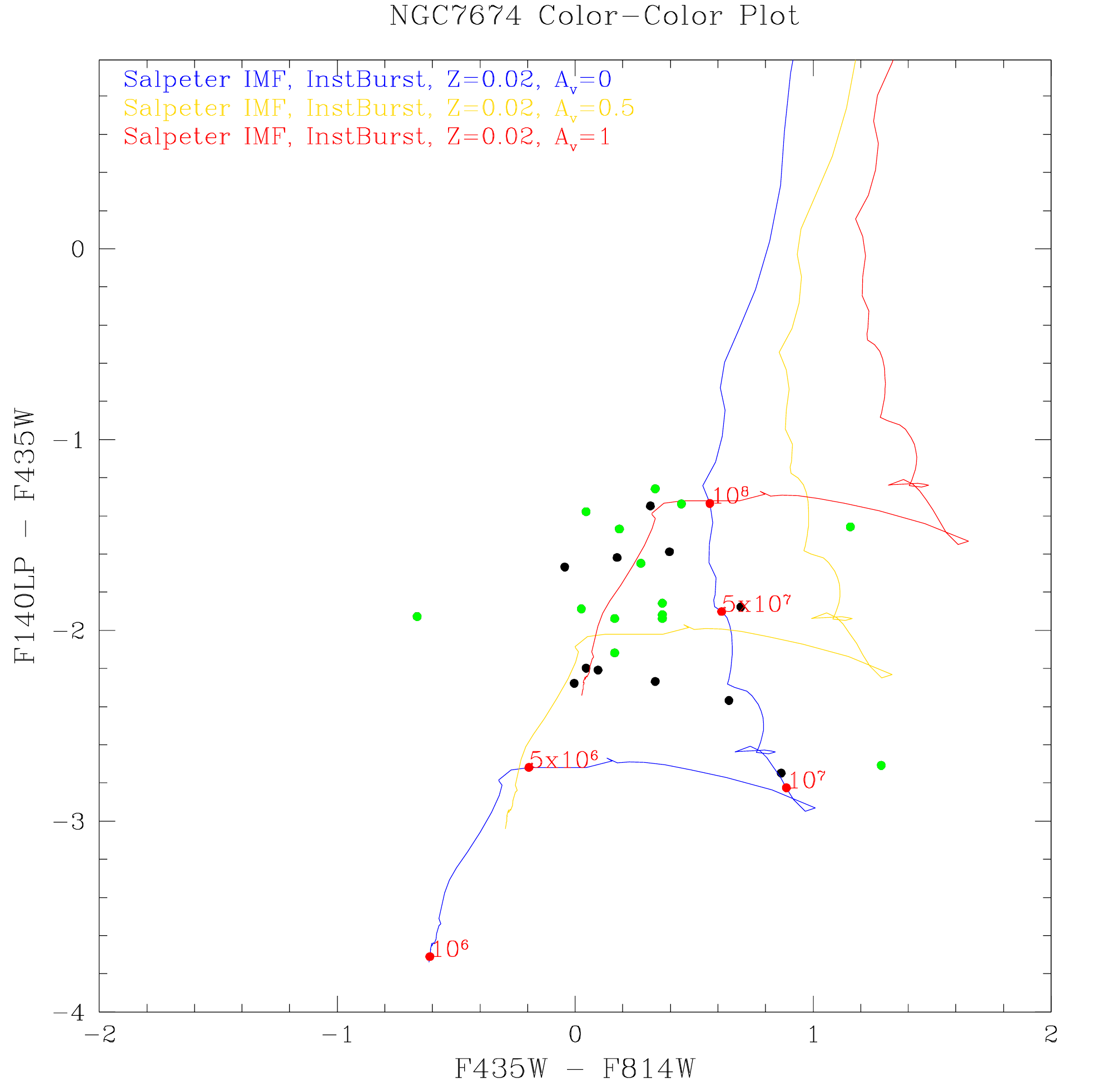}
\caption{Color-Color plot of all star clusters identified in NGC 7674 in F814W, F435W, and F140LP plotted against SSP models with various amount of visual extinction. The green points correspond to the clusters found in dustier regions of the galaxy in Figure 57}
\end{figure*}

\begin{deluxetable*}{ccccccccc}
\tabletypesize{\footnotesize}
\tablewidth{0pt}
\tablecaption {Observed Properties of Star Clusters in NGC 7674}
\tablehead{
\colhead{ID} & \colhead{RA} & \colhead{Dec} & \colhead{$M_{B}$} & \colhead{$\sigma_{B}$} & \colhead{$M_{I}$} & \colhead{$\sigma_{I}$} & \colhead{$M_{FUV}$} & \colhead{$\sigma_{FUV}$}} \\
\startdata
1 & 351.9831239 & 8.778349264 & -12.20 & 0.02 & -12.57 & 0.03 & -14.06 & 0.02 \\
2 & 351.9849596 & 8.772096618 & -11.55 & 0.02 & -11.73 & 0.04 & -13.17 & 0.05 \\
3 & 351.9850026 & 8.772116709 & -11.44 & 0.02 & -11.76 & 0.03 & -12.79 & 0.08 \\
4 & 351.9832985 & 8.779441129 & -10.77 & 0.11 & -11.11 & 0.13 & -13.04 & 0.06 \\
5 & 351.9830552 & 8.780545473 & -10.91 & 0.03 & -11.08 & 0.05 & -12.85 & 0.04 \\
6 & 351.9840117 & 8.778513551 & -12.29 & 0.04 & -12.74 & 0.03 & -13.63 & 0.04 \\
7 & 351.9849762 & 8.780046055 & -12.87 & 0.02 & -12.92 & 0.03 & -15.07 & 0.02 \\
8 & 351.9855017 & 8.779670342 & -12.42 & 0.04 & -12.61 & 0.09 & -13.89 & 0.06 \\
9 & 351.9862949 & 8.779049017 & -17.60 & 0.01 & -18.76 & 0.01 & -19.06 & 0.01 \\
10 & 351.9882912 & 8.773938539 & -12.62 & 0.01 & -13.32 & 0.01 & -14.50 & 0.02 \\
11 & 351.9885463 & 8.773640234 & -11.47 & 0.03 & -12.12 & 0.05 & -13.84 & 0.03 \\
12 & 351.9857724 & 8.780029971 & -11.03 & 0.05 & -10.37 & 0.20 & -12.96 & 0.07 \\
13 & 351.9869669 & 8.777906039 & -11.45 & 0.04 & -11.50 & 0.06 & -12.83 & 0.05 \\
14 & 351.9868119 & 8.778402264 & -13.02 & 0.01 & -13.05 & 0.03 & -14.91 & 0.01 \\
15 & 351.9875291 & 8.777102907 & -13.70 & 0.03 & -13.66 & 0.03 & -15.37 & 0.02 \\
16 & 351.9886887 & 8.774122684 & -11.76 & 0.03 & -11.76 & 0.06 & -14.04 & 0.05 \\
17 & 351.9874886 & 8.7769875 & -11.39 & 0.19 & -12.26 & 0.16 & -14.14 & 0.04 \\
18 & 351.9864376 & 8.77992672 & -11.09 & 0.04 & -11.46 & 0.07 & -13.01 & 0.06 \\
19 & 351.9873013 & 8.778022066 & -10.96 & 0.06 & -11.33 & 0.13 & -12.90 & 0.07 \\
20 & 351.9867238 & 8.779687958 & -12.50 & 0.03 & -12.67 & 0.04 & -14.62 & 0.01 \\
21 & 351.9877438 & 8.778029429 & -12.90 & 0.02 & -13.24 & 0.04 & -14.16 & 0.02 \\
22 & 351.988047 & 8.779603842 & -10.36 & 0.13 & -11.65 & 0.09 & -13.07 & 0.03 \\
23 & 351.9895457 & 8.777038636 & -10.67 & 0.05 & -10.77 & 0.09 & -12.88 & 0.07 \\
24 & 351.9886045 & 8.779833873 & -12.05 & 0.03 & -12.33 & 0.05 & -13.70 & 0.03 \\
25 & 351.9823723 & 8.778916581 & -11.27 & 0.03 & -11.67 & 0.04 & -12.86 & 0.07
\enddata
\end{deluxetable*}

\begin{deluxetable*}{ccccccc}
\tabletypesize{\footnotesize}
\tablewidth{0pt}
\tablecaption {Derived Properties of Star Clusters in NGC 7674}
\tablehead{
\colhead{ID} & \colhead{Log(Age)} & \colhead{$\sigma_{Age}$} & \colhead{Log($M/M_{\odot}$)} & \colhead{$\sigma_{M}$} & \colhead{$A_{V}$} & \colhead{$\sigma_{A_{V}}$}} \\
\startdata  
1 & 6.74 & 0.03 & 5.48 & 0.21 & 0.70 & 0.11 \\
2 & 6.52 & 0.02 & 5.50 & 0.19 & 1.30 & 0.07 \\
3 & 6.52 & 0.03 & 5.56 & 0.19 & 1.50 & 0.08 \\
4 & 6.72 & 0.28 & 4.87 & 0.27 & 0.60 & 0.17 \\
5 & 6.52 & 0.08 & 5.14 & 0.22 & 1.10 & 0.13 \\
6 & 6.44 & 0.10 & 6.14 & 0.24 & 1.70 & 0.16 \\
7 & 6.42 & 0.10 & 6.09 & 0.22 & 1.10 & 0.14 \\
8 & 6.52 & 0.05 & 5.90 & 0.20 & 1.40 & 0.09 \\
9 & 6.80 & 0.03 & 7.78 & 0.17 & 0.90 & 0.03 \\
10 & 7.00 & 0.07 & 5.97 & 0.17 & 0.70 & 0.04 \\
11 & 7.00 & 0.10 & 5.46 & 0.17 & 0.60 & 1.12 \\
12 & 6.52 & 0.01 & 5.08 & 0.20 & 0.90 & 0.09 \\
13 & 6.52 & 0.01 & 5.51 & 0.18 & 1.40 & 0.04 \\
14 & 6.52 & 0.01 & 5.98 & 0.17 & 1.10 & 0.03 \\
15 & 6.52 & 0.01 & 6.31 & 0.17 & 1.20 & 0.03 \\
16 & 6.50 & 0.08 & 5.39 & 0.22 & 0.90 & 0.13 \\
17 & 7.00 & 0.43 & 5.43 & 0.27 & 0.60 & 0.06 \\
18 & 6.74 & 0.18 & 5.04 & 0.27 & 0.70 & 0.20 \\
19 & 6.74 & 0.19 & 4.98 & 0.29 & 0.70 & 0.22 \\
20 & 6.58 & 0.27 & 5.70 & 0.27 & 0.90 & 0.20 \\
21 & 6.54 & 0.01 & 6.14 & 0.18 & 1.50 & 0.05 \\
22 & 6.82 & 0.10 & 4.76 & 0.22 & 0.60 & 1.12 \\
23 & 6.68 & 0.11 & 4.90 & 0.24 & 0.70 & 0.16 \\
24 & 6.52 & 0.28 & 5.70 & 0.24 & 1.30 & 0.16 \\
25 & 6.66 & 0.30 & 5.41 & 0.30 & 1.20 & 0.23
\enddata
\end{deluxetable*}

\end{document}